\documentclass[12pt]{article}
\pdfoutput=1
\usepackage{mheck}
\usepackage{multirow}
\usepackage{multicol}
\usepackage{adjustbox}
\usepackage{subfig}
\usepackage{mathtools}
\usepackage{float}

\newcommand{\massG}{\mathbf{m}_G}
\newcommand{\massF}{\mathbf{m}_F}
\newcommand{\mass}{\mathbf{m}}
\newcommand{\hh}[2]{$\mathbf{h}_{#1}^{#2}$}
\newcommand{\two}[2]{\mathbf{h}_{#1}^{#2}}

\newcommand{\anomal}[1]{{\mathcal A}_{#1}}
\newcommand{\six}[3]{{\bf #1}_{#2}^{#3}}
\newcommand{\strich}{\mathbf{S}}

\institution{SCGP}{\ Simons Center for Geometry and Physics, SUNY, Stony Brook, NY, 11794-3636 USA}
\institution{UvA}{\   Institute for Theoretical Physics, University of Amsterdam,  Amsterdam, The Netherlands}

\title{Universal Features of BPS Strings in Six-dimensional SCFTs}

\authors{Michele Del Zotto \worksat{\SCGP}\footnote{e-mail: {\tt delzotto@scgp.stonybrook.edu}} and Guglielmo Lockhart \worksat{\UvA}\footnote{e-mail: {\tt lockhart@uva.nl}}}

\abstract{In theories with extended supersymmetry the protected observables of UV superconformal fixed points are found in a number of contexts to be encoded in the BPS solitons along an IR Coulomb-like phase. For six-dimensional SCFTs such a role is played by the BPS strings on the tensorial Coulomb branch. In this paper we develop a uniform description of the worldsheet theories of a BPS string for rank-one 6d SCFTs. These strings are the basic constituents of the BPS string spectrum of arbitrary rank six-dimensional models, which they generate by forming bound states. Motivated by geometric engineering in F-theory, we describe the worldsheet theories of the BPS strings in terms of topologically twisted 4d $\mathcal{N}=2$ theories in the presence of $1/2$-BPS 2d $(0,4)$ defects. As the superconformal point of a 6d theory with gauge group $G$ is approached, the resulting worldsheet theory flows to an $\mathcal{N}=(0,4)$ NLSM with target the moduli space of one $G$ instanton, together with a nontrivial left moving bundle characterized by the matter content of the six-dimensional model. We compute the anomaly polynomial and central charges of the NLSM, and argue that the 6d flavor symmetry $F$ is realized as a current algebra on the string, whose level we compute. We find evidence that for generic theories the $G$ dependence is captured at the level of the elliptic genus by characters of an affine Kac-Moody algebra at negative level, which we interpret as a subsector of the chiral algebra of the BPS string worldsheet theory. We also find evidence for a spectral flow relating the R--R and NS--R elliptic genera. These properties of the string CFTs lead to constraints on their spectra, which in combination with modularity allow us to determine the elliptic genera of a vast number of string CFTs, leading also to novel results for 6d and 5d instanton partition functions.}

\begin{document}
\maketitle
\phantom{.}
\tableofcontents

\newpage
\section{Introduction and summary}

\noindent Recently much progress has been made in understanding six-dimensional $\cn=(1,0)$ supersymmetric theories and their compactifications thanks to the development of novel techniques relating to their holographic description, to their geometric engineering in F-theory, and to 6d field theory itself \cite{Apruzzi:2013yva,Heckman:2013pva,Gaiotto:2014lca,Ohmori:2014pca,DelZotto:2014hpa,Heckman:2014qba,Ohmori:2014kda,Intriligator:2014eaa,DelZotto:2014fia,Heckman:2015bfa,Bhardwaj:2015xxa,Apruzzi:2015wna,DelZotto:2015isa,Ohmori:2015pua,DelZotto:2015rca,Heckman:2015ola,Louis:2015mka,Cordova:2015fha,Heckman:2015axa,Ohmori:2015pia,Ohmori:2015tka,Bhardwaj:2015oru,Cremonesi:2015bld,Heckman:2016ssk,Cordova:2016xhm,Font:2016odl,Morrison:2016nrt,Morrison:2016djb,Benvenuti:2016dcs,Bena:2016oqr,Shimizu:2016lbw,Anderson:2016cdu,Heckman:2016xdl,Apruzzi:2016nfr,Razamat:2016dpl,Cordova:2016emh,Bobev:2016phc,Mekareeya:2016yal,Kim:2017xan,Yankielowicz:2017xkf,DelZotto:2017pti,Chang:2017xmr,Mekareeya:2017jgc,Mekareeya:2017sqh,Dibitetto:2017klx,Apruzzi:2017iqe,Heckman:2017uxe,Kim:2017toz,Mayrhofer:2017nwp,Ganor:2017rrz,Merkx:2017jey,Apruzzi:2017nck,Nazzal:2018brc,Hanany:2018uhm,Anderson:2018heq,Kim:2018bpg,Cordova:2018cvg,Apruzzi:2018oge,Lee:2018ihr,Hanany:2018vph}. Six-dimensional theories have tensorial Coulomb-like phases in which the tensor dynamics abelianizes and can be more easily understood. In particular, a feature of this `tensor branch' of the moduli space is that a spectrum of BPS strings is generated (see e.g. \cite{Seiberg:1996vs}) which possess 2d $(0,4)$ supersymmetry on their worldsheet; approaching the origin of the tensor branch corresponds to a renormalization group flow on the strings' worldsheet to an IR (0,4) CFT. Considerable advances have been made in characterizing these BPS strings and in understanding their elliptic genera \cite{Lockhart:2012vp,Haghighat:2013gba,Haghighat:2013tka,Hohenegger:2013ala,Haghighat:2014pva,Hosomichi:2014rqa,Kim:2014dza,Cai:2014vka,Haghighat:2014vxa,Huang:2015sta,Honda:2015yha,Gadde:2015tra,Sugimoto:2015nha,Hayashi:2015zka,Kim:2015fxa,Iqbal:2015fvd,Nieri:2015dts,Yun:2016yzw,Hayashi:2016abm,Haghighat:2016jjf,Kim:2016foj,DelZotto:2016pvm,Gu:2017ccq,Hayashi:2017jze,Haghighat:2017vch,Choi:2017vtd,Bastian:2017jje,Kim:2017jqn,Choi:2017luj,DelZotto:2017mee,Zhu:2017ysu,Kim:2018gak,Kim:2018gjo},\footnote{ This type of analysis has also been carried out in the related contexts of 5d supersymmetric field theories \cite{Harvey:2014nha,Haghighat:2015coa,Hohenegger:2015cba}, little string theories \cite{Hohenegger:2015cba,Kim:2015gha,Hohenegger:2015btj,Hohenegger:2016eqy,Hohenegger:2016yuv,Ahmed:2017hfr,Bastian:2017ing,Bastian:2017ary}, and 6d (1,0) supergravity theories \cite{Haghighat:2015ega,Lawrie:2016axq,Couzens:2017way}.} the main motivation for such studies being the conjecture that the $\Omega$-background partition function for six-dimensional theories localizes on contributions from the BPS strings, which can also be exploited to reconstruct the corresponding superconformal index and other partition functions \cite{Haghighat:2013gba,Haghighat:2013tka,Kim:2013nva,Kim:2016usy,Hayling:2017cva,Hayling:2018fmv}.\footnote{ In analogy with the case of 4d $\cn=2$ SCFTs \cite{Cecotti:2010fi,Cecotti:2011rv,Cecotti:2011gu,DelZotto:2011an,Alim:2011kw,Iqbal:2012xm,Cecotti:2012jx,Cecotti:2013lda,Cecotti:2014zga,Cordova:2015nma,Cecotti:2015hca,Cecotti:2015lab,Caorsi:2016ebt,Cordova:2017ohl,Cirafici:2017iju,Cordova:2017mhb,Caorsi:2017bnp}, one might ask whether it is possible to recover some, if not all, BPS properties of the 6d SCFTs from the knowledge of the corresponding BPS string CFTs.} The currently available methods, which often involve UV realizations of the strings' worldsheet theories in which RG flow invariant quantities can more easily be computed, are extremely powerful and can frequently be employed to compute at once the elliptic genera of arbitrary bound states of BPS strings. A downside of relying on UV techniques however is that they tend to be tailored to specific classes of 6d SCFTs, and as a consequence certain features which are common to the BPS strings of all 6d (1,0) theories tend to be somewhat obscured. Motivated by this, in this paper we take a different route and seek to reformulate the various known features of the BPS strings directly in the language of 2d conformal field theory. The payoff of this approach is that we will find a very natural and uniform picture for how the global symmetries of the string are realized at the level of the CFT, which also turns out to be a quite powerful asset in computing the strings' elliptic genera. In particular, this will allow us to determine the elliptic genera for the one string sectors of essentially all rank one 6d SCFTs, which also leads to several new results concerning 6d and 5d Nekrasov partition functions. Along the way, we develop a description of the BPS string CFTs in terms of twisted compactifications of 4d $\cn=2$ theories on $\PP^1$ in the presence of 2d (0,4) surface defects.\\

\noindent In this paper we focus on models that admit a geometric engineering in F-theory without frozen singularities \cite{Tachikawa:2015wka}, and therefore whenever referring to an F-theory geometry we implicitly make such an assumption.\footnote{\label{ft:frozen} At the time of this writing progress is being made towards understanding the frozen phase of F-theory; a very nice account of these recent advances can be found in the recent talks given by Alessandro Tomasiello at the Banff 2018 and at Madrid 2018 F-theory conferences, which are available online.} The relevant six-dimensional F-theory geometric backgrounds have been classified \cite{Heckman:2013pva,DelZotto:2014hpa,Heckman:2015bfa}; in this geometric setup the BPS strings are realized as stacks of D3-branes wrapped on combinations of intersecting rational curves in the F-theory base. For almost all 6d SCFTs these rational curves also support wrapped seven-branes which lead to nontrivial gauge groups in the SCFT. The Hilbert spaces of these models along the tensor branch are divided in superselection sectors labeled by the BPS string charges. In particular, the one-string subsector of the Hilbert space can always be identified with the one-string subsector of specific rank-one theories.\footnote{ We define the rank of a 6d SCFT as the dimension of its tensor branch, in analogy with the definition of the rank of a 4d $\cn=2$ SCFT.} For this reason, in this work we choose to focus on the spectrum of BPS strings of rank-one 6d SCFTs. A review of the F-theory geometries  and of various properties of rank-one theories can be found in section \ref{sec:Ftheory}.\\

\noindent The tensor branches of rank-one 6d SCFTs are realized geometrically in terms of an elliptically fibered Calabi-Yau such that the base of the fibration has local model given by the total space of a line bundle $\mathcal{O}(-n)\to\mathbb{P}^1 $ \cite{Heckman:2013pva}. A specific model is characterized by two pieces of data: an integer $n$ between 1 and 12 specifying the degree of the line bundle, and a Lie group $G$. The former is interpreted in the SCFT as the Dirac pairing of an elementary BPS string with itself; the latter, which is determined geometrically from the structure of the elliptic fiber along the base $\PP^1$, specifies the 6d gauge symmetry. The BPS string of the 6d SCFT can be viewed as an instanton for $G$, and indeed the Green-Schwarz term in the 6d tensor branch Lagrangian identifies the BPS string charge with the instanton charge for the gauge group $G$. From anomaly cancellation it follows that not all pairs $n$ and $G$ are allowed; in particular, given a pair $(n,G)$ the corresponding matter content and flavor symmetry $F$ is often uniquely determined (with only a few exceptions in which more than one choice of matter is allowed \cite{Grassi:2011hq}). We adopt the notation $\six{n}{G}{}$ for the corresponding six-dimensional SCFT. It is well known that D3-brane probes of seven-branes in F-theory give rise to four-dimensional theories with $\cn=2$ supersymmetry \cite{Banks:1996nj,Douglas:1996js,Fayyazuddin:1998fb,Aharony:1998xz}. Therefore, exploiting an adiabatic approximation, we can view the worldsheet theories for the 6d BPS strings as twisted compactifications of such 4d $\cn=2$ theories on the base $\PP^1$. This strategy has been adopted in \cite{DelZotto:2016pvm} to describe the BPS string instantons for the minimal 6d SCFTs without matter. In presence of matter, the geometry is modified by introducing transverse seven-branes that wrap noncompact curves intersecting the base $\PP^1$ at points. These extra seven-branes are interpreted as surface defects preserving 2d $(0,4)$ supersymmetry for the corresponding 4d $\cn=2$ theories on the base $\PP^1$, which generalize the chiral defects studied by Martucci in \cite{Martucci:2014ema} in the context of duality-twisted compactifications of 4d $\cn=4$ SYM. The geometric engineering setup allows us to deduce several properties of these generalized chiral defects as well as of the corresponding BPS string CFTs, as we discuss in section \ref{sec:BPSstrings}.\\

\noindent The BPS string and its bound states are interesting probes into the physics of the six-dimensional SCFT. This is true first and foremost at the level of the Nekrasov partition function on the tensorial Coulomb branch, which, specializing to theories with one tensor multiplet, is given by
\begin{equation}\label{eq:znek6d}
Z_{\text{Nekrasov}} = Z_{\text{pert}}\cdot Z_{\text{inst}}, \qquad Z_{\text{inst}} = \sum_{k=0}^\infty Q^kZ_{k\text{ inst}}.
\end{equation}
The 6d Nekrasov partition function is an elliptic generalization of the 4d and 5d Nekrasov partition functions \cite{Nekrasov:2002qd}, which are given respectively in terms of rational and trigonometric functions of various fugacities. The $Z_{pert}$ factor includes contributions to the partition function from the BPS particles, which do not carry instanton charge, whereas $Z_{\text{inst}}$ is the contribution of the instantons, which are identified with BPS strings wrapped on $T^2$ in the $T^2\times\mathbb{R}^4$ omega background. Indeed, $Z_{k\text{ inst}}$ is given by the Ramond-Ramond elliptic genus of the worldsheet theory of a bound state of $k$ strings \cite{Haghighat:2013gba}, with periodic boundary conditions on the left movers. This interpretation holds for all 6d SCFTs including the E-- and M-- string SCFTs which may be viewed respectively as 6d SCFTs with an $Sp(0)$ or $SU(1)$ gauge group. We review elliptic genera of the BPS strings and their relation to the 6d Nekrasov partition function in section \ref{sec:ellgen}.\newline

\noindent In the following sections we also encounter other ways in which the CFT of a string carries nontrivial information about the 6d SCFT. For example, the $\six{2}{SU(2)}{}$ SCFT has $SO(8)$ flavor symmetry on the tensor branch, but only $SO(7)$ flavor symmetry at the origin \cite{Ohmori:2015pia}. This is reflected in the spectrum of the string, which as we will see in section \ref{sec:excep} can be organized in terms of $SO(8)_1$ affine characters (with a certain specialization of fugacities), but only has an $SO(7)$ adjoint  representation worth of chiral currents rather than a full $SO(8)$ adjoint representation. Moreover we will find evidence that some features of 6d SCFTs that emerge in their F-theory classification have a natural explanation from the perspective of the string worldsheet CFT. For example, F-theory predicts that 6d SCFTs with unpaired tensors (i.e. with a tensor multiplet that is not paired up to a gauge algebra via the Green-Schwarz mechanism) only exist for $n=1 $ and $2$. From the perspective of the string worldsheet CFT, this is a consequence of unitarity: the requirement that the left moving central charge
\begin{equation}
c_L=6\,h^\vee_G-6\,n+12\geq 0
\end{equation}
for $n\geq 3$ is only possible if $h^\vee_G > 0$. \newline

\noindent Motivated by these observations, in sections \ref{sec:1spn} and \ref{sec:univ} we seek a uniform description for the BPS string CFTs; though several aspects of our discussion apply to bound states of arbitrary number of strings, for brevity we choose to restrict our attention to the CFTs of a single string. We find that these CFTs share a number of universal features which strongly constrain their spectrum and the form of their elliptic genera. We choose to take an empirical approach, relying on known results for various BPS string CFTs (such as computation of the elliptic genera by localization or by other methods) and inferring general features of the CFTs from the existing results. We first focus in section \ref{sec:1spn} on the BPS strings of the $\six{1}{Sp(n)}{}$ SCFTs, which are essentially free theories (since the reduced moduli space of one $Sp(N)$ instanton is just a $\mathbb{Z}_2$ orbifold of $\mathbb{C}^{2N}$), and then rephrase our findings in full generality in section \ref{sec:univ}.\newline

\noindent The CFT describing a BPS string consists of two components: a center of mass piece, and an interacting piece on which we focus, which we denote by \hh{n}{G} for the string of the $\six{n}{G}{}$ 6d SCFT. The interacting piece can be understood as a $\mathcal{N}=(0,4)$ nonlinear sigma model with target space the reduced moduli space of one $G$ instanton, $\widetilde{\mathcal{M}}_{G,1}$. The information on the 6d gauge symmetry is carried by scalar (bosonic) superfields, whereas the 6d flavor symmetry $F$ which acts on the matter content is coupled to spinor (fermionic) superfields, which are sections of a chiral vector bundle on $\widetilde{\mathcal{M}}_{G,\,1}$. We find that the NLSM consists of the following numbers of bosonic and fermionic components (where we further distinguish between L (chiral) and R (anti-chiral) degrees of freedom):
\begin{center}
\begin{tabular}{|c|c|c|}
\hline
&L&R\\\hline
\# noncompact bosons &$4(h^\vee_G-1)$&$4(h^\vee_G-1)$\\
\# fermions &$4h^\vee_G-12(n-2)$&$4(h^\vee_G-1)$\\\hline
\end{tabular}
\end{center}
From this data one recovers the central charges for \hh{n}{G} 
\begin{align}
c_L &= 6h^\vee_G-6n+8;\\
c_R &= 6h^\vee_G-6.
\end{align}
Since the flavor symmetry $F$ couples chirally to the fermionic superfields, we are immediately led to the conclusion that it is realized in the CFT as a chiral current algebra (that is, as the chiral half of a WZW model). This is of course already a very well known fact for the E-string CFT, whose flavor symmetry is captured by the level 1 $E_8$ current algebra \cite{Ganor:1996mu}; in this paper we will see very concretely how the statement generalizes to all \hh{n}{G} CFTs. The level can be read off from the anomaly polynomial as the coefficient of the corresponding `t Hooft anomaly. In turn, the string's anomaly polynomial can be read off from the anomaly polynomial of the $\six{n}{G}{}$ SCFT by anomaly inflow, following the approach of \cite{Kim:2016foj,Shimizu:2016lbw}. Anticipating our discussion from section \ref{sec:wzwunivii}, in table \ref{tab:flavint} we list the current algebras realizing the flavor symmetry of all BPS string CFTs. We only encounter subtleties for certain theories with $G=SO(11)$ or $SO(12)$, which we address in section \ref{sec:wzwunivii}.\newline

\begin{table}[p!]
\begin{center}
\begin{multicols}{2}
\scalebox{0.84}{\begin{tabular}{|c|l|l|}\hline
$n$&G&F
\\\hline\hline
12&$E_8$&$-$\\\hline
8&$E_7$&$-$\\\hline
7&$E_7$&$-$\\\hline
6&$E_6$&$-$\\
6&$E_7$&$SO(2)_{12}$\\\hline
5&$F_4$&$-$\\
5&$E_6$&$U(1)_6$\\
5&$E_7$&$SO(3)_{12}$\\\hline
4&$SO(N),\,N\geq 8$&$Sp(N-8)_1$\\
4&$F_4$&$Sp(1)_3$\\
4 &$E_6$&$SU(2)_{6}\times U(1)_{12}$\\
4&$E_7$&$SO(4)_{12}$\\\hline
3&$SU(3)$&$-$\\
3&$SO(7)$&$Sp(2)_1$\\
3&$SO(8)$&$Sp(1)_1^3$\\
3&$SO(9)$&$Sp(2)_1\times Sp(1)_2$\\
3&$SO(10)$&$Sp(3)_1\times (SU(1)_4\times U(1)_4)$\\
3&$SO(11)$&$Sp(4)_1\times \text{Ising}$\\
3&$SO(12)$&$Sp(5)_1$\\
3&$G_2$&$Sp(1)_1$\\
3&$F_4$&$Sp(2)_3$\\
3 &$E_6$&$SU(3)_{6}\times U(1)_{18}$\\
3&$E_7$&$SO(5)_{12}$\\\hline
2&$SU(1)$&$SU(2)_1$\\
2&$SU(2)$&$SO(8)_1\to SO(7)_1\times\text{Ising}$\\
2&$SU(N), N>2$&$SU(2N)_1$\\
2&$SO(7)$&$Sp(1)_1\times Sp(4)_1$\\
\hline
\end{tabular}}
\columnbreak
\scalebox{0.84}{\begin{tabular}{|c|l|l|}\hline
$n$&G&F\\\hline\hline
2&$SO(8)$&$Sp(2)_1^3$\\
2&$SO(9)$&$Sp(3)_1\times Sp(2)_2$\\
2&$SO(10)$&$Sp(4)_1\times (SU(2)_4\times U(1)_{8})$\\
2&$SO(11)$&$Sp(5)_1\times ???$\\
2&$SO(12)_a$&$Sp(6)_1\times SO(2)_8$\\
2&$SO(12)_b$&$Sp(6)_1\times \text{Ising}\times \text{Ising}$\\
2&$SO(13)$&$Sp(7)_1$\\
2&$G_2$&$Sp(4)_1$\\
2&$F_4$&$Sp(3)_3$\\
2 &$E_6$&$SU(4)_{6}\times U(1)_{24}$ \\
2&$E_7$&$SO(6)_{12}$\\\hline
1&$Sp(0)$&$(E_8)_1$\\
1&$Sp(N),N\geq 1$&$SO(4N+16)_1$\\
1&$SU(3)$&$SU(12)_1$\\
1&$SU(4)$&$SU(12)_1\times SU(2)_1$\\
1 &$SU(N),N\geq 4$&$SU(N\!+\!8)_1\!\times\! U(1)_{2N(N-1)(N+8)}$\\
1&$SU(6)_*$&$SU(15)_1$\\
1&$SO(7)$&$Sp(2)_1\times Sp(6)_1$\\
1&$SO(8)$&$Sp(3)_1^3$\\
1&$SO(9)$&$Sp(4)_1\times Sp(3)_2$\\
1&$SO(10)$&$Sp(5)_1\times (SU(3)_4\times U(1)_{12})$\\
1 &$SO(11)$&$Sp(6)_1\times ???$\\
1&$SO(12)_a$&$Sp(7)_1\times SO(3)_8$\\
1&$SO(12)_b$&$Sp(7)_1\times ???$\\
1&$G_2$&$Sp(7)_1$\\
1&$F_4$&$Sp(4)_3$\\
1 &$E_6$&$SU(5)_{6}\times U(1)_{30}$ \\
1&$E_7$&$SO(7)_{12}$\\\hline
\end{tabular}}
\end{multicols}
\end{center}
\caption{Current algebra associated to the flavor symmetry $F$ of the $\six{n}{G}{}$ 6d SCFTs. The $???$ indicate cases for which we do not have a good understanding of the worldsheet realization of the flavor symmetry. The notation $SO(12)_{a,b}$ and $SU(6)_*$ is explained in section \ref{sec:higgsingchains}.}
\label{tab:flavint}
\end{table}

\noindent At the level of the spectrum of the CFT, for $F=\prod_{i=1}^{n_F} F_i$ a product of simple and abelian factors, this implies that the Hilbert space factorizes as
\begin{equation}
H_n^{G} = \bigoplus_{\vec{\lambda}} \left(\bigotimes_{i=1}^{n_F}H^{WZW_{F_i}}_{\lambda_i}\right) \otimes H^{residual}_{\vec{\lambda}},
\end{equation}
where $\vec{\lambda}=(\lambda_1,\dots,\lambda_{n_F})$ labels highest weights corresponding to integrable highest weight representations of the $F_i$ WZW models. The residual factor of the Hilbert space includes both chiral and anti-chiral degrees of freedom that depend on $G$ and $v$. Correspondingly, the dependence of the elliptic genus also factorizes:
\begin{equation}\label{eq:ellgfi}
\mathbb{E}_{n}^{G}(\massG,\massF,v,q) = \sum_{\vec\lambda}\big(\prod_{i=1}^{n_F}\widehat\chi^{F_i}_{\lambda_i}(\mass_{F_i},q)\big)\xi_{\vec\lambda}^{n,G}(\massG,v,q),
\end{equation}
where $\widehat \chi^{F_i}_{\lambda_i}(\mass_{F_i},q)$ are WZW$_{F_i}$ characters, and $\xi_{\vec\lambda}^{n,G}(\massG,v,q)$ are the holomorphic contributions of $H^{residual}_{\vec{\lambda}}$ to the elliptic genus. In these expressions, $\massF$ and $\massG$ are respectively exponentiated fugacities for $F$ and $G$, while $v=e^{2\pi i \epsilon_+}$ is a fugacity coupling to $J^3_R+J^3_I$, where $J^3_R$ and $J^3_I$ are respectively Cartan generators for the $SU(2)_R$ subgroup of the $SO(4)\sim SU(2)_L\times SU(2)_R$ isometry group of $\mathbb{R}^4$ and for the $SU(2)_I$ R-symmetry group of the 6d SCFT, which is also identified with the superconformal R-symmetry group of the 2d (0,4) CFT (consistent with the fact that the latter cannot act on the non-compact target space \cite{Witten:1997yu}). \newline

\noindent Rewriting the elliptic genus as in equation \eqref{eq:ellgfi} reveals a number of interesting features which we ultimately interpret as properties of the chiral algebra of the $(0,4)$ CFT. First of all, we find that for \hh{n}{G} theories with $n\neq h^\vee_G$ the functions $\xi_{\vec\lambda}^{n,G}$ admit the following expansion:
\begin{equation}\label{eq:fgvin}
\begin{tabular}{c}
$ \displaystyle{\xi_{\vec\lambda}^{n,G}(\massG,v,q) =\hspace{4.2in}}$\\
$\displaystyle{\sum_{\nu\in \text{Rep}(G)}\,\sum_{\ell=-2\vert n-h^\vee_G\vert+1}^{0}\,\sum_{m\in\mathbb{Z}} n^{\vec\lambda}_{\nu,\ell,m}\times \frac{q^{-\frac{c_G}{24}+h^G_\nu}\chi^G_\nu(\massG)}{\widetilde\Delta_{G}(\mass_{G},q)}\times \frac{q^{-\frac{c_v}{24}+h^v_{\ell,m}}v^{\ell+2(n-h^\vee_G)m}}{\prod_{j=1}^\infty(1-q^j)^{}},}$\\
\end{tabular}
 \end{equation}
where the $n^{\vec\lambda}_{\nu,\ell,m}$ are integer coefficients, and $c_G,$ $h^G_\nu$ (resp. $c_v=1$, and $h^v_{\ell,m}$) are the central charges and conformal dimensions of operators of the level $-n$ $G$ Kac-Moody algebra (resp. $U(1)$ Kac-Moody algebra at $R^2\sim n-h^\vee_G$). By analyzing the elliptic genus in specific examples, we find evidence that the $G$ dependence indeed organizes itself in terms of irreducible characters of the $-n$ $G$ Kac-Moody algebra, where the level $k_G = -n$ is in agreement with the expectation from the anomaly polynomial of the string.\newline

\noindent The central charge of the left-moving part of the CFT can be written as follows:
\begin{equation}
c_L = c_F+c_G+c_v-24\frac{(h^\vee_G-1)^2}{4(n-h^\vee_G)},
\end{equation}
where $c_F$ is the total Sugawara central charge of the WZW sector capturing the 6d flavor symmetry $F$. The last term arises because the $G$-- and $F$--neutral vacuum in the chiral algebra carries nonzero $U(1)_v$ charge $h^\vee_G-1$ and therefore sits in a Verma module of $U(1)_v$ which is distinct from the vacuum Verma module. \newline

\noindent The diagonal subgroup $SU(2)_v$ of $SU(2)_R\times SU(2)_I$, to which the chemical potential $v$ couples, does not act chirally on the full spectrum of the CFT; however, it can be thought of as a chiral symmetry once we restrict to the chiral algebra underlying the elliptic genus; we find that shifting its fugacity $v\to q^{1/2}/v$ implements a spectral flow which leads to a relation between the Ramond--Ramond elliptic genus $\mathbb{E}_{n}^G$ and the Neveu-Schwarz--Ramond elliptic genus $\mathcal{E}_n^G$:
\begin{equation}\label{eq:elrel}
\mathcal{E}_n^{G}(\mass_{G},\mass_{F},v,q) = q^{\frac{n-h^\vee_G}{4}}v^{-(n-h^\vee_G)} \mathbb{E}_n^{G}(\mass_{G},\mass_{F},q^{1/2}/v,q).
\end{equation}
This fact turns out to be quite convenient: whereas the low energy spectrum that contributes to the Ramond--Ramond elliptic genus is complicated due to the presence of fermionic zero modes on top of the bosonic zero modes, the low energy spectrum contributing to the Neveu-Schwarz--Ramond elliptic genus is much simpler. At the zero energy level only the bosonic generators of the moduli space of one $G$ instanton contribute; their contribution is given by:
\begin{align}\label{eq:int0e}
\mathcal{E}_n^{G}(\mass_{G},\mass_{F},v,q)\bigg\vert_{q^{-\frac{c_L}{24}}} = v^{h^\vee_G-1} \sum_{k=0}^\infty v^{2k}\chi_{k\cdot \theta_G}^{G}(\mass_{G}),
\end{align}
where up to the overall factor of $v^{h^\vee_G-1}$ the right hand side coincides with the Hilbert series of $\widetilde{\mathcal{M}}_{G,1}$ \cite{Benvenuti:2010pq}.\newline

\noindent At the first excited level, if the 6d matter fields transform in a direct sum of representations 
\begin{equation}
\bigoplus_{i=1}^r\, (R^G_i,R^F_i),
\end{equation}
we find the following set of contributions:
\begin{equation}\label{eq:int1e}
\mathcal{E}_n^{G}(\mass_{G},\mass_{F},v,q)\bigg\vert_{q^{-\frac{c_L}{24}+\frac{1}{2}}} =-v^{h^\vee_G-1}\sum_{k=0}^\infty v^{2k+1}\sum_{i=1}^r\chi_{\lambda^G_i+k\cdot\th_G}^{G}(\mass_{G})\chi_{\lambda^F_i}^{F}(\mass_F),
\end{equation}
where $\lambda_i^{G}$ and $\lambda_i^{F}$ are respectively the highest weights of the representations $R^G_i$ and $R^F_i$, $\theta_G$ is the highest weight of the adjoint representation of $G$, and $\chi^G_\lambda$, $\chi^F_\lambda$ are Lie algebra characters. Using the spectral flow \eqref{eq:elrel}, equations \eqref{eq:int0e} and \eqref{eq:int1e} determine an infinite number of coefficients in the Ramond--Ramond elliptic genus as well.\newline

\noindent In section \ref{sec:5dlim} we discuss in some detail the five-dimensional limit of the 6d Nekrasov partition function \eqref{eq:znek6d}. Geometric considerations suggest three different behaviors according to whether $n\geq 3$, $n=2$, or $n=1$. For 6d  SCFTs with $n\geq 3$ the five-dimensional limit we consider is a 5d $\mathcal{N}=1$ theory with the gauge group and matter content obtained by the naive dimensional reduction of the 6d field content; for $n=2$ one in addition finds free decoupled states (see also \cite{Hayashi:2013qwa}); finally for $n=1$ one simply obtains the 5d $\mathcal{N}=1$ theory of one free hypermultiplet. This distinction between the three cases is reflected in the $\tau\to i\infty$ behavior of the Ramond--Ramond elliptic genus: for $n=1$ at lowest energy one finds a single state, while for $n\geq2 $ one obtains the one instanton piece $Z_{1 \text{ inst}}$ of the 5d Nekrasov partition function with same gauge symmetry and matter content as the 6d theory, plus additional extra states for $n=2$. The $n\geq 2$ case is of particular interest, since for many of the 6d SCFTs that we study the 5d Nekrasov partition function is not known, and the elliptic genus of the BPS string can be used to obtain new information about $Z_{1\text{ inst}}$ in 5d.\newline 

\noindent In section \ref{sec:modul} we translate these features of the \hh{n}{G} CFTs into a series of constraints on their Ramond-Ramond elliptic genera. For convenience, we switch off the $F$ and $G$ fugacities $\massF,\massG$; then the elliptic genus can be expressed as the following meromorphic Jacobi form:
\begin{equation}
\mathbb{E}_n^G(v,q) = \frac{\mathcal{N}_n^G(v,q)}{\eta(q)^{12(n-2)-4+24\,\delta_{n,1}}\varphi_{-2,1}(v^2,q)^{h^\vee_G-1}}.
\end{equation}
The numerator $\mathcal{N}_n^G(v,q)$ is now a holomorphic Jacobi form of even weight
\begin{equation}
m_n^G = 6n-2h^\vee_G-12+12\,\delta_{n,1}\quad\in\quad 2\,\mathbb{Z}
\end{equation}
and index
\begin{equation}
k_n^G=n+3\,h^\vee_G-4\quad\in\quad \mathbb{Z}_+
\end{equation}
with respect to $\epsilon_+$. According to a structure theorem for the bi-graded ring of even weight holomorphic Jacobi forms, the vector space of forms of given weight and index is finite-dimensional and has a known basis, given by products of  the standard Jacobi and modular forms  $\varphi_{0,1}(v,q),$ $\varphi_{-2,1}(v,q)$, $E_4(q),$ and $E_6(q)$. We find that the constraints from section \ref{sec:univ} are strong enough to uniquely determine the elliptic genus for 59 out of the 72 CFTs for which $\text{rank}(G)\leq 7$. In particular, we are able to compute the elliptic genus for a number of CFTs for which the elliptic genus was not previously known.\newline

\noindent  In section \ref{sec:excep} we discuss two BPS string CFTs which fall outside of the general discussion of section \ref{sec:univ}: the theories \hh{2}{SU(2)} and \hh{3}{SU(3)}, for which $n=h^\vee_G$ (the only other CFT belonging to this class is the E-string CFT \hh{1}{Sp(0)} which however is trivially given by the $E_8$ current algebra at level 1). For these theories the expansion \eqref{eq:fgvin} is not valid, but we are able to find simple alternative series expansions in $v$. Moreover, for these two CFTs the level of the Kac-Moody algebra implied from the anomaly polynomial is the critical level $k=-h^\vee_G$. The irreducible highest-weight modules of Kac-Moody algebras at critical level have markedly different structure from the noncritical case. Interestingly, for these two CFTs the $G$ dependence of the elliptic genus does not seem to be captured in terms of the corresponding irreducible characters.\newline

\noindent In section \ref{sec:concl} we present our conclusions and formulate a number of questions which we leave to future research. The appendices are organized as follows: in appendix \ref{sec:Lie} we briefly review simple and affine Lie algebras and set up our notation. In section \ref{sec:appmod} we review the properties of modular and Jacobi forms we make use of in the main text. In appendix \ref{sec:catalogue} we collect a list of references to other works where the elliptic genera of various BPS string CFTs are presented in a form which may be readily compared to our results. In appendix \ref{sec:soodd} we compute the elliptic genera of the strings of the $\six{4}{SO(2M+1)}{}$ SCFTs (with $M\geq 4$), which have not previously appeared in the literature. In appendix \ref{sec:WZW} we recall basic properties of chiral WZW models that we make use of in the text. In appendix \ref{app:xiexample} we give a detailed description of the $\xi^{n,G}_\lambda$ functions that appear in equation \eqref{eq:ellgfi} for a specific choice of $n, G$, and we also show how the form of these functions leads to constraints on the elliptic genus. In appendix \ref{app:bjk} we provide extensive tables of coefficients of the series expansions of the elliptic genera which we determined by exploiting modularity in section \ref{sec:modul}. Finally, in appendix \ref{sec:5dapp} we discuss the computation of the one-instanton component of 5d Nekrasov partition functions starting from the elliptic genera of the \hh{n}{G} theories, and provide our results for a number of theories with $n\geq 2$, several of which have not previously appeared in the literature.

\section{Review of 6d rank-one SCFTs}\label{sec:Ftheory}
In this section we review the geometric engineering of 6d SCFTs in the context of F-theory, highlighting a number of aspects that will be relevant to describing their BPS strings. We begin in section \ref{sec:6df} with a broad overview of the geometric engineering of 6d SCFTs. In section \ref{sec:dtypeeng} we discuss in more detail the example of 6d D-type conformal matter, whose BPS strings we will study in some detail in section \ref{sec:1spn}. In section \ref{sec:anomf} we review how 6d anomaly polynomials are encoded in the F-theory geometry. Finally in section \ref{sec:higgsingchains} we discuss in more detail the 6d SCFTs of rank 1 which are the focus of this paper.
\subsection{Six-dimensional SCFTs from F-theory}\label{sec:6df}
Consider an F-theory compactification to six dimensions, defined by a Calabi-Yau threefold $X$ that is elliptically fibered over a base $B$, which is a complex (K\"ahler) surface \cite{Vafa:1996xn,Morrison:1996na,Morrison:1996pp,Bershadsky:1996nh,Katz:2011qp}. As we are interested in the geometric engineering of SCFTs decoupled from gravity, we consider a local models such that $B$ has infinite volume. Moreover, any geometry corresponding to an SCFT has no scales in it, which leaves us with bases that typically have the form
\be
B \simeq \mathbb C^2 / \Gamma\,,
\ee
where $\Gamma$ is a discrete subgroup of $U(2)$ under whose action the only fixed point is the origin $(0,0)\in \mathbb C^2$. In case the elliptic fibration is trivial
\be
X \simeq B \times T^2 \,,
\ee
the F-theory background has a perturbative type IIB interpretation, and the CY condition on $X$ forces $\Gamma \subset SU(2)$. This background preserves a higher amount of supersymmetry, and the corresponding 6d SCFTs are the ADE $(2,0)$ theories. Otherwise, one obtains backgrounds preserving $(1,0)$ supersymmetry. In this case, the elliptic fibration can degenerate along a codimension-one locus in the base $B$, along a curve which is called the discriminant locus. The complex structure parameter $\tau$ of the elliptic fiber of $X$ is interpreted in IIB as the axio-dilaton field, and a nontrivial discriminant signals that $\tau$ undergoes monodromies which are sourced by seven-branes; hence, the discriminant locus is interpreted in F-theory backgrounds as a curve of coalesced stacks of wrapped seven-branes.\footnote{ In this paper we consider only singularities that are not frozen. Frozen singularities correspond to bound states of seven-branes that involve an $O7^+$ \cite{Tachikawa:2015wka}, see also the remark in footnote \ref{ft:frozen}.} Requiring a geometry that has no scales in it forces the possible components of the discriminant to be noncompact curves through the origin, which gives rise to (generalized) flavor symmetries for the SCFT. The noncompact flavor curves, being of codimension one in the F-theory base, support degenerate elliptic fibers that obey the Kodaira classification; this has to be contrasted with the behavior of the elliptic fiber at the origin, which is a codimension-two locus: there, the elliptic fiber can degenerate into a non-Kodaira type fiber.\\

\noindent To determine the tensor branch geometry, one resolves the origin of the base $B$ by successive blow-ups until no further codimension-two components in the discriminant support singularities of non-Kodaira type. Indeed, a non-Kodaira singularity in codimension-two signals the presence of tensionless strings in the geometry, corresponding to a partially unresolved tensor branch, which can be resolved by further blow-ups. By this process of repeatedly blowing up one obtains a curve $\Sigma$ which can have several compact irreducible rational components intersecting transversally; we denote these components by $\Sigma_I$, with $I=1,...,R$, where $R$ is the rank of the corresponding tensor branch. This can be understood by considering the corresponding IIB reduction. Let $\omega_I$ be harmonic forms such that 
\begin{equation}
\int_{\Sigma_J} \omega_I = \delta_{IJ}.
\end{equation}
In the decomposition of the IIB RR potential
\begin{equation}
C_4^{+} = \sum_{I=1}^R B^I \wedge \omega_I 
\end{equation}
one obtains $R$ anti-self-dual two-forms $B^I$ that are part of the 6d $\cn=(1,0)$ tensor multiplets. The scalar components $\phi^I$ of the tensor multiplets correspond to
the periods of the K\"ahler form of $\widehat B$:
\begin{equation}
J = \sum_{I=1}^R \phi^I \omega_I.
\end{equation}
Hence the vacuum expectation values of the scalar fields $\phi^I$ correspond to the K\"ahler moduli given by the volumes of the rational curves $\Sigma_I$. In particular, the superconformal fixed point at the origin of the tensor branch is attained by setting all such volumes to zero, shrinking the curve $\Sigma$ to a point. This is possible iff $\Sigma \cdot \Sigma < 0$ by the Artin-Grauert criterion \cite{Artin, Grauert}. Whenever the curve has multiple irreducible components, this implies that the matrix 
\be\label{eq:AM}
A_{IJ} \equiv -\Sigma_I \cdot \Sigma_J
\ee
is positive definite. This is not a surprise because the matrix in equation \eqref{eq:AM} gives the kinetic terms for the effective action of the scalars along the tensor branch $A_{IJ} \partial_\mu \phi^I \partial^\mu \phi^J$.\\

\begin{table}
\begin{center}
\begin{tabular}{|ccc|c|c|c|}
\hline
ord($f$)	&ord($g$)		&ord($\Delta$)		&type			&singularity			&non-abelian algebra \\
\hline
\hline
$\geq0$	&$\geq0$		&0				&I${}_0$			&none				& none \\
$0$		&$0$			&1				&I${}_1$			&none				& none \\
$0$		&$0$			&$n\geq2$		&I${}_n$			&$A_{n-1}$			&$su_n$ or $sp_{[n/2]}$ \\
$\geq1$	&$1$			&2				&II				&none				& none \\
$1$		&$\geq2$		&3				&III				&$A_1$				&$su_2$\\
$\geq2$	&$2$			&4				&IV				&$A_2$				&$su_3$ or $su_2$ \\
$\geq2$	&$\geq3$		&6				&I${}_0^\ast$		&$D_4$				&$so_8$ or $so_7$ or $g_2$ \\
$2$		&$3$			&$n\geq7$		&I${}_{n-6}^\ast$	&$D_{n-2}$			&$so_{2n-4}$ or $so_{2n-5}$\\
$\geq3$	&$4$			&8				&IV${}^\ast$		&$E_6$		&$e_6$ or $f_4$ \\
$3$		&$\geq5$		&9				&III${}^\ast$		&$E_7$		&$e_7$ \\
$\geq4$	&$5$			&10				&II${}^\ast$		&$E_8$		&$e_8$ \\
\hline
\hline
$\geq4$	&$\geq6$		&$\geq12$		&non-Kodaira		&-					&- \\
\hline
\end{tabular}

\vspace{1cm}

\begin{tabular}{|c|c|} \hline
Type  & Monodromy Cover equation\\ \hline\hline
I${}_m$, $m\ge3$ \phantom{$\Bigg|$} & $\psi^2+(9g/2f)|_{z=0}$ \\ \hline 
IV \phantom{$\Bigg|$} &$\psi^2-(g/z^2)|_{z=0}$  \\ \hline
I${}_0^*$ \phantom{$\Bigg|$}  & $\psi^3+(f/z^2)|_{z=0}\cdot\psi+(g/z^3)|_{z=0}$ \\ \hline
I${}_{2n-5}^*$, $n\ge3$ \phantom{$\Bigg|$} &$\psi^2+\frac14(\Delta/z^{2n+1})(2zf/9g)^3|_{z=0}$ \\ \hline 
I${}_{2n-4}^*$, $n\ge3$ \phantom{$\Bigg|$} &$\psi^2+(\Delta/z^{2n+2})(2zf/9g)^2|_{z=0}$ \\ \hline 
IV${}^*$ \phantom{$\Bigg|$} & $\psi^2-(g/z^4)|_{z=0}$ \\ \hline 
\end{tabular}

\end{center}
\caption{\textsc{Top:} summary of Kodaira singularities and corresponding non-abelian gauge algebras for F-theory seven-branes. \textsc{Bottom:} monodromy covers for $\Sigma$ using adapted coordinates in which $\Sigma$ is the locus $\{z = 0\}$ in the F-theory base.}
\label{tab:sing}
\end{table}

\noindent Along the tensor branch, we can use the standard dictionary relating Kodaira singularities to coalesced seven-brane stacks. To make this dictionary explicit in the discussion that follows, we are going to assume that the elliptic fibration of the resolved CY $\widehat X$ has a section.\footnote{ This is not strictly necessary for general F-theory engineerings of six-dimensional models, see \cite{Morrison:2014era}. For some recent application to six-dimensional models decoupled from gravity see e.g. \cite{Bhardwaj:2015oru,Monnier:2017oqd,Anderson:2018heq,Lee:2018ihr}.} In this case, the elliptic fibration can be described by a local Weierstrass model
\begin{equation}
y^{2} = x^{3}+f x + g \,,
\end{equation}
where $f$ and $g$ are local functions on $\widehat B$ that globally are sections respectively of $-4K$ and $-6K$, $K$ being the canonical bundle of $\widehat B$. The discriminant of the elliptic fibration is the local function
\begin{equation}
\Delta = 4f^{3} + 27g^{2}
\end{equation}
which is a section of $-12K$. By an abuse of notation, we also denote by $\Delta$ the discriminant locus, which is the curve $\Delta=0$. If the discriminant $\Delta$ has several irreducible components $\Delta_\alpha$, such that the order of vanishing of $\Delta$ along such irreducible components is
\be
\text{ord}(\Delta \big|_{\Delta_\alpha}) = N_\alpha > 0 \, ,
\ee
this signals that a configuration of seven-branes with RR charge $N_\alpha$ is wrapped on the curve $\Delta_\alpha$. Since the base $\widehat B$ is noncompact there are two possibilites: the curve $\Delta_\alpha$ itself can be either compact or noncompact. In the latter case the divisor is a flavor divisor, corresponding to a global symmetry. In the former case, the corresponding stack of seven-branes contributes a gauge sector to the model with coupling
\be\label{eq:gaugecoup}
1/g_\alpha^2 \sim \text{vol}(\Delta_\alpha).
\ee
The gauge and global symmetries corresponding to the various components of $\Delta_\alpha$ are determined by the structure of the corresponding Kodaira singularity (see the top part of table \ref{tab:sing}). In particular, the Calabi-Yau condition on $\widehat X$ is satisfied provided that
\begin{equation}\label{eq:c1b}
c_1(B) = - \frac{1}{12} \sum_\alpha N_\alpha \, \delta(\Delta_\alpha) \ ,
\end{equation}
in obvious notation. The irreducible compact components $\Sigma_I$ of $\Sigma$ that are part of the discriminant correspond to gauge theory subsectors of the model; we denote by $\mathfrak{g}_{\Sigma_I}$ the Lie algebra of the corresponding gauge group. This is computed in terms of two pieces of data: the first is the order of vanishing of $(f,g,\Delta)$ along $\Sigma_I$, and the second is the monodromy of the fibration, which determines which gauge algebra occurs for a given
singularity type, according to Tate's algorithm. Tate's algorithm assigns to each curve $\Sigma_I$ a monodromy cover which is captured by an equation in an auxiliary variable $\psi$, valued in a line bundle over $\Sigma_I$. Explicit equations for these monodromy covers are given in table \ref{tab:sing}: \textsc{Bottom}, where we have adapted locally the base coordinates so that $\Sigma_I$ is identified with locus $z=0$.  For all cases except I${}_0^*$, the equation of the monodromy cover takes the form
\be
\psi^2 - P(f,g,z) = 0 \, ,
\ee
where $P(f,g,z)$ is a Laurent polynomial in $f$, $g$ and $z$. The cover splits (leading to no monodromy) if $P$ is a perfect square. In the I${}_0^*$ case, the monodromy cover equation defines a degree $3$ cover of $\Sigma_I$, and one must analyze this system further to determine whether the cover is
irreducible ($\mathfrak{g}_\Sigma = g_2$), splits into two components ($\mathfrak{g}_\Sigma = so(7)$),
or splits into three components ($\mathfrak{g}_\Sigma = so(8)$).\footnote{ We are being very explicit here since later on we will be interpreting these geometric structures in terms of surface defects of the theory describing D3-branes wrapped on two-cycles in $\widehat B$.}\\

\begin{figure}
\begin{center}
\includegraphics[scale=0.6]{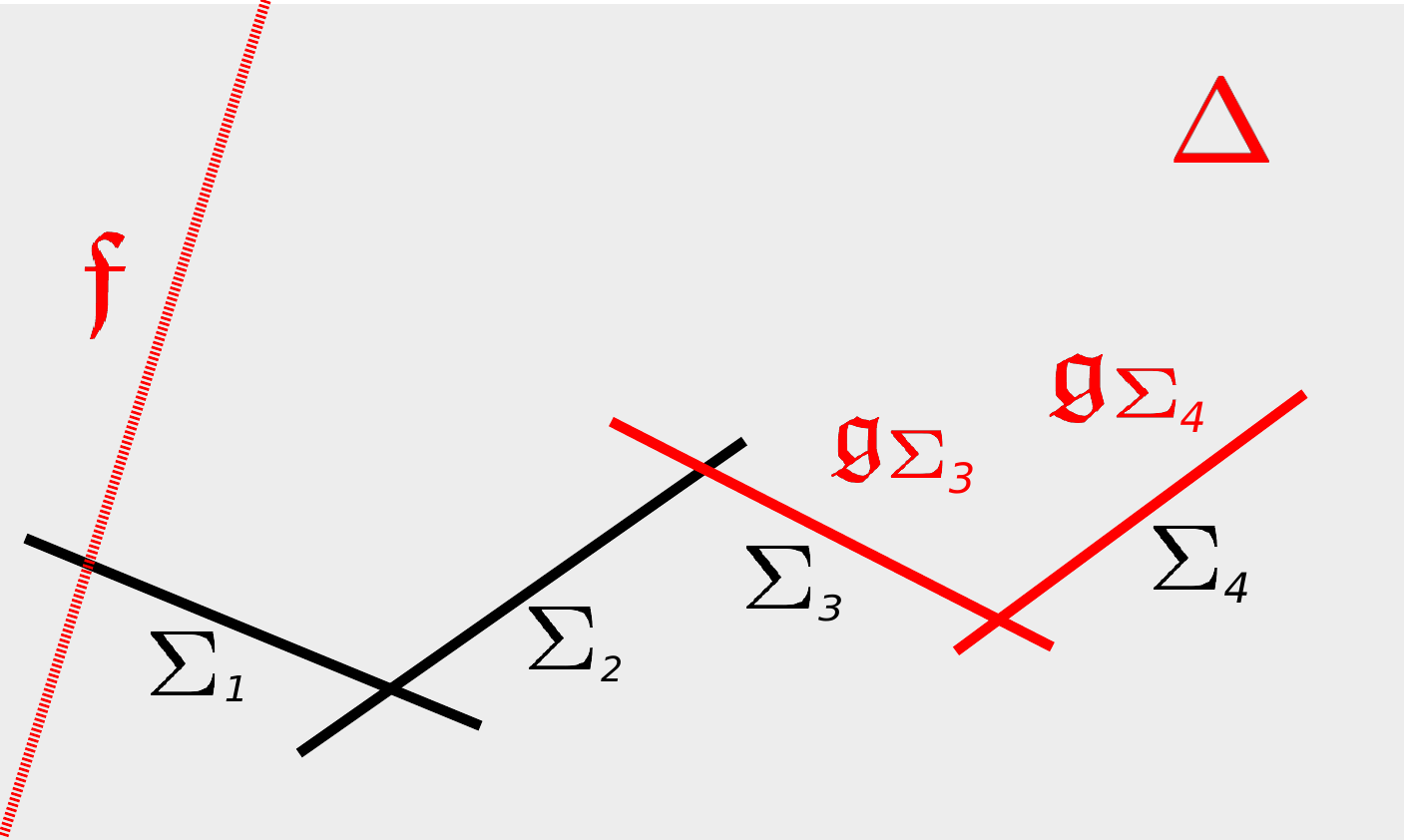}
\vspace{1.3cm}
\caption{Schematic description of the geometric engineering of the tensor branch of a 6d SCFT in F-theory. The rational curves $\Sigma_I$ are in one-to-one correspondence with tensor multiplets, whose scalars' vacuum expectation values coincide with $\text{vol}(\Sigma_I)$. In red we have depicted the discriminant locus, corresponding to the divisor $\Delta$ of the F-theory base. In this example, we have a flavor divisor intersecting the curve $\Sigma_1$, carrying flavor symmetry $\mathfrak{f}$, while the compact curves $\Sigma_3$ and $\Sigma_4$ both support non-abelian gauge groups.}\label{fig:afbase}
\end{center}
\end{figure}

\noindent A schematic description of the typical geometries for the 6d tensor branches can be found in figure \ref{fig:afbase}. The Calabi-Yau condition on $\widehat X$ gives several restrictions both at the level of the base geometries and at the level of the corresponding elliptic fibrations. In particular, the base geometries have to satisfy \cite{Heckman:2013pva}
\begin{itemize}
\item[1.)] For all $I = 1,..., R$ the self-intersection numbers are bounded:
\be
1 \leq - \Sigma_I \cdot \Sigma_I \leq 12\, ;
\ee
\item[2.)] Whenever for some $I\in \{1,...,R\}$
\be
 - \Sigma_I \cdot \Sigma_I \geq 3\,,
\ee
the rational curve $\Sigma_I$ is forced to be part of the discriminant by the CY condition;
\item[3.)] Whenever
\be
\Sigma_I \cdot \Sigma_J \neq 0,
\ee
the self intersections of $\Sigma_I$ and $\Sigma_J$ are constrained; for instance, compact rational curves with with self-intersection smaller than $3$ never intersect in $\widehat B$.\footnote{ The precise rules for composition of the bases can be found in \cite{Heckman:2013pva,Heckman:2015bfa}. We do not dwell on these details here because our focus in this work are the rank-one theories.}
\end{itemize}

\begin{table}[t]
\begin{center}
\begin{tabular}{|c||c|c|c|c|c|c|c|c|c|}
	\hline
	$- \Sigma_I \cdot \Sigma_I$ & $3$ & $4$ & $5$ & $6$ & $7 - 8$ & $9-12$ \\
\hline
Kodaira type & $IV$ & $I_0^*$ & $IV^*_{ns}$ & $IV^*$& $III^*$ & $II^*$ \\
    \hline
    gauge symmetry & $\mathfrak{su}_3$ & $\mathfrak{so}_8$ & $\mathfrak{f}_4$ & $\mathfrak{e}_6$ & $\mathfrak{e}_7$  & $\mathfrak{e}_8$ \\
    \hline
\end{tabular}

\vspace{1cm}

\caption{Minimal gauge algebras for rational curves with $- \Sigma_I \cdot \Sigma_I \geq 3$. For the rational curve with  $- \Sigma_I \cdot \Sigma_I  = 7$ the theory also has matter in the $\tfrac{1}{2}{\bf 56}$, while for the rational curves with $- \Sigma_I \cdot \Sigma_I  = 12 - k $ and $k=1,2,3$ one has $k$ additional E-string subsectors.}\label{tab:minimali}
\end{center}
\end{table}
\noindent By property 2.), the elliptic fibration has to degenerate along the curves with self-intersections smaller than $-3$. For each such rational curve the minimal degeneration gives the gauge algebras in Table \ref{tab:minimali}, that can be further enhanced by tuning the Weierstrass model \cite{Morrison:2012np}. In particular, for a given base geometry, characterized by a collection of compact rational curves with intersection matrix $A_{IJ}$, we have a plethora of possible compatible F-theory models, each characterized by a different structure of the discriminant. These models have to satisfy further consistency conditions dictated by the geometry of $\widehat X$. The resulting geometries are related to each other by Higgs branch RG flows, which amount to deformations of the complex structure of $\widehat X$ (that can be translated into deformations of the complex structure of $X$). The effect of such Higgs branch RG flows is to make the singularities of the elliptic fiber less and less severe. In IIB language, the Higgs branch deformations correspond to separating the stacks of coalesced seven-branes and sending some at infinity, thus lowering their effective number. At the bottom of a Higgs deformation chain sits a minimal model that is characterized solely by the intersection matrix $A_{IJ}$, which automatically determines the corresponding minimal gauge algebras and matter content as in table \ref{tab:minimali}. At the intersection of the curves that are part of the discriminant the corresponding singularity enhances in codimension two, giving rise to matter charged under the gauge groups supported by the curves in the base geometry \cite{Katz:1996xe}.\\

\begin{figure}[t]
\begin{center}
\includegraphics[scale=0.8]{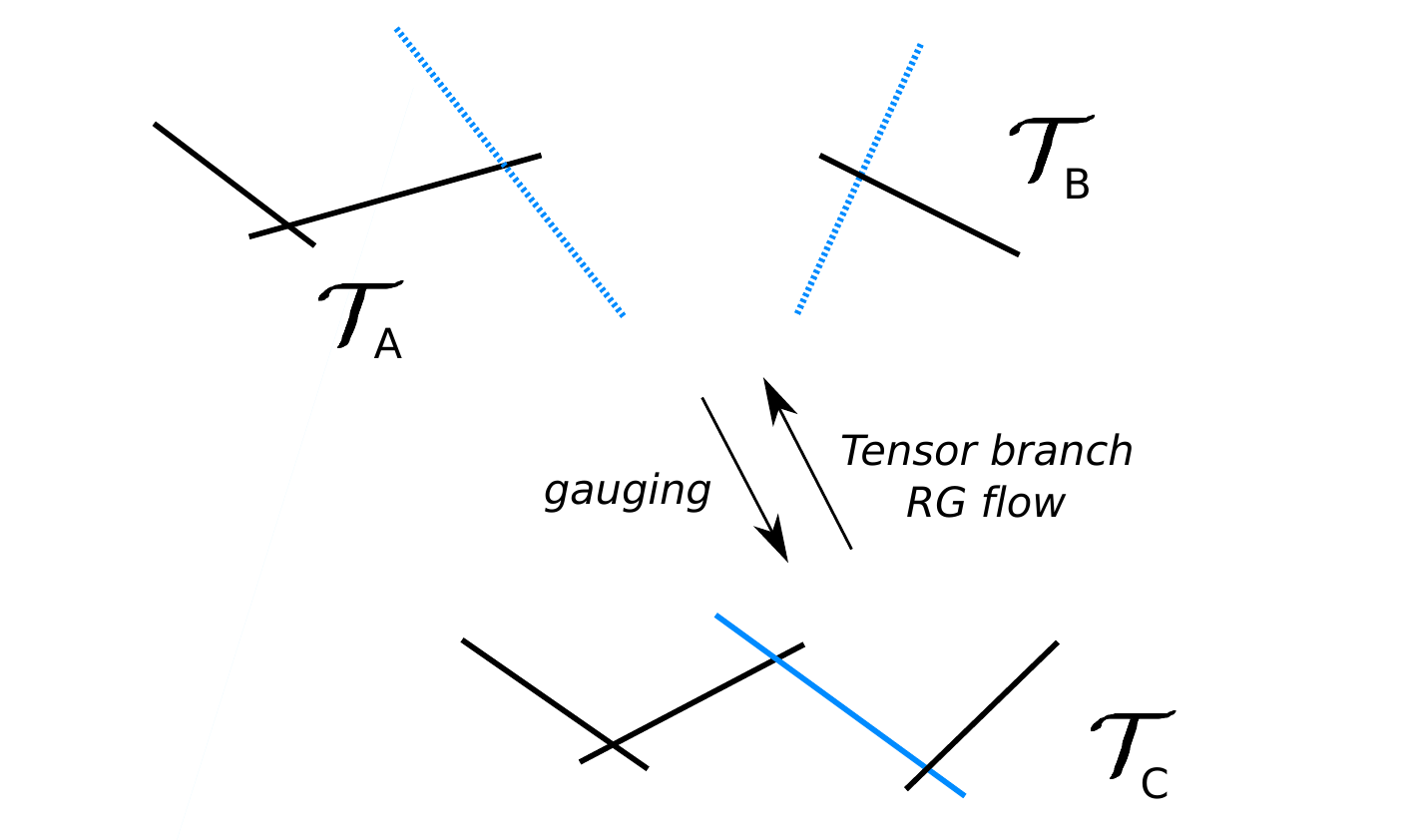}
\vspace{1.3cm}
\caption{Schematic description of the gauging of a (generalized) flavor symmetry in 6d SCFTs. Two models ${\mathcal T}_A$ and ${\mathcal T}_B$ that share the same flavor divisor can give rise to a consistent theory ${\mathcal T}_C$ by making the corresponding noncompact flavor curve (dotted blue line in the figure) into a compact one.}\label{fig:gauging}
\end{center}
\end{figure}

\noindent  Another salient aspect of the base geometries is that they give a natural geometric mechanism for gauging a 6d flavor symmetry: gauging simply amounts to making a flavor divisor compact, as illustrated in figure \ref{fig:gauging}. The precise restrictions imposed by the geometries on this type of mechanism are implicit in the classification result. At any rate, the lesson one learns from this mechanism is that all higher rank 6d SCFTs obtained from F-theory (without frozen singularities) can be obtained by coupling together rank-one building blocks. Below we will see the counterpart of this phenomenon at the level of the BPS strings of these models.

\subsection{Example: conformal matter of D-type}\label{sec:dtypeeng}

Let us discuss now a simple example corresponding to the rank-one conformal matter of D-type. In this case the base $B$ is smooth, and we have a Weierstrass model for $X$ that in a neighborhood of the origin is given by
\be\label{eq:weier}
f(s,t) = -3 \alpha^2 s^2 t^2 \qquad g(s,t) = (2 \alpha^3 + \beta s^{k - 6} t^{k - 6}) s^3 t^3 \qquad k \geq 6 \,,
\ee
where $\alpha,\beta \in \C$. The corresponding discriminant is given by
\be
\Delta = - 432 \, \beta  s^{k-6} t^{k-6}  (4 \alpha^3 s^6 t^6 + \beta s^k t^k).
\ee
We clearly see that we have a pair of noncompact flavor curves given by $D_s \equiv \{t=0\}$ and $D_t \equiv \{s=0\}$ that intersect at the origin. These are part of the discriminant locus and support singularities of type $I^*_{k-6}$ (where $k\geq 6$). This can be easily seen because the order of vanishing 
\be
\text{ord}(f,g,\Delta)\big|_{D_s}  =  (2,3,k-6) = \text{ord}(f,g,\Delta)\big|_{D_t}
\ee
away from the origin, and in each case the monodromy cover of Table \ref{tab:sing} factorizes, which corresponds to a singularity of split type. The origin $D_s \cdot D_t$, however, has a non-Kodaira type singularity, where 
\be
\text{ord}(f,g,\Delta)\big|_{D_s \cdot D_t} = (4,6,2k) \, .
\ee 
By a single blow-up, we obtain an exceptional rational curve $\Sigma$ with self intersection $-\Sigma \cdot \Sigma = 1$, along which the order of vanishing of $(f,g,\Delta)$ is $(0,0,2k - 12)$. The corresponding monodromy cover in this case is non-split, and we obtain an $I_{2k-12}^{ns}$ fiber, corresponding to the gauge algebra $\mathfrak{sp}_{k-6}$. Higgsing in this case corresponds to deforming the polynomial for $g(s,t)$ in equation \eqref{eq:weier} in such a way that $k \mapsto k-1$. The endpoint for this chain of Higgsings is the E-string theory, with trivial gauge algebra $\mathfrak{sp}_{0}$.\\

\noindent Now consider taking two of these models and compactifying a common flavor divisor with gauge algebra $\mathfrak{so}_{8+k}$ corresponding to the $I^*_{k-6}$ fiber. The corresponding intersection pairing is easily determined from the matter content of these models by requiring gauge anomaly cancellation (see discussion below). We obtain
\begin{equation}
A_{IJ} = \left(\begin{matrix} 1 & -1 & 0 \\ -1 & 4 & -1 \\ 0 & -1 & 1 \end{matrix}\right)
\end{equation}
where 
\be
\mathfrak{g}_{\Sigma_1} = \mathfrak{g}_{\Sigma_3} = \mathfrak{sp}_k \qquad\qquad \mathfrak{g}_{\Sigma_2} = \mathfrak{so}_{8+k}.
\ee
Blowing such rational curves down, one obtains a Weierstrass model over a singular base. In this case, the base is no longer $\C^2$, but rather $\C^2/\Z_2$, where $\Z_2$ acts on $(s,t)$ as the cyclic discrete subgroup of $SU(2)$. Proceeding similarly, considering $N$ rank one D-type theories one obtains a Weierstrass model over $\C^2/\Z_N$, where $\Z_N$ acts as
\be
(s,t) \to (\omega s, \omega^{-1} t) \qquad\qquad \omega^N = 1.
\ee
One straightforwardly sees that the Weierstrass model in equation \eqref{eq:weier} is compatible with this action. To proceed, one first has to blow-up the base, giving rise to a collection of $-2$ curves intersecting transverally according to an A-type Dynkin graph; at each intersection, one finds a local model of the type discussed above, which can then be resolved along similar lines \cite{DelZotto:2014hpa}.

\subsection{Anomaly polynomials}\label{sec:anomf}

\begin{table}
\begin{center}

\begin{tabular}{|c|c|c|c|} \hline
$\mathfrak{g}$  \phantom{\Big|}& $R$ & $\text{\,tr}_R F^2$ & $\text{\,tr}_R F^4 $ \\ 
\hline\hline
 $\mathfrak{su}_N $ \phantom{\Big|}& $\text{Adj}$ &$2N \text{\,tr} F^2$&$(6+N)(\text{\,tr} F^2)^2$ \\
$N=2,3$ \phantom{\big|} & $\Lambda$ &$\text{\,tr} F^2$& $\frac12(\text{\,tr} F^2)^2$ \\ \hline
$\mathfrak{su}_N $  \phantom{\Big|} & $\text{Adj}$ &$2N \text{\,tr} F^2$&$6(\text{\,tr} F^2)^2+2N \text{\,tr} F^4  $ \\
$N\ge4$      \phantom{\Big|}    &      $\Lambda$ &$\text{\,tr} F^2$& $0(\text{\,tr} F^2)^2+\text{\,tr} F^4$ \\
     \phantom{\Big|}  & $\Lambda^2$ &$(N-2)\text{\,tr} F^2$&$3(\text{\,tr} F^2)^2+(N-8)\text{\,tr} F^4  $  \\
\hline 
 $\mathfrak{su}_6$ \phantom{\Big|}   &$\Lambda^3$ & $6 \text{\,tr} F^2$ &$6(\text{\,tr} F^2)^2 - 6\text{\,tr} F^4  $  \\
 \hline
 $\mathfrak{so}_8$	 \phantom{\Big|}&$\text{Adj}$ 		&$12 \text{\,tr} F^2$ 		&$4 \text{\,tr}_{V} F^4 + 4 \text{\,tr}_{S_+} F^4 + 4 \text{\,tr}_{S_-} F^4$\\
\phantom{\Big|} &$\Lambda=V,S_+,S_-$ & $2 \text{\, tr}_\Lambda F^2$ & $\sum_{\Lambda=V,S_+,S_-} \text{\,tr}_\Lambda F^4 = 3 (\text{\,tr} F^2)^2$\\
\hline

$\mathfrak{so}_N$  \phantom{\Big|}	& $\text{Adj}$ 		&$(2N-4)\text{\,tr} F^2$			&$12(\text{\,tr} F^2)^2+(2N-16)\text{\,tr} F^4  $ \\
$N\geq 7 $       		 \phantom{\Big|}& $V$ 		&$2\text{\,tr} F^2$				&$0(\text{\,tr} F^2)^2+2 \text{\,tr} F^4$ \\
$N\neq 8$          	 \phantom{\Big|}& $S_*$ 			& $\dim(S_*)(\frac14 \text{\,tr} F^2)$ 	& $\dim(S_*)(\frac3{16} (\text{\,tr} F^2)^2-\frac18 \text{\,tr} F^4 ) $ \\ 
\hline
 $\mathfrak{sp}_N$ 	 \phantom{\Big|}&$\text{Adj}$ 		&$(2N+2)\text{\,tr} F^2$	&$3(\text{\,tr} F^2)^2+(2N+8)\text{\,tr} F^4$\\
     $N\ge2$       	 \phantom{\Big|}&$\Lambda$ 		&$\text{\,tr} F^2$		&$0(\text{\,tr} F^2)^2+\text{\,tr} F^4$\\
     				 \phantom{\Big|}&$\Lambda^2_{irr}$ &$(2N-2)\text{\,tr} F^2$	&$3 (\text{\,tr} F^2)^2+(2N-8)\text{\,tr} F^4  $  \\ 
\hline
$\mathfrak{g}_2$ \phantom{\Big|}& $\text{Adj}$ &$8\text{\,tr} F^2$&$10(\text{\,tr} F^2)^2$ \\
        & $\mathbf{7}$ &$2\text{\,tr} F^2$&$(\text{\,tr} F^2)^2$ \\ \hline
$\mathfrak{f}_4$ \phantom{\Big|}& $\text{Adj}$ &$18\text{\,tr} F^2$&$15(\text{\,tr} F^2)^2$\\
       & $\mathbf{26}$ &$6\text{\,tr} F^2$&$3 (\text{\,tr} F^2)^2$\\ \hline
 $\mathfrak{e}_6$ \phantom{\Big|}& $\text{Adj}$ &$24\text{\,tr} F^2$&$18(\text{\,tr} F^2)^2$ \\
       & $\mathbf{27}$ &$6\text{\,tr} F^2$&$3 (\text{\,tr} F^2)^2$\\ \hline
 $\mathfrak{e}_7$ \phantom{\Big|}& $\text{Adj}$ &$36\text{\,tr} F^2$&$24(\text{\,tr} F^2)^2$\\
       & $\mathbf{56}$ &$12\text{\,tr} F^2$&$6 (\text{\,tr} F^2)^2$\\ \hline
 $\mathfrak{e}_8$ \phantom{\Big|}& $\text{Adj}$ &$60\text{\,tr} F^2$&$36(\text{\,tr} F^2)^2$\\
\hline

\end{tabular}

\end{center}
\smallskip
\caption{Anomaly coefficients in our normalization \cite{Grassi:2000we,Grassi:2011hq}.}\label{tab:anomtab}
\end{table}

\noindent It is well known that there is a deep interplay between the intersection theory of the F-theory base and anomaly cancellation in 6d \cite{Sadov:1996zm,Grassi:2000we,Grassi:2011hq}. The invariant field strengths for the tensor fields $B^I$ in the theory are given by
\be
H^I = dB^I + L^I_3
\ee
where $L^I_3$ is a Chern-Simons three-form associated to the 4-form
\be
\label{EqDefCSGamma}
d L^I_3 = X^I_4 = \tfrac{1}{4} \alpha^I \, p_1(M_6) + \sum_\ell \beta^I_\ell \, c_2(\mathfrak{g}_\ell) + \tfrac12 \sum_{i \, j} \beta^I_{ij} \, c_1(U(1)^i) c_1(U(1)^j)  \,;
\ee
$L^I_3$ can be interpreted as an effective degree-3 abelian gauge field with field strength $X^I_4$, coupled to the tensor multiplets through the Green-Schwarz terms. Moreover, $p_1(M_6)$ is the first Pontryagin class of the tangent bundle of the 6d worldvolume $M_6$, while $c_1(U(1)^i)$ and $c_2(\mathfrak{g}_\ell)$ are respectively the first Chern class of the $i$-th $U(1)$ bundle and the second Chern class associated to $\mathfrak{g}_\ell$, \emph{i.e.} a degree-4 characteristic form associated to the quadratic Casimir of $\mathfrak{g}_\ell$. The vectors $ \mathbf{a},\, \mathbf{b}_\ell, \, \mathbf{b}_{ij}$ with components respectively $\alpha^I, \, \beta^I_\ell $ and $\beta^I_{ij}$ are the anomaly coefficients of the theory; $ \mathbf{a}$ is a gravitational anomaly coefficient, while $\mathbf{b}_\ell$ and $\mathbf{b}_{ij}$ are respectively non-abelian and abelian anomaly coefficients. It is convenient to group the $X^I_4$ into a vector of four-forms $\mathbf{X}_4$.\newline

\noindent Let us first consider the case of a compact elliptic threefold. In order for the local gravitational, gauge and mixed anomalies of the theory to be canceled via the Green-Schwarz mechanism, the following anomaly cancellation condition must be satisfied \cite{Green:1984bx, Sagnotti:1992qw, Sadov:1996zm}: the anomaly polynomial ${\mathcal A}_8$ has to factor:
\be\label{eq:polyfactor}
{\mathcal A}_8 =  \tfrac12 \mathbf{X}_4 \cdot  \mathbf{X}_4 = \tfrac12  A_{I J} \, X^I_4 \wedge X^J_4 \,,
\ee
where $A_{IJ}$ is the intersection pairing. If this is the case, the gauge and gravitational anomalies can be cured with local counterterms of the form $B_I \wedge X^I_4$ in the six-dimensional effective action. The cancellation conditions that do not involve the abelian anomaly coefficients are as follows. The cancellation of the gravitational anomaly requires
\begin{equation}
0=H-V+29T-273 \qquad \qquad \mathbf{a} \cdot \mathbf{a} = 9-T \,,
\end{equation}
where $H$ is the total number of hypermultiplets in the theory, $V$ is the total number of vector multiplets, and $T$ is the total number of tensor multiplets. The mixed gauge-gravitational anomaly gives:\footnote{ We are adopting the same conventions as section 9 of  \cite{Grassi:2000we} with $\mathbf{a}|_{\text{here}}=-K_B|_{\text{there}}$ and $\mathbf{b}|_{\text{here}} = \Sigma|_{\text{there}}$.}
\begin{equation}
\mathbf{a} \cdot \mathbf{b}_\ell = \frac{1}{6} \left(- A_{\text{Adj}_\ell} + \sum_{R_\ell} n^\ell_{R_\ell} A_{R_\ell}\right) \,.
\end{equation}
The cancellation of non-abelian gauge anomalies requires the following constraint on the matter representations
\begin{equation}\label{eq:GSSWnab1}
0= B_{\text{Adj}_\ell} - \sum_{R_\ell} n^\ell_{R_\ell} B_{R_\ell} , 
\end{equation}
and moreover 
\begin{equation}\label{eq:GSSWnab2}
\mathbf{b}_\ell \cdot \mathbf{b}_\ell = \frac{1}{3} \left(-C_{\text{Adj}_\ell} + \sum_{R_\ell} x^\ell_{R_\ell} C_{R_\ell}\right) \quad\text{and}\quad \mathbf{b}_\ell \cdot \mathbf{b}_{\ell'} = \sum_{R_\ell, R_{\ell'}} n^{\ell,\ell'}_{R_\ell, R_\ell'} A_{R_\ell} A_{R_{\ell'}} ~(\ell \neq \ell')
\end{equation}
need to be satisfied. The anomaly cancellation conditions that depend on the abelian anomaly coefficients are the following:
\begin{align}
\mathbf{a} \cdot \mathbf{b}_{ij} &=
\frac{1}{6}
\sum_{R_\ell} n^{i,j}_{q_i, q_j} q_i q_j  \label{eq:GSSWabelian1} \\
0&= \sum_{R_\ell, q_i} n^{\ell,i}_{R_\ell,q_i}  q_i  E_{R_\ell}\,,\label{eq:GSSWabelian2} \\
\mathbf{b}_\ell \cdot \mathbf{b}_{ij} &= \sum_{R_\ell q_i,q_j} n^{\ell,i,j}_{R_\ell,q_i,q_j} q_i q_j  A_{R_\ell}\,, \label{eq:GSSWabelian3}\\
\mathbf{b}_{ij} \cdot \mathbf{b}_{kp}+\mathbf{b}_{ip} \cdot \mathbf{b}_{kj}+\mathbf{b}_{ik} \cdot \mathbf{b}_{jp}
&= \sum_{q_i,q_j,q_k,q_p} n^{i,j,k,p}_{q_i,q_j,q_k,q_p} q_i q_j q_k q_p \, \label{eq:GSSWabelian4}.
\end{align}
The Lie algebra coefficients $A$, $B$, $C$ and $E$ are defined by
\bea\label{eq:ABCE}
\text{tr}_{R_\ell} F^2 &= A_{R_\ell} \text{tr} F^2 \,, &
\text{tr}_{R_\ell} F^4 &= B_{R_\ell} \text{tr} F^4 +
C_{R_\ell} (\text{tr} F^2)^2  \,, &
\text{tr}_{R_\ell} F^3 &= E_{R_\ell} \text{tr} F^3
\eea
for $F \in \mathfrak{g}_i$. For $\mathfrak{g}_\ell$ with rank $\leq 2$, $B_{R_i}$ is defined to vanish for any $R_i$. The coefficients $n^{\ell_1,\cdots,\ell_k, i_1,\cdots,i_K}_{R_{\ell_1}, \cdots,R_{\ell_k},  q_{i_1}, \cdots, q_{i_K}}$ are defined to be the number of hypermultiplets simultaneously in representation $R_{\ell_1}$ of $\mathfrak{g}_{\ell_1}$, ..., $R_{\ell_k}$ of $\mathfrak{g}_{\ell_k}$, $q_{i_1}$ of $u(1)_{i_1}$, ... and $q_{i_K}$ of $u(1)_{i_K}$. In table \ref{tab:anomtab} we give the conventions we use to compute all possible traces needed for this paper \cite{Sadov:1996zm,Grassi:2000we,Grassi:2011hq}. \\

\noindent The link to the algebraic geometry of the F-theory base is as follows \cite{Sadov:1996zm,Grassi:2000we,Grassi:2011hq}: the $\mathbf{a}$ coefficient can be identified with the class $-K$, while the $\mathbf{b}_\ell$ anomaly coefficients can be identified with the curve classes $\Sigma_I$.\footnote{ Here we are assuming that all the curves $\Sigma_I$ belong to the discriminant, which is the case for all the rank-one models that appear in this paper except for the $E$-string theory.} On the other hand, the $U(1)$ anomaly coefficients can be identified with appropriate curve classes in the F-theory base (see e.g. \cite{Lee:2018ihr} for a detailed review).\\

\noindent Decoupling gravity amounts to taking the volume of the F-theory base to infinity, and focusing on a subset of curves which remains compact, see e.g. the discussions in \cite{DelZotto:2014fia,Bhardwaj:2015oru,Lee:2018ihr}. This results in some curves becoming noncompact; the noncompact curves which intersect the compact ones are interpreted as flavor divisors. Note that there is no known six-dimensional theory with a $U(1)$ gauge symmetry, but there are several examples that have $U(1)$ global symmetries. Along the way all of the above conditions can be dropped except for the subset of equations \eqref{eq:GSSWnab1} and  \eqref{eq:GSSWnab2} that corresponds to gauge anomaly cancellation. Nevertheless the anomaly coefficients $\mathbf{a}$, as well as the ones corresponding to flavor divisors, give rise to `t Hooft anomalies for our models, which can be determined by using the second equation in \eqref{eq:GSSWnab2} as well as equation \eqref{eq:GSSWabelian3}.

\subsection{Rank one 6d SCFTs and their Higgsing trees}\label{sec:higgsingchains}

There are only nine F-theory bases that can give rise to rank-one SCFTs. These are given by  Hirzebruch-Jung singularities, denoted by $HJ_{p,q}$, that can be resolved by blowing up the origin leading to a single rational curve $\Sigma \subset \widehat B$. Recall that a singularity of type $HJ_{p,q}$  is the orbifold of $\C^2$
\be
HJ_{p,q}\,\,\colon \qquad (z_1,z_2) \to (\omega z_1, \omega^q z_2) \qquad \omega^p = 1.
\ee
The rank one theories correspond to Hirzebruch-Jung orbifold singularities of type
\be
(p,q)=(n,1) \quad \text{with}\quad n=1,2,3,4,5,6,7,8,12 \,.
\ee
Performing a single blow up in the base results in a single rational curve $\Sigma$ of self-intersection 
\begin{equation}
\Sigma \cdot \Sigma = -n \, .
\end{equation}
The resolved base is
\begin{equation}\label{eq:normbundle}
\widehat B = \text{Tot}\left(\mathcal{O}(-n) \to \mathbb{P}^1\right) \,,
\end{equation} 
where the K\"ahler class of the base $\mathbb{P}^1$ corresponds to the vev of the tensor multiplet scalar parametrizing the 6d tensorial Coulomb branch. For $n> 2$, all these curves are forced to be part of the discriminant for $\widehat X$ and have minimal singularities associated to them which we list in table \ref{tab:HJ}. For a generic F-theory model, the curves with $n=1,2$ in the base are not necessarily part of the discriminant. For 6d SCFTs with $(1,0)$ supersymmetry, the situation is slightly more constrained. Indeed, the minimal singularity associated to $n=2$ corresponds to the M-string theory is a type $I_1$ fiber. On the other hand, in the minimal case for $n=1$ the generic fiber on the $\mathbb{P}^1$ is nondegenerate, corresponding to the E-string CFT. By tuning the corresponding Weierstrass models, one can obtain higher Kodaira singularities for each of these bases, giving rise to higher rank gauge symmetries at the price of introducing flavor divisors. At the locus where a flavor divisor meets a gauge divisor, the singularity of the fiber enhances and one obtains localized matter.   \\

\begin{table}
\begin{center}
\begin{tabular}{|c|ccccccccc|}
\hline
$HJ_{n,1}$& $HJ_{1,1}$ &$HJ_{2,1}$&$HJ_{3,1}$&$HJ_{4,1}$&$HJ_{5,1}$&$HJ_{6,1}$&$HJ_{7,1}$&$HJ_{8,1}$&$HJ_{12,1}$\phantom{$\Big|$}\\ 
\hline
fiber & $I_0$ & $I_1$  & $IV$ & $I_0^*$ & $IV^*_{ns}$ & $IV^*$ & $III^*$ & $III^*$ & $II^*$\phantom{$\Big|$}\\ 
$\mathfrak{g}_{min}$ & none & none &$\mathfrak{su}_3$&$\mathfrak{so}_8$&$\mathfrak{f}_4$&$\mathfrak{e}_6$&$\mathfrak{e}_7\oplus \tfrac{1}{2}\mathbf{56}$&$\mathfrak{e}_7$&$\mathfrak{e}_8$\phantom{$\Big|$}\\
\hline
\end{tabular}
\end{center}\caption{ $HJ_{p,q}$ singularities and corresponding minimal gauge groups for rank one 6d SCFTs.}\label{tab:HJ}
\end{table}

\begin{table}
\begin{center}
{\scalebox{0.85}{
\begin{tabular}{|r|r|r|r|}
\hline
\phantom{$\Big|$} $n$ & $G_\Sigma$ & $R_\Sigma$& $F_\Sigma$\\
\hline
\phantom{$\Big|$}1,2& $SU(2)$&$(32 - 12 n) \tfrac{1}{2} \Lambda$&$SO(32 -12n)$\\
\phantom{$\Big|$}1,2,3&$SU(3)$&$(18 - 6 n)\Lambda$&$SU(18-6n)$\\
\phantom{$\Big|$}1,2& $SU(4)$&$(16-4n)\Lambda\oplus(2-n)\Lambda^2$&$SU(16-4n)\otimes Sp(2-n)$\\
\phantom{$\Big|$}1,2& $SU(5)$&$(16-3n)\Lambda\oplus(2-n)\Lambda^2$&$SU(16-3n)\otimes SU(2-n) \otimes U(1)^{2-n}$\\
\phantom{$\Big|$}1,2& $SU(6)$&$(16- 2n)\Lambda\oplus(2-n)\Lambda^2$&$SU(16-2n)\otimes SU(2-n) \otimes U(1)^{2-n}$\\
\phantom{$\Big|$}1&$SU(6)_\ast$&$(16-n)\Lambda\oplus\tfrac{1}{2}(2-n)\Lambda^3$&$SU(16-n)\otimes SO(2-n)$\\
\phantom{$\Big|$} 1,2&$SU(N), \ N\geq7$&$(16+(N-8)n)\Lambda\oplus(2-n)\Lambda^2$&$SU(16\!+\!(N-8)n)\!\otimes\! SU(2\!-\!n)\!\otimes\! U(1)^{2-n}$\\
\phantom{$\Big|$} 1& $Sp(N), \ N\geq2$&$(16+4N)\tfrac{1}{2} \Lambda$ &$SO(16+4N)$\\
\phantom{$\Big|$} 1,2,3& $SO(7)$&$(3-n)V\oplus 2(4-n)S$&$Sp(3-n)\otimes Sp(8-2n)$\\
\phantom{$\Big|$} 1,2,3,4& $SO(8)$&$(4-n)(V\oplus S_+ \oplus S_-)$&$Sp(4-n)\otimes Sp(4-n)\otimes Sp(4-n)$\\
\phantom{$\Big|$} 1,2,3,4&$SO(9)$&$(5-n)V\oplus(4-n)S$ & $Sp(5-n)\otimes Sp(4-n)$\\
\phantom{$\Big|$} 1,2,3,4&$SO(10)$&$(6-n)V\oplus(4-n)S_+$&$Sp(6-n)\otimes SU(4-n)\otimes U(1)$\\
\phantom{$\Big|$} 1,2,3,4&$SO(11)$&$(7-n)V\oplus(4-n)\tfrac{1}{2} S$&$Sp(7-n)\otimes SO(4-n)$\\
\phantom{$\Big|$} 1,2,3,4&$SO(12)$&$(8-n)V\oplus(4-n)\tfrac{1}{2} S_\pm$&$Sp(8-n)\otimes SO(4-n)$\\
\phantom{$\Big|$} 2,4& $SO(13)$&$(9-n)V\oplus(2-\tfrac{1}{2}n)\tfrac{1}{2} S$&$Sp(9-n)\otimes SO(2-\tfrac{1}{2}n)$\\
\phantom{$\Big|$} 4& $SO(N),\ N\geq14$	&$(n-8)V$	&$Sp(n-8)$\\
\phantom{$\Big|$} 1,2,3&$G_2$&$(10-3n)\mathbf{7}$&$Sp(10-3n)$\\
\phantom{$\Big|$} 1$\dots$5&$F_4$&$(5-n)\mathbf{26}$&$Sp(5-n)$\\
\phantom{$\Big|$} 1$\dots$6&$E_6$&$(6-n)\mathbf{27}$	&$SU(6-n)\otimes U(1)$\\
\phantom{$\Big|$} \!\!\!\!\!1$\dots$8&$E_7$&$(8-n)\tfrac{1}{2} \mathbf{56}$&$SO(8-n)$\\
\phantom{$\Big|$} 12&$E_8$&none&none\\

\hline
\end{tabular}
}}
\caption{Global symmetries and matter of rank one 6D SCFTs. For $SU(N)$ and $Sp(N)$, $\Lambda$ is the fundamental representation, and $\Lambda^k$ its exterior powers. For $SO(N)$, $V$ denotes the vector representation; for $N$ odd, $S$ denotes the spinor representation, while for $N$ even we denote respectively by $S_+$ and $S_-$ the spinor and conjugate spinor representations. Recall that $k$ fundamentals in the $SU(N)$ case contribute a factor of $SU(k)$ to the flavor symmetry group, instead of $U(k)$, see e.g. \cite{Hanany:1997gh,Gadde:2015tra}. There are two theories with $G=SU(6);$ we use the terminology $G=SU(6)_*$ to distinguish the one with $F_\Sigma=SU(15)$ from the other.}\label{tab:sapori}
\end{center}
\end{table}

\noindent We introduce the following special notation for rank one models:
\be
\six{n}{\,G}{F}
\ee
where ${\bf n}$ refers to the Hirzebruch-Jung singularity type for the F-theory base, $G$ labels the gauge group corresponding to the seven-brane stack wrapping the resolved base curve $\Sigma \subset \widehat B$, whose self intersection is $n$, and $F$ is the flavor symmetry for the corresponding model. In particular, in our notation the M-string corresponds to the model $\six{2}{SU(1)}{SU(2)}$, while the E-string is the model $\six{1}{Sp(0)}{E_8}$. The flavor symmetry for a given SCFT is encoded by the matter content of the corresponding model, which is in turn uniquely determined (with a few exceptions) by the requirement of anomaly cancellation once $n$ and $G$ are specified. Therefore we often drop the flavor symmetry label. We refer the reader to table \ref{tab:sapori} for a summary of the rank one models and their flavor symmetries \cite{Bertolini:2015bwa}. In consulting the table, it is useful to recall that $N$ hypermultiplets in a complex representation of the gauge group support an $U(N)$ symmetry group, which is enhanced to $SO(2N)$ for quaternionic ({\it i.e.}, pseudoreal) representations, and to $Sp(N)$ for real representations. In the quaternionic case, the underlying complex representation is a representation by half-hypermultiplets, so that $N$ is allowed to be a half-integer. The fundamental (resp. anti-symmetric) representation of $SU(N)$ is quaternionic for $N=2$ and is complex for $N>2$ (resp. real for $N=4$ and complex for $N>4$);  the fundamental representation of $Sp(N)$ is always quaternionic; the spinor representations of $SO(N)$ are real for $N=0,1,7 \mod 8$, complex for $N=2,6 \mod 8$, and quaternionic for $N=3,4,5 \mod 8$. It is interesting to remark that table \ref{tab:sapori} respects the exceptional isomorphisms of Lie algebras \cite{Bertolini:2015bwa}. Extending the $SO(N)$ formula to the representation of $SO(6)$, one obtains $(2-n)V + 4(4-n)S_\pm$, which agrees with the result for $SU(4)$ as $(V,S_\pm) \simeq (\Lambda^2,\Lambda)$. Similarly, for $SO(5)$, one obtains $(1-n)V+8(4-n)\tfrac{1}{2}S$, which implies that $n=1$ and $V$ does not occur; this agrees with the $Sp(2)$ case as $S \simeq \Lambda$.\newline

\noindent For the theories $\six{n}{SO(12)}{}$ with $\mathbf{n} = 1,2$ additional remarks are in order. In these cases the matter sector includes respectively $3$ or $2$ half-spinors. Recall that spinors of different chirality are Casimir equivalent, so one may entertain making different choices of spinorial matter. Consider the case $\mathbf{n}=1$ first; choosing the spinorial matter to be in the $\tfrac32 S_+$ or $\tfrac32 S_-$ representation gives rise to a flavor symmetry $SO(3)$, while the choices $\tfrac12 S_+ \oplus  S_-$ and $S_+ \oplus  \tfrac12 S_-$ give $SO(1) \times SO(2) \simeq U(1)$ flavor symmetry; these choices lead naively to different 6d SCFTs, as the flavor symmetry groups are different. In later sections we also use the terminology $SO(12)_a$ and $SO(12)_b$ to distinguish between the $G=SO(12)$ theories with $F=SO(3)$ and $F=SO(1)\times SO(2)$.  Similarly for $\mathbf{n}=2$ one can obtain a flavor symmetry $SO(2) \simeq U(1)$ or $SO(1) \times SO(1)$ associated different choices of chirality for the spinorial matter. We again use the terminology $SO(12)_a$ and $SO(12)_b$ to distinguish between these two possibilities. Notice that these possibilities do not arise for the models  $\six{n}{SO(10)}{}$ because the representations $S_+$ and $S_-$ are each other's complex conjugate. Naively one would think that a similar phenomenon might arise also for $SO(8)$ as the representations $V=\mathbf{8}_v$, $S_+ = \mathbf{8}_s$ and $S_-=\mathbf{8}_c$ are rotated by triality, but this is not the case, as these representations are not Casimir equivalent --- see Appendix C of \cite{Grassi:2011hq} for a proof.\footnote{ We thank D. R. Morrison for pointing this out to us.}\\

\noindent In the remainder of this section we review the classification of rank-one SCFTs \cite{Erler:1993zy,Danielsson:1997kt} from the F-theory perspective \cite{Heckman:2015bfa}. A convenient organizing principle is to distinguish nine different classes of theories based on the value of $n$; in each class, one can arrange the theories in `Higgsing trees' where each node of the tree corresponds to a specific rank 1 SCFT, and arrow connecting two theories indicates that it is possible to Higgs the first theory to obtain the second. In drawing the Higgsing trees, since $n$ is fixed, we denote the Higgsable SCFTs by the alternative notation $G\oplus R_G$ emphasizing the representation of $G$ in which the matter transforms.

\subsubsection{Models with finite-length Higgsing trees} 

The rank-one models corresponding to F-theory bases with 
\be
n=3,5,6,7,8,12
\ee 
all have finite-length Higgsing trees. The rank one theories 
\be
\six{12}{E_8}{}\,, \qquad \six{8}{E_7}{}\,, \qquad \six{7}{E_7}{}
\ee 
are isolated, meaning that these theories are not part of any rank-one Higgsing tree that includes higher rank gauge groups. They can nevertheless be obtained by Higgsing higher rank 6d SCFTs, obtained by coupling the $\six{12}{E_8}{}$ with several copies of $\six{1}{Sp(0)}{E_8}$ theories, gauging the diagonal $E_8$ flavor symmetry of these models \cite{Kim:2016foj}.\newline

\noindent The rank one theory $\six{6}{E_6}{}$ has a Higgsing tree of length one:
\be
E_7 \oplus {\bf 56}  \longrightarrow \six{6}{E_6}{},
\ee
where the theory $\six{6}{E_7}{U(1)}$ has matter charged with respect to $E_7$ transforming as a hypermultiplet in the ${\bf 56}$ representation. The theory $\six{5}{F_4}{}$ has a Higgsing tree of length two,
\be\label{eq:Higgsingtree5}
E_7 \oplus \tfrac{3}{2} {\bf 56} \longrightarrow E_6 \oplus {\bf 27} \longrightarrow \six{5}{F_4}{}.
\ee
The richest finite-length Higgsing tree is the one for the theory $\six{3}{SU(3)}{}$, given by
\be\label{eq:alberodiHiggs3}
\xymatrix{\six{3}{SU(3)}{}&\\
G_2 \oplus {\bf 7}\ar[u]&\\
SO(7) \oplus S^{\oplus 2}\ar[u]&\\
SO(8) \oplus V \oplus S_+ \oplus S_-\ar[u]&\\
F_4 \oplus {\bf 26}^{\oplus 2} \ar[u]& SO(9) \oplus V^{\oplus 2} \oplus S\ar[ul]\\
E_6  \oplus {\bf 27}^{\oplus 3} \ar[u]& SO(10) \oplus V^{\oplus 3} \oplus S_+\ar[u] \\
E_7  \oplus (\tfrac12 {\bf 56})^{\oplus 5}\ar[u] & SO(11)\oplus V^{\oplus 4} \oplus \tfrac12 S\ar[u] \\
 & SO(12)\oplus V^{\oplus 5} \oplus \tfrac12 S_\pm\ar[u]}
\ee

\subsubsection{Models with infinite-length Higgsing trees}

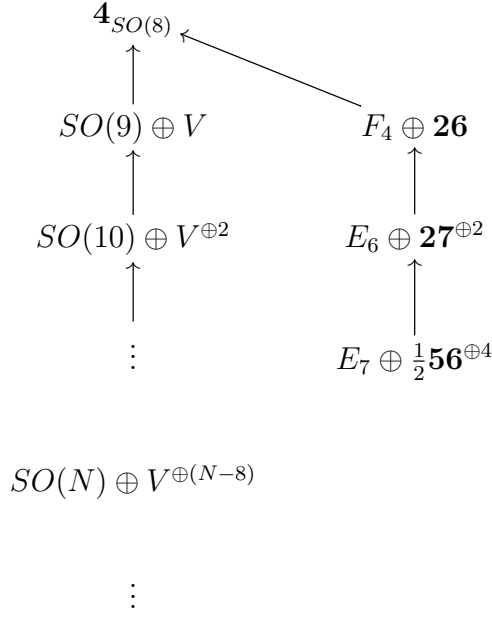
\begin{figure}
$$
\begin{gathered}
\xymatrix{
\six{4}{SO(8)}{}&\\
SO(9)\oplus V \ar[u]&F_4\oplus {\bf 26}\ar[ul]\\
SO(10)\oplus V^{\oplus 2}\ar[u]&E_6 \oplus {\bf 27}^{\oplus 2}\ar[u]\\
\vdots\ar[u]& E_7 \oplus \tfrac{1}{2}{\bf 56}^{\oplus 4}\ar[u]\\
SO(N)  \oplus V^{\oplus (N-8)}&\\
\vdots
}
\end{gathered}
$$
\caption{The Higgsing tree for $\six{4}{SO(8)}{}$.}\label{fig:alberodihiggsD}
\end{figure}

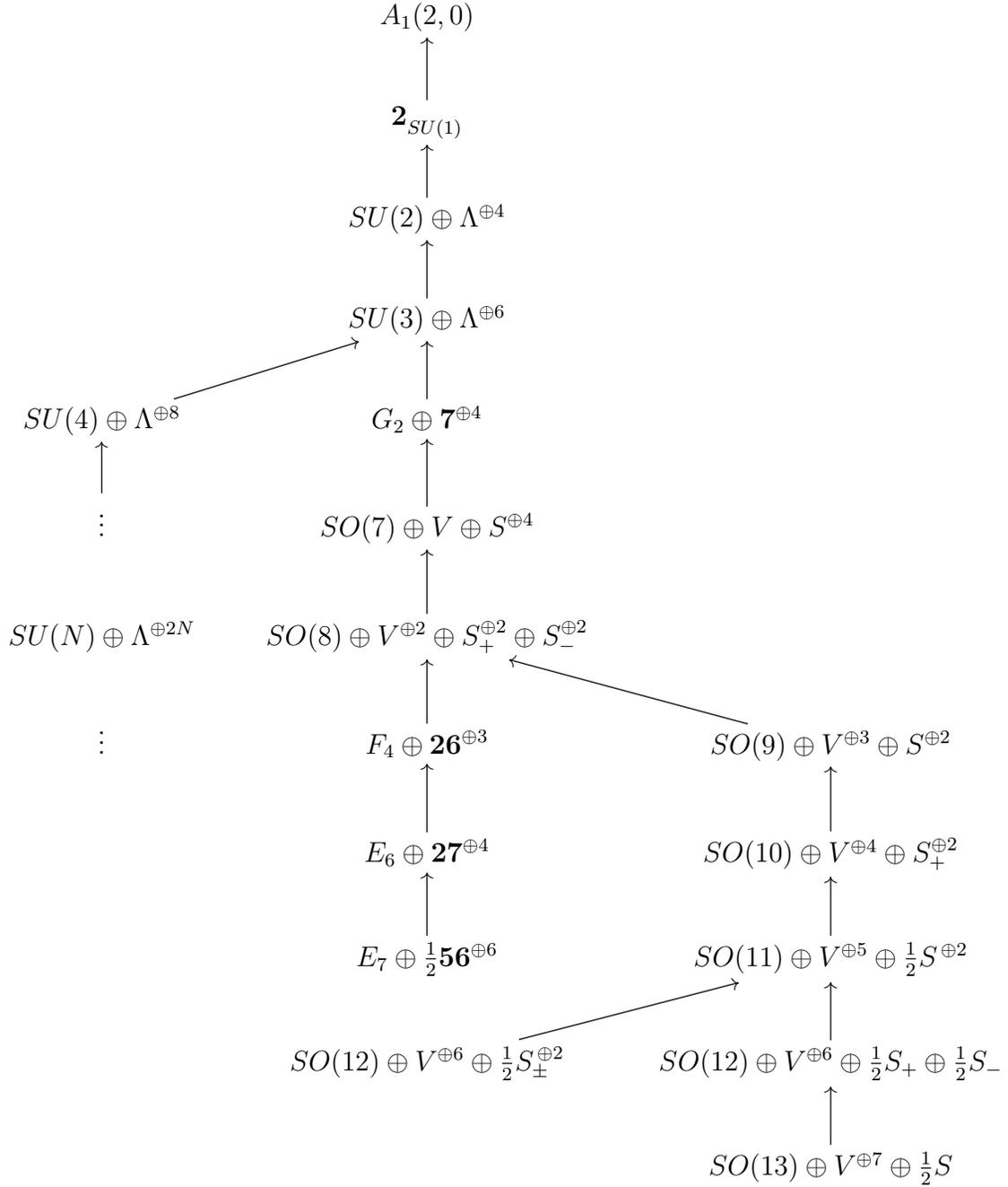
\begin{figure}
$$
\begin{gathered}
\xymatrix{
&A_1 (2,0)&\\
&\six{2}{SU(1)}{}\ar[u]&\\
&SU(2) \oplus \Lambda^{\oplus 4}\ar[u]&\\
&SU(3) \oplus \Lambda^{\oplus 6}\ar[u]&\\
SU(4) \oplus \Lambda^{\oplus 8}\ar[ur]&G_2 \oplus {\bf 7}^{\oplus 4}\ar[u]&\\
\vdots\ar[u]&SO(7) \oplus V \oplus S^{\oplus 4}\ar[u]&\\
SU(N) \oplus \Lambda^{\oplus 2N}&SO(8) \oplus V^{\oplus 2} \oplus S_+^{\oplus 2}\oplus S_-^{\oplus 2}\ar[u]&\\
\vdots&F_4\oplus {\bf 26}^{\oplus 3}\ar[u]&SO(9) \oplus V^{\oplus 3} \oplus S^{\oplus 2}\ar[ul]\\
&E_6 \oplus {\bf 27}^{\oplus 4}\ar[u]&SO(10) \oplus V^{\oplus 4} \oplus S_+^{\oplus 2} \ar[u]\\
&E_7 \oplus \tfrac{1}{2}{\bf 56}^{\oplus 6}\ar[u]&SO(11) \oplus V^{\oplus 5} \oplus \tfrac{1}{2} S^{\,\oplus 2}\ar[u]\\
&SO(12) \oplus V^{\oplus 6} \oplus \tfrac{1}{2} S_\pm^{\, \oplus 2}\ar[ur]&SO(12) \oplus V^{\oplus 6} \oplus \tfrac{1}{2} S_+ \oplus \tfrac12 S_-\ar[u]\\
&&SO(13) \oplus V^{\oplus 7} \oplus \tfrac12 S \ar[u]\\
}
\end{gathered}
$$
\caption{The M-string Higgsing tree.}\label{fig:alberodihiggsM}
\end{figure}

\begin{figure}
{\footnotesize
$$
\begin{gathered}
\xymatrix{
&\six{1}{Sp(0)}{E_8}&&\\
&Sp(1) \oplus \Lambda^{\oplus 10}\ar[u]&&\\
&SU(3) \oplus \Lambda^{\oplus 12}\ar[u]&&\\
Sp(2) \oplus \Lambda^{\oplus 14}\ar[uur]&SU(4) \oplus \Lambda^{\oplus 12}\oplus \Lambda^2\ar[u]&&G_2 \oplus {\bf 7}^{\oplus 7}\ar[ull]\\
Sp(3) \oplus \Lambda^{\oplus 16}\ar[u]& SU(5) \oplus \Lambda^{\oplus 13}\oplus \Lambda^2\ar[u] &&SO(7) \oplus V^{\oplus 2} \oplus S^{\oplus 6\ar[u]}\\
\vdots\ar[u]&SU(6) \oplus \Lambda^{\oplus 14}\oplus \Lambda^2\ar[u] &SU(6) \oplus \Lambda^{\oplus 15}\oplus \tfrac{1}{2} \Lambda^3\ar[ul]&SO(8) \oplus V^{\oplus 3} \oplus S_+^{\oplus 3}\oplus S_-^{\oplus 3} \ar[u]\\
Sp(N) \oplus \Lambda^{\oplus 8+2N}&\vdots\ar[u]&SO(9) \oplus V^{\oplus 4} \oplus S^{\oplus 3}\ar[ur]&F_4\oplus {\bf 26}^{\oplus 4}\ar[u]\\
\vdots &SU(N) \oplus \Lambda^{\oplus \, 8 + N} \oplus \Lambda^2& SO(10) \oplus V^{\oplus 5} \oplus S_+^{\,\oplus 3}\ar[u]&E_6 \oplus {\bf 27}^{\oplus 5}\ar[u]\\
&\vdots&SO(11) \oplus V^{\oplus 6} \oplus \tfrac{1}{2} S^{\,\oplus 3}\ar[u]&E_7 \oplus \tfrac{1}{2}{\bf 56}^{\oplus 7}\ar[u]\\
&&SO(12) \oplus V^{\oplus 7} \oplus \tfrac{1}{2} S_\pm^{\, \oplus 3}\ar[u]&SO(12) \oplus V^{\oplus 7} \oplus \tfrac{1}{2} S_\pm \oplus \tfrac12 (S_\mp)^{\oplus 2}\ar[ul]
}
\end{gathered}
$$
}
\caption{The E-string Higgsing tree.}\label{fig:alberodihiggsE}
\end{figure}
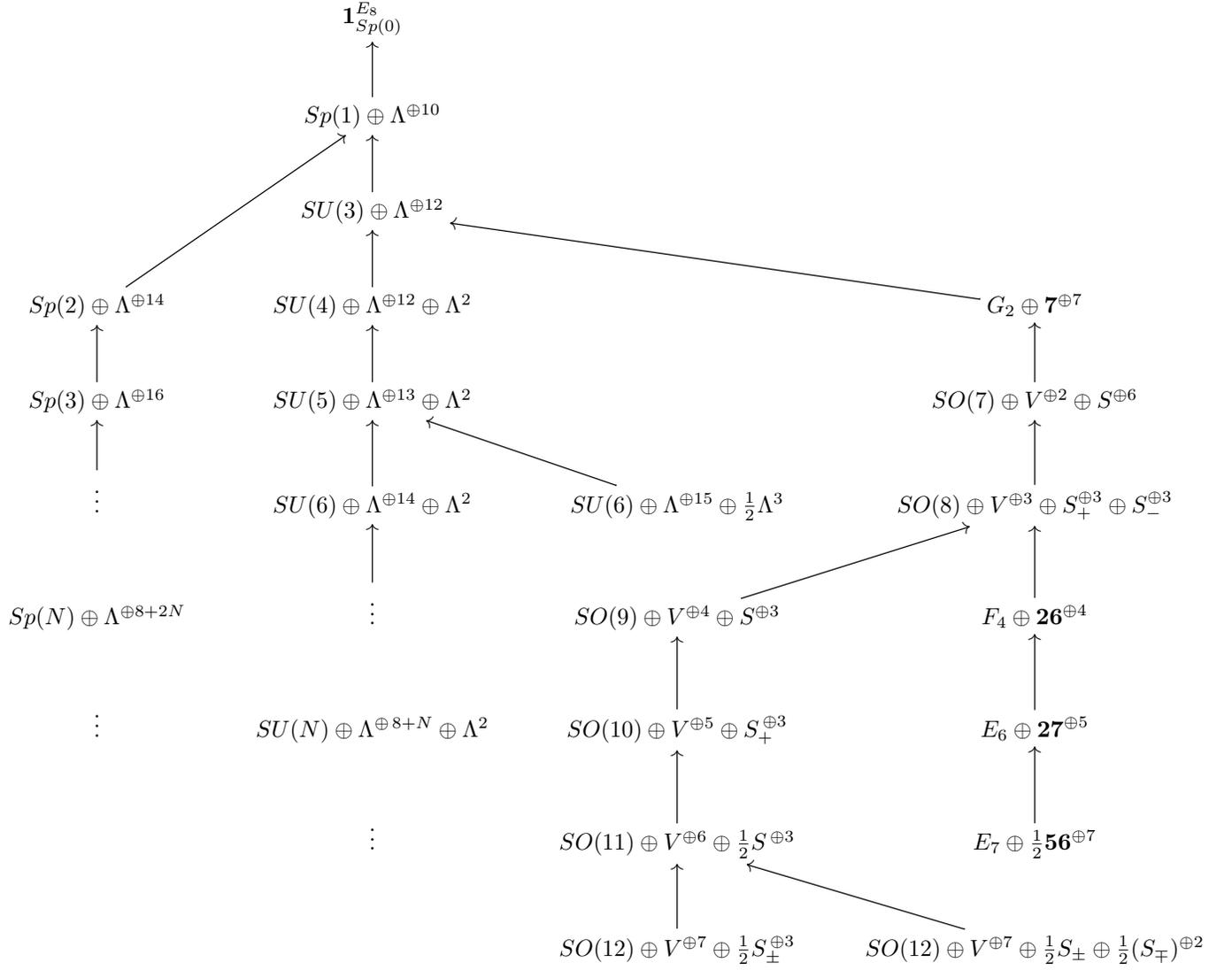

The remaining bases with
\be
n=1,2,4
\ee 
all have infinitely long Higgsing trees, at the bottom of which one finds the theories
\be
\six{1}{Sp(0)}{E_8} \qquad \six{2}{SU(1)}{} \qquad \six{4}{SO(8)}{}
\ee
We draw the corresponding Higgsing trees in figures \ref{fig:alberodihiggsD}, \ref{fig:alberodihiggsM}, and \ref{fig:alberodihiggsE}. \\

\noindent The M-strings and E-strings deserve special attention. The first is in fact technically a Higgsable theory. In the F-theory setup this is related to the fact that the discriminant has an $I_1$ singular fiber along the base curve $\Sigma$, this explains our notation $\six{2}{SU(1)}{}$ for this model. This can be deformed to a smooth configuration with fiber $I_0$ and no monodromy. In this case we have an enhancement of supersymmetry and we obtain the  $A_1$ (2,0) theory. This can also be readily understood from the perspective of type IIA string theory, where the M-string is realized by the following brane configuration:
\be
\begin{tabular}{c|cccccccccc}
&$0$&$1$&$2$&$3$&$4$&$5$&$6$&$7$&$8$&$9$\\
\hline
NS5$_1$& X & X & X & X & X & X & \{$p_0$\} &--&--&--\\ 
NS5$_2$& X & X & X & X & X & X & \{$p_1$\} &--&--&--\\
D6 & X & X &X & X & X & X & X &-- & --& --
\end{tabular}
\ee
In this IIA setup, Higgsing amounts to initiating an RG flow by moving the D6 brane away from the NS5 branes along a direction transverse to them. In the IR limit, the D6 brane is moved infinitely far away from the NS5 branes, and the 6d theory reduces to the $A_1$ $\mathcal{N}=(2,0)$ SCFT.\\

\noindent A similar picture holds for the E-string SCFT \cite{Hanany:1997gh}. One can consider the following brane configuration
\be
\begin{tabular}{c|cccccccccc}
&$0$&$1$&$2$&$3$&$4$&$5$&$6$&$7$&$8$&$9$\\
\hline
O8$^-$+8 D8 & X & X & X & X & X & X & \{$p_0$\} &X& X &X\\ 
$\tfrac12$NS5$_1$& X & X & X & X & X & X & \{$p_0$\} &--&--&--\\
NS5$_2$& X & X & X & X & X & X & \{$p_1$\} &--&--&--\\
D6 & X & X &X & X & X & X & X &-- & --& --
\end{tabular}
\ee
which gives a brane realization of the $G=`SU(1)$' theory, sitting at the end of the $SU(n)$ Higgsing chain.  Alternatively, we can consider the following brane configuration:
\be
\begin{tabular}{c|cccccccccc}
&$0$&$1$&$2$&$3$&$4$&$5$&$6$&$7$&$8$&$9$\\
\hline
O8$^-$+8 D8 & X & X & X & X & X & X & \{$p_0$\} &X& X &X\\ 
NS5& X & X & X & X & X & X & \{$p_1$\} &--&--&--\\
\end{tabular}
\ee
obtained by removing the D6 brane and also moving away the 1/2-NS5 brane along $x_{6,7,8,9}$. This provides a realization of the $G=`Sp(0)$' theory. The only difference between the two theories is that the latter describes the E-string theory, while the former describes the E-string theory plus a decoupled free (1,0) hypermultiplet. Going from the former to the latter can be viewed as a quite trivial Higgs branch RG flow. As the E-string's 2d CFT is not sensitive to the presence of the extra decoupled hypermultiplet \cite{Gadde:2015tra}, in the remainder of the paper we do not distinguish between the two theories and only consider the $G=`Sp(0)$' theory.

\section{BPS strings and wrapped D3 branes}\label{sec:BPSstrings}

In this section we discuss the geometric engineering of the BPS strings of rank-one 6d SCFTs. In section \ref{sec:BPSstringsuniv} we point out that these sectors are the basic elements of the BPS spectrum of 6d SCFTs of arbitrary rank. In section \ref{sec:defects} we discuss some of the features of the (0,4) surface defects that appear in our construction. In section \ref{sec:anominflow} we use anomaly inflow from 6d to determine the anomaly polynomial of the 2d worldsheet theories of the strings. In section \ref{sec:gaiotto} we discuss about an analogy between the structure of the singular elliptic fibers and UV curves for wrapped M5 branes.

\subsection{Universality of rank-one BPS strings}\label{sec:BPSstringsuniv}
Along the tensor branch, six-dimensional theories have BPS strings that are engineered in F-theory by D3 branes wrapping the rational curves $\Sigma_I$. By an abuse of notation, we denote the classes of the curves $\Sigma_I$ with the same symbol used for the curves themselves. The integer lattice
\begin{equation}
\Gamma \equiv \sum_{I=1}^R \, \mathbb{Z} \, \Sigma_I  \subseteq H_{2,c}(B,\mathbb{Z})
\end{equation}
is identified with the string charge lattice of the model. A BPS string configuration with charge 
\be
\strich = N_1 \, \Sigma_1 + \dots + N_R \, \Sigma_R \,,
\ee
where $N_i \geq 0$, is engineered  as a bound state of $N_1$ D3 branes wrapping the curve $\Sigma_1$, $N_2$ D3 branes wrapping the curve $\Sigma_2$, etcetera. The configuration is irreducible only if the support of the $N_I$ is on curves that mutually intersect. We denote the cone of positive charges corresponding to the $\Sigma_I$ generators as $\Gamma_+$. All the resulting worldsheet theories preserve $(0,4)$ supersymmetry.\\

\noindent Let us restrict momentarily to 6d SCFT that do not possess abelian global symmetries. For a given charge vector $\strich$, let us denote by ${\mathcal I}(\strich) \subset \{1,\dots, R\}$ the support of the charge vector, e.g. the $I \in \{1,\dots, R\}$ for which $N_I \neq 0$. Let us also denote by ${\mathcal I}_G(\strich) \subset {\mathcal I}(\strich)$ the index set of those curves in the support of the charge vector that are also part of the discriminant. Let us further denote by ${\mathcal I}_F(\strich)$ an index set running over those compact and noncompact irreducible components of the discriminant that are not in the support of the charge vector but intersecting the compact curves $\Sigma_I$ such that $I\in {\mathcal I}(\strich)$. For all curves $\{\Sigma_I\}_{I \in {\mathcal I}_G(\strich)}$ (resp. $\{\Delta_\alpha\}_{ 
\alpha \in {\mathcal I}_F(\strich)}$) we denote by $G_I$ (resp. $F_\alpha$) the gauge group associated to the corresponding seven-brane stack. The $(0,4)$ model corresponding to a BPS string of charge $\strich \in \Gamma_+$ has global symmetry
\be\label{eq:2dsyms}
SU(2)_L \times SU(2)_R \times SU(2)_I \times \prod_{I\in {\mathcal I}_G(\strich)} G_I \times \prod_{\alpha \in {\mathcal I}_F(\strich)} F_\alpha.
\ee
The subgroup $SU(2)_L \times SU(2)_R$ arises from the $SO(4)$ symmetry rotating the ${\mathbb R}^4$ plane transverse to the D3 branes within ${\mathbb R}^{1,5}$; $SU(2)_I$ is the R-symmetry of the small $(0,4)$ superconformal algebra governing the IR fixed point, which is also identified with the $Sp(1)$ R-symmetry of the 6d SCFT. Of course, for 6d SCFTs that have global $U(1)$ symmetries one can modify equation \eqref{eq:2dsyms} by adding a factor $\prod_a U(1)_a$ to the global symmetry of the model.\footnote{ Since $U(1)$ symmetries in F-theory are read off from the MW group of the elliptic fibration, they are a global property of the base; it has been suggested to one of us by J.J. Heckman that as a consequence they cannot give rise to gauge symmetries, but only to global ones. While this paper was in preparation \cite{Lee:2018ihr} appeared which contains a proof of this conjecture for models that can be coupled to gravity.} Both gauge and global symmetries of the 6d SCFT give rise to global symmetries for the 2d $(0,4)$ worldsheet SCFTs, but as we shall see below they are realized differently in the BPS string worldsheet theory: the 6d flavor symmetry is realized chirally, while the 6d gauge symmetry is not. \\

\noindent From this description of the BPS string sector of a given 6d SCFT, it follows that the subsectors with unit total BPS string charge \be \strich = \delta^{IJ} \Sigma_J \ee are universal, and can always be described as strings of rank-one models. To characterize these universal subsectors, therefore, it is enough to focus on the rank-one 6d SCFTs, whose salient features we have described above. The main topic of this paper is to probe the physics of rank-one 6d SCFTs exploiting their one-string BPS subsectors. Moreover, understanding the structure of these subsectors is a necessary first step towards the a general understanding of multi-string BPS sectors (though various results have already been obtained for various specific instances of higher rank SCFTs \cite{Haghighat:2013gba,Haghighat:2013tka,Hohenegger:2013ala,Gadde:2015tra,Kim:2015fxa,Kim:2016foj,Haghighat:2017vch,Kim:2018gak,Kim:2018gjo}).

\subsection{BPS strings of rank-one models and surface defects}\label{sec:defects}

\begin{table}
\begin{center}
\begin{tabular}{| c | p{8.0pc} | c |c|}
\hline
seven-brane & D3 probe & $\Delta$ & $H^1_G$\\
\hline
\hline
\phantom{$\Bigg|$} $I_n$ & $U(1)$ with $n$ charge-one hypers & 1 & $H^1_{SU(n)}$ \\
\hline
\phantom{$\Bigg|$} $I_n^*$ &  $SU(2)$ with $n+4$ fundamentals & 2 & $H^1_{SO(2n+8)}$\\
\hline
\phantom{$\Bigg|$} $ II $ & $(A_1,A_2)$ AD  & 6/5 & $H^1_{\varnothing}$ \\
\hline
\phantom{$\Bigg|$} $III$ & $(A_1,A_3)$ AD & 4/3 & $H^1_{SU(2)}$ \\
\hline
\phantom{$\Bigg|$} $IV$ & $(A_1,D_4)$ AD & 3/2& $H^1_{SU(3)}$\\
\hline
\phantom{$\Bigg|$} $IV^*$ & $E_6$ MN  & 3 & $H^1_{E_6}$ \\
\hline
\phantom{$\Bigg|$} $III^*$ & $E_7$ MN & 4 & $H^1_{E_7}$ \\
\hline
\phantom{$\Bigg|$} $II^*$ & $E_8$ MN & 6 & $H^1_{E_8}$\\
\hline
\end{tabular}

\vspace{1cm}

\caption{Rank-one 4d $\cn=2$ theories of D3-brane probes of F-theory seven-branes. Each entry of this table must be supplemented with a free center of mass hypermultiplet.}\label{tab:koda}
\end{center}
\end{table}

\noindent In this subsection we discuss the BPS string worldsheet theories from the perspective of geometric engineering in F-theory. We find that the BPS strings are obtained by reduction on $\PP^1$ of the 4d $\mathcal{N}=2$ worldvolume theory of an instanton of an 8d SYM, with the insertion of appropriate classes of surface defects. These defects are generalizations of the chiral defects studied by Martucci \cite{Martucci:2014ema}. More precisely, in eight dimensions for any simple, simply laced gauge group there there exists a corresponding type of seven-brane. The instantons on the seven-brane worldvolume are engineered by parallel stacks of D3-branes; for $A_n$ and $D_n$ gauge groups the worldvolume theories of the D3 branes are determined by the corresponding ADHM quivers, while for exceptional gauge groups they are given by the Minahan-Nemeschansky SCFTs \cite{Minahan:1996fg,Minahan:1996cj} and their higher rank generalizations \cite{Fayyazuddin:1998fb,Aharony:1998xz,Benini:2009gi}. While the chiral defects studied in \cite{Martucci:2014ema} are defined for $\cn=4$ SYM which is a Lagrangian theory, our chiral defects arise in $\mathcal{N}=2$ theories that generally do not admit a Lagrangian description, which happens for exceptional gauge groups. In these cases the complex structure of the elliptic fiber is frozen, hence these are not strictly speaking duality defects in the sense of \cite{Martucci:2014ema}. Nevertheless, these support chiral fermionic degrees of freedom, which are responsible for shifting the central charges of the IR CFT on the BPS string, as we will see below. In addition to these chiral defects, in order to realize the BPS strings of 6d theories with non-simply laced gauge group one is also led to consider a different class of `folding' defects, which implement the projection from a suitable simply-laced gauge group to the desired non-simply laced one. In this section we only discuss basic properties of these classes of defects which we infer from geometric engineering, while a more detailed study is left to future work \cite{DZLfuture}.\\ 

\noindent For rank-one models, the D3 brane configurations considered in section \ref{sec:BPSstringsuniv} simplify, as the BPS string charge lattice is one dimensional. We denote the corresponding base curve $\Sigma$. The global symmetry for the worldvolume theory of a stack of $Q$ D3 branes in a generic 6d (1,0) SCFT is
\be\label{eq:oneglobal}
SU(2)_L \times SU(2)_R \times SU(2)_I \times G \times \prod_{\alpha \in I_F} F_\alpha \times \prod_a U(1)_a.
\ee
The worldsheet theory describing a bound state of BPS strings with charge $\strich = Q \Sigma$ for a rank-one 6d SCFT arises from of a stack of $Q$ D3-branes wrapping the rational curve $\Sigma$. When $\Sigma$ is not part of the discriminant, one can think of the corresponding string worldsheet theory as a twisted compactification of the 4D $\cn =4$ $U(Q)$ SYM theory that lives on the D3-branes worldvolume \cite{Martucci:2014ema,Haghighat:2015ega,Lawrie:2016axq}. However, this is not the situation that one encounters for the generic rank-one 6d SCFT; rather, for a typical member of this class of theories $\Sigma$ is part of the discriminant, the sole exception being the E-string SCFT $\six{1}{Sp(0)}{E_8}$. \\

\noindent More precisely, the twisted compactification of the D3 branes' worldvolume theory leads to a 2d UV description of the strings with $\mathcal{N}=(0,4)$ supersymmetry whose properties we study in this section. This UV theory flows in the IR to the $\mathcal{N}=(0,4)$ CFT which is the subject of later sections.\newline

\begin{figure}
\begin{center}
\includegraphics[scale=1]{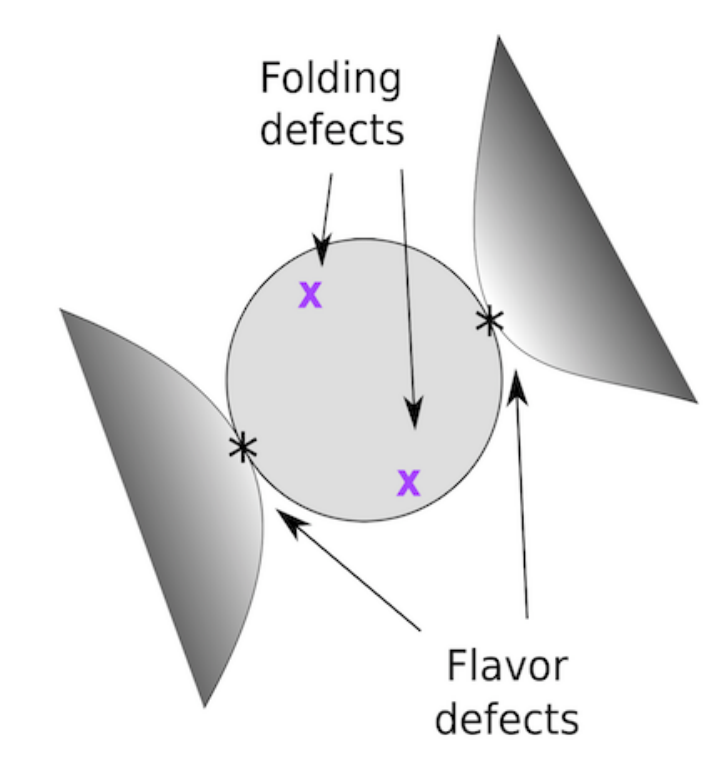}
\end{center}
\caption{Generic geometry for a rank-one F-theory base.}\label{fig:fgeneric}
\end{figure}

\noindent Consider a collection of F-theory seven-branes in flat space. The worldvolume theories of D3 branes that are placed on top of the seven-branes enjoy $\cn=2$ supersymmetry. For seven-brane stacks corresponding to Kodaira singularities of type $I_N$ and $I_N^*$ the worldvolume theories governing the low energy behavior of a stack of parallel $Q$ D3-branes are ADHM theories: for $I_N$ singularities we have the 4d $\cn=2$ theory
\be
H^Q_{SU(N)} \equiv U(Q) \oplus \Lambda^{\oplus N} 
\ee
where $\Lambda$ is the fundamental representation, while for $I_N^*$ singularities with $N\geq1$ we have the 4d $\cn=2$ theory
\be
H^Q_{SO(N+8)} \equiv Sp(Q) \oplus \Lambda^{\oplus N+8} \oplus \Lambda^2_\text{irr}
\ee
where $\Lambda^2_\text{irr}$ denotes the irreducible component of the antisymmetric representation of the $Sp(Q)$ gauge group. Notice that with the exception of the singularity $I_0^*$ the corresponding 4d $\cn=2$ theories are not conformal and often they are not asymptotically free either - this is the case for instance for a single D3-brane probe of the $I_N$ seven-brane stack, which is a $U(1)$ theory. For the exotic seven-branes of type $II,III,IV$ and $II^*,III^*,IV^*$ the corresponding $\cn=2$ models are strongly interacting, non-Lagrangian SCFTs, that we denote $H^Q_G$, which have flavor symmetry $SU(2)_L \times G$ where $G=\varnothing, SU(2),SU(3),E_6,E_7,E_8\equiv E_{n=0,1,2,6,7,8}$ respectively.\\

\noindent In considering six-dimensional F-theory backgrounds, by invoking an adiabatic approximation we can identify the worldsheet theory on the BPS strings with the twisted compactification on $\Sigma$ of precisely the 4D models governing the worldvolumes of D3-brane stacks, along the lines of \cite{Bershadsky:1995qy}. If $\Sigma$ belongs to the discriminant, the resulting worldvolume theory is the twisted compactification of one of the 4D $\cn=2$ theories we introduced above. We refer the reader to table \ref{tab:koda} for the list of the corresponding $\cn=2$ theories for the case of a single D3-brane, which is the case we address in detail in this paper. This strategy was exploited in \cite{DelZotto:2016pvm} for the theories $\six{3}{SU(3)}{}$, $\six{4}{SO(8)}{}$, $\six{6}{E_6}{}$, $\six{8}{E_7}{}$, and $\six{12}{E_8}{}$, building upon the results of \cite{Putrov:2015jpa}. The relevant twisted compactification in these cases is the $\beta$-twist on $\mathbb{R}^{1,1} \times {\mathbb P}^1$ by the $U(1)_R$ R-symmetry of the corresponding 4d $\cn=2$ SCFTs \cite{Kapustin:2006hi}.  In the language of \cite{Festuccia:2011ws,Dumitrescu:2012ha,Dumitrescu:2012at,Closset:2013vra}, one couples the 4d $\cn=2$ theories to a rigid supersymmetry background of the form $\mathbb{R}^{1,1} \times S^2$ or $T^2 \times S^2$ by switching on a unit monopole flux on the $S^2$ for the R-symmetry background gauge field \cite{Closset:2013sxa,Nishioka:2014zpa,Honda:2015yha,Benini:2015noa,Gadde:2015wta}. Notice that this is defined for theories with a $U(1)_R$ R-symmetry such that the corresponding R-charges are quantized over the integers. For a review, see e.g. \cite{DelZotto:2016pvm} and references therein.  Most of the theories in table \ref{tab:koda} are not superconformal and typically have an anomalous $U(1)_R$ symmetry. Nevertheless, in the geometric engineering picture one sees that each time such a non-conformal system of branes arises the compactification on $\Sigma$ requires the insertion of additional flavor divisors intersecting the D3 worldvolume along the directions transverse to $\Sigma$. On the one hand, these defects are responsible for compensating the $U(1)_R$ anomaly, thus rendering the background consistent; on the other hand, they introduce extra degrees of freedom which make the corresponding 2d $(0,4)$ models consistent. The models $\six{3}{SU(3)}{}$, $\six{4}{SO(8)}{}$, $\six{6}{E_6}{}$, $\six{8}{E_7}{}$, and $\six{12}{E_8}{}$ are special in that the 6d gauge anomaly polynomial coefficient multiplying $\text{ tr } F^4$ cancels automatically, with no need for extra matter; correspondingly, these theories have no flavor divisors. For all other theories, in addition to the gauge seven-branes wrapped on the base curve $\Sigma$ the background involves extra components of the discriminant that are noncompact and intersect $\Sigma\simeq S^2$, corresponding to flavor seven-branes. Schematically, the corresponding geometry is  the following:
\be
\begin{tabular}{c|cccccccccc}
IIB & \multicolumn{6}{c}{$\substack{\displaystyle{\,\,\mathbb{R}^{1,5}} \\ \vspace{-0.05in}  \\\overbrace{\hspace{1.2in}} \\  \vspace{-0.01in}}$}& \multicolumn{4}{c}{$\substack{\displaystyle{\,\,\widehat{B}} \\ \vspace{-0.05in} \\\overbrace{\hspace{.9in}} \\  \vspace{-0.01in}}$} \\
background & \multicolumn{4}{c}{$\substack{\displaystyle{\,\,\mathbb{C}^2_\parallel}\\\vspace{-.04in}\\\overbrace{\hspace{.9in}}}$} & \multicolumn{2}{c}{$\substack{\displaystyle{\,\,\mathbb{R}^{1,1}}\\\vspace{-.04in}\\\overbrace{\hspace{.4in}}}$} & \multicolumn{2}{c}{$\substack{\displaystyle{\,\,\Sigma}\\\vspace{-.04in}\\\overbrace{\hspace{.4in}}}$} & &\\  
 & 0 &1 & 2 & 3 & 4 & 5 & 6 & 7 & 8 & 9\\
\hline
seven-branes & X & X&  X & X & X & X & X & X & - & - \\
flavor seven-branes & X & X&  X & X & X & X & - & - & X & X \\ 
D3 branes & - & - & - & - & X & X & X & X & - & - \\
\end{tabular}
\ee
\noindent From the adiabatic perspective of the theory on the D3-brane, it is manifest that these additional flavor seven-branes behave like surface defects which preserve (0,4) supersymmetry. The insertion of these defects alters the $S^2$ metric on $\Sigma$ by introducing conical defects, which lead to an effective shift of the $U(1)_R$ symmetry of the model. This is a key feature that allows for a twisted compactification even for the 4D $\cn=2$ theories in Table \ref{tab:koda} that do not have an R-symmetry to begin with. We denote the corresponding systems as 
\be\label{eq:flavordefectnotation}
\left(H_G^Q\,,\,\mathfrak{D}^{R_{\mathfrak{D}}}_F\right)_{S^2} \,,
\ee
where $H_G^Q$ is the theory on the worldvolume of $Q$ D3-branes probing a stack of seven-branes of type $G$, while $\mathfrak{D}^{R_\mathfrak{D}}_F$ is a defect preserving 2d (0,4) supersymmetry, which supports a flavor symmetry $F$ and contributes an effective shift to the $R$-charge by $R_\mathfrak{D}$ units. The subscript $S^2$ on the right hand side of equation \eqref{eq:flavordefectnotation} is a reminder that these surface defects are inserted at points on the $S^2$ used for the compactification. In the absence of defects, the direction normal to the rational curve $\Sigma$ has the structure of the fiber of an $\mathcal O (-n)$ bundle, where
\begin{equation}\label{eq:DeltaG}
n = R_{G} = 2 \Delta_G,
\end{equation}
where $R_{G}$ is the R-charge of the Coulomb branch operator for the $H^1_G$ theories. Whenever the corresponding generalized chiral defects have a nonzero effective $R_\mathfrak{D}$ charge, this formula gets modified to
\be\label{eq:effectiveRshift}
n =  2 \Delta_G - R_\mathfrak{D}.
\ee
This allows one to determine the effective R-charge shift $R_\mathfrak{D}$ associated to a defect by inspection of the geometric engineering setup.\\

\begin{figure}
\begin{center}
\begin{tabular}{ccc}
\includegraphics[scale=0.5]{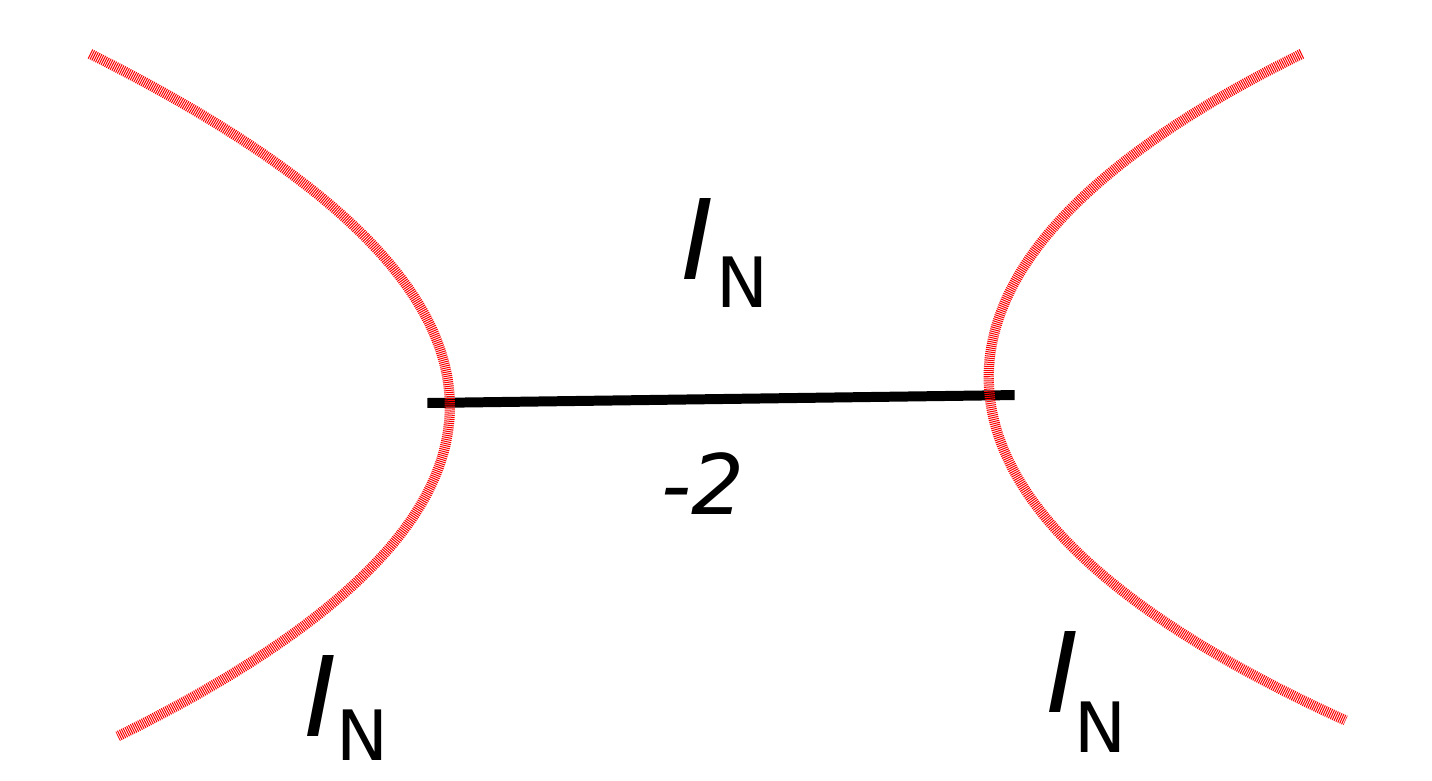}&$\qquad$&\includegraphics[scale=0.5]{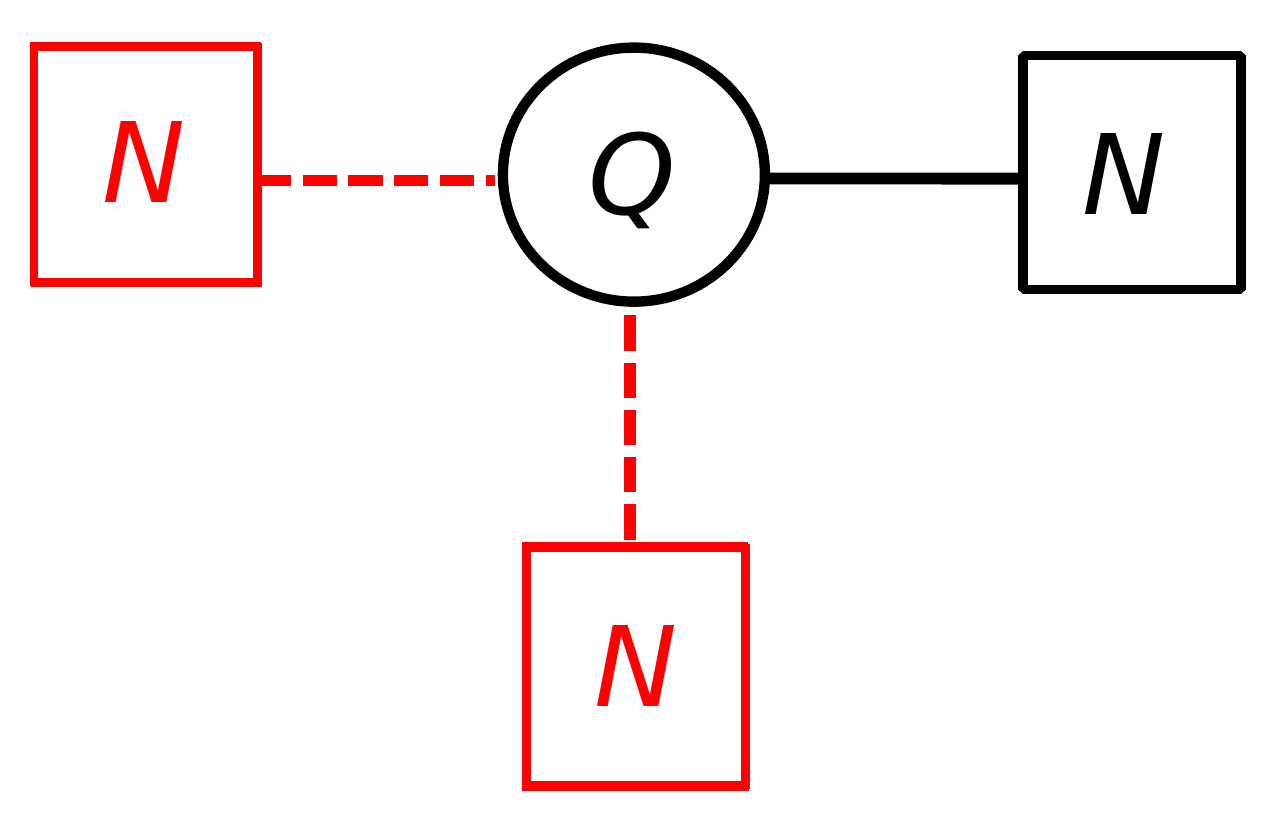}\\
\end{tabular}
\end{center}
\caption{\textsc{left:} Schematic description of the geometry of the F-theory model correspoding to the $\six{2}{SU(N)}{}$ theory. \textsc{right:} 2d quiver theory for a bound state of $Q$ BPS strings \cite{Haghighat:2013tka}.}\label{fig:SUN_example}
\end{figure}

\noindent As an example, let us consider the six-dimensional models $\six{2}{SU(N)}{}$ which involve seven-branes of $I_N$ type. The corresponding geometry is shown in figure \ref{fig:SUN_example}. Proceeding naively, reducing the 4d $\cn=2$ theories $H^Q_{SU(N)}$ on $S^2$ gives rise to 2d $(0,4)$ theories with field content analogous to that of the 4d theory, since the $S^2$ reduction gives a 2d $(0,4)$ vector multiplet (resp. hypermultiplet) for each 4d $\cn=2$ vector multiplet (resp. hypermultiplet). This is of course not the complete description of the worldsheet degrees of freedom. On the left of the figure one can see the F-theory geometry for the corresponding rank-one 6d tensor branch, which involves two extra noncompact flavor seven-branes of $I_N$ type, corresponding to generalized chiral defects that make the corresponding compactification consistent by contributing extra degrees of freedom. In other words, we have a compactification of the $H^Q_{SU(N)}$ theory with two defects:
\be
\left(H_{SU(N)}^Q, \mathfrak{D}^0_{SU(N)} \oplus \mathfrak{D}^0_{SU(N)}\right)_{S^2} \approx \left(H_{SU(N)}^Q, \mathfrak{D}^0_{SU(2N)}\right)_{S^2}
\ee
where on the right hand side we have emphasized the fact that the two defects can be brought on top of each other and merged, leading to a single defect supporting the larger flavor symmetry $SU(2N)$. Notice that for these models the effective $R$-charge shift $R_{\mathfrak{D}}$ is null, where to determine the effective $R$-charge shift $R_{\mathfrak{D}}$ we read off the dimension of $\Delta_{SU(N)} = 1$ from table \ref{tab:koda}, which implies that
\begin{equation}
n = 2 = 2 \Delta_G,
\end{equation}
so that $R_{\mathfrak{D}}=0$ in equation \eqref{eq:effectiveRshift}. This has a natural explanation: in this case, the effective shift is used up to compensate for the $U(1)_R$ anomaly.  The resulting two-dimensional $(0,4)$ theory admits a quiver description, which is shown on the right hand side of figure \ref{fig:SUN_example}. In the figure the degrees of freedom arising from the 4d $\cn=2$ theory $H^Q_{SU(N)}$ on $S^2$ are shown in black, while the degrees of freedom corresponding to the defect insertions are shown in red. Each $ \mathfrak{D}^0_{SU(N)}$ defect supports a $(0,4)$ bifundamental Fermi multiplet and therefore contributes chirally to the quiver theory. \\

\noindent As an example without a Lagrangian description, consider the models with gauge group $E_7$ and matter hypermultiplets in the $\tfrac12{\bf 56}$. We claim that the following configurations for the $H_{E_7}^Q$ theory involving defects,
\be\label{eq:flavordefectex}
\left(H_{E_7}^Q \,,\, \underbrace{\mathfrak{D}_{SO(1)}^{1} \oplus \cdots \oplus \mathfrak{D}_{SO(1)}^{1}}_{m \text{  times}}\right)_{S^2} \approx \left(H_{E_7}^Q,\mathfrak{D}_{SO(m)}^{m}\right)_{S^2},
\ee
give rise to the BPS strings for the entire class of 6d SCFTs
\be
\six{(8-m)}{E_7}{SO(\ell)} \qquad m = 0,1,...,7
\ee
whose gauge symmetry on the tensor branch is $E_7$.\\

\begin{figure}
\begin{center}
\begin{gather*}
\begin{gathered}
\xymatrix{
&&&& \bullet \ar@{-}[dl]\ar@{..>}@/^1.5pc/[dd]\\
\bullet \ar@{-}[r] &\cdots \ar@{-}[r] &\bullet \ar@{-}[r]& \bullet\\
&&&& \bullet \ar@{-}[ul]
}
\end{gathered}
\qquad  \longrightarrow\ 
\begin{gathered}
\xymatrix{ 
\bullet \ar@{-}[r] &\cdots \ar@{-}[r] &\bullet \ar@{-}[r]& \bullet\ar@{=>}[r]&\bullet}\end{gathered}\\
\qquad\\
\begin{gathered}
\xymatrix{& \bullet \ar@{-}[r] &\cdots \ar@{-}[r]& \bullet \ar@{-}[r] &\bullet\ar@{..>}@/^1.5pc/[dd]\\
 \bullet \ar@{-}[ur]\ar@{-}[dr] \\
&\bullet \ar@{-}[r] &\cdots \ar@{-}[r]& \bullet \ar@{-}[r] &\bullet
}
\end{gathered}
\qquad  \longrightarrow\ 
\begin{gathered}
\xymatrix{ 
\bullet \ar@{=>}[r] &\bullet \ar@{-}[r] &\cdots \ar@{-}[r] & \bullet\ar@{-}[r]&\bullet}\end{gathered}\\
\qquad\\
\begin{gathered}
\xymatrix{& \bullet\ar@{-}[dl]  \ar@{..>}@/^1pc/[d]\\
 \bullet \ar@{-}[r] &\bullet \\
&\bullet \ar@{-}[ul] \ar@{..>}@/_1pc/[u]}\end{gathered}
\qquad  \longrightarrow\ 
\begin{gathered}
\xymatrix{ 
\bullet \ar@3{->}[r] &\bullet
}
\end{gathered}\\
\\
\begin{gathered}
\xymatrix{
&& \bullet \ar@{-}[r] &\bullet\ar@{..>}@/^1.5pc/[dd]\\
\bullet \ar@{-}[r]& \bullet \ar@{-}[ur]\ar@{-}[dr] \\
&&\bullet \ar@{-}[r] &\bullet
}
\end{gathered}
\qquad  \longrightarrow\ 
\begin{gathered}
\xymatrix{ 
\bullet \ar@{-}[r] &\bullet \ar@{=>}[r] & \bullet\ar@{-}[r]&\bullet
}
\end{gathered}
\end{gather*}
\caption{Dynkin diagram foldings.}
\label{fig:foldynk}
\end{center}
\end{figure}
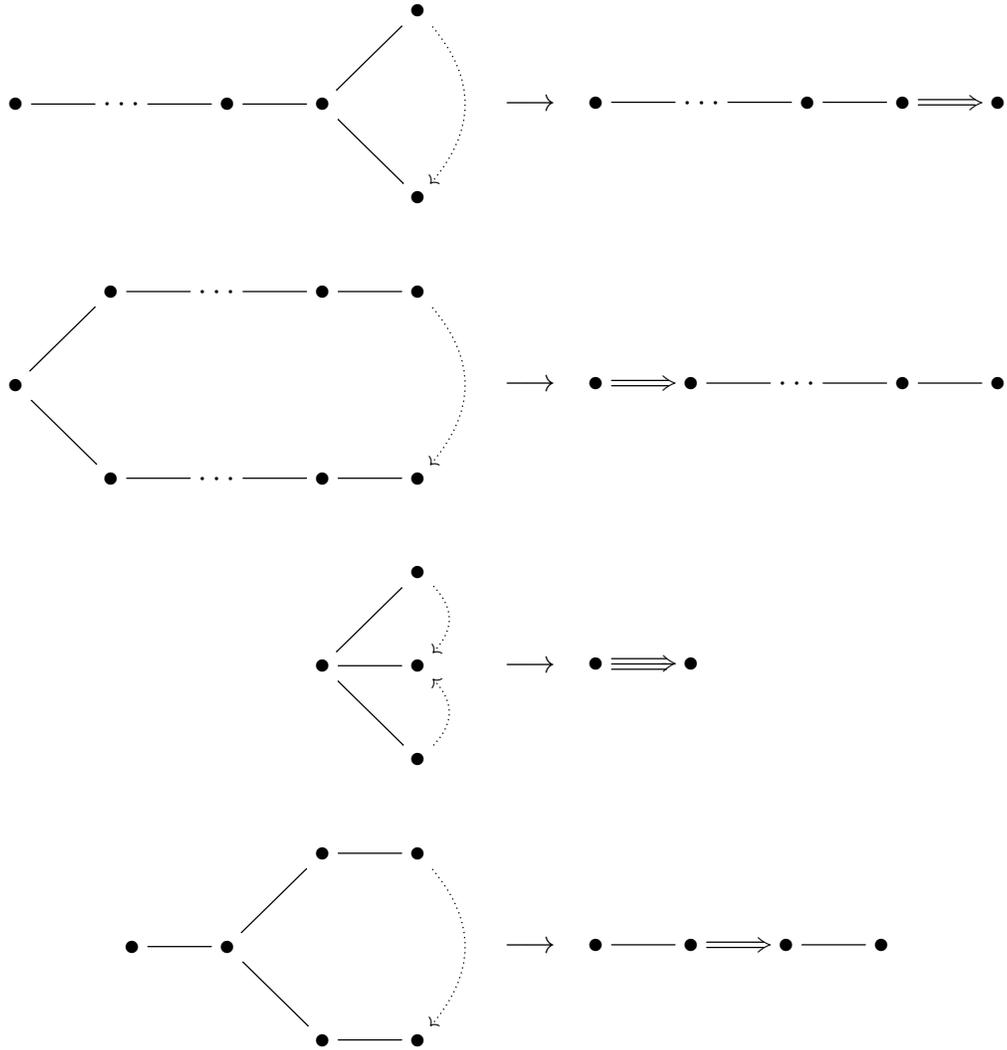

\noindent If an effective tensor branch gauge group is not simply laced one must include additional (0,4) defects that implement the relevant outer automorphism twist on the flavor symmetry (see figure \ref{fig:foldynk}), which we will refer to as folding defects. We denote backgrounds with folding defects as
\be
\left(H_G^Q,\mathfrak{F}_{\mathbb G}^{R_\mathfrak{F}}\right)_{S^2}
\ee
where $\mathbb G$ is the outer automorphism that the folding defect implements on the flavor symmetry of the model $H_G^Q$ and $R_\mathfrak{F}$ is the shift to the $U(1)_R$ charge induced by the defect. Crucially, the folding defects do not act as orbifold projections on the degrees of freedom of the BPS string, which would correspond to gauging the corresponding discrete symmetry. Rather, their role is to restrict the algebra of operators of the theory $H_G^Q$ on $S^2$ to a smaller closed consistent subalgebra in which only the non-simply laced flavor symmetry $G/\mathbb{G}$ is manifest.\\

\noindent A simple example of folding occurs for the BPS strings of the $\six{1}{Sp(N)}{SO(16+4N)}$ theories. In that context, we need to insert two folding defects to get from $SU(2N)$ to $Sp(N)$. Each folding defect produces an effective shift of the $R$-charge by $1/2$. The resulting models are given by
\be
\begin{aligned}
\two{1}{Sp(N)}&= \left(H_{SU(2N)}^Q,\mathfrak{F}_{\mathbb{Z}_2}^{1/2} \oplus \mathfrak{F}_{\mathbb{Z}_2}^{1/2} \oplus \mathfrak{D}^0_{SO(2N+8)}\oplus \mathfrak{D}^0_{SO(2N+8)} \right)_{S^2} \\
&= \left(H_{SU(2N)}^Q,\mathfrak{F}_{\mathbb{Z}_2}^{1/2} \oplus \mathfrak{F}_{\mathbb{Z}_2}^{1/2} \oplus \mathfrak{D}^0_{SO(4N+16)}\right)_{S^2}\,.
\end{aligned}
\ee
We draw the corresponding quiver theory in figure \ref{fig:spneng}(b). The insertion of the folding defects projects the $SU(2N)$ ADHM theory to the $Sp(N)$ one, which has a gauge group $O(Q)$ with an anti-symmetric multiplet as well as a bifundamental hypermultiplet carrying $Sp(N)$ flavor symmetry. The $\mathfrak{D}^0_{SO(4N+16)}$ flavor defect on the other hand supports an ortho-symplectic bifundamental $(0,4)$ Fermi multiplet, which couples to the $Sp(N)$ gauge symmetry.\\

\begin{figure}
\begin{center}
\begin{tabular}{ccc}
\includegraphics[scale=0.5]{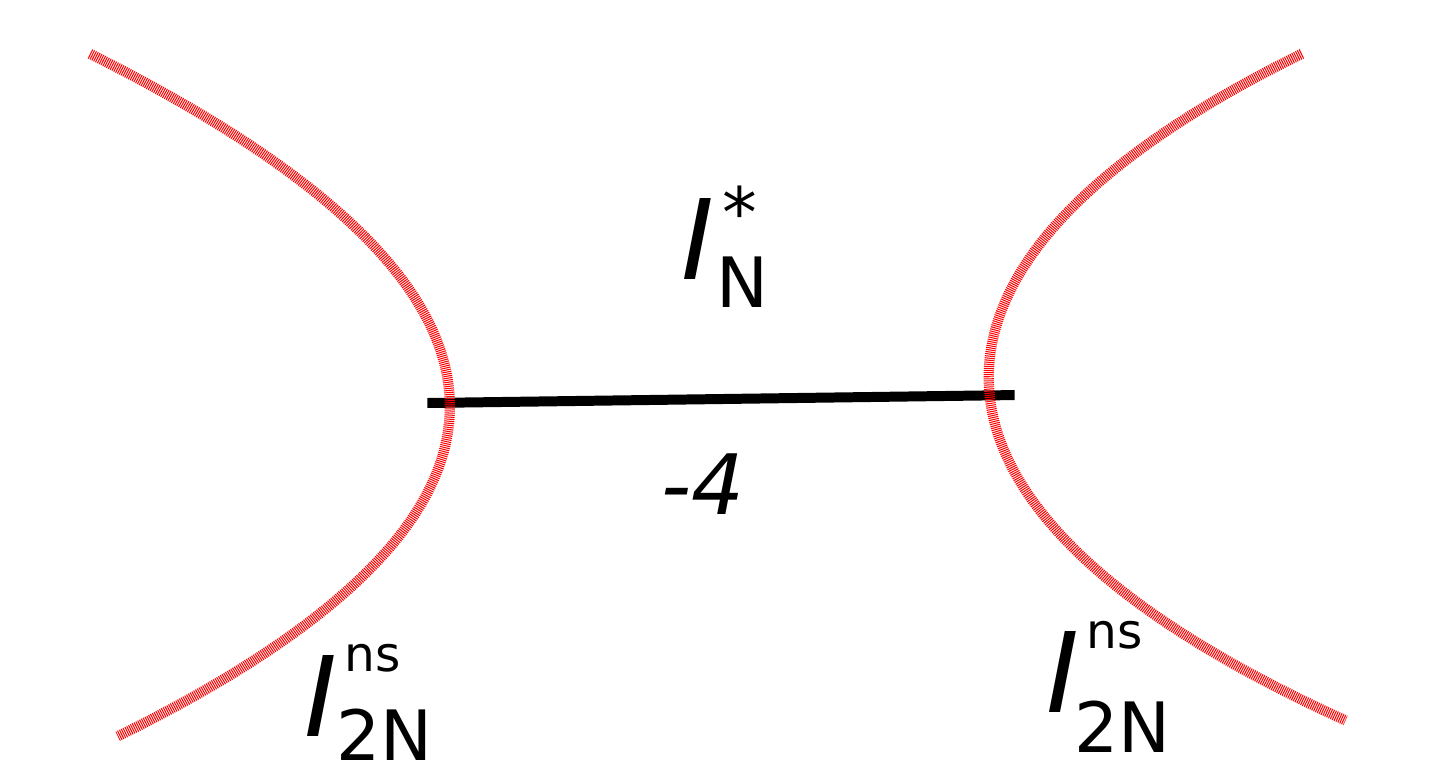}&$\qquad$&\includegraphics[scale=0.5]{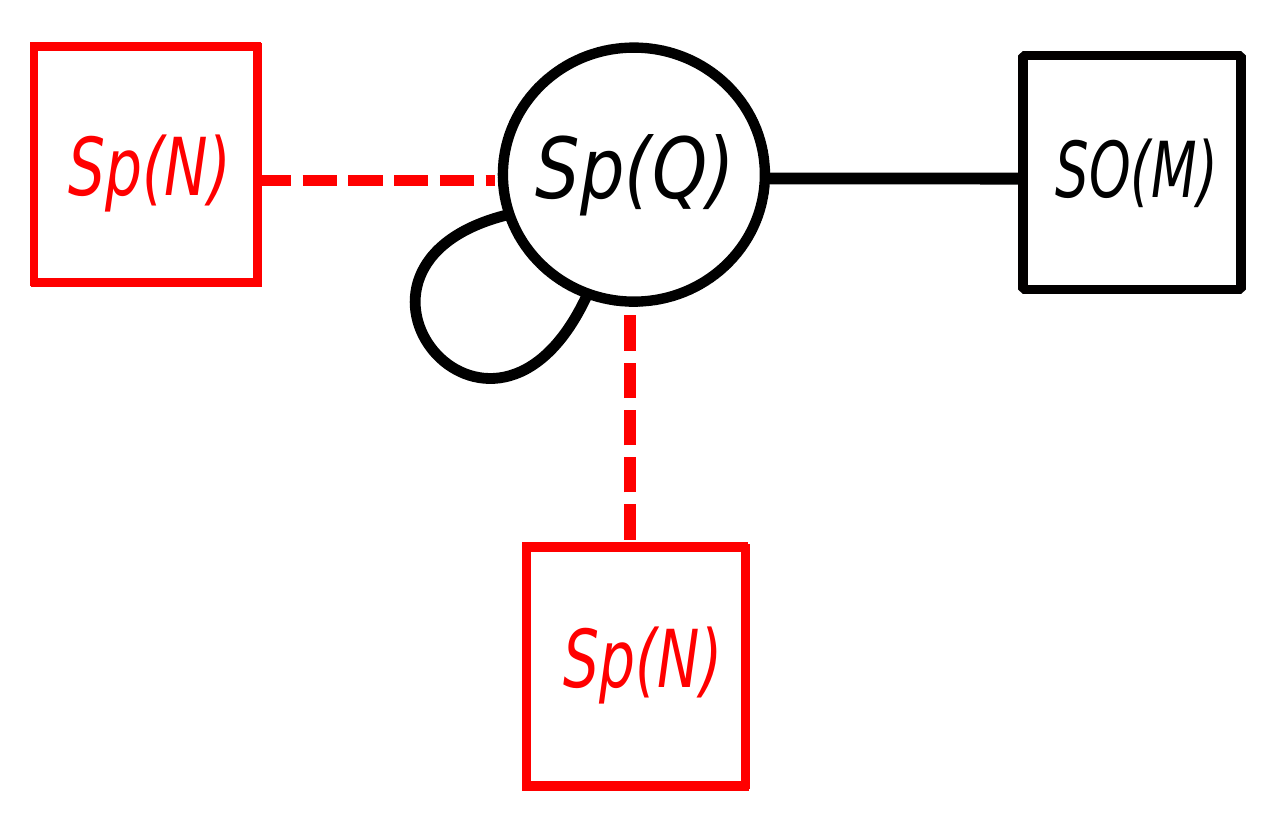}\\
\end{tabular}
\end{center}
\caption{\textsc{left:} Schematic description of the geometry of the F-theory model correspoding to the $\six{4}{SO(2N+8)}{}$ theory. \textsc{right:} 2d quiver theory for a bound state of $Q$ BPS strings, here $M=2N+8$ \cite{Gadde:2015tra}.}\label{fig:SON_example}
\end{figure}

\noindent One can similarly interpret the strings for the models $\six{4}{SO(8+2N)}{}$ as arising from compactifications of the theory $H^Q_{SO(8+2N)}$ on $S^2$ with $(0,4)$-defect insertions 
\be
\two{4}{SO(8+2N)} = \left(H^Q_{SO(8+2N)}\,,\,\mathfrak{D}^0_{Sp(N)}\oplus \mathfrak{D}^0_{Sp(N)}\right)_{S^2} \,.
\ee
We draw the resulting quiver in Figure \ref{fig:SON_example}. Again here we see that the 2d $(0,4)$ defects support Lagrangian degrees of freedom (bifundamental ortho-symplectic half-hypermultiplets). The $SO(8+2N)$ ADHM quiver of the theory without defects is clearly visible in black in the figure.\\

\noindent Next, we provide some additional examples that further clarify the nature of folding and generalized chiral defects.

\subsubsection{Folding defects: a detailed example}

\begin{figure}
\begin{center}
\includegraphics[scale=0.7]{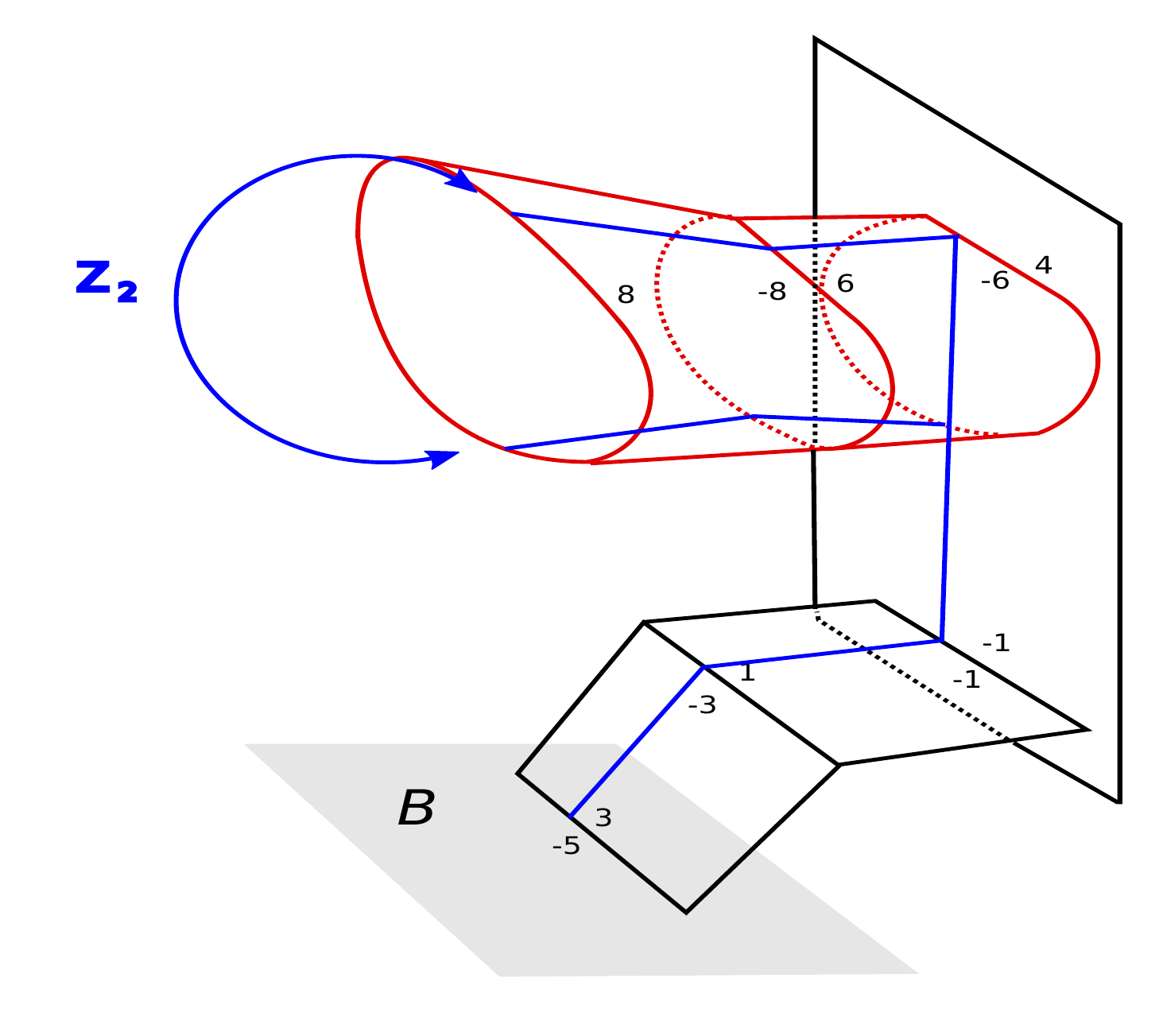}
\end{center}
\caption{Geometry of the $\six{5}{F_4}{}$ theory \cite{DelZotto:2017pti}.}\label{fig:5F4}
\end{figure}

\noindent The theory $\six{5}{F_4}{}$ is the only example of a 6d SCFT of rank one with non-simply laced group and no matter, and therefore involves folding defects but no generalized chiral defects. The BPS strings for this theory are realized starting from an $S^2$ compactification of the theory $H^Q_{E_6}$ with the insertion of folding defects corresponding to the $\mathbb{Z}_2$ outer automorphism
\be
\mathbb{Z}_2 \colon E_6 \to F_4.
\ee
We claim that the model
\be\label{eq:F4HE6}
\left(H_{E_6}^Q \, , \, \mathfrak{F}_{{\mathbb Z}_2}^{1/2}\oplus \mathfrak{F}_{{\mathbb Z}_2}^{1/2}\right)_{S^2}
\ee
\noindent corresponds to the worldsheet theory of $Q$ BPS strings of the $\six{5}{F_4}{}$ theory. The fact that two insertions are needed is manifest in the geometry  (see figure \ref{fig:5F4}), which has been studied in detail in \cite{DelZotto:2017pti}, (see also \cite{Esole:2017rgz}) and is due to the compactness of the gauge divisor in the fully resolved CY geometry. The generic fiber along the $-5$ curve in the F-theory base is indeed a fiber of type $IV^*$, which correspond to an exotic $E_6$ seven-brane stack. A stack of $Q$ D3 branes probing the seven-branes supports a generalized rank $Q$ MN theory of $E_6$ type, denoted $H^{\,Q}_{E_6}$, which has flavor symmetry $SU(2)_L\times E_6$. There are two points over the base $\mathbb P^1$ where two legs of the affine  $E_6$ Dynkin diagram `merge', which corresponds to the `folding' of the surfaces in red in Figure \ref{fig:5F4}. From the perspective of the $H^Q_{E_6}$ theory supported on the stack of D3-branes, these two points  on the $S^2$ are the locations of the folding defects. The $R$-charge of the folding defects is determined by the order of the monodromy cover equation in Table \ref{tab:sing}. From the table we see that there are only two options: a monodromy cover of order two implies an $R$-charge of $\tfrac12$, while a monodromy cover of order three (which occurs only in the $G_2$ case) implies an $R$-charge of $\tfrac13$. These considerations result in the 4d-2d system of equation \eqref{eq:F4HE6} giving rise to the BPS strings of the $\six{5}{F_4}{}$ theory.\newline

\begin{figure}
\begin{center}
$$
\begin{gathered}
\xymatrix{\six{6}{E_6}{} \ar[dd]_{\text{add matter}}&&\left(H_{E_6}^Q\, ,\, \varnothing \right)_{S^2} \ar[dd]^{\text{flavor defects}}\ar@/^6.5pc/[dddd]^{\text{folding defects}}\\
&&\\
\six{5}{E_6}{U(1)}\ar[dd]_{\text{Higgs}}&&\left(H_{E_6}^Q\, ,\, \mathfrak{D}^1_{U(1)} \right)_{S^2}\ar[dd]^{\text{Higgs}} \\
&&\\
\six{5}{F_4}{}&&\left(H_{E_6}^Q\, ,\, \mathfrak{F}^{1/2}_{{\mathbb Z}_2} \oplus  \mathfrak{F}^{1/2}_{{\mathbb Z}_2} \right)_{S^2}}\\
\end{gathered} 
$$ 
\caption{Higgsing tree and folding defects: the case of $\six{5}{F_4}{}$.}\label{fig:FoldingDefectsHiggs}
\end{center}
\end{figure}
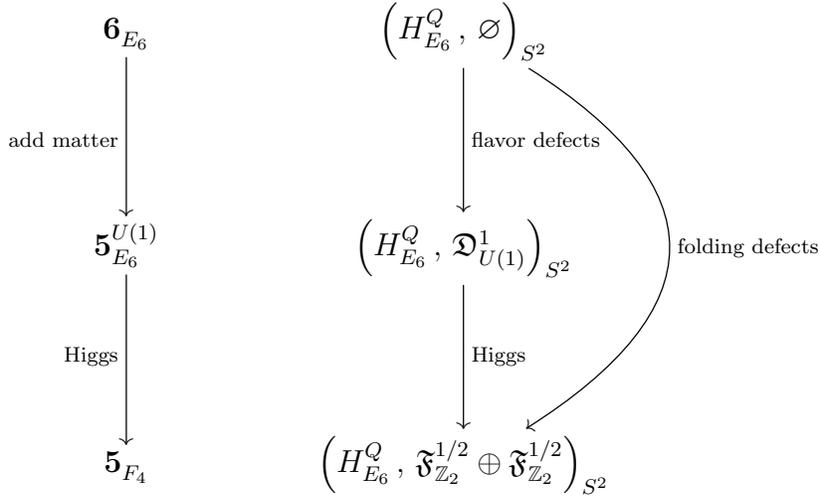

\noindent We remark that an alternative characterization for the $\six{5}{F_4}{}$ BPS strings worldsheet theories can be achieved by first adding generalized chiral defects to the theory \be \left( H^Q_{ E_6} \,,\,\varnothing \right)_{S^2}\ee and then Higgsing to the $\six{5}{G}{F}$ theory along the Higgsing three of equation \eqref{eq:Higgsingtree5}. We summarize this in Figure \ref{fig:FoldingDefectsHiggs}. From this perspective it is clear that the role of the folding defects is to project out degrees of freedom from the worldsheet 2d (0,4) SCFT.\\

\noindent The authors of \cite{Dey:2016qqp} have discussed in an example how one may directly obtain the Hilbert series of the moduli spaces of instantons for a non-simply laced gauge group $G/\mathbb{G}$ by projecting out states from the Hilbert series of $\mathcal{M}_{G,1}$. It would be interesting to see whether such an explicit procedure for folding can also be implemented at the level of the elliptic genera.\\

\subsubsection{Global anomalies and surface defects}
Six-dimensional theories may suffer from global anomalies that are the analogue of the famous Witten anomaly constraining the number of half-hypers in $SU(2)$ theories in four-dimensions \cite{Witten:1982fp}. These Bershadsky-Vafa anomalies occur for groups $SU(2)$, $SU(3)$ and $G_2$ only \cite{Bershadsky:1997sb}. In this section we are going to study the interplay of these anomalies with the structure of chiral defects in the $G_2$ case.\\ 

\begin{figure}
\begin{center}
\includegraphics[scale=1]{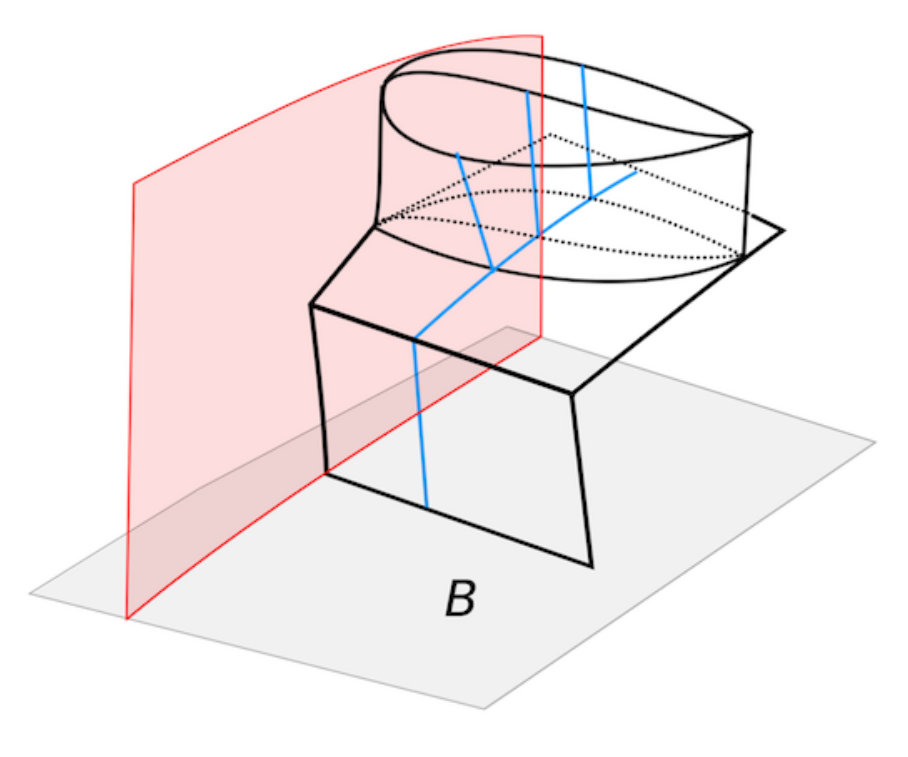}
\end{center}
\caption{Geometry corresponding to a non-split singularity for the model $\six{3}{G_2}{Sp(1)}$. We interpret the folding from $D_4$ to $G_2$ in this geometry as monodromy defects for the D3 branes; in addition to this, non compact flavor divisors give rise to matter in the ${\bf 7}$ of $G_2$ localized at the intersection with the compact divisor.}\label{fig:G2geom}
\end{figure}

\noindent The geometries of the $G_2$ models involve both flavor divisors and non-split singularities, and therefore the BPS strings are described by compactifications of D3 brane configurations that involve both flavor and folding surface defects. We claim that the models
\be
\left(H_{SO(8)}^Q \,,\,\mathfrak{F}_{{\mathbb Z}_3}^{1/3}\oplus \mathfrak{F}_{{\mathbb Z}_3}^{1/3}\oplus \mathfrak{D}_{Sp(1)}^{1/3} \oplus \underbrace{ \mathfrak{D}_{Sp(1)}^{1/3} \oplus \cdots \oplus \mathfrak{D}_{Sp(1)}^{1/3}}_{3m \text{  times}}\right)_{S^2} \qquad m=0,1,2 
\ee
give rise to the strings of the family of six-dimensional (1,0) theories
\be
\six{(3-m)}{G_2}{Sp(1+3m)} \qquad m=0,1,2.
\ee
In 6d the Bershadsky-Vafa global anomaly requires the matter to consist of either $1, 4,$ or $7$ fundamental hypermultiplets. From the point of view of the D3 brane probe, this corresponds to the fact that the R-charges must be quantized over the integers in order for the compactification to 2d to be consistent. The folding defects $\mathfrak{F}_{{\mathbb Z}_3}^{1/3}$ have fractional $R_{\mathfrak{F}}$ effective charges which are dictated by the order of the corresponding monodromy cover; this in turn forces the charges of the defects to be multiples of $1/3$ and their overall number to be equal to 1 mod 3. In the M-theory picture, the $G_2$ symmetry is realized in terms of a non-split Kodaira fiber, see figure \ref{fig:G2geom} (details about $G_2$ fibers can also be found in \cite{Esole:2017qeh}); the surface defects are located at the the trivalent points in the fibration of the $D_4$ singularity along the base ${\mathbb P}^1$. Only in the $G_2$ case one encounters a third order non-split fiber; in all other instances where the 6d gauge algebra is non simply-laced the order of the monodromy cover is $2$, and correspondingly the effective $R$-charges are half-integers.

\subsection{BPS string anomaly inflow}\label{sec:anominflow}
The aim of this section is to determine certain coefficients in the anomaly polynomial of the worldsheet theories of the BPS strings, which will be used in sections \ref{sec:1spn} and \ref{sec:univ}  to make various statements about the structure of their IR CFTs. In particular, we first focus on the anomalies associated to both abelian and non-abelian global symmetries, and then compute gravitational and R-symmetry anomaly coefficients in term of which the central charges of the IR CFT of the string are determined.\newline

\noindent In the presence of BPS strings the Bianchi identities for the six-dimensional anti-self-dual three-forms\footnote{ In the F-theory setup the tensor fields of the 6d theory arise from the reduction of $C_4^+$ and it is natural to adopt conventions in which they are anti-self-dual. From the perspective of the 6d field theory the opposite convention is most often chosen and the BPS strings are thus commonly referred to as self-dual.} which are the curvatures of the two-form fields in the corresponding tensor multiplets get modified \cite{Blum:1993yd}. It is convenient to group the three-form curvatures into a single vector, which we denote by
\be
\mathbf{H} = (H^1,\dots, H^R)\,,
\ee
where $H^I = d B^I$ and $R$ is the rank of the six-dimensional SCFT. The modified Bianchi identities read as follows:
\be\label{eq:BIANCHI}
d \,\mathbf{H} = \mathbf{X}_4 + \mathbf{J}_s
\ee
where $\mathbf{J}_s$ denotes the BPS string current. Setting $\mathbf{J}_s=0$ above, the GSSW mechanism \cite{Green:1984bx,Sagnotti:1992qw} entails that whenever the anomaly polynomial for the 6d theory factors as
\be
\anomal{8} = \tfrac12 \mathbf{X}_4 \cdot \mathbf{X}_4\,,
\ee
the gauge anomalies can be canceled. However, in case $\mathbf{J}_s \neq 0$ the contribution to the anomaly due to BPS strings needs to be canceled via an anomaly inflow mechanism. This was studied in \cite{Kim:2016foj,Shimizu:2016lbw} (see also \cite{Freed:1998tg,Henningson:2005hd,Kim:2012wc}) whose results we adapt to our discussion in what follows. The anomaly inflow mechanism implies that the anomalies of the 2d worldsheet theories of the BPS strings are given by
\be
\anomal{4} = - \int_{N_4} \Big(\mathbf{J}_s \cdot \mathbf{X}_4 + \tfrac12 \mathbf{J}_s \cdot \mathbf{J}_s\Big) \,,
\ee
where $N_4$ is a manifold transverse to the worldsheet of a BPS string within the worldvolume of the 6d SCFT. In particular, for a configuration with BPS string charge $\mathbf{S}$ one obtains that
\be
\int_{N_4} \mathbf{J}_s \cdot \mathbf{X}_4 = \mathbf{S} \cdot \mathbf{X}_4 \qquad\text{ and }\qquad \int_{N_4} \mathbf{J}_s \cdot \mathbf{J}_s = \mathbf{S}\cdot \mathbf{S} \, \chi_4 (N_4).
\ee
This implies that
\be\label{eq:anomal2d}
\anomal{4} = -  \tfrac12 \mathbf{S} \cdot \Big( 2 \, \mathbf{X}_4 + \mathbf{S} \, ( c_2(SU(2)_L) - c_2(SU(2)_R)) \Big)
\ee
where we have used the identity
\be
\chi_4(N_4) = c_2(SU(2)_L) - c_2(SU(2)_R)
\ee
for the Euler class of the $SO(4) = SU(2)_L \times SU(2)_R$ bundle $TN_4$. At this point, consider a bound state of BPS strings of charge 
\be
\strich = \sum_{I=1}^R Q^I \Sigma_I\,.
\ee
The anomaly polynomial of the corresponding 2d $(0,4)$ BPS string worldsheet theory is completely determined by inflow from the anomaly of the 6d $(1,0)$ theory. In particular, we have a contribution from
\be
\strich \cdot \mathbf{X}_4 = \sum_{I=1}^R Q^I \Sigma_I \cdot \left(\tfrac14 \mathbf{a} \, p_1(M_6) +  \sum_{\ell}\mathbf{b}_\ell c_2(G^\ell) + \tfrac12 \sum_{i \, j}\mathbf{b}_{ij} c_1(U(1)^i)c_1(U(1)^j)\right),
\ee
where we refer to section \ref{sec:anomf} for the notation. These intersections can be evaluated exploiting the equations in Section \ref{sec:anomf} and table \ref{tab:anomtab}.
For a rank-one model with gauge symmetry $G$ such as the ones we consider, 
\be \strich = Q\, \Sigma = Q\, \mathbf{b}_G \,, \ee and these equations simplify. One finds that the coefficients associated to the `t Hooft anomaly of $U(1)$ flavor symmetry factors are given by
\be
\strich \cdot \mathbf{b}_{ij} = Q \, \Sigma \cdot \mathbf{b}_{ij} = Q \sum_R n_{R,ij} A_R q^i q^j
\ee
while the coefficients for the non-abelian flavor symmetry `t Hooft anomalies are given by
\be
\strich \cdot \mathbf{b}_\alpha = Q \, \Sigma \cdot \mathbf{b}_\alpha = Q \sum_{R,R_\alpha} n_{R,R_\alpha} A_R A_{R_\alpha}\,.
\ee

\noindent Let us illustrate this more concretely by looking at some examples; we begin with the 6d theories involving abelian flavor symmetries. The models $\six{(6-k)}{E_6}{}$ always have a $U(1)$ factor in the flavor symmetry for $k\geq 1$. The $U(1)$ charges can be determined by exploiting the Higgsing
\be
\six{(6-k)}{E_7}{} \to \six{(6-k)}{E_6}{}
\ee
where $E_7$ is broken to $E_6$ leaving a global $U(1)$. The $U(1)$ charge for the $\mathbf{27}$ of $E_6$ is determined by the branching rules
\be
\mathbf{133} \to \mathbf{78}_0 \oplus \mathbf{27}_2 \oplus \mathbf{27}_{-2} \oplus \mathbf{1}_0 \qquad\qquad \mathbf{56} \to \mathbf{27}_1 \oplus \overline{\mathbf{27}}_{-1}
\ee
which gives a fundamental hyper in the $\mathbf{27}$ with $U(1)$ charge $1$ for each $\mathbf{56}$ which survives the Higgsing. This allows us to compute
\be
\Sigma  \cdot \mathbf{b}_{U(1)} = k \times 6 \times 1 \times 1 = 6 k \quad\text{for}\quad\six{(6-k)}{E_6}{}, \quad k\geq 1.
\ee
where we have used $A_{\mathbf{27}} = 6$ from Table \ref{tab:anomtab}. This is twice the coefficient which multiplies the term $c_1(U(1))^2$ in the anomaly polynomial of the 2d string.\newline

\noindent Similarly, for the $\six{(4-k)}{SO(10)}{}$ theories we determine the $U(1)$ charges by exploiting the Higgsing\\
\be
\six{(4-k)}{SO(12)}{}\to \six{(4-k)}{SO(10)}{}
\ee
where $SO(12)$ is broken to $SO(10)$ leaving a global $U(1)$. From the branching rules
\be
\mathbf{66} \to \mathbf{45}_0 \oplus \mathbf{10}_2 \oplus \mathbf{10}_{-2} \oplus \mathbf{1}_0 \qquad \mathbf{12} \to \mathbf{10}_0\oplus \mathbf{1}_2 \oplus \mathbf{1}_{-2} \qquad \mathbf{32} \to \mathbf{16}_{-1} \oplus \overline{\mathbf{16}}_{1}\,,
\ee
we obtain
\be
\Sigma  \cdot \mathbf{b}_{U(1)} = k \times 4 \times 1 \times 1  = 4k \quad\text{for}\quad \six{(4-k)}{SO(10)}{} \quad k\geq 1, 
\ee
using also the fact that the vectors of $SO(10)$ are neutral with respect to the $U(1)$.\newline

\noindent For the model $\six{1}{SU(N)}{}$, as remarked by \cite{Kim:2015fxa} only a mixed $U(1)$ survives from the flavor symmetry arising from the $U(1)\times U(1)$ factor which rotates the $N+8$ fundamental multiplets and the anti-symmetric multiplet. With respect to this mixed $U(1)$, we claim that the charges of the fundamental and anti-symmetric multiplets are given respectively by $4-N$ and $N+8$, which gives
\bea
\Sigma  \cdot \mathbf{b}_{U(1)} &= (N+8) \times 1 \times (4-N)^2 + 1 \times (N-2) \times (N+8)^2\\& = 2N(N-1)(N+8),
\eea
where we have used that $A_{\Lambda^2} = N-2$ from Table \ref{tab:anomtab}. We find that this assignment of charges is required in order for the CFT describing the BPS strings to be consistent with the results of section \ref{sec:univ} below.\\

\noindent To illustrate the calculation of anomaly coefficients for non-abelian flavor symmetries, consider the theories $\six{(8-k)}{E_7}{SO(k)}$ with $k\geq 3$. The $SO(k)$ anomaly coefficient receives contributions from the bifundamental $(\tfrac12 \mathbf{56},\mathbf{k})$, leading to the coefficient
\be
\Sigma \cdot \mathbf{b}_{SO(k)} = \frac{1}{2} \times 12 \times 2 = 12.
\ee
Similarly, for $\six{(5-k)}{F_4}{Sp(k)}$ the bifundamental matter $(\mathbf{26},\tfrac12\mathbf{2k})$ leads to the anomaly coefficient
\be
\Sigma \cdot \mathbf{b}_{SO(k)} =  6 \times \tfrac12 = 3.
\ee
The calculation for other 6d SCFTs proceeds analogously.\newline

\noindent In order to compute the anomalies of the BPS strings' worldsheet theory under the remaining global symmetries $SU(2)_L $ and $ SU(2)_R$ one can exploit the following identity:
\be
p_1(M_6) = p_1(N_4) + p_1(M_2)
\ee
where we have split the 6d $p_1(M_6)$ characteristic class in its component along the 2d worldsheet $M_2$ and its component along the four normal directions $N_4$. Then the following identity can be used to recover the coefficients of the global anomalies for the factors $SU(2)_L$ and $SU(2)_R$ of the global symmetry:
\be
p_1(N_4) = - 2 c_2(SU(2)_L) - 2 c_2(SU(2)_R).
\ee
\noindent Specializing the anomaly polynomial to rank one theories one obtains \cite{Kim:2016foj,Shimizu:2016lbw}
\begin{equation}
\begin{aligned}\label{eq:anpol}
\anomal{4} &= - \tfrac12 \mathbf{S} \cdot \Big( 2 \, \mathbf{X}_4 + \mathbf{S} \, \left( c_2(SU(2)_L) - c_2(SU(2)_R)\right) \Big)\\
&= - \tfrac12 \mathbf{S} \cdot \mathbf{S} \,\Big(c_2(SU(2)_L) - c_2(SU(2)_R)\Big)  \\
&\qquad - \tfrac14 \mathbf{S} \cdot \mathbf{a} \,\Big(p_1(M_2) - 2 c_2(SU(2)_L) - 2  c_2(SU(2)_R)\Big)\\
& \qquad - \mathbf{S} \cdot \mathbf{b}_{SU(2)_I} \, c_2(I) - \mathbf{S} \cdot \mathbf{b}_G \, c_2(G) \\
&\qquad - \mathbf{S} \cdot \mathbf{b}_{F_\alpha} \, c_2(F_{\alpha}) - \tfrac12 \sum_{i\, j} \mathbf{S} \cdot \mathbf{b}_{ij} \,c_1(U(1)^i)c_1(U(1)^j)
\end{aligned}
\end{equation}

\noindent As we have reviewed in section \ref{sec:anomf}, the F-theory geometric engineering requires to identify $\mathbf{a}$ with the class of $ - K$ in $H_2(B,\mathbb{Z})$. In this case a straightforward application of the Riemann-Roch formula gives
\be
- \Sigma \cdot (\Sigma + K) = 2 \quad \Rightarrow \quad  \Sigma \cdot \mathbf{a} = 2 + \Sigma \cdot \Sigma = 2 - n \quad \Rightarrow \quad \mathbf{S} \cdot \mathbf{a} = Q \, (2 - n).
\ee
A conjecture of \cite{Ohmori:2014kda} determines the $SU(2)_I$ anomaly coefficient as:
\be
- \mathbf{S} \cdot \mathbf{b}_{SU(2)_I} = Q h^\vee_{G}.
\ee
Finally, the gravitational anomaly $k$ of these models, which is a measure of the difference among the central charges $c_L$ and $c_R$, can be read off from the coefficient of $-\tfrac{1}{24} \,p_1(M_2)$. In our normalization, we obtain:
\begin{equation}
 k = c_R - c_L= 6\, Q \, (2 - n),
\end{equation}
while the coefficient of $c_2(SU(2)_I)$ is given by the 2d $SU(2)_I$ R-symmetry anomaly coefficient $k_R$, which is related to $c_R$ by a $(0,4)$ Ward identity, namely
\begin{equation}
c_R = 6 k_R.
\end{equation}
This determines the central charges of the 2d $\cn=(0,4)$ infrared CFT describing a bound state of $Q$ BPS strings of a rank-one 6d SCFT to be the following:
\begin{equation}\label{eq:central_charges}
\begin{tabular}{|cc|}
\hline 
$ c_R =  6 Q \, h^\vee_G\qquad$ & $c_L = c_R + 6 Q (2 - n )$ \\ & $\phantom{c_L\,\,}=6 Q (h^\vee_G - n + 2)$. \\
\hline
\end{tabular}
 \end{equation}

\noindent Notice that the unitarity of the left moving sector requires, in particular, that 
\begin{equation}
h^\vee_G + 2 \geq n.
\end{equation}
For $n>2$, this implies $h^\vee_G>0$. This is a manifestation from the BPS string perspective of the fact that such theories necessarily support a nontrivial gauge algebra \cite{Morrison:2012np}. Of course, equation \eqref{eq:central_charges} relies on the validity of the conjectural expression for the 6d (1,0) anomaly polynomial.

\subsection{An analogy with 4d UV curves}\label{sec:gaiotto}

\begin{figure}
\begin{center}
\includegraphics[scale=0.7]{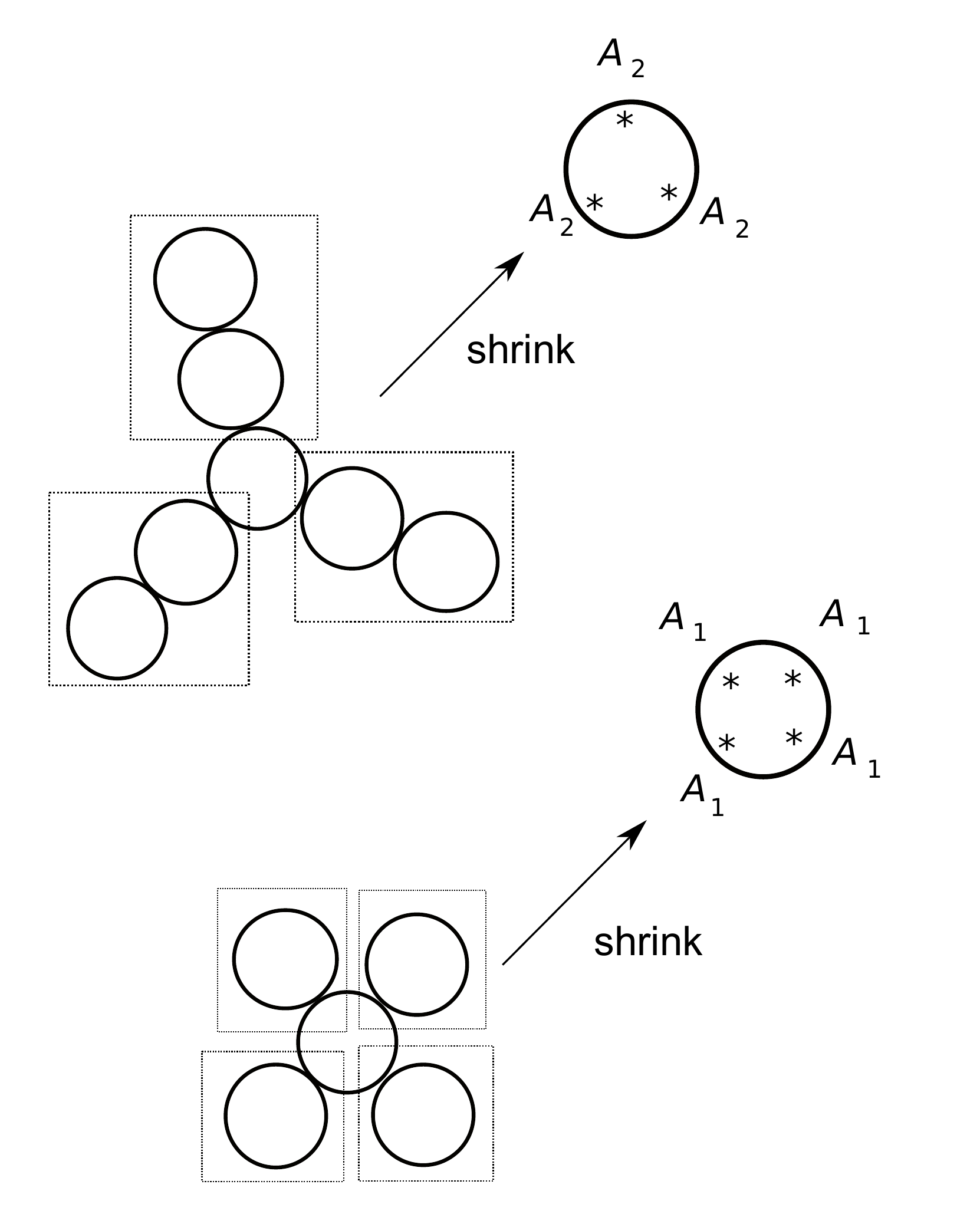}
\end{center}
\caption{Analogy with Gaiotto curves. In the top part of the figure, a Kodaira  fiber of type $IV^*$ is shrunk to a sphere with three punctures reminiscent of the Gaiotto curve of the $E_6$ Minahan-Nemeschansky theory. In the bottom part, the analogous procedure is shown for a fiber of type $I_0^*$, which shrinks to the four-punctured sphere reminiscent of the Gaiotto curve for the $\mathcal{N}=2$ SCFT with $G=SU(2)$ and $N_f=4$.}
\label{fig:gaiottocurve}
\end{figure}

At low energies the worldvolume theory of a stack of $N$ D3-branes is described by $U(N)$ $\cn=4$ SYM theory. If we view F-theory as the limit of M-theory on an elliptically fibered Calabi-Yau where the volume of the elliptic fiber is sent to zero, the D3-branes wrapping a divisor in the base of the Calabi-Yau arise from M5-branes wrapping a divisor of the Calabi-Yau $X$ which also includes the elliptic fiber. Consistent with this, the coupling $\tau$ of the theory governing the D3 branes is identified with the complex structure $\tau_E$ of the elliptic fiber. This was discussed in the context of the E-string e.g. in references \cite{Douglas:1996xp,Minahan:1998vr}, where it was found that the E-string arises as a compactification of the $\cn=4$ SYM in presence of a network of chiral defects, encoded in the geometry of the elliptic fibration of a del Pezzo dP$_9$ surface, which can be realized as a $\PP^2$ surface blown up at 9 points.\\

\noindent If on the other hand a stack of M5 branes is compactified on a curve ${\mathcal C}$ with appropriate boundary conditions at punctures on the curve, it is well known that one obtains a $\mathcal{N} = 2$ SQFT whose properties are encoded by the topology of ${\mathcal C}$ and by the boundary conditions at the punctures \cite{Gaiotto:2009we,Gaiotto:2009hg}. Moreover, the punctures can be viewed as originating from collections of transverse M5 branes intersecting the ${\mathcal C}$.  In the IR Coulomb phase, the stack of M5 branes wrapping the curve $\mathcal C$ recombines to a single M5 wrapping the Seiberg-Witten curve $\Sigma \subset T^* {\mathcal C}$.\\

\noindent For the rank-one 6d (1,0) theories without matter $\six{4}{SO(8)}{},$ $\six{6}{E_6}{},$ $\six{8}{E_7}{},$ and $\six{12}{E_8}{}$, we find it worthwhile to point out that the D3 branes are described by $\mathcal{N}=2$ SCFTs whose UV curve $\mathcal{C}$ resembles a limit of the exceptional Kodaira fiber associated to the two-cycle on which the D3 branes are wrapped, as illustrated in figure \ref{fig:gaiottocurve}.\\

\noindent Let us see how this works out explicitly building upon the analysis in  \cite{Morrison:2016nrt} for the various 6d $(1,0)$ SCFTs without matter. In the $\six{4}{SO(8)}{}$ case, the $I^*_0$ fiber can be mapped to a sphere with 4 $A_1$ singularities; correspondingly, the theory on a wrapped D3 brane is the 4d $\mathcal{N}=2$ CFT with $SU(2)$ gauge group and $N_f=4$, which has a Gaiotto curve of class S$[A_1]$ given by a sphere with 4 punctures, each carrying an $SU(2)$ global symmetry (cfr. Figure \ref{fig:gaiottocurve} (bottom)). In the $\six{6}{E_6}{}$ case, the Kodaira type $IV^*$ fiber depicted in Figure \ref{fig:gaiottocurve} (top) degenerates to a ${\mathbb P}^1$ with three $A_2$ singularities, consistent with the fact that the D3 brane is described by the $E_6$ Minahan-Nemeschansky theory, which is the class S$[A_2]$ theory associated to a sphere with 3 maximal punctures each carrying a flavor symmetry of type $SU(3)$. In the $\six{8}{E_7}{}$ case, the Kodaira type $III^*$ fiber degenerates to a ${\mathbb P}^1$ with singularities of type $A_3$, $A_3$ and $A_1$, while the worldvolume theory of the D3 brane is the $E_7$ MN theory, which is identified with a Sicilian theory of class S$[A_4]$ that has two maximal puncturs with symmetry $SU(4)$ and one puncture of a different kind supporting $SU(2)$ global symmetry. Finally, in the $\six{12}{E_8}{}$ case the Kodaira type $II^*$ fiber degenerates to a ${\mathbb P}^1$ with singularities of type $A_5$, $A_2$ and $A_1$, and correspondingly the D3 brane worldvolume theory is the $E_8$ MN theory, which can be realized as a Sicilian theory of class S$[A_5]$ with three punctures with global symmetries $SU(2), SU(3)$ and $SU(5)$ respectively.\\

\section{Elliptic genera and 6d $T^2\times\mathbb{R}^4$ partition functions}\label{sec:ellgen}
By subjecting a (1,0) SCFT to the 6d $\Omega$-background $T^2\times \mathbb{R}^4_{\epsilon_1,\epsilon_2}$ \cite{Losev:2003py,Hollowood:2003cv} one can compute a partition function that receives two distinct kinds of contributions arising from:
\begin{enumerate}
\item Towers of BPS particles arising from the 6d tensor, vector, and hypermultiplets, with KK momentum along the 6d circle, and
\item BPS strings wrapped around the 6d circle, which couple to the various tensor multiplets of the 6d SCFT.
\end{enumerate}
These two classes of contributions organize themselves as follows
\begin{equation}\label{eq:z6d}
Z^{6d}_{T^2\times\mathbb{R}^4}(\vec\varphi,\mathbf{m},x,v,q) = Z_{\text{pert}}(\mass,x,v,q)\times \bigg(\sum_{\,\,\,\strich \in \Gamma^+} e^{-\vec\Phi \, \cdot \, \strich} \, \mathbb{E}_{\,\strich}(\mass,x,v,q)\bigg).
\end{equation}
The BPS particles contribute to the factor $Z_{\text{pert}}$ , while the strings contribute via their elliptic genera to the remaining factor\cite{Haghighat:2013gba}. More precisely, $\mathbb{E}_{\,\strich}$ is the elliptic genus of the 2d $\mathcal{N}=(0,4)$ theory that describes a bound state of strings with charge $\strich \in \Gamma^+$. In particular, 
\be
\mathbb{E}_{\bf 0} = 1.
\ee
\noindent Let us define
\begin{equation}
\vec\varphi = \vec\phi+ i \int_{T^2} \vec B,
\end{equation}
where $\vec\phi$ are the vevs of the scalar fields in the tensor multiplets, while the imaginary components are fluxes on the torus of the two-form fields belonging to the tensor multiplets. If the $I$-th tensor multiplet is coupled to a gauge multiplet transforming under the gauge algebra $G_I$, the Yang-Mills coupling for $G_I$ is identified with $\phi_I$ as follows:
\begin{equation}
\phi_I = \frac{8\pi^2}{g_{G_I}^2}.
\end{equation}
The chemical potentials $\vec\Phi$ that appear on the right hand side of equation \eqref{eq:z6d} and couple to the BPS string charges are identified with $\vec\varphi$, up to shifts by linear combinations of the parameters $\tau$ and $\mass$ (see e.g. \cite{Haghighat:2014vxa,Hayashi:2017jze}).\\

\noindent The partition function depends  on several further parameters: on the fugacities $\mass$ for the flavor and gauge symmetries of the 6d SCFT, on $q= e^{2\pi i \tau}$, where $\tau$ is the complex modulus of $T^2$, and on the $\Omega$-background parameters $x=e^{2\pi i \epsilon_-},v=e^{2\pi i \epsilon_+}$, which couple respectively to the Cartan generator of $SU(2)_L$ and to the linear combination $J^3_R+J^3_I$ of the Cartan generators $J^3_R$ and $J^3_I$ of $SU(2)_R$ and $SU(2)_I$, where $SU(2)_L\times SU(2)_R = SO(4)$ is the rotation group of $\mathbb{R}^4$ and $SU(2)_I$ is the R-symmetry group of the 6d (1,0) SCFT.\newline

\noindent We focus on the single-string sector of the BPS spectrum of any 6d SCFT, for which $\strich = (0,\dots,1,\dots,0)$. The theory that describes a single string coupled to the $I$-th tensor multiplet depends on the following data: the Dirac self-pairing of the string $n_I$, the gauge group $G_I$ (which couples via a Green-Schwarz term to the $I$-th tensor multiplet), and the effective flavor symmetry group for the string, which was discussed in section \ref{sec:BPSstringsuniv}. The catalogue of theories of a single BPS string mirrors the list of rank-one 6d SCFTs of section \ref{sec:higgsingchains}, and Higgsing the 6d (1,0) SCFT corresponds from the perspective of its BPS string to introducing a deformation that triggers an RG flow to the CFT of the BPS string of the Higgsed 6d SCFT. As discussed in section \ref{sec:higgsingchains}, most of the allowed combinations of $n_I$ and $G_I$ allow for a single choice of flavor group $F$, the exceptions being 
\be
(n,G) = (1,SU(6))
\ee
for which $F$ is either $SU(14)\times U(1)$ or $SU(15)$,
\be
(n,G)=(2,SO(12))
\ee
for which $F$ is either $SO(2)$ or $SO(1)\times SO(1)$, and 
\be
(n,G)=(1,SO(12))
\ee
for which $F$ is either $SO(3)$ or possibly $SO(2)\times SO(1)$; see section \ref{sec:wzwunivii} for further details on these theories.\newline

\noindent For a single string, we can split the fugacities $\mass$ into two sets: a first set, $\mass_{G_I}$, which couples to the Cartan of the 6d gauge algebra, and a second set, $\mass_{F_I}$, which couples to the Cartan of the flavor symmetry group.\footnote{ We collect basic facts of Lie algebras and their representations, as well as our notational conventions, in appendix \ref{sec:finit}.} The Ramond-Ramond elliptic genus of the string associated to the $I$-th tensor multiplet is defined as the following trace over the Hilbert space of the theory:
\begin{align}
\mathbb{E}_{\, \Sigma_I}(&\mass,x,v,q)=\nonumber\\
&Tr_{RR}(-1)^{F_L+F_R}e^{2\pi i \tau H_L}e^{2\pi i \bar\tau H_R}x^{J^3_L}v^{J^3_R+J^3_I}\prod_{j=1}^{\text{rank}(F_I)}(m^j_{F_I})^{K_{F_I}^j}\prod_{j=1}^{\text{rank}(G_I)}(m^j_{G_I})^{K_{G_I}^j}.
\end{align}
where $K_{F_I}^j,K_{G_I}^j$ are respectively Cartan generators for the Lie algebras $F_I$ and $G_I$, and for a Lie group $G$ the chemical potential $\mathbf{\mu}_G$, which takes values in the complexification of the dual of the Cartan of $G$, is related to the exponentiated fugacity $\mathbf{m}_G$ as follows:
\begin{equation}
\mathbf{m}_G = e^{2\pi i \mathbf{\mu}_G}.
\end{equation}
In explicit computations, we find it convenient to pick a specific basis in which to expand these chemical potentials. The choices we make for various groups are explained in section \ref{sec:finit}.\newline

\noindent The elliptic genus of the theories under consideration consists of two factors: 
\begin{equation}
\mathbb{E}_{\, \Sigma_I}(\mass,x,v,q) = \mathbb{E}_{c.m.}(x,v,q) \mathbb{E}_{n_I}^{G_I}(\mass_{G_I},\mass_{F_I},v,q).
\end{equation}
The first factor captures the degrees of freedom of the string associated to its motion in $\mathbb{R}^4_{\epsilon_1,\epsilon_2}$ and is given by\footnote{ We refer to appendix \ref{sec:appmod} for our notation for modular and Jacobi forms.}
\begin{equation}
\mathbb{E}_{c.m.}(x,v,q) = \frac{\eta(q)^2}{\theta_1(v\, x,q)\theta_1(v/x,q)}.
\end{equation}
The second factor in the elliptic genus captures the `internal' dynamics of the string that describe its propagation on the reduced moduli space $\widetilde{\mathcal{M}}_{G_I,1}$ of one $G_I$ instanton. For a single string, this component turns out to be independent of $x$ but does depend on $v$. \newline

\noindent The Ramond boundary conditions on the left movers are the relevant ones for computing the 6d Nekrasov partition function, but we will also find it useful to consider the elliptic genus with Neveu-Schwarz boundary conditions on the left-movers; we will denote the latter elliptic genus by
\begin{equation}
\mathcal{E}_{n_I}^{G_I}(\mass_{G_I},\mass_{F_I},v,q).
\end{equation}\\

\noindent The theory of a single BPS string flows in the infrared to an interacting fixed point described by a $\mathcal{N}=(0,4)$ CFT which we denote by \hh{n_I}{G_I}. At the time of writing of this paper, the elliptic genera for a number of BPS string worldsheet theories have been computed, but a sizable number of cases remains for which the elliptic genera are not known. Our approach in the next sections will be to analyze the elliptic genera of the BPS string worldsheet theories from the perspective of the infrared CFT and to identify features shared by the various CFTs. In appendix \ref{sec:catalogue} we provide a summary of the results available in the literature on which we rely in the coming sections.

\section{A motivating example: strings of $D$-type conformal matter SCFTs}\label{sec:1spn}

In this section we study in some detail the \hh{1}{Sp(N)} CFTs describing the strings of the $\six{1}{Sp(N)}{}$ 6d SCFTs (which correspond to conformal matter of $(D_{4+N}, D_{4+N})$ type). The \hh{1}{Sp(N)} CFTs are particularly simple, which allows for a very detailed analysis. This brings to light several interesting properties of this class of CFTs, which in section \ref{sec:univ} we will argue are also shared by other \hh{n}{G} theories. We begin in section \ref{sec:spneng} by reviewing the brane construction of the $\six{1}{Sp(N)}{}$ SCFTs and give explicit formulas for elliptic genera of the \hh{1}{Sp(N)} CFTs; in section \ref{sec:spnwzw} we argue that the 6d $SO(16+4N)$ flavor symmetry is realized in the CFT as a level 1 current algebra; in section \ref{sec:spngv} we discuss how the 6d gauge symmetry $G$ is realized in the string CFT, and find evidence that its contribution to the elliptic genus is captured by irreducible characters of the affine Kac-Moody algebra for $Sp(N)$ at level $-1$; finally, in section \ref{sec:spnns} we discuss the NS-R elliptic genus of these theories and argue for the existence of a spectral flow that relates it to the R-R elliptic genus; we also discuss the spectrum of operators of low conformal weight counted by the NS-R elliptic genus.

\subsection{Brane engineering and elliptic genus}\label{sec:spneng}
\begin{figure}[t]
\begin{center}
\subfloat[\phantom{x}]{\includegraphics[width=0.45\textwidth]{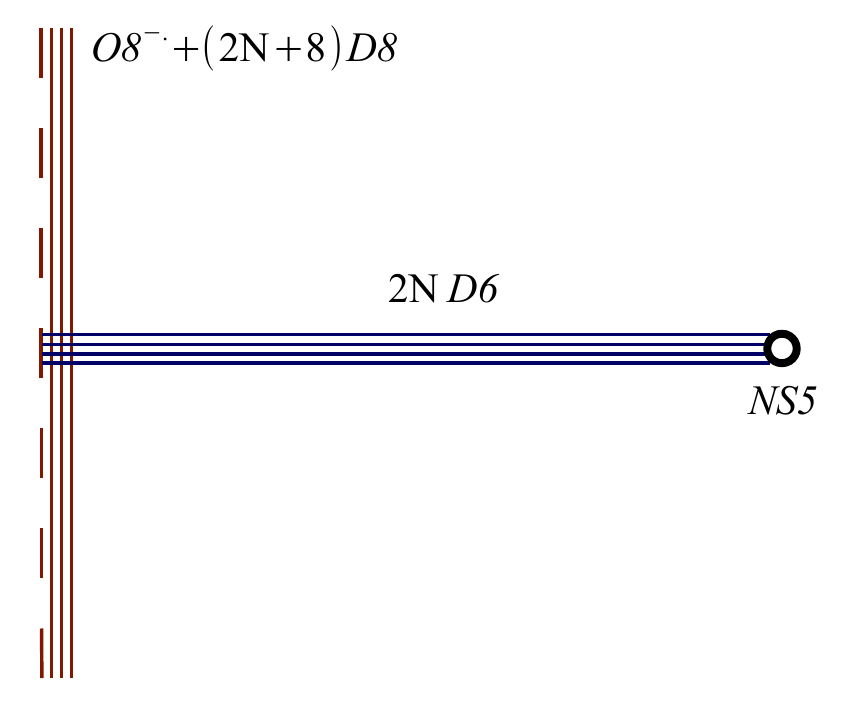}}
\subfloat[\phantom{x}]{\includegraphics[width=0.35\textwidth]{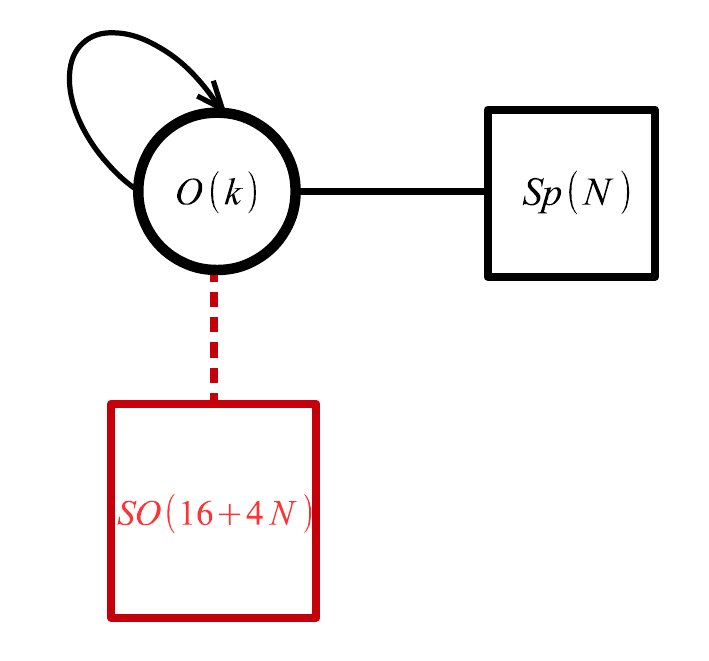}}
\caption{(a): Type IIA brane engineering of the $\six{1}{Sp(N)}{}$ 6d SCFTs, and (b): 2d (0,4) quiver theory for $k$ BPS strings of the SCFT. The dashed and continuous lines symbolize respectively bifundamental Fermi and hypermultiplets; the gauge node supports an $O(k)$ adjoint vector multiplet and a hypermultiplet in the symmetric representation of $O(k)$ represented by the arrow.}
\label{fig:spneng}
\end{center}
\end{figure}

Recall that six-dimensional (1,0) SCFTs with orthogonal or symplectic groups can be engineered by means of massive Type IIA brane constructions involving orientifold planes \cite{Brunner:1997gf,Hanany:1997gh}. In particular, the SCFT of interest to us can be engineered from the brane configuration of figure \ref{fig:spneng}(a), which involves one $O8^-$ orientifold plane and $(8+2N)$ D8 branes, as well as a NS5 brane and $N$ D6 branes stretching between the NS5 brane and its mirror under the orientifold. This brane engineering is closely related to the ones considered in \cite{Kim:2014dza,Yun:2016yzw}. The strings are realized in terms of D2 branes suspended between the NS5 brane and its mirror image (not shown in the figure).\\

\noindent The theories for $k$ BPS strings with $G=Sp(N)$ admit a UV Lagrangian description that can be read off from the brane construction presented above; the Lagrangian theory involves (0,4) multiplets coupled to a $O(k)$ gauge group \cite{Kim:2014dza,Yun:2016yzw}. This 2d gauge theory is summarized by the quiver of figure \ref{fig:spneng}(b). The theory \hh{1}{Sp(N)} of a single string in particular is extremely simple: the gauge group is just $O(1) = \mathbb{Z}_2$, which does not have an adjustable gauge coupling, so the Lagrangian realization describes directly the BPS string CFT. The spectrum of the \hh{1}{Sp(N)} theory, expressed in terms of $\mathcal{N}=(0,4)$ consists of $4+N$ Fermi multiplets, transforming under the vector representation of $SO(16+4N)$, and $N$ hypermultiplets transforming under the $2N$-dimensional representation of $Sp(N)$. All multiplets are charged under the $O(1)$ gauge group. From the CFT point of view, the $O(1)$ gauge symmetry can be viewed as implementing a GSO projection.\newline

\noindent The elliptic genus of the \hh{1}{Sp(N)} theory is given by\cite{Yun:2016yzw}:
\begin{align}
&\mathbb{E}_1^{Sp(N)}(\mass_{Sp(N)},\mass_{SO(16+4N)},v,q) =\nonumber\\
&\frac{1}{2}\sum_{a\in\{0,\frac{1}{2},\frac{\tau}{2},\frac{1+\tau}{2}\}}\left(\prod_{i=1}^{8+2N}\frac{\th_1(e^{2\pi i a}m^i_{SO(16+4N)},q)}{\eta(q)}\right)\left(\prod_{i=1}^{N}\frac{\eta(q)^2}{\th_1(e^{2\pi i a}v\,m^i_{Sp(N)},q)\th_1(e^{2\pi i a}v/m^i_{Sp(N)},q)}\right),\label{eq:spnel}
\end{align}
where the chemical potentials for $Sp(N)$ and $SO(16+4N)$ that appear as arguments in the right hand side given by equations \eqref{eq:mso2n} and \eqref{eq:mso2n}. The four sectors labeled by different values of $a$ correspond to the four possible choices of $O(1)$ holonomy around the two circles of the $T^2$. We can alternatively use the Jacobi theta functions $\theta_\ell(z,q), \, \ell=1,\dots,4$ (see appendix \ref{sec:appmod} for our conventions) to rewrite the elliptic genus as
\begin{align}
\mathbb{E}_1^{Sp(N)}(&\mass_{Sp(N)},\mass_{SO(16+4N)},v,q) =\nonumber\\
&\qquad \frac{1}{2}\sum_{\ell=1}^4\left(\prod_{i=1}^{8+2N}\frac{\th_\ell(m^i_{SO(16+4N)},q)}{\eta(q)}\right)\left(\prod_{i=1}^{N}\frac{\eta(q)^2}{\th_\ell(v\,m^i_{Sp(N)},q)\th_\ell(v/m^i_{Sp(N)},q)}\right).\label{eq:spnellge}
\end{align}

\noindent The simplicity of this class of CFTs reflects the fact that the moduli space of one $Sp(N)$ instanton is just
\begin{equation} \mathcal{M}_{Sp(N),1} = \mathbb{H}^N / \mathbb{Z}_2 = \mathbb{C}^{2N}/\mathbb{Z}_{2},
\end{equation}
where the $\mathbb{Z}_2$ acts on the $\mathbb{C}^{2N}$ coordinates $z_{1},\dots,z_{2N}$ simultaneously as: 
\begin{equation} z_{n}\to -z_{n},\qquad n=1,\dots, 2N.
\end{equation}
The bosonic degrees of freedom that capture the target space of the BPS string CFT  consist simply of $2N$ free complex bosons and give rise to the second term on the right hand side of equation \eqref{eq:spnellge}, with the orbifolding by $\mathbb{Z}_2$ being reflected by the sum over $\ell=1,\dots,4$.\\

\begin{figure}[t]
\begin{center}
\includegraphics[width=0.8\textwidth]{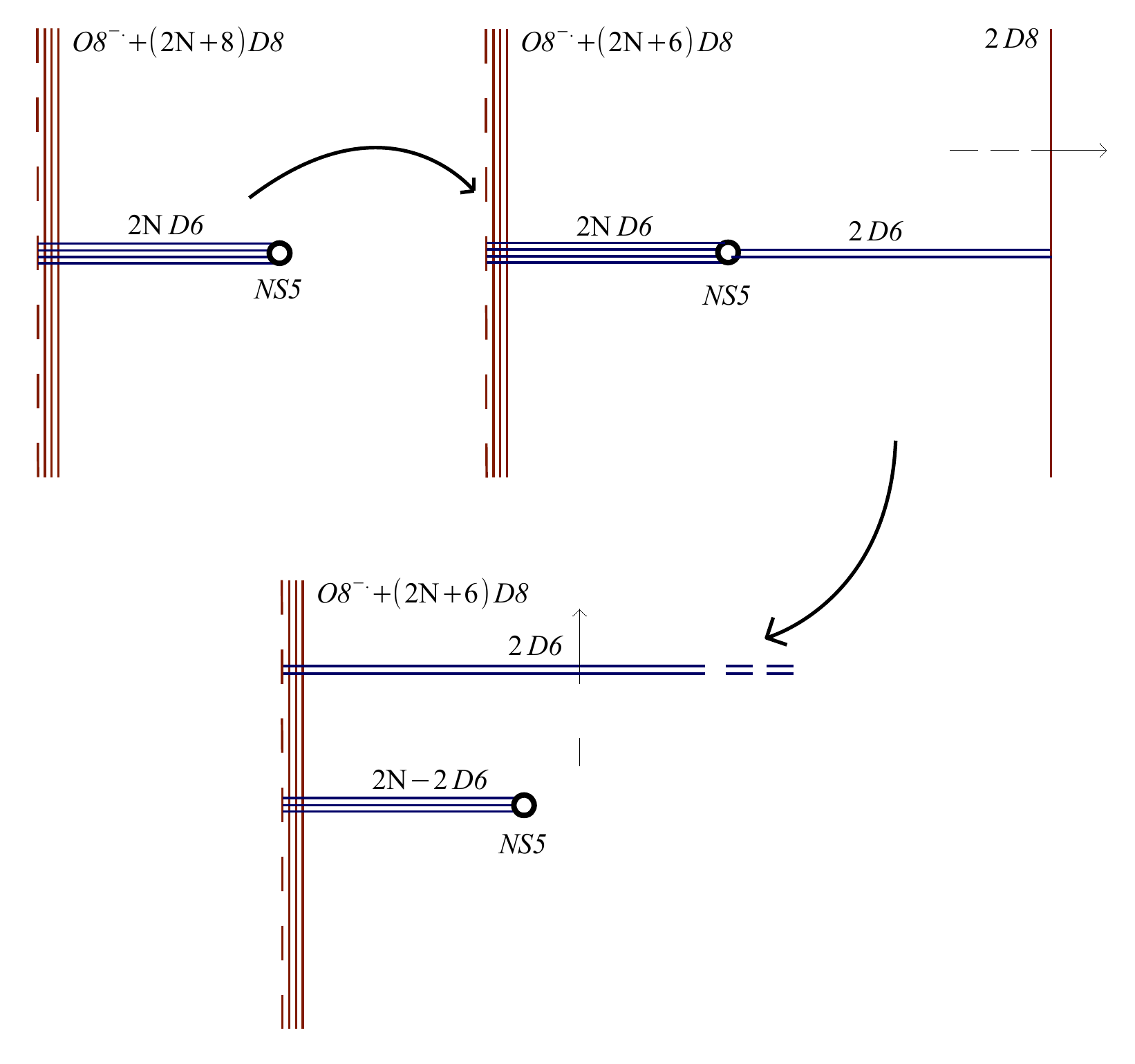}
\caption{Higgsing of the $\six{1}{Sp(N)}{}$ theory to the $\six{1}{Sp(N-1)}{}$ theory. Higgsing can be performed by moving a pair of D8 branes to infinity past the NS5 brane, which creates two additional D6 branes via a Hanany-Witten transition \cite{Hanany:1996ie}. The new D6 branes combine with two of the already existent D6 branes and are also moved off to infinity, lowering the overall number of D6 branes to $2(N-1)$.}
\label{fig:spNHiggs}
\end{center}
\end{figure}

\noindent Incidentally, it is straightforward to see the effect of Higgsing the $G=Sp(N)$ 6d theory to the $G=Sp(N-1)$ theory at the level of the elliptic genus: the former reduces to the latter when one sets \footnote{ Notice the appearance of a shift of the flavor symmetry fugacities by $v=e^{2\pi i \epsilon_+}$. See \cite{Okuda:2010ke} for a related discussion.} 
\be
m_{SO(4N+16)}^{2N+8} = v\, m_{Sp(N)}^N \qquad m_{SO(4N+16)}^{2N+7} = v/m_{Sp(N)}^N \,,
\ee
which leads to cancellations between the contributions to the elliptic genus from one Fermi and one hypermultiplet. For this specialization of values, it is possible to simultaneously give a mass to the Fermi multiplet and to the hypermultiplet and remove them from the spectrum of the CFT. In the brane construction, the Higgsing is realized by removing a pair of D6 branes from the stack of $2N$ D6 branes as in figure \ref{fig:spNHiggs}.\\

\noindent By successively removing all D6 branes, the 6d gauge group can be completely Higgsed, and one is left with the E-string (or `$Sp(0)$') theory which consists of four (0,4) Fermi multiplets (or, in other words, 16 real fermions). The elliptic genus is given by:
\begin{equation}\label{eq:ellestring}
\mathbb{E}_1^{\text{E-string}}(\mass_{SO(16)},v,q) = \frac{1}{2}\sum_{\ell=1}^4\left(\prod_{i=1}^{8}\frac{\th_\ell(m^i_{SO(16)},q)}{\eta(q)}\right).
\end{equation}

\subsection{Current algebra realization of $F=SO(16+4N)$}\label{sec:spnwzw}

In this section we show that the flavor symmetry $F$ of the \hh{1}{Sp(N)} theories is realized in terms of a Kac-Moody algebra, a statement which we will eventually extend to all \hh{n}{G} theories in section \ref{sec:wzwuniv}.\\

\noindent Recall that the CFT of one E-string, equation \eqref{eq:ellestring}, is given by  the $E_8$ Kac-Moody algebra at level 1. Indeed, the elliptic genus of one E-string coincides with the character of the unique level 1 character of $E_8$,\footnote{ See appendices \ref{sec:appaff} and \ref{sec:WZW} for our conventions on affine algebras and their characters.}
\begin{equation}\label{eq:ellestring2}
\mathbb{E}_1^{\text{E-string}} = \widehat\chi^{E_8}_{\bf{1}}(\mass_{E_8},q),
\end{equation}
where the regular embedding of $SO(16)$ into $E_8$ fixes the precise relation between $\mass_{E_8}$ and $\mass_{SO(16)}$ fugacities. The appearance of the $E_8$ level 1 chiral algebra is understood since the early work of \cite{Ganor:1996mu}: the E-string can be realized in M-theory as a M2 brane suspended between an M5 brane and an M9 brane; the latter carries $E_8$ gauge degrees of freedom to which the M2 brane couples chirally, giving rise to a current algebra. Generalizing to multiple E-strings, it was found in \cite{Minahan:1998vr} that the theory of $k$ E-strings involves a $E_8$ at level $k$ sector, describing the M2-M9 boundary degrees of freedom, coupled to a residual piece of the CFT describing propagation of the bound state of strings in $\mathbb{R}^4$.\\

\noindent  In the Type IIA brane construction at weak coupling only the $SO(16)$ maximal subgroup of $E_8$ is visible \cite{Polchinski:1996fm}. The elliptic genus for $k$ strings may also be expressed in terms of characters of level $k$ modules of the Kac-Moody algebra associated to this $SO(16)$ subgroup. For example, for one string the elliptic genus receives contributions from the level 1 $SO(16)$ WZW primaries associated to the vacuum and spinor representations:
\begin{equation}\label{eq:e8so16}
\mathbb{E}_1^{\text{E-string}}(\mass_{SO(16)},v,q) = \widehat\chi^{SO(16)}_{\bf{1}}(\mass_{SO(16)},q)+\widehat\chi^{SO(16)}_{\bf{s}}(\mass_{SO(16)},q),
\end{equation}
where\footnote{ For $G=SO(2N)$ we adopt the shorthand notation $\bf{1},\bf{v},\bf{s},\bf{c}$ to indicate respectively the level 1 WZW primaries corresponding to the trivial, vector, spinor, and conjugate spinor representations of $SO(2N)$; for a WZW model of specified level $k$ we label the WZW primaries in terms of the Dynkin label $\lambda$ of the corresponding finite representation. Characters of the WZW model are denoted by a hat (e.g. $\widehat{\chi}^G_{\lambda}(\mass_G,q)$), as opposed to the characters of the simple Lie group $G$ (e.g. $\chi^G_\lambda(\massG)$). We sometimes omit the arguments of these functions when they are clear from the context. We refer to appendices \ref{sec:Lie} and \ref{sec:WZW} for our conventions for Lie algebras and WZW models.}
\begin{align}
\widehat\chi^{SO(16)}_{\bf{1}}(\mass_{SO(16)},q) &= \frac{1}{2}\left(\prod_{i=1}^{8}\frac{\theta_3(m^i_{SO(16)},q)}{\eta(q)}+\prod_{i=1}^{8}\frac{\theta_4(m_{SO(16)}^i,q)}{\eta(q)}\right),\\
\widehat\chi^{SO(16)}_{\bf{s}}(\mass_{SO(16)},q) &= \frac{1}{2}\left(\prod_{i=1}^{8}\frac{\theta_2(m^i_{SO(16)},q)}{\eta(q)}+\prod_{i=1}^{8}\frac{\theta_1(m^i_{SO(16)},q)}{\eta(q)}\right).
\end{align}
The fact that the elliptic genus can also be interpreted in terms of a SO(16) Kac-Moody algebra is of course hardly surprising: the theory of figure \ref{fig:spneng} with $N=0$ consists simply of 16 free chiral fermions with identical boundary conditions, transforming in the vector representation of the flavor symmetry $SO(16)$; this free CFT indeed provides a free field realization of the $SO(16)$ current algebra at level 1, which explains the appearance of the affine flavor symmetry in the elliptic genus.\newline

\noindent The situation for the \hh{1}{Sp(N)} theories is only slightly more complicated: the chiral half of the \hh{1}{Sp(N)} CFT again consists of a theory of free fermions in the vector representation of $F=SO(16+4N)$, all with identical periodicities, now coupled to a CFT of noncompact free bosons charged under $SU(2)_R\times Sp(N)$. The only coupling between the fermionic and bosonic components of the CFT arises through the assigment of identical periodicities to the two sectors. So in analogy with the E-string case one also expects the \hh{1}{Sp(N)} CFT to involve a $SO(16+4N)$ current algebra at level one realized in terms of the free chiral fermions, which arise from the D2-D8 strings in the brane setup of figure \ref{fig:spneng}; only, for $N>0$ we expect this sector to be coupled nontrivially to an additional sector of the CFT describing the dependence on the 6d gauge algebra.\\

\noindent Recall that for a WZW model the level is proportional to $\tfrac12$ of the anomaly coefficient for the term in the anomaly polynomial that is associated to the corresponding 't Hooft anomaly. In our case this is precisely the coefficient of the term $c_2(F)$ within $\Sigma \cdot \mathbf{X}_4$. For the models $\six{1}{Sp(N)}{}$ the matter transforms as a bifundamental half hypermultiplet $\tfrac12( \mathbf{2N}, \mathbf{16+4N})$, and so we obtain
\bea
\Sigma \cdot \mathbf{b}_{SO(16+4N)} &=  \tfrac12 \times A_{Sp(N),\mathbf{2N}} \times A_{SO(4N+16), \mathbf{16+4N}}\\
&= \tfrac12 \times 1  \times 2 = 1.
\eea

\noindent The occurrence of the level 1 $SO(16+4N)$ current algebra coupled to a gauge sector is easily seen at the level of the elliptic genus, which can be rewritten in terms of the four characters 
\be
\widehat\chi^{SO(16+4N)}_{\bf{1}},\quad \widehat\chi^{SO(16+4N)}_{\bf{v}},\quad \widehat\chi^{SO(16+4N)}_{\bf{s}}, \quad\text{and }\widehat\chi^{SO(16+4N)}_{\bf{c}}
\ee
of the current algebra as follows:
\be\label{eq:EspNwzw}
\mathbb{E}_1^{Sp(N)} = \sum_{\lambda = \mathbf{1}, \mathbf{v}, \mathbf{s}, \mathbf{c}} \widehat\chi^{SO(16+4N)}_{\lambda}(\mass_{SO(16+4N)},q) \, \, \xi^{1,Sp(N)}_{\lambda}(\mass_{Sp(N)},v,q)
\ee
where the functions
\begin{align*} \xi^{1,Sp(N)}_{\bf{1}}(\mass_{Sp(N)},v&, q) =\nonumber\\
 &\hspace{-0.2in}\frac{1}{2}\left(\prod_{i=1}^{N}\frac{\eta(q)}{\theta_3(v\, m_{Sp(N)}^i,q)}\frac{\eta(q)}{\theta_3(v/ m_{Sp(N)}^i,q)}+\prod_{i=1}^{N}\frac{\eta(q)}{\theta_4(v\, m_{Sp(N)}^i,q)}\frac{\eta(q)}{\theta_4(v /m_{Sp(N)}^i,q)}\right),
 \end{align*}
 \begin{align*}
 \xi^{1,Sp(N)}_{\bf{v}}(\mass_{Sp(N)},v,& q)= \nonumber\\
&\hspace{-0.2in} \frac{1}{2}\left(\prod_{i=1}^{N}\frac{\eta(q)}{\theta_3(v\, m_{Sp(N)}^i,q)}\frac{\eta(q)}{\theta_3(v/ m_{Sp(N)}^i,q)}-\prod_{i=1}^{N}\frac{\eta(q)}{\theta_4(v\, m_{Sp(N)}^i,q)}\frac{\eta(q)}{\theta_4(v /m_{Sp(N)}^i,q)}\right),
  \end{align*}
 \begin{align*}
 \xi^{1,Sp(N)}_{\bf{s}}(\mass_{Sp(N)},v,& q)=\nonumber\\
&\hspace{-0.2in} \hspace{-0.5in} \frac{1}{2}\left(\prod_{i=1}^{N}\frac{\eta(q)}{\theta_2(v\, m_{Sp(N)}^i,q)}\frac{\eta(q)}{\theta_2(v/ m_{Sp(N)}^i,q)}+(-1)^N\prod_{i=1}^{N}\frac{\eta(q)}{\theta_1(v\, m_{Sp(N)}^i,q)}\frac{\eta(q)}{\theta_1(v /m_{Sp(N)}^i,q)}\right),
  \end{align*}
 \begin{align*}
\xi^{1,Sp(N)}_{\bf{c}}(\mass_{Sp(N)},v,& q)=\nonumber\\
&\hspace{-0.2in} \hspace{-0.5in} \frac{1}{2}\left(\prod_{i=1}^{N}\frac{\eta(q)}{\theta_2(v\, m_{Sp(N)}^i,q)}\frac{\eta(q)}{\theta_2(v/ m_{Sp(N)}^i,q)}-(-1)^N\prod_{i=1}^{N}\frac{\eta(q)}{\theta_1(v\, m_{Sp(N)}^i,q)}\frac{\eta(q)}{\theta_1(v /m_{Sp(N)}^i,q)}\right),
 \end{align*}
 capture the contribution to the elliptic genus of $2N$ free noncompact complex bosons with both periodic and anti-periodic boundary conditions, which reflects the fact that the moduli space of one $Sp(N)$ instanton $\mathcal{M}_{Sp(N),1}$ is the orbifold space $\mathbb{C}^{2N}/\mathbb{Z}_2$.\\
 
\noindent In the $N=0$ case, corresponding to the E-string, one obtains $\xi^{1,Sp(0)}_{\bf{1}}=\xi^{1,Sp(0)}_{\bf{s}}=1$ and $\xi^{1,Sp(0)}_{\bf{v}}=\xi^{1,Sp(0)}_{\bf{c}}=0$, which neatly recovers expression \eqref{eq:e8so16}.\\

\noindent At the level of central charges, the current algebra of $F=SO(16+4N)$ contributes 
\begin{equation}
c^F_L = 8+2N\qquad\text{and}\qquad c^F_R = 0
\end{equation}
while the residual piece of the CFT contributes
\begin{equation}
c^{residual}_L=4N\qquad\text{and}\qquad  c^{residual}_R =6N,
\end{equation}
so that the total central charge of the \hh{1}{Sp(N)} CFT adds up to 
\begin{equation}
c_L= 8+6N\qquad\text{and}\qquad c_R = 6N.
\end{equation}
Finally, adding the contributions $(c_L,c_R)=(4,6)$ from the decoupled center of mass hypermultiplet which we have suppressed in this section, one obtains
\begin{equation}
c_L\to 6(N+2)\qquad\text{and}\qquad c_R \to 6(N+1),
\end{equation}
consistent with the values \eqref{eq:central_charges} computed from the anomaly polynomial.\\

\subsection{Worldsheet realization of $G=Sp(N)$}\label{sec:spngv}
In this section we turn to a discussion of the 6d gauge symmetry $G=Sp(N)$ from the perspective of the \hh{1}{Sp(N)} CFT. The analysis of the $G$-dependent sector of the CFT is complicated by the non-compactness of the target space, on which both $G$ and $SU(2)_R$ act as isometries in a nonchiral fashion. Nonetheless, we find that the elliptic genus is captured in terms of a $Sp(N)$ Kac-Moody algebra at level $-1$, coupled to a $c=1$ theory that encodes the dependence on $v$. Eventually, we will provide a rationale for this structure as being part of the chiral algebra of the theory, but we begin by taking an agnostic approach and studying the $G$ dependence of the series expansion of the $\xi^{1,Sp(N)}_\lambda$ functions.\newline

\begin{table}[t!]
$
\begin{tabular}{c|cccccccccccccccccc}
$j\backslash k$&\!-8\!&\!-7\!&\!-6\!&\!-5\!&\!-4\!&\!-3\!&\!-2\!&\!-1\!&\!0\!&\!1\!&\!2\!&\!3\!&\!4\!&\!5\!&\!6\!&\!7\!&\!8\!\\\hline
0&      &   &   &   &   &   &   &   & $\chi_{(0)}$ &   &   &   &   &   &   &   &     \\
1&      &   &   &   &   &   & $\chi_{(2)}$ &   &   &   & $\chi_{(2)}$ &   &   &   &   &   &     \\
2&      &   &   &   & $\chi_{(4)}$ &   &   &   & $\chi_{(2)}$ &   &   &   & $\chi_{(4)}$ &   &   &   &     \\
3&      &   & $\chi_{(6)}$ &   &   &   &   &   &   &   &   &   &   &   & $\chi_{(6)}$ &   &      \\
4&    $\chi_{(8)}$ &   &   &   &   &   &   &   &   &   &   &   &   &   &   &   & $\chi_{(8)}$    \\
5&      &   &   &   &   &   & $\chi_{(4)}$ &   &   &   & $\chi_{(4)}$ &   &   &   &   &   &     \\
\end{tabular}$\caption{Nonvanishing coefficients $a^{(1)}_{j,k}(\mass_{Sp(1)})$ in the series expansion of $\xi_{\bf{1}}^{1,Sp(1)}$, written in terms of the $Sp(1)$ characters $\chi_{(n)}=\chi^{Sp(1)}_{(n)}(\mass_{Sp(1)})$.}
\label{tab:sp1xi1}

\vspace{0.3in}

$
\begin{tabular}{c|cccccccccccccccccc}
$j\backslash k$&\!-8\!&-7&\!-6\!&-5&\!-4\!&-3&\!-2\!&-1&\!0\!&1&\!2\!&3&\!4\!&5&\!6\!&7&\!8\!\\\hline
0 &         &   &   &   &   &   &   & -$\chi_{(1)}$ &   & -$\chi_{(1)}$ &   &   &   &   &   &   &   \\
1&   &   &   &   &   & -$\chi_{(3)}$ &   &   &   &   &   & -$\chi_{(3)}$ &   &   &   &   &   \\
2&   &   &   & -$\chi_{(5)}$ &   &   &   &   &   &   &   &   &   &-$\chi_{(5)}$ &   &   &   \\
3&   & -$\chi_{(7)}$ &   &   &   &   &   & -$\chi_{(3)}$ &   & -$\chi_{(3)}$ &   &   &   &   &   & -$\chi_{(7)}$ &   \\
4&   &  &   &   &   &   &   &  &   &  &   &   &   &   &   &  &   \\
5&   &  &   &   &   &   &   &  &   &  &   &   &   &   &   &  &   \\
\end{tabular}$
\caption{Nonvanishing coefficients $a^{(2)}_{j,k}(\mass_{Sp(1)})$ in the series expansion of $\xi_{\bf{v}}^{1,Sp(1)}$, written in terms of the $Sp(1)$ characters $\chi_{(n)}=\chi^{Sp(1)}_{(n)}(\mass_{Sp(1)})$.} \vspace{0.3in}\label{tab:sp1xi2}
\end{table}

\begin{table}[t]
$
\begin{tabular}{c|cccccccccccccccccc}
$j\backslash k$&-8&-7&-6&-5&-4&-3&-2&-1&0&1&2&3&4&5&6&7&8\\\hline
0&   &   &   &   &   &   &   &   &   & $\chi_{(0)}$ &   & $\chi_{(2)}$ &   & $\chi_{(4)}$ &   & $\chi_{(6)}$ &   \\
1&   &   &   &   &   &   &   &   &   &   &   &   &   &   &   &   &   \\
2&   &   &   &   &   &   &   & $\chi_{(2)}$ &   & $\chi_{(2)}$ &   &   &   &   &   &   &   \\
3&   &   &   &   &   &   &   &   &   &   &   &   &   &   &   &   &   \\
4&   &   &   &   &   & $\chi_{(4)}$ &   &   &   &   &   & $\chi_{(4)}$ &   &   &   &   &   \\
5&   &   &   &   &   &   &   &   &   &   &   &   &   &   &   &   &   \\
\end{tabular}$
\caption{Nonvanishing coefficients $a^{(3)}_{j,k}(\mass_{Sp(1)})$ in the series expansion of $\xi^{1,Sp(1)}_{\bf{s}}$, written in terms of the $Sp(1)$ characters $\chi_{(n)}=\chi^{Sp(1)}_{(n)}(\mass_{Sp(1)})$.} \vspace{0.3in}
\label{tab:sp1xi3}
$
\begin{tabular}{c|cccccccccccccccccc}
$j\backslash k$&-8&-7&-6&-5&-4&-3&-2&-1&0&1&2&3&4&5&6&7&8\\\hline
0&   &   &   &   &   &   &   &   &   &   & -$\chi_{(1)}$ &   & -$\chi_{(3)}$ &   & -$\chi_{(5)}$ &   & -$\chi_{(7)}$ \\
1&   &   &   &   &   &   &   &   & -$\chi_{(1)}$ &   &   &   &   &   &   &   &   \\
2&   &   &   &   &   &   &   &   &   &   &   &   &   &   &   &   &   \\
3&   &   &   &   &   &   & -$\chi_{(3)}$ &   &   &   & -$\chi_{(3)}$ &   &   &   &   &   &   \\
4&   &   &   &   &   &   &   &   & -$\chi_{(3)}$ &   &   &   &   &   &   &   &   \\
5&   &   &   &   & -$\chi_{(5)}$ &   &   &   &   &   &   &   & -$\chi_{(5)}$ &   &   &   &   \\
\end{tabular}$ 
\caption{Nonvanishing coefficients $a^{(4)}_{j,k}(\mass_{Sp(1)})$ in the series expansion of $\xi^{1,Sp(1)}_{\bf{c}}$, written in terms of the $Sp(1)$ characters $\chi_{(n)}=\chi^{Sp(1)}_{(n)}(\mass_{Sp(1)})$.}
\label{tab:sp1xi4}
\end{table}

\begin{table}[ht]
\begin{equation*}
\begin{tabular}{c|cccccccccccccccc}
$j\backslash k$&-7&-6&-5&-4&-3&-2&-1&0&1&2&3&4&5&6&7\\\hline
0&  &    &   &   &   &   &   & $\chi_{(00)}$ &   &   &   &      &   &   &\\
1&  &    &   &   &   & $\chi_{(20)}$  &   & $\chi_{(01)}$ &   & $\chi_{(20)}$  &   &   &   &      &\\
2&   &   &   & $\chi_{(40)}$ &   & -$\chi_{(02)}$ &   &   &   & -$\chi_{(02)}$ &   & $\chi_{(40)}$ &   &      &\\
3&  &  $\chi_{(60)}$ &   & -$\chi_{(22)}$ &   &   &   & -$\chi_{(21)}$ &   &   &   & -$\chi_{(22)}$ &   & $\chi_{(60)}$ &\\
\end{tabular}\end{equation*}\caption{Nonvanishing coefficients $a^{(1)}_{j,k}(\mass_{Sp(2)})$ in the series expansion of $\xi_{\mathbf{1}}^{1,Sp(2)}$, written in terms of the $Sp(2)$ characters $\chi_{(n_1n_2)}=\chi^{Sp(2)}_{(n_1n_2)}(\mass_{Sp(2)})$.}
\label{tab:sp2xi1}

\vspace{0.3in}

\begin{equation*}
\begin{tabular}{c|cccccccccccccccc}
$j\backslash k$&-7&\!-6\!&-5&\!-4\!&-3&\!-2\!&-1&\!0\!&1&\!2\!&3&\!4\!&5&\!6\!&7\\\hline
0&      &   &   &   &   &   & -$\chi_{(10)}$ &   & -$\chi_{(10)}$ &   &   &   &   &   &      \\
1&      &   &   &   & -$\chi_{(30)}$  &   &   &   &   &   & -$\chi_{(30)}$     &   &   &   &   \\
2&      &   & -$\chi_{(50)}$ &   & $\chi_{(12)}$  &   &   &   &   &   & $\chi_{(12)}$  &   & -$\chi_{(50)}$    &   &   \\
3&    -$\chi_{(70)}$  &   & $\chi_{(32)}$  &   &   &   & $\chi_{(12)}$  &   & $\chi_{(12)}$  &   &   &   & $\chi_{(32)}$  &   & -$\chi_{(70)}$    \\
\end{tabular}\end{equation*}
\caption{Nonvanishing coefficients $a^{(2)}_{j,k}(\mass_{Sp(2)})$ in the series expansion of $\xi_{\mathbf{v}}^{1,Sp(2)}$, written in terms of the $Sp(2)$ characters $\chi_{(n_1n_2)}=\chi^{Sp(2)}_{(n_1n_2)}(\mass_{Sp(2)})$.} \label{tab:sp2xi2}
\end{table}

\noindent  In order to obtain a clearer understanding of the $G$-dependence of the $\xi_\lambda^{1,Sp(N)}$ functions, it turns out to be helpful to first rescale them by a factor of
\begin{equation} \eta(q)\cdot \widetilde\Delta_{Sp(N)}(\mass_{Sp(N)},q),\end{equation}
where
\begin{equation}
\widetilde\Delta_{G}(\mass_G,q) = \prod_{j=1}^
\infty (1-q^j)^{\text{rank}(G)}\prod_{\alpha\in \Delta^G_+}(1-q^j m_\alpha)(1-q^j m_\alpha^{-1})
\end{equation}
is closely related to the Weyl-Kac determinant for the affine Lie algebra $G$, equation \eqref{eq:kdet}. Here, $\Delta^G_+$ denotes the set of positive roots of the Lie algebra of $G$, and to any root $\alpha$ we associate a fugacity $m_\alpha$ as in equation \eqref{eq:malpha}. We then expand the rescaled $\xi^{1,Sp(N)}_\lambda$ functions in terms of $q$ and $v$; this requires care, as the dependence on these variables is meromorphic. We choose to expand in the region $0<q<v$, and furthermore require that  $v< m_\alpha$  and $v<m_\alpha^{-1}$ for any positive root $\alpha$; this allows us to first expand the $\xi_\lambda^{1,Sp(N)}$ in powers of $q$ and next perform a series expansion in $v$, leading to the following expressions:
\begin{align}
\xi^{1,Sp(N)}_{\bf{1}}(\mass_{Sp(N)},v,q)&= \frac{q^{N/12}}{\eta(q)\widetilde\Delta_{Sp(N)}(\mass_{Sp(N)},q)}\sum_{j,k=0}^\infty a^{(1)}_{jk}(\mass_{Sp(N)})q^jv^k,\label{eq:xi1exp}\\
\xi^{1,Sp(N)}_{\bf{v}}(\mass_{Sp(N)},v,q)&= \frac{q^{N/12+1/2}}{\eta(q)\widetilde\Delta_{Sp(N)}(\mass_{Sp(N)},q)}\sum_{j,k=0}^\infty a^{(2)}_{jk}(\mass_{Sp(N)})q^jv^k,\\ 
\xi^{1,Sp(N)}_{\bf{s}}(\mass_{Sp(N)},v,q)&= \frac{q^{-N/6}}{\eta(q)\widetilde\Delta_{Sp(N)}(\mass_{Sp(N)},q)}\sum_{j,k=0}^\infty a^{(3)}_{jk}(\mass_{Sp(N)})q^jv^k,\\
\xi^{1,Sp(N)}_{\bf{c}}(\mass_{Sp(N)},v,q)&= \frac{q^{-N/6}}{\eta(q)\widetilde\Delta_{Sp(N)}(\mass_{Sp(N)},q)}\sum_{j,k=0}^\infty a^{(4)}_{jk}(\mass_{Sp(N)})q^jv^k.
\end{align}
This expansion turns out to be quite natural, as is indicated by the fact that many of the coefficients $a^{(i)}_{jk}$ turn out to vanish, and the nonvanishing coefficients are simple linear combinations of characters $\chi^{Sp(N)}_{\lambda}(\mass_{Sp(N)})$ of irreducible representations $\lambda$ of $Sp(N)$. We illustrate this by looking at some explicit examples. For $G=Sp(1)$ we display the first several nonvanishing coefficients of $\xi^{1,Sp(1)}_{\bf{1}},\xi^{1,Sp(1)}_{\bf{v}},\xi^{1,Sp(1)}_{\bf{s}},$ and $\xi^{1,Sp(1)}_{\bf{c}}$ in tables \ref{tab:sp1xi1}-\ref{tab:sp1xi4}; in tables \ref{tab:sp2xi1} and \ref{tab:sp2xi2} we display a similar set of coefficients for the $G=Sp(2)$ functions $\xi^{1,Sp(2)}_{\bf{1}}$ and $\xi^{1,Sp(2)}_{\bf{v}}$ (from which tables of coefficients for $\xi^{1,Sp(2)}_{\bf{s}},\xi^{1,Sp(2)}_{\bf{c}}$ may easily be obtained by spectral flow, as discussed in the next section). For $Sp(N), N>2$ one obtains similar tables of coefficients, though increasingly higher numbers of representations of $Sp(N)$ appear.\newline

\noindent We find that we can fully account for the $Sp(N)$ dependence of the elliptic genus if we assume that the functions $\xi^{1,Sp(N)}_\lambda$ take the following form:

\begin{align}\label{eq:spngans}
\xi_{\lambda}^{1,Sp(N)}&(\mass_{Sp(N)},\mass_{SO(16+4\,N)},v,q) =\nonumber\\
& \sum_{\nu\in \text{Rep}(Sp(N))}\,\sum_{\ell=-2N+1}^{0}\,\sum_{m\in\mathbb{Z}} n^\lambda_{\nu,\ell,m}\times \frac{q^{-\frac{c_{Sp(N)}}{24}+h^{Sp(N)}_\nu}\chi^{Sp(N)}_\nu(\mass_{Sp(N)})}{\widetilde\Delta_{Sp(N)}(\mass_{Sp(N)},q)}\times \frac{q^{-\frac{c_v}{24}+h^v_{\ell,m}}v^{\ell-2Nm}}{\prod_{j=1}^\infty(1-q^j)^{}},
\end{align}
where $\chi^{Sp(N)}_\nu(\mass_{Sp(N)})$ is the character of the irreducible representation $\nu$ of $Sp(N)$, and the quantities
\begin{align}\label{eq:cspnGv}
c_{Sp(N)}=-(2N+1)\qquad \text{ and } \qquad h^{Sp(N)}_{\nu} = \frac{\langle \nu,\nu+2\rho_{Sp(N)}\rangle}{2N}
\end{align}
are respectively the central charge of the (chiral half of the) $Sp(N)$ WZW model at level $k=-1$, and the conformal dimension of a WZW primary
of this theory. On the other hand, the $v$-depedence of the elliptic genus is suggestive of a CFT at $c=1$ (consistent with the fact that in the $q,v$ expansion of the elliptic genus only a $U(1)_v$ generated by the Cartan of the $SU(2)_R\times SU(2)_I$ global symmetry is visible), where $ h^v_{\ell,m}$ is the conformal dimension of a $U(1)_{-N}$ WZW primary of charge $\ell-2\,N\, m$:
\begin{equation}
h^v_{\ell,m} = -\frac{(\ell-2N m)^2}{4N}.
\end{equation}

\noindent The only data that is not fixed in equation \eqref{eq:spngans} are the coefficients $n^\lambda_{\nu,\ell,m}$, which by inspection all turn out to be either $1$, $0$, or $-1$ \footnote{Though for the theories considered in section \ref{sec:univ} other values also occur.}. For fixed $\lambda,\nu,\ell$, only a finite number of coefficients are nonvanishing due to the requirement that $H_L$ be bounded from below. Moreover, the level-matching condition for the spectrum of the CFT on the torus implies that $n^\lambda_{\nu,\ell,m} = 0$ for the choices of $\lambda,\nu,\ell,m$ for which $H_L+\frac{c_L-c_{R}}{12}$ is not an integer.\newline

\noindent Equation \eqref{eq:spngans} turns out to be quite constraining, as one sees by looking at an example. We consider the case $G=Sp(2)$ and for definiteness focus on the function $\xi^{1,Sp(2)}_{\mathbf{1}}(\mass_{Sp(2)},v,q)$. We compute the first several coefficients in equation \eqref{eq:spngans}, only keeping terms consistent with $H_L$ being a positive integer, and expand the resulting expression as in equation \eqref{eq:xi1exp}. The nonvanishing coefficients are displayed in table \ref{tab:sp1acoef} (we do not display the ones with $k>0$ since they are specular to the ones with $k<0$). Comparing with table \ref{tab:sp2xi1}, one finds that of the nine undetermined coefficients $n^{\bf{0}}_{\nu,\ell,m}$, five turn out to be equal to 1:
\begin{equation}
n^{\bf{0}}_{(00),0,0} = n^{\bf{0}}_{(20),-2,0}=n^{\bf{0}}_{(40),0,1}=n^{\bf{0}}_{(60),-2,1}=n^{\bf{0}}_{(01),0,0}=1;
\end{equation}
four turn out to be equal to $-1$:
\begin{equation}
n^{\bf{0}}_{(02),-2,0} = n^{\bf{0}}_{(22),0,1}=n^{\bf{0}}_{(21),0,0}=-1,
\end{equation}
and one vanishes:
\begin{equation}
n^{\bf{0}}_{(23),-2,1} = 0.
\end{equation}
For $G=Sp(N)$ with other values of $N$ we have similarly performed extensive checks that \eqref{eq:spngans} is consistent with the known expressions for the elliptic genus.\newline

\begin{table}[t!]
\begin{tabular}{c|cccccccccccc}
$j\backslash k$&\!-7\!\!&\!\!-6\!\!&\!\!-5\!\!&\!\!-4\!\!&\!\!-3\!\!&\!\!-2\!\!&\!\!-1\!\!&\!\!0\!\!\\\hline
0&    &&&&&   &   & $n^{\bf{0}}_{(00),0,0}\chi_{(00)}$ \\
1&    &  &      && &   $n^{\bf{0}}_{(20),-2,0}\chi_{(20)}$   &  & $n^{\bf{0}}_{(01),0,0}\chi_{(01)}$          \\
2& &     & &   $n^{\bf{0}}_{(40),0,1}\chi_{(40)}$ &&    $n^{\bf{0}}_{(02),-2,0}\chi_{(02)}$ &&            \\
3&&    $n^{\bf{0}}_{(60),-2,1}\chi_{(60)}+n^{\bf{0}}_{(23),-2,1}\chi_{(23)}$ &&    $n^{\bf{0}}_{(22),0,1}\chi_{(22)}$ &&      &&    $n^{\bf{0}}_{(21),0,0}\chi_{(21)}$ \end{tabular}
\caption{Potential nonvanishing coefficients $a^{(1)}_{j,k}(\mass_{Sp(2)})$ in the series expansion of $\xi_{\mathbf{1}}^{1,Sp(2)}$, obtained from equation \eqref{eq:spngans}. The coefficients are written in terms of the $Sp(2)$ characters $\chi_{(n_1n_2)}=\chi^{Sp(2)}_{(n_1n_2)}(\mass_{Sp(2)})$.}\label{tab:sp1acoef}
\end{table}

\noindent While $F=SO(16+4N)$ is truly a chiral symmetry of the \hh{1}{Sp(N)} CFT, the flavor symmetry $G=Sp(N)$ is not. Nevertheless, one can argue on quite general grounds that flavor symmetries of 2d theories with $(0,2)$ supersymmetry are realized as affine Kac-Moody subalgebras of the chiral algebra \cite{Dedushenko:2015opz}. In particular, flavor symmetries that act on noncompact degrees of freedom lead to current algebras at negative level. \\

\noindent This is of course consistent with the string worldsheet anomaly. Indeed, by the same argument that fixed the level of the 6d flavor symmetry contribution $SO(16+4N)_1$, one can see that the coefficient of $c_2(G)$ in the string worldsheet anomaly is given by 
\be
\Sigma \cdot \mathbf{b}_{Sp(N)} = \Sigma \cdot \Sigma = -1
\ee
because of the identification of the anomaly coefficient $\mathbf{b}_{Sp(N)}$ with the unit string charge vector of the corresponding BPS string instanton $\Sigma$.\\

\noindent As a consequence, we expect the $G$ dependence of the elliptic genus to be given in terms of irreducible characters of the level $k=-1$ affine $Sp(N)$ algebra. In the remainder of the section we provide empirical evidence that this is indeed the case, while a more thorough study of the chiral algebra will be undertaken in a future work \cite{wip}.\\
\begin{table}[t!]\begin{center}
\begin{tabular}{|r|l|}\hline
$\nu$ &$\widehat{\chi}^{Sp(2)}_\nu(\mass_{Sp(2)},q)\cdot \widetilde\Delta_{Sp(2)}(\mass_{Sp(2)},q)$\\\hline
$(00)$& $q^{\frac{5}{24}}\left[\chi^{Sp(2)}_{(00)}-\chi^{Sp(2)}_{(22)} q^5+\chi^{Sp(2)}_{(41)} q^6+\chi^{Sp(2)}_{(04)}q^7-\chi^{Sp(2)}_{(61)}q^{10}+\chi^{Sp(2)}_{(80)}q^{12}+\mathcal{O}(q^{16})\right]$\\
$(10)$& $q^{\frac{5}{24}+\frac{1}{2}}\left[\chi^{Sp(2)}_{(10)}-\chi^{Sp(2)}_{(12)} q^3+\chi^{Sp(2)}_{(50)}q^5+\chi^{Sp(2)}_{(14)}q^{8}-\chi^{Sp(2)}_{(52)}q^{10}\right.$\\
&$\quad\quad\,\,\left.+\chi^{Sp(2)}_{(90)}q^{14}-\chi^{Sp(2)}_{(16)}q^{15}+\mathcal{O}(q^{16})\right]$\\
$(01)$& $q^{\frac{5}{24}+1}\left[\chi^{Sp(2)}_{(01)}-\chi^{Sp(2)}_{(21)} q^2+\chi^{Sp(2)}_{(40)}q^3+\chi^{Sp(2)}_{(05)}q^{9}-\chi^{Sp(2)}_{(62)}q^{12}-\chi^{Sp(2)}_{(25)}q^{13}\right.$\\
&$\quad\quad\,\,\left.+(\chi^{Sp(2)}_{(44)}+\chi^{Sp(2)}_{(81)})q^{14}+\mathcal{O}(q^{15})\right]$\\
$(20)$& $q^{\frac{5}{24}+1}\left[\chi^{Sp(2)}_{(20)}-\chi^{Sp(2)}_{(02)} q-\chi^{Sp(2)}_{(03)}q^3+(\chi^{Sp(2)}_{(60)}+\chi^{Sp(2)}_{(23)})q^{6}-\chi^{Sp(2)}_{(42)}q^{7}+\chi^{Sp(2)}_{(24)}q^{9}\right.$\\
&$\quad\quad\,\,\left.-\chi^{Sp(2)}_{(43)}q^{10}-\chi^{Sp(2)}_{(06)}q^{12}+\mathcal{O}(q^{15})\right]$\\
$(30)$& $q^{\frac{5}{24}+\frac{3}{2}}\left[\chi^{Sp(2)}_{(30)}-\chi^{Sp(2)}_{(12)} q+\chi^{Sp(2)}_{(14)}q^6+\chi^{Sp(2)}_{(70)}q^{7}-\chi^{Sp(2)}_{(52)}q^{8}-\chi^{Sp(2)}_{(16)}q^{13}+\mathcal{O}(q^{15})\right]$\\
$(40)$& $q^{\frac{5}{24}+2}\left[\chi^{Sp(2)}_{(40)}-\chi^{Sp(2)}_{(22)} q+\chi^{Sp(2)}_{(04)}q^3+\chi^{Sp(2)}_{(05)}q^{6}+\chi^{Sp(2)}_{(80)}q^{8}-\chi^{Sp(2)}_{(62)}q^{9}-\chi^{Sp(2)}_{(25)}q^{10}\right.$\\
&$\quad\quad\,\,\left.+\chi^{Sp(2)}_{(44)}q^{11}+\mathcal{O}(q^{14})\right]$\\
$(50)$& $q^{\frac{5}{24}+\frac{5}{2}}\left[\chi^{Sp(2)}_{(50)}-\chi^{Sp(2)}_{(32)} q+\chi^{Sp(2)}_{(14)}q^3+\chi^{Sp(2)}_{(90)}q^{9}-(\chi^{Sp(2)}_{(72)}+\chi^{Sp(2)}_{(16)})q^{10}\right.$\\
&$\quad\quad\,\,\left.+\chi^{Sp(2)}_{(54)}q^{12}+\mathcal{O}(q^{14})\right]$\\
$(60)$& $q^{\frac{5}{24}+3}\left[\chi^{Sp(2)}_{(60)}-\chi^{Sp(2)}_{(42)} q+\chi^{Sp(2)}_{(24)}q^3-\chi^{Sp(2)}_{(06)}q^{6}+(\chi^{Sp(2)}_{(10,0)}-\chi^{Sp(2)}_{(07)})q^{10}\right.$\\
&$\quad\quad\,\,\left.-\chi^{Sp(2)}_{(82)}q^{11}+\mathcal{O}(q^{13})\right]$\\
$(70)$& $q^{\frac{5}{24}+\frac{7}{2}}\left[\chi^{Sp(2)}_{(70)}-\chi^{Sp(2)}_{(52)} q+\chi^{Sp(2)}_{(34)}q^3-\chi^{Sp(2)}_{(16)}q^{6}+\chi^{Sp(2)}_{(11,0)}q^{11}-\chi^{Sp(2)}_{(92)}q^{12}+\mathcal{O}(q^{13})\right]$\\
$(80)$& $q^{\frac{5}{24}+4}\left[\chi^{Sp(2)}_{(80)}-\chi^{Sp(2)}_{(62)} q+\chi^{Sp(2)}_{(44)}q^3-\chi^{Sp(2)}_{(26)}q^{6}+\chi^{Sp(2)}_{(08)}q^{10}+\mathcal{O}(q^{12})\right]$\\
$(90)$& $q^{\frac{5}{24}+\frac{9}{2}}\left[\chi^{Sp(2)}_{(90)}-\chi^{Sp(2)}_{(72)} q+\chi^{Sp(2)}_{(54)}q^3-\chi^{Sp(2)}_{(36)}q^{6}+\chi^{Sp(2)}_{(18)}q^{10}+\mathcal{O}(q^{12})\right]$\\
$(10,0)$& $q^{\frac{5}{24}+5}\left[\chi^{Sp(2)}_{(10,0)}-\chi^{Sp(2)}_{(82)} q+\chi^{Sp(2)}_{(64)}q^3-\chi^{Sp(2)}_{(46)}q^{6}+\chi^{Sp(2)}_{(28)}q^{10}+\mathcal{O}(q^{11})\right]$\\
$(11,0)$& $q^{\frac{5}{24}+\frac{11}{2}}\left[\chi^{Sp(2)}_{(11,0)}-\chi^{Sp(2)}_{(92)} q+\chi^{Sp(2)}_{(74)}q^3-\chi^{Sp(2)}_{(56)}q^{6}+\chi^{Sp(2)}_{(38)}q^{10}+\mathcal{O}(q^{11})\right]$\\
\dots&\dots\\
$(n,0)$&$q^{\frac{5}{24}+\frac{n}{2}}\left[\chi^{Sp(2)}_{(n,0)}-\chi^{Sp(2)}_{(n-2,2)}q+\chi^{Sp(2)}_{(n-4,4)}q^3-\chi^{Sp(2)}_{(n-6,6)}q^6+\mathcal{O}(q^{10})\right]\qquad (n>12)$\\\hline
\end{tabular}\end{center}
\caption{Irreducible characters of $Sp(2)$ at level $-1$ (rescaled by an overall factor of $\widetilde\Delta_{Sp(2)}(\mass_{Sp(2)},q)$), computed up to $\mathcal{O}(q^{\frac{5}{24}+16})$. For $n$ sufficiently high, the leading order terms in the $\nu=(n,0)$ representation behave regularly, as indicated in the last entry; we have checked the validity of this formula up to $\mathcal{O}(q^{\frac{5}{24}+16})$ for $n\leq 31$.}\label{tab:sp2ch}
\end{table}

\noindent The Kazhdan-Lusztig conjecture, which has been proven for affine Kac-Moody algebras at level $k>h^\vee_G$ in \cite{kashiwara1990kazhdan,casian1990kazhdan}, provides a prescription to determine the irreducible modules of the Kac-Moody algebra and their characters by evaluating a certain collection of Kazhdan-Lusztig polynomials \cite{kazhdan1979representations}; see e.g. appendix C of \cite{Beem:2013sza} for an explanation of this relation. We have implemented the computation of the characters with the aid of the \texttt{Mathematica} and \texttt{Sage} platforms and the \texttt{Coxeter} program \cite{du2002computing}. We illustrate our results for the two cases $G=Sp(1)$ and $Sp(2)$. The first case turns out to be trivial: the irreducible $Sp(1)_{-1}$ characters turn out to be given simply by
\begin{equation}
\widehat\chi^{Sp(1)}_{(n)}(\mass_{Sp(1)},q) = q^{-\frac{c_{Sp(1)}}{24}+h^{Sp(1)}_\nu}\frac{\chi^{Sp(1)}_{(n)}(\mass_{Sp(1)})}{\widetilde\Delta_{Sp(1)}(\mass_{Sp(1)},q)},
\end{equation}
which are exactly the quantities appearing in equation \eqref{eq:spngans}, and the $\xi^{1,Sp(N)}_{\lambda}$ functions can be written in terms of them as:
\begin{align}
\xi_{\bf{1}}^{1,Sp(1)}(\mass_{Sp(1)},v,q) &=  \sum_{n=0}^\infty \widehat\chi^{Sp(1)}_{(2n)}(\mass_{Sp(1)},q)\frac{\sum_{m=-n}^nq^{-m^2}v^{2m}}{\eta(q)},\\
\xi_{\bf{v}}^{1,Sp(1)}(\mass_{Sp(1)},v,q) &= -\sum_{n=0}^\infty \widehat\chi^{Sp(1)}_{(2n+1)}(\mass_{Sp(1)},q)\frac{\sum_{m=-n-1}^nq^{-\frac{1}{4}(-1-2m)^2}v^{-1-2m}}{\eta(q)},
\end{align}
and 
\begin{align}
\xi_{\bf{s}}^{1,Sp(1)}(\mass_{Sp(1)},v,q) &=\xi_{\bf{1}}^{1,Sp(1)}(\mass_{Sp(1)},q^{1/2}/v,q),\\
\xi_{\bf{c}}^{1,Sp(1)}(\mass_{Sp(1)},v,q) &=\xi_{\bf{v}}^{1,Sp(1)}(\mass_{Sp(1)},q^{1/2}/v,q).
\end{align}
For $G=Sp(2)$, on the other hand, we find:
\begin{align}
\xi_{\bf{1}}^{1,Sp(2)}(\mass_{Sp(2)},v,q) &=  \left(\widehat\chi^{Sp(2)}_{(0,0)}(\mass_{Sp(2)},q)+\widehat\chi^{Sp(2)}_{(0,1)}(\mass_{Sp(2)},q)\right)\frac{1}{\eta(q)}\nonumber\\
&+\sum_{n=1}^\infty \widehat\chi^{Sp(2)}_{(2n,0)}(\mass_{Sp(2)},q)\frac{q^{-n^2}(v^{-2n}+v^{2n})}{\eta(q)},\\
\xi_{\bf{v}}^{1,Sp(2)}(\mass_{Sp(2)},v,q) &= -\sum_{n=0}^\infty \widehat\chi^{Sp(2)}_{(2n+1,0)}(\mass_{Sp(2)},q)\frac{q^{-\frac{1}{4}(2n+1)^2}(v^{-2n-1}+v^{2n+1})}{\eta(q)}.
\end{align}
We have verified these expressions up to $\mathcal{O}(q^{\frac{1}{6}+15})$ and $\mathcal{O}(q^{\frac{2}{3}+15})$ respectively. The relevant level $-1$ $Sp(2)$ characters, computed to sufficiently high order, are reported in table \ref{tab:sp2ch}. One also finds that
\begin{align}
\xi_{\bf{s}}^{1,Sp(2)}(\mass_{Sp(2)},v,q) &=\xi_{\bf{1}}^{1,Sp(2)}(\mass_{Sp(2)},q^{1/2}/v,q),\\
\xi_{\bf{c}}^{1,Sp(2)}(\mass_{Sp(2)},v,q) &=\xi_{\bf{v}}^{1,Sp(2)}(\mass_{Sp(2)},q^{1/2}/v,q).
\end{align}

\noindent In section \ref{sec:gengauge} we will discuss the generalization of the results of this section to arbitrary \hh{n}{G} theories.

\subsection{NS--R elliptic genus, spectral flow, and low energy spectrum}\label{sec:spnns}
In this section we compute the NS--R elliptic genus\footnote{ Note that for this particular class of 2d CFTs the terminology `NS--R elliptic genus' is somewhat of a misnomer, since the sum over $O(1)=\mathbb{Z}_2$ holonomies leads automatically to summing over different fermion and boson periodicities.} of the \hh{1}{Sp(N)} theories and analyze the low energy states that contribute to it, remarking on their interpretation from the point of view of the parent 6d (1,0) SCFT. We also comment on the existence of a spectral flow that relates the NS--R and R--R elliptic genera. \newline

\noindent The NS--R elliptic genus is obtained by imposing anti-periodic boundary conditions  on the chiral fermions:
\begin{align}\label{eq:spnnsell}
&\mathcal{E}_1^{Sp(N)}(\mass_{Sp(N)},\mass_{SO(16+4N)},v,q) =\nonumber\\
& \frac{1}{2}\!\!\!\sum_{a\in\{0,\frac{1}{2},\frac{\tau}{2},\frac{1+\tau}{2}\}}\!\!\!\left(\prod_{i=1}^{8+2N}\frac{\th_1(e^{2\pi i a}q^{1/2}m^i_{SO(16+4N)},q)}{\eta(q)}\right)\!\!\left(\prod_{i=1}^{N}\frac{\eta(q)^2}{\th_1(e^{2\pi i a}v\,m^i_{Sp(N)},q)\th_1(e^{2\pi i a}v/m^i_{Sp(N)},q)}\right)\!.
\end{align}
The NS boundary conditions on the fermions are reflected in the factor of $q^{1/2}=e^{\pi i \tau}$ in the argument of the theta functions in the first factor of the elliptic genus. As the same time, the sum over $O(1)$ holonomies also instructs us to sum over different boundary conditions. For the $N=0$ case of the E-string, for example, the choice of NS versus R boundary conditions is immaterial since the shift by $q^{1/2}$ can be readsorbed by shifting $a\to a-\tau/2$ in the sum. On the other hand, for $N\geq 1$ the choice of boundary conditions does matter, since the sum over holonomies also affects the hypermultiplets in the second term on the right hand side of equation \eqref{eq:spnnsell}. One can again absorb the factor of $q^{1/2}$ by a shift in the holonomy $a$, but this must be compensated by simultaneously shifting $v\to v\, q^{1/2}$.\footnote{ In fact, we find it more convenient to combine this shift with the transformation $v\to 1/v$, which only changes the elliptic genus by an overall phase, so we express the spectral flow as $v\to q^{1/2}/v$.} This leads to the following relation between NS--R and R--R elliptic genera (valid up to an overall minus sign):
\begin{align}
\mathcal{E}_1^{Sp(N)}(\mass_{Sp(N)},\mass_{SO(16+4N)},v,q) = q^{\frac{k_{\epsilon_+}}{4}}v^{-k_{\epsilon_+}} \mathbb{E}_1^{Sp(N)}(\mass_{Sp(N)},\mass_{SO(16+4N)},q^{1/2}/v,q),
\end{align}
where $ k_{\epsilon_+} = 1-h^\vee_{Sp(N)}$ is the $\epsilon_+$-index of the elliptic genus, which can be read off from the $\epsilon_+^2$ coefficient in the anomaly polynomial \eqref{eq:anpol} \cite{DelZotto:2016pvm,DelZotto:2017mee},  upon setting $c_2(R)=c_2(I)=-\epsilon_+^2$ and subtracting the contributions to the anomaly polynomial from the center of mass hypermultiplet. Decomposing the elliptic genus as in equation \eqref{eq:EspNwzw}, one finds that the effect of the transformation $v\to q^{1/2}/v$ is simply to exchange 
\begin{align}\xi^{1,Sp(N)}_{\bf{1}}(\mass_{Sp(N)},\mass_{SO(16+4\,N)},v,q)&\leftrightarrow\xi^{1,Sp(N)}_{\bf{s}}(\mass_{Sp(N)},\mass_{SO(16+4\,N)},v,q),\\
\xi^{1,Sp(N)}_{\bf{v}}(\mass_{Sp(N)},\mass_{SO(16+4\,N)},v,q)&\leftrightarrow\xi^{1,Sp(N)}_{\bf{c}}(\mass_{Sp(N)},\mass_{SO(16+4\,N)},v,q),
\end{align}
which is evident from looking at e.g. tables \ref{tab:sp1xi1}-\ref{tab:sp1xi4} for $G=Sp(1)$. We will discuss this spectral flow in more detail in section \ref{sec:genspec}, where we will find it to be a property of all \hh{n}{G} theories.\\

\noindent We next turn to a discussion of the low energy spectrum in the NS--R sector. This sector has a unique fermionic vacuum tensored with a degenerate set of bosonic vacua which arise from the zero energy modes of the scalars $q_i,\widetilde q_i$, $1=1,\dots,N$ in the hypermultiplets. In the Ramond sector one finds a set of fermionic zero modes, which originate from the Fermi multiplets charged under the 6d flavor symmetry group $SO(16+4N)$; on the other hand, since in the Neveu-Schwarz sector the fermionic vacuum is unique, the zero energy sector is captured simply by the ADHM quantum mechanics of one pure $Sp(N)$ instanton, represented by the following 1d $\mathcal{N}=2$ quiver:
\begin{center}
\includegraphics[width=0.6\textwidth]{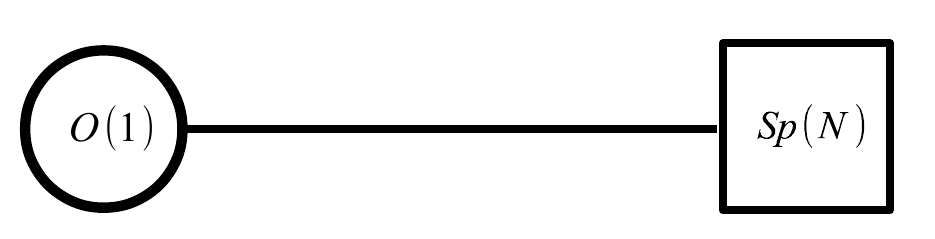}
\end{center}
Accordingly, we find that the lowest energy component of the NS--R elliptic genus is given by%
\begin{align}\label{eq:ellspnq0}
\mathcal{E}_1^{Sp(N)}(\mass_{Sp(N)},\mass_{SO(16+4N)},v,q)\bigg\vert_{q^{-\frac{1}{3}-\frac{N}{4}}} &= \frac{1}{2}\sum_{s=\pm 1}\prod_{j=1}^N\frac{v}{(1-s\, v\, m^j_{Sp(N)})(1-s\,v/m^j_{Sp(N)})}.
\end{align}
This may be expanded in a power series in $v$ as follows:\footnote{ Note that only the $U(1)_v$ Cartan subgroup of the diagonal subgroup $SU(2)_v$ of $SU(2)_R\times SU(2)_I$ is visible in this expansion.}
\begin{equation}\label{eq:spnhilb}
\mathcal{E}_1^{Sp(N)}(\mass_{Sp(N)},\mass_{SO(16+4N)},v,q)\bigg\vert_{q^{-\frac{1}{3}-\frac{N}{4}}} = v^N \sum_{k=0}^\infty v^{2k}\chi_{k\cdot \theta_{Sp(N)}}^{Sp(N)}(\mass_{Sp(N)}),
\end{equation}
which coincides with the Hilbert series of the moduli space of one $Sp(N)$ instanton \cite{Benvenuti:2010pq} up to an overall factor of $v^N$; in this expression, $\theta_{Sp(N)}$ is the highest weight of the adjoint representation of $Sp(N)$,
\begin{equation}
\theta_{Sp(N)} = (20\dots0).
\end{equation}

\noindent The contribution at $k=0$ indicates that the $Sp(N)$-neutral vacuum is assigned $U(1)_v$ charge $+N = h^\vee_{Sp(N)}-1$ (see sections \ref{sec:genspec} and \ref{sec:n35d} for further remarks). The $k=1$ terms correspond to mesonic operators constructed out $O(1)$ gauge-invariant quadratic combinations of the zero modes $q_i^{(0)},\widetilde q_i^{(0)}$ of the hypermultiplet scalars; these terms appear with an accompanying power of $v^{2}$, consistent with the assignment of $U(1)_v$ charge $+1$ to the fields $q_i^{(0)},\widetilde q_i^{(0)}$. The $k>1$ terms count operators constructed out of combinations of higher numbers the $q_i^{(0)},\widetilde q_i^{(0)}$, taking into account the relations arising from F-terms in the quantum mechanics.\newline

\noindent We can also easily construct a set of operators that appear at the first excited level $H_L= -c_L/24+ 1/2$ in the elliptic genus: these involve gauge-invariant combinations of the scalar zero modes $q_i^{(0)},\widetilde q_i^{(0)}$ as well as the first excited component of a fermionic operator, $\psi_i^{(1/2)}$, which contributes $L_0=1/2$ to the energy. The simplest such combination consists of the gauge-invariant operators 
\begin{equation}
q_i^{(0)}\psi_j\qquad\text{ and }\qquad \widetilde q_i^{(0)}\psi_j,
\end{equation}
which transform together in the $(\bf{2N,16+4N})$ bifundamental representation of $G\times F$ = $Sp(N)\times SO(16+4N)$, which coincides with the representation in which the matter content of the 6d (1,0) SCFT on the tensor branch transforms. The fermion $\psi^i$ is neutral under $U(1)_v$, and therefore these fields contribute to the $q^{-c_L/24+1/2}v^{N+1}$ term in the $v$-expansion of the elliptic genus.\newline

\noindent At the same energy level one can form further states by considering combinations of multiple of $q^{(0)}_i$ and $\widetilde q^{(0)}_i$ with a single $\psi^{(1/2)}_i$. Such fields will still transform in the $16+4N$-dimensional vector representation of $F$, but the precise representation of $G$ will be affected by the existence of relations. By examining the explicit expressions for the elliptic genus, we find the first excited level terms take the following form:
\begin{align}\label{eq:1stexcspn}
\mathcal{E}_1^{Sp(N)}(\mass_{Sp(N)},\mass_{SO(16+4N)},v,q)&\bigg\vert_{q^{-\frac{1}{3}-\frac{N}{4}+\frac{1}{2}}} =\nonumber\\
&v^N\sum_{k=0}^\infty v^{2k+1}\chi_{\omega_1}^{SO(16+4N)}(\mass_{SO(16+4N)})\chi_{\omega_1+k\cdot\th_{Sp(N)}}^{Sp(N)}(\mass_{Sp(N)}),
\end{align}
where for both $SO(16+4N)$ and $Sp(N)$ $\omega_1$ indicates the highest weight $(10\dots 0)$ of the fundamental representation, which is respectively $16+4N$ and $2N$-dimensional.\newline

\noindent In other words, the ground state and first excited level contributions to the elliptic genus take a very simple form which is determined in terms of basic data of the 6d (1,0) SCFT: its gauge symmetry and matter content. As we will discuss in section \ref{sec:genspec} this feature generalizes to all BPS string CFTs \hh{n}{G}.\newline

\noindent On the other hand, at the lowest energy level of the Ramond elliptic genus we find a single state:
\begin{equation}
\mathbb{E}_1^{Sp(N)}(\mass_{Sp(N)},\mass_{SO(16+4N)},v,q) =\bigg\vert_{q^{-1/3}}=1.
\end{equation}
This is a peculiarity of the theories \hh{n}{G} with $n=1$; the theories with $n>1$, on the other hand, have a rich set of lowest-energy states in the Ramond sector. We will give a more detailed analysis of the low energy states in the Ramond sector in section \ref{sec:5dlim}.
\section{Universal features of BPS string CFTs}\label{sec:univ}

In this section we extend our discussion in the previous section of the \hh{1}{Sp(N)} theories to arbitrary \hh{n}{G} theories. We find several features which are common to all theories of BPS strings. This section is organized as follows: in section \ref{sec:sigma} we discuss the interpretation of the \hh{n}{G} CFTs as nonlinear sigma models on the moduli space of one $G$-instanton, and use this realization to compute their central charges. In section \ref{sec:wzwuniv} we argue that the 6d flavor symmetry $F$ is realized on the string as a current algebra whose level we determine. In section \ref{sec:gengauge} we discuss the realization of the 6d gauge symmetry in the \hh{n}{G} theory, and find evidence that it contributes to the elliptic genus in terms of irreducible characters of the affine $G$ Kac-Moody algebra at level $-n$. Finally, in section \ref{sec:genspec} we comment on the existence of a spectral flow relating the states that contribute to the NS--R and R--R elliptic genera.

\subsection{BPS string CFTs as $\mathcal{N}=(0,4)$ NLSMs}\label{sec:sigma}

An $\mathcal{N}=(0,4)$ NLSM can be described concisely in (0,1) superspace $\Sigma^{(0,1)}$, which consists of two bosonic coordinates $z,\overline{z}$ and one fermionic coordinate $\theta^-$, as follows \cite{Howe:1988cj,Papadopoulos:1993mf}: the field content for a generic $\mathcal{N}=(0,1)$ nonlinear sigma model consists of $D$ bosonic superfields $\phi^i(z,\overline{z},\theta^-),$  $i=1,\dots,D$, defining a map 
\begin{equation}
\phi: \Sigma^{(0,1)}\rightarrow \mathcal{M}
\end{equation}
into a Riemaniann manifold $\mathcal{M}$,\footnote{ Equivalently, the bosonic component of $\phi$ defines a map from $\mathbb{R}^2$ to $\mathcal{M}$ and the fermionic component defines a section of $S_-\otimes\phi^*(T\mathcal{M})$.} and $n$ fermionic superfields $\psi^a_+(z,\overline{z},\theta^-)$ defining a section $\psi$ of the bundle $S_+\otimes\phi^* E$, where $S_+$ is the spinor bundle over $\Sigma^{(0,1)}$ and $E$ is an $n$-dimensional vector bundle. The sigma model depends on the metric $g_{ij}$ and B-field $B_{ij}$ on $\mathcal{M}$, as well a connection $\Omega$ and a covariantly constant metric $h_{ab}$ on $E$. The action
\begin{equation}
S = \int \mathrm{d}z\mathrm{d}\overline{z}\, \mathrm{d}\theta^- \bigg[D_-\phi^i\partial_+\phi^j(g_{ij}+B_{ij})+i\psi^a_+\nabla_-\psi^b_+ h_{ab}\bigg],
\end{equation}
where $D_-=i\partial_{\theta^-}+\theta^-\partial_{\overline{z}}$ and $\nabla_-\psi_+^b=(D_-\psi^b_-+D_-\phi^i\Omega{}_i^b{}_c\psi^c_+)$, defines a nonlinear sigma model with $\mathcal{N}=(0,1)$ supersymmetry.\newline

\noindent The requirement that the NLSM display $\mathcal{N}=(0,4)$ supersymmetry imposes additional constraints on the data of the model. First of all, the target manifold $\mathcal{M}$ must be a hyperk\"ahler manifold of quaternionic dimension $K=D/4$. The metric $g_{ij}$ must be Hermitian with respect to each of the three complex structures. Furthermore, the holonomy of the connection
\begin{equation}
\Gamma_{ij}^{(+)K} = \hat\Gamma_{ij}^K+H_{ij}^K,
\end{equation}
where $\hat\Gamma_{ij}^K$ is the Christoffel connection on $\mathcal{M}$ and $H_{ijk}=\frac{3}{2}\partial_{[i}B_{jk]}$, must be a subgroup of $Sp(K)$. Finally, the bundle $E\otimes \mathbb{C}$ must be holomorphic with respect to all three complex structures.\newline

\noindent The NLSMs describing the \hh{n}{G} theories consist of two separate components. The first component describes the propagation of the string in $\mathbb{R}^4$ and is given by a $\mathcal{N}=(0,4)$ NLSM with $\mathcal{M}=\mathbb{R}^4$ and trivial chiral fermion bundle; this is just the theory of one free $(0,4)$ hypermultiplet and has $c_L=4,c_R=6$. The second component describes the `internal' degrees of freedom of the string and has as target space the (reduced) moduli space of one $G$-instanton, $\widetilde{\mathcal{M}}_{G,\,1}$. The information about the 6d matter content is reflected by the choice of chiral spinor bundle (for instance, for the \hh{1}{Sp(N)} theories the $F=SO(16+4N)$ flavor symmetry that rotates the 6d hypermultiplets is carried by $16+4N$ fermionic superfields).\newline

\noindent Two issues afflict the latter component of the nonlinear sigma model. First, the target space is noncompact, which leads to a continuum of bosonic vacua. This is cured by turning on chemical potentials for the $G\times SU(2)_R$ isometries of $\widetilde{\mathcal{M}}_{G,\, 1}$, where $SU(2)_R$ acts as hyperk\"ahler rotations on the three complex structures of  $\widetilde{\mathcal{M}}_{G,\, 1}$. This leads to a still infinitely degenerate, but discrete, set of vacua. Second, the moduli space of one $G$-instanton has a singularity corresponding to the instanton shrinking to zero size. Lifting the ADHM quantum mechanics to 2d suggests a UV description flowing to the NLSM in the IR which is reliable at finite instanton size, but is potentially ill-behaved at the small-instanton singularity \cite{Witten:1994tz}; however, alternative UV completions have been found whose low energy behavior is identical to the ADHM model everywhere except at the small-instanton singularity, where one finds additional localized degrees of freedom \cite{Kim:2015fxa} leading to a consistent UV description of the NLSM.\newline

\noindent Regardless of these complications, we will be able to infer several nontrivial properties of the BPS strings and their elliptic genera from basic properties of the sigma model. We begin in this section by re-deriving the \hh{n}{G} central charges \eqref{eq:central_charges}. It is easiest to begin by considering the maximally Higgsed theories for any $n$. First of all, recall that for $n=3,4,5,6,8,12$ the strings are instantons in theories without matter. The chiral fermion bundle is therefore  trivial, and the only contributions come from the $4(h^\vee_G-1)$ noncompact bosons describing the target space $\widetilde{\mathcal{M}}_{G,\, 1}$ and an equal number of anti-chiral spinors \cite{DelZotto:2016pvm}. Furthermore, using the relation $h^\vee_G = 3(n-2)$ which holds for these theories, we obtain
\begin{equation}
c_L = 4\left(3n-7\right) \qquad \text{and}\qquad c_R = 6\left(3n-7\right).
\end{equation}
The M-string and E-string cases are different: as explained in section \ref{sec:higgsingchains} they can respectively be thought of as theories with $G=SU(1)$ and $Sp(0)$, for which $h^\vee_G = 1$ and the moduli space of instantons is trivial. Nevertheless, both theories support a bundle of chiral fermions. In the M-string case this consists of four fermions \cite{Haghighat:2013gba}, leading to $c_L=2$ and $c_R=0$ (i.e. $c_L=c_R=6$ upon including the center of mass hypermultiplet contribution). The E-string CFT on the other hand consists of 16 chiral fermions, leading to $c_L=8$ and $c_R=0$.\newline

\noindent The central charges for Higgsable theories can be computed from the ones of the maximally Higgsed theories as follows: the relevant deformation of the (0,4) CFT of the Higgsable theory that initiates the Higgs RG flow involves giving mass to an equal number of chiral fermionic superfields $\psi^a_+$ and of bosonic superfields $\phi^i$ (in particular, this is consistent with the fact that the gravitational anomaly is invariant under the RG flow). In particular, the difference between the rank of the chiral fermion bundle and the dimension of the target space the same before and after Higgsing. The dimension of the target space is $4(h^\vee_G-1)$, and therefore upon Higgsing from the \hh{n}{G} theory to the \hh{n}{G_0} theory one has
\begin{equation}
c_L \to c_L-6(h^\vee_G-h^\vee_{G_0}),\qquad c_R \to c_R-6(h^\vee_G-h^\vee_{G_0}).
\end{equation}

\begin{table}[t!]
\begin{center}
\begin{tabular}{|c|c|c|}
\hline
&L&R\\\hline
\# noncompact bosons &$4(h^\vee_G-1)$&$4(h^\vee_G-1)$\\
\# fermions &$4h^\vee_G-12(n-2)$&$4(h^\vee_G-1)$\\\hline
\end{tabular}
\caption{Degrees of freedom of the \hh{n}{G} NLSM.}
\label{tab:ferbos}
\end{center}
\end{table}

\noindent This fixes the number of bosonic and fermionic degrees of freedom of the BPS string CFTs; the results are summarized in table \ref{tab:ferbos}. The only loophole in the argument above is for the \hh{7}{E_7} theory, which is maximally Higgsed but has nontrivial matter content. However, the number of chiral fermions given in table \ref{tab:ferbos} depends linearly on $n$ (or equivalently on the number of defects in the D3 brane worldvolume from the perspective of section \ref{sec:BPSstrings}); each defect contributes a fixed number of chiral fermions; this in particular also holds for $G=E_7$ for $n\neq 7$, so by extension this shows that table \ref{tab:ferbos} also applies to the \hh{7}{E_7} theory. \\

\noindent From the data of table \ref{tab:ferbos} one recovers immediately the correct central charges for the nonlinear sigma models:
\begin{align}
c_L &= 6(h^\vee_G-n+2)-4,\\
c_R &= 6h^\vee_G-6,
\end{align}
consistent with equation \eqref{eq:central_charges} (with $Q=1$). For a generic number $Q$ of strings, the same line of reasoning leads to the following prediction for the numbers of bosonic and fermionic degrees of freedom in the NLSM:\newline

\begin{table}[h!]
\begin{center}
\begin{tabular}{|c|c|c|}
\hline
&L&R\\\hline
\# noncompact bosons &$4(Qh^\vee_G-1)$&$4(Qh^\vee_G-1)$\\
\# fermions &$4Qh^\vee_G-12Q(n-2)$&$4(Qh^\vee_G-1)$\\\hline
\end{tabular}
\caption{Degrees of freedom of the NLSM describing a bound states of $Q$ strings for the 6d (1,0) SCFT with gauge algebra $G$ and string Dirac pairing $-n$.}
\label{tab:ferbosQ}
\end{center}
\end{table}

\noindent From table \ref{tab:ferbos} one can also read off the ground state energies in the Ramond-Ramond sector of the \hh{n}{G} NLSM:\footnote{ See also the recent paper \cite{Kim:2018gjo} for a similar discussion.} 
\begin{equation}
E^{Ramond}_{L,R}=-\frac{1}{24}(\#\text{bosons}_{L,R}-\#\text{fermions}_{L,R});
\end{equation}
this gives of course $E^{Ramond}_{R}=0$ consistent with supersymmetry, whereas on the chiral side one finds
\begin{equation}\label{eq:eramondr}
E^{Ramond}_{L} = \frac{1}{6}-\frac{n-2}{2}.
\end{equation}
This determines the leading order power of $q$ appearing in the Ramond elliptic genus, with the single exception of the \hh{1}{G} theories, as we will discuss in section \ref{sec:5dlim}. The general statement will turn out to be that for an arbitrary \hh{n}{G} theory
\begin{equation}\label{eq:ellimo}
\mathbb{E}_{n}^{G}(\mass_{G},\mass_{F},v,q) = q^{E_L^{Ramond}}\left(\delta_{n,1}q^{-1}+f_{n}^{G}(\mass_G,\mass_F,v)\,q^0+\mathcal{O}(q)\right),
\end{equation}
where $f_{n}^{G}(\mass_G,\mass_F,v)$ includes a combination of Ramond ground state contributions and, for $n=1$, excitations above the $L_0=-1$ singlet state (see section \ref{sec:5dlim}).\newline

\noindent Similarly, for $Q$ strings one finds from table \ref{tab:ferbosQ} that the ground state energy is given by
\begin{equation}\label{eq:eramondr2}
E^{Ramond}_{R} = \frac{1}{6}-Q\frac{n-2}{2}.
\end{equation}

\subsection{Current algebra realization of $F$ (I)}\label{sec:wzwuniv}
In this section we discuss the realization of the flavor symmetry of the 6d SCFT in terms of a current algebra in the \hh{n}{G} CFTs. Indeed, the bosonic superfields describe the propagation on the one $G$-instanton moduli space, and are not charged under the flavor symmetry $F$; the only fields that are charged under it are the fermionic superfields $\psi^a_+$, which contribute chirally to the CFT. This is also consistent with the F-theory picture where the flavor symmetry is carried by the (0,4) defects on the D3 worldvolume, which arise from seven-branes intersecting the D3 brane along a codimension 1 locus and therefore contribute chirally to the topologically twisted theory on the D3 brane.\newline 

\noindent In particular, this suggests that the 6d flavor symmetry $F$, which is also a flavor symmetry of the \hh{n}{G} theories, is realized in terms of chiral current algebras. Let us write $F$ as a product of $n_F$ simple or abelian factors, $F=\prod_{i=1}^{n_F} F_i$. Then, the currents $j^a_{F_i}(z),$ $a=1,\dots,\,\text{rank}(F_i)$ obey the OPE
\begin{equation}
j^a_{F_i}(z)j^b_{F_i}(w) \sim \frac{k_i \delta^{ab}}{(z-w)^2} +\sum_c i f^{F_i}_{abc} \frac{j_{F_i}^c(w)}{z-w},
\end{equation}
where $f^{F_i}_{abc}$ are the structure constants of $F_i$, and the coefficient of the most singular term determines the level $k$.\footnote{ For abelian factors, which correspond to a boson of compact radius $R$, $k_i$ is replaced by $R^2$.} The level can be read off from the anomaly polynomial, and it is given by 
\be
\Sigma \cdot \mathbf{b}_F \qquad\text{or}\qquad \Sigma \cdot \mathbf{b}_{U(1)}.
\ee

\noindent The WZW model corresponding to each simple factor has central charge
\begin{equation}
c_{F_i} = \frac{\text{dim}(F_i)\, k}{h^\vee_{F_i}+k}, 
\end{equation}
and a set of WZW primaries which are in one-to-one correspondence with the integrable level $k$ representations of $F$; abelian factors have $c_{U(1)}=1$ and a set of WZW primaries detailed in section \ref{sec:WZW}. The total central charge associated to $F$ is given by:
\begin{equation}\label{eq:cFcharge}
\boxed{c_F = \sum_{i|\,F_i\text{ simple}}\frac{\text{dim}(F_i)k_i}{h^\vee_{F_i}+k_i}+\sum_{i|\,F_i\text{ abelian}}1.}
\end{equation}
This proposal implies that the Hilbert space of the \hh{n}{G} CFT decomposes as:
\begin{equation}\label{eq:hilbdec}
\boxed{H_n^{G} = \bigoplus_{\vec{\lambda}} \left(\bigotimes_{i=1}^{n_F}H^{WZW_{F_i}}_{\lambda_i}\right) \otimes H^{residual}_{\vec{\lambda}},}
\end{equation}
where the $i$-th entry of $\vec\lambda=(\lambda_1,\dots,\lambda_{n_F})$ runs over the primaries of the WZW$_{F_i}$ model. While the WZW sector of the CFT is chiral, the spectrum of the residual part of the CFT includes both chiral and anti-chiral components and couples to $SU(2)_R$ and $G$.\newline

\noindent From \eqref{eq:hilbdec} it follows moreover that the R--R elliptic genus for any BPS string CFT decomposes in terms of characters $\widehat\chi^{F_i}_{\lambda_i}(\mass_{F_i},q)$ of the WZW model for $F_i$ in the following way:
\begin{equation}\label{eq:elldecomp}
\boxed{\mathbb{E}_{n}^{G}(\massG,\massF,v,q) = \sum_{\vec\lambda}\big(\prod_{i=1}^{n_F}\widehat\chi^{F_i}_{\lambda_i}(\mass_{F_i},q)\big)\xi_{\vec\lambda}^{n,G}(\massG,v,q).}
\end{equation}

\noindent The $\widehat\chi_\lambda^{F_i}(\mass_{F_i},q)$ are vector-valued Jacobi forms with modular parameter $\tau$ and (exponentiated) elliptic parameters $\mass_{F_i}$, and therefore the functions $\xi_{\vec\lambda}^{n,G}(\massG,v,q)$ must also transform as a vector-valued Jacobi form with modular parameter $\tau$ and (exponentiated) elliptic parameters $\mass_G $ and $v$. The modular $S$-matrix for this Jacobi form can be easily determined by requiring that the elliptic genus transform as a scalar Jacobi form with the correct modular anomaly.\newline 

\noindent We now turn to a discussion of the $G\times SU(2)_R$-dependent part of the CFT, and will return to a more in-depth discussion of the flavor symmetry $F$ in section \ref{sec:wzwunivii}.

\subsection{Worldsheet realization of $G$}\label{sec:gengauge}
In section \ref{sec:spngv} we found that the 6d gauge symmetry $G$ contributes to the elliptic genus according to equation \eqref{eq:spngans}, which we furthermore refined by showing that the $Sp(N)$ representations that appear organize themselves in terms of irreducible characters of the affine $Sp(N)$ Kac-Moody algebra at level $-1$. In this section we discuss the generalization to other \hh{n}{G} theories. For the theories for which $n\neq h^\vee_G$ we discuss a realization of the 6d gauge symmetry in terms of an affine $G$ Kac-Moody algebra at level $-n$, in agreement with the expectation from the coefficient of $c_2(G)$ in the anomaly polynomial of the \hh{n}{G} theory, equation \eqref{eq:anpol}.\newline

\noindent The representation theory of affine Kac-Moody algebras is different for the three cases $k<-h^\vee_G,$ $k=-h^\vee_G,$ and $k>-h^\vee_G$. A survey of the \hh{n}{G} theories shows that the case $k<-h^\vee_G$ occurs only for the M-string CFT \hh{2}{SU(1)}, the critical case $k=-h^\vee_G$ occurs for the three theories \hh{3}{SU(3)}, \hh{2}{SU(2)}, and \hh{1}{Sp(0)}, while all other theories belong to the case $k>-h^\vee_G$. The critical case is subtle and requires a separate discussion, which we postpone to section \ref{sec:excep}. On the other hand, for $k \neq -h^\vee_G$ we find the following generalization of equation \eqref{eq:spngans}:

\begin{equation}\label{eq:FGv}
\begin{tabular}{|c|}
\hline 
$ \displaystyle{\xi_{\vec\lambda}^{n,G}(\massG,v,q) =\hspace{4.2in}}$\\
$\displaystyle{\sum_{\nu\in \text{Rep}(G)}\,\sum_{\ell=-2\vert n-h^\vee_G\vert+1}^{0}\,\sum_{m\in\mathbb{Z}} n^{\vec\lambda}_{\nu,\ell,m}\times \frac{q^{-\frac{c_G}{24}+h^G_\nu}\chi^G_\nu(\massG)}{\widetilde\Delta_{G}(\mass_{G},q)}\times \frac{q^{-\frac{c_v}{24}+h^v_{\ell,m}}v^{\ell+2(n-h^\vee_G)m}}{\prod_{j=1}^\infty(1-q^j)^{}},}$\\
\hline
\end{tabular}
 \end{equation}
which we have verified by comparing against known results for the elliptic genera of a large number of theories. The sum runs over all representations of $G$, and
\begin{align} \label{eq:cGv}
c_G &= \frac{\text{dim}(G)(-n)}{h^\vee_G-n},\qquad h^G_{\nu} = \frac{\langle \nu,\nu+2\rho_G\rangle}{2(h^\vee_G-n)},\nonumber\\
c_v &= 1,\qquad\qquad\qquad\,\,\,\,\, h^v_{\ell,m} = \frac{(\ell+2(n-h^\vee_G)m)^2}{4(n-h^\vee_G)},
\end{align}
are the central charges and conformal dimensions for the $G_{-n}$ and $U(1)_{n-h^\vee_G}$ WZW models (see appendix \ref{sec:WZW}), suggestive that the 6d gauge symmetry $G$ is again realized in terms of an affine Kac-Moody algebra at level $k=-n$, and the $U(1)_v$ symmetry generated by $J^3_R+J^3_I$ is realized in terms of a $U(1)_{n-h^\vee_G}$ algebra; we will return to this shortly. \newline

\noindent Analogously to the $n=1,G=Sp(N)$ case, the $n^{\vec\lambda}_{\nu,\ell,m}$ coefficients must obey various constraints:
\begin{itemize}
\item The $n^{\vec\lambda}_{\nu,\ell,m}$ must be integers, which follows immediately from the integrality of the coefficients of the elliptic genus.
\item A look at equations \eqref{eq:FGv} and \eqref{eq:cGv} reveals that for $n<h^\vee_G$ only a finite number of terms in the sum over the integer $ m$ would correspond to states of energy above the ground state energy, since,  $n<h^\vee_G$ implies that $h^v_{\ell,m}$ is negative and grows quadratically with $m$. As the elliptic genus describes a unitary theory whose energy is bounded from below, for fixed $\lambda, \nu,$ and $\ell$ all but a finite number of coefficients $n^{\vec\lambda}_{\nu,\ell,m}$ can be nonvanishing.
\item The level-matching condition for the CFT on the torus forces the energy of any state that contributes to the elliptic genus to differ from the Ramond ground state energy $-\frac{(c_R-c_L)}{12}$ by an  integer. For many choices of $\nu,\ell,$ and $m$, the coefficient of $n^{\vec\lambda}_{\nu,\ell,m}$ in the elliptic genus must therefore vanish, since the corresponding term in equation \eqref{eq:FGv}  would violate the level-matching condition.
\end{itemize}

\noindent We next turn to a few examples. First of all, the M-string CFT \hh{2}{SU(1)} is the only case for which $h^\vee_G < n$, and it is interesting to see how equation \eqref{eq:FGv} captures the elliptic genus. As the group $G=SU(1)$ is trivial, only the $\bf 1$ representation contributes. Moreover since the dependence on $v$ is holomorphic one expects the $\xi^{2,SU(1)}_\lambda$ functions to be expressed in terms of affine characters for $SU(2)_R$ at level $2-h^\vee_{SU(1)}= 1$. Indeed, the M-string elliptic genus
\begin{equation}
\mathbb{E}_2^{SU(2)}(\mass_{SU(2)},v,q) = \frac{\theta_1(v\, m^1_{SU(2)},q)\theta_1(v/m^1_{SU(2)},q)}{\eta(q)^2},
\end{equation}
can be written in terms of level 1 characters of $F=SU(2)$ and of $SU(2)_v$ as follows:\footnote{ Since the elliptic genus is holomorphic in $v$ rather than meromorphic, it can be expressed in terms of representations of the $SU(2)_v$ diagonal subgroup of $SU(2)_R\times SU(2)_I$, and not solely of its Cartan subgroup $U(1)_v$.} 
\begin{equation}
\mathbb{E}_2^{SU(2)}(\mass_{SU(2)},v,q) = \widehat{\chi}_{\bf 1}^{F}(\massF,q)\widehat{\chi}_{\bf 2}^{SU(2)_R}(v,q)-\widehat{\chi}_{\bf 2}^{F}(\massF,q)\widehat{\chi}_{\bf 1}^{SU(2)_R}(v,q).
\end{equation}
On the other hand equation \eqref{eq:FGv} reads:
\begin{equation}\label{eq:FAAGv}
\xi_\lambda^{2,SU(1)}(\massG,v,q) =\sum_{\ell=-1}^{0}\,\sum_{m\in\mathbb{Z}} n^\lambda_{\mathbf{1},\ell,m}\frac{q^{(\ell/2+m)^2}v^{\ell+2\,m}}{\eta(q)}.
 \end{equation}
Comparing this with equations \eqref{eq:su2ch} and \eqref{eq:su2ch2} for the SU(2) level 1 characters, one finds that
\begin{equation}
n^{\bf 1}_{\mathbf{1},1,m}=-n^{\bf 2}_{\mathbf{1},0,m}=1\qquad \text{and}\qquad n^{\bf 1}_{\mathbf{1},0,m}=n^{\bf 2}_{\mathbf{1},1,m}=0\qquad \forall\, m\in\mathbb{Z}.
\end{equation}\\

\begin{table}[t!]\begin{center}
\begin{tabular}{|r|r|}\hline
$\nu$ &$\widehat{\chi}^{SO(8)}_\nu(\mass_{SO(8)},q)\cdot \widetilde\Delta_{SO(8)}(\mass_{SO(8)},q)$\\\hline
$(0000)$& $q^{\frac{56}{24}}\left[\chi^{SO(8)}_{(0000)}-2\chi^{SO(8)}_{(0100)} q^3+(\chi^{SO(8)}_{(2000)}+\chi^{SO(8)}_{(0020)}+\chi^{SO(8)}_{(0002)}) q^4+3\chi^{SO(8)}_{(0200)}q^7\right.$\\
&$\quad\quad\,\,\left.-2(\chi^{SO(8)}_{(2100)}+\chi^{SO(8)}_{(0120)}+\chi^{SO(8)}_{(0102)})q^{8}+(\chi^{SO(8)}_{(2020)}+\chi^{SO(8)}_{(2002)}+\chi^{SO(8)}_{(0022)})q^{9}\right.$\\
&$\quad\quad\,\,\left.+(\chi^{SO(8)}_{(4000)}+\chi^{SO(8)}_{(0040)}+\chi^{SO(8)}_{(0004)})q^{10}+\mathcal{O}(q^{12})\right]$\\
$(0100)$& $q^{\frac{56}{24}+3}\left[\chi^{SO(8)}_{(0100)}-(\chi^{SO(8)}_{(2000)}+\chi^{SO(8)}_{(0020)}+\chi^{SO(8)}_{(0002)})  q+\chi^{SO(8)}_{(1011)}q^3-3\chi^{SO(8)}_{(0200)}q^{4}\right.$\\
&$\quad\quad\,\,\left.+2(\chi^{SO(8)}_{(2100)}+\chi^{SO(8)}_{(0120)}+\chi^{SO(8)}_{(0102)})q^{5}-2(\chi^{SO(8)}_{(2020)}+\chi^{SO(8)}_{(2002)}+\chi^{SO(8)}_{(0022)})q^{6}\right.$\\
&$\quad\quad\,\,\left.-(\chi^{SO(8)}_{(4000)}+\chi^{SO(8)}_{(0040)}+\chi^{SO(8)}_{(0004)})q^{7}+\mathcal{O}(q^{9})\right]$\\
$(0200)$& $q^{\frac{56}{24}+7}\left[\chi^{SO(8)}_{(0200)}-(\chi^{SO(8)}_{(2100)}+\chi^{SO(8)}_{(0120)}+\chi^{SO(8)}_{(0102)})q+(\chi^{SO(8)}_{(2020)}+\chi^{SO(8)}_{(2002)}+\chi^{SO(8)}_{(0022)})q^2\right.$\\
&$\quad\quad\,\,\left.+(\chi^{SO(8)}_{(4000)}+\chi^{SO(8)}_{(0040)}+\chi^{SO(8)}_{(0004)})q^{3}+\mathcal{O}(q^{5})\right]$\\
\hline
\end{tabular}\end{center}
\caption{Irreducible characters of $SO(8)$ at level $-4$ (rescaled by an overall factor of $\widetilde\Delta_{SO(8)}(\mass_{SO(8)},q)$), for a set of highest-weight representations that contribute in the elliptic genus.}\label{tab:so8ch}
\end{table}

\noindent Next, we consider the \hh{4}{SO(8)} theory, for which $n>-h^\vee_G$ and we expect the affine $SO(8)$ Kac-Moody algebra at level $-4$ to appear. As $F$ is trivial, the elliptic genus is given by the function $\xi^{4,SO(8)}_{\bf 1}$; we find that this function can be written in terms of $SO(8)_{-4}$ irreducible representations as follows:
\begin{align}
\xi_{\bf{1}}^{4,SO(8)}(\mass_{SO(8)},v,q) &\!=\!\! \sum_{n=0}^\infty \widehat\chi^{SO(8)}_{(0n00)}(\mass_{SO(8)},q)\sum_{m=0}^2a_m\!\!\left[\!\frac{v^{-b_{m,n}+1}q^{-\frac{(b_{m,n}-1)^2}{8}}}{\eta(q)}-\frac{v^{b_{m,n}+1}q^{-\frac{(b_{m,n}+1)^2)}{8}}}{\eta(q)}\!\right]\!\!,
\end{align}
where $a_0=a_2=1$, $a_1=2$, and $b_{m,n}=2m+2n$. This expression is indeed also consistent with equation \eqref{eq:FGv}. In table \ref{tab:so8ch} we list the first few terms in the $q$ expansion of the irreducible characters of the relevant representations of $SO(8)_{-4}$, which we obtained in terms of Kazhdan-Lusztig polynomials as outlined in section \ref{sec:spngv}.\\

\begin{table}[p!]\begin{center}
\begin{tabular}{|r|r|}\hline
$\nu$ &$\widehat{\chi}^{G_2}_\nu(\mass_{G_2},q)\cdot \widetilde\Delta_{G_2}(\mass_{G_2},q)$\\\hline
$(00)$& $q^{\frac{7}{4}}\left[\chi^{G_2}_{(00)}-\chi^{G_2}_{(11)} q^7+\chi^{G_2}_{(30)}q^8+\mathcal{O}(q^{18})\right]$\\
$(01)$& $q^{\frac{7}{4}+4}\left[\chi^{G_2}_{(01)}-\chi^{G_2}_{(11)} q^3+\chi^{G_2}_{(02)}q^6-\chi^{G_2}_{(12)} q^{10}+\chi^{G_2}_{(31)}q^{11}+\mathcal{O}(q^{14})\right]$\\
$(02)$& $q^{\frac{7}{4}+10}\left[\chi^{G_2}_{(02)}-\chi^{G_2}_{(40)} q^2-\chi^{G_2}_{(12)}q^4+\chi^{G_2}_{(31)} q^{5}+\chi^{G_2}_{(03)}q^{8}+\mathcal{O}(q^{13})\right]$\\
$(03)$& $q^{\frac{7}{4}+18}\left[\chi^{G_2}_{(03)}-\chi^{G_2}_{(41)} q^2+\chi^{G_2}_{(60)}q^4-\chi^{G_2}_{(13)}q^5+\chi^{G_2}_{(32)} q^{6}+\chi^{G_2}_{(04)}q^{10}-\chi^{G_2}_{(42)}q^{12}+\mathcal{O}(q^{16})\right]$\\
$(04)$& $q^{\frac{7}{4}+28}\left[\chi^{G_2}_{(04)}-\chi^{G_2}_{(42)} q^2+\chi^{G_2}_{(61)}q^4-\chi^{G_2}_{(14)}q^6+\chi^{G_2}_{(33)} q^{7}-\chi^{G_2}_{(71)}q^{11}+\chi^{G_2}_{(05)}q^{12}+\mathcal{O}(q^{14})\right]$\\
$(05)$& $q^{\frac{7}{4}+40}\left[\chi^{G_2}_{(05)}\!-\!\chi^{G_2}_{(43)} q^2+\chi^{G_2}_{(62)}q^4-\chi^{G_2}_{(15)}q^7+\chi^{G_2}_{(34)} q^{8}-\chi^{G_2}_{(10,0)}q^{10}-\chi^{G_2}_{(72)}q^{12}+\mathcal{O}(q^{14})\right]$\\
$(06)$& $q^{\frac{7}{4}+54}\left[\chi^{G_2}_{(06)}-\chi^{G_2}_{(44)} q^2+\chi^{G_2}_{(63)}q^4-\chi^{G_2}_{(16)}q^8+\chi^{G_2}_{(35)} q^{9}-\chi^{G_2}_{(10,1)}q^{10}+\mathcal{O}(q^{13})\right]$\\
$(07)$& $q^{\frac{7}{4}+70}\left[\chi^{G_2}_{(07)}-\chi^{G_2}_{(45)} q^2+\chi^{G_2}_{(64)}q^4-\chi^{G_2}_{(17)}q^9+(\chi^{G_2}_{(36)}-\chi^{G_2}_{(10,2)})q^{10}+\mathcal{O}(q^{14})\right]$\\
$(08)$& $q^{\frac{7}{4}+88}\left[\chi^{G_2}_{(08)}-\chi^{G_2}_{(46)} q^2+\chi^{G_2}_{(65)}q^4-(\chi^{G_2}_{(18)}+\chi^{G_2}_{(10,3)})q^{10}+\chi^{G_2}_{(37)}q^{11}+\mathcal{O}(q^{14})\right]$\\
$(09)$& $q^{\frac{7}{4}+108}\left[\chi^{G_2}_{(09)}-\chi^{G_2}_{(47)} q^2+\chi^{G_2}_{(66)}q^4-\chi^{G_2}_{(10,4)}q^{10}-\chi^{G_2}_{(19)}q^{11}+\chi^{G_2}_{(3,8)}q^{12}+\mathcal{O}(q^{14})\right]$\\
$(0,\!10)$& $q^{\frac{7}{4}+130}\left[\chi^{G_2}_{(0,10)}+\mathcal{O}(q^{2})\right]$\\
$(0,\!11)$& $q^{\frac{7}{4}+154}\left[\chi^{G_2}_{(0,11)}+\mathcal{O}(q^{2})\right]$\\
$(10)$& $q^{\frac{7}{4}+2}\left[\chi^{G_2}_{(10)}-\chi^{G_2}_{(01)} q^2-\chi^{G_2}_{(02)}q^8+\chi^{G_2}_{(40)}q^{10}+\chi^{G_2}_{(12)} q^{12}+\mathcal{O}(q^{13})\right]$\\
$(11)$& $q^{\frac{7}{4}+7}\left[\chi^{G_2}_{(11)}-\chi^{G_2}_{(30)} q-\chi^{G_2}_{(02)} q^3+\chi^{G_2}_{(12)}q^7-2\,\chi^{G_2}_{(03)}q^{11}+\mathcal{O}(q^{13})\right]$\\
$(12)$& $q^{\frac{7}{4}+14}\left[\chi^{G_2}_{(12)}-\chi^{G_2}_{(31)} q-\chi^{G_2}_{(03)} q^4+\chi^{G_2}_{(41)}q^6+\chi^{G_2}_{(13)}q^{9}-\chi^{G_2}_{(32)}q^{10}+\mathcal{O}(q^{14})\right]$\\
$(13)$& $\!\!q^{\frac{7}{4}+23}\!\left[\!\chi^{G_2}_{(13)}\!-\!\chi^{G_2}_{(32)} q\!+\!(\chi^{G_2}_{(70)}\!-\!\chi^{G_2}_{(04)}) q^5\!+\!\chi^{G_2}_{(42)}q^7\!-\!\chi^{G_2}_{(61)}q^{9}\!+\!\chi^{G_2}_{(14)}q^{11}\!-\!\chi^{G_2}_{(33)}q^{12}\!+\!\mathcal{O}(q^{17})\right]$\\
$(14)$& $q^{\frac{7}{4}+34}\left[\chi^{G_2}_{(14)}-\chi^{G_2}_{(33)} q+\chi^{G_2}_{(71)} q^5-\chi^{G_2}_{(05)}q^6+(\chi^{G_2}_{(43)}-\chi^{G_2}_{(90)})q^{8}-\chi^{G_2}_{(62)}q^{10}+\mathcal{O}(q^{13})\right]$\\
$(15)$& $q^{\frac{7}{4}+47}\left[\chi^{G_2}_{(15)}-\chi^{G_2}_{(34)} q+\chi^{G_2}_{(72)} q^5-\chi^{G_2}_{(06)}q^7-\chi^{G_2}_{(91)}q^{8}+\chi^{G_2}_{(44)}q^{9}-\chi^{G_2}_{(63)}q^{11}+\mathcal{O}(q^{15})\right]$\\
$(16)$& $q^{\frac{7}{4}+62}\left[\chi^{G_2}_{(16)}-\chi^{G_2}_{(35)} q+\chi^{G_2}_{(73)} q^5-(\chi^{G_2}_{(92)}+\chi^{G_2}_{(07)})q^8+\chi^{G_2}_{(45)}q^{10}-\chi^{G_2}_{(64)}q^{12}+\mathcal{O}(q^{16})\right]$\\
$(17)$& $q^{\frac{7}{4}+79}\left[\chi^{G_2}_{(17)}-\chi^{G_2}_{(36)} q+\chi^{G_2}_{(74)} q^5-\chi^{G_2}_{(93)}q^8-\chi^{G_2}_{(08)}q^{9}+\chi^{G_2}_{(46)}q^{11}+\mathcal{O}(q^{13})\right]$\\
$(18)$& $q^{\frac{7}{4}+98}\left[\chi^{G_2}_{(18)}-\chi^{G_2}_{(37)} q+\chi^{G_2}_{(75)} q^5-\chi^{G_2}_{(94)}q^8-\chi^{G_2}_{(09)}q^{10}+\chi^{G_2}_{(47)}q^{12}+\mathcal{O}(q^{14})\right]$\\
$(19)$& $q^{\frac{7}{4}+119}\left[\chi^{G_2}_{(19)}-\chi^{G_2}_{(38)} q+\mathcal{O}(q^{5})\right]$\\
$(1,\!10)$& $q^{\frac{7}{4}+142}\left[\chi^{G_2}_{(1,10)}+\mathcal{O}(q)\right]$\\
\hline
\end{tabular}\end{center}
\caption{Irreducible level $-3$ characters of $G_2$  (rescaled by an overall factor of $\widetilde\Delta_{G_2}(\mass_{G_2},q)$), for a set of highest-weight representations that contribute to the elliptic genus. The expansion is given to an order sufficient to compute the elliptic genus to $\mathcal{O}(q^{13})$ and $\mathcal{O}(v^{22})$.}\label{tab:g2ch}
\end{table}

\noindent Finally, let us examine the \hh{3}{G_2} theory, for which $F=Sp(1)$ at level 1. In this case we expect the affine $G_2$ Kac-Moody algebra at level $-3$ to appear. The elliptic genus can be written as
\begin{align}
\mathbb{E}_{3}^{G_2}(\mass_{G_2},\mass_{Sp(1)},v,q) &= \widehat\chi_{(0)}^{Sp(1)}(\mass_{Sp(1)},q)\xi_{(0)}^{3,\,G_2}(\mass_{G_2},v,q)\nonumber\\
&+\widehat\chi_{(1)}^{Sp(1)}(\mass_{Sp(1)},q)\xi_{(1)}^{3,\,G_2}(\mass_{G_2},v,q).
\end{align}
We find that the $\mass_{G_2}$ dependent pieces of the elliptic genus admit the following expansions:
\begin{align}
\xi_{(0)}^{3,G_2}(\mass_{G_2},v,q) &= \sum_{n=0}^\infty \widehat\chi^{G_2}_{(0n)}(\mass_{G_2},q)\left[\frac{v^{-2(n+1)}q^{-(n+1)^2}}{\eta(q)}-\frac{v^{2n}q^{-n^2}}{\eta(q)}\!\right]\nonumber\\
&+ \sum_{n=0}^\infty \widehat\chi^{G_2}_{(1n)}(\mass_{G_2},q)\left[\frac{v^{-2n}q^{-n^2}}{\eta(q)}-\frac{v^{2(n+2)}q^{-(n+2)^2}}{\eta(q)}\right]
\end{align}
and
\begin{align}
\xi_{(1)}^{3,G_2}(\mass_{G_2},v,q) &= \sum_{n=0}^\infty \widehat\chi^{G_2}_{(0n)}(\mass_{G_2},q)\left[\frac{v^{-1+2(n+2)}q^{-\frac{(-1+2(n+2))^2}{4}}}{\eta(q)}-\frac{v^{-1-2(n-1)}q^{-\frac{(-1-2(n-1))^2}{4}}}{\eta(q)}\right]\nonumber\\
&+ \sum_{n=0}^\infty \widehat\chi^{G_2}_{(1n)}(\mass_{G_2},q)\left[\frac{v^{-1+2(n+1)}q^{-\frac{(-1+2(n+1))^2}{4}}}{\eta(q)}-\frac{v^{-1-2(n+1)}q^{-\frac{(-1-2(n+1))^2}{4}}}{\eta(q)}\right],
\end{align}
which we have verifed to hold to $\mathcal{O}(q^{13})$ and $\mathcal{O}(v^{22})$. The level $-3$ characters that appear in these expressions, which we have computed by making use of Kazhdan-Lusztig polynomials, are reported in table \ref{tab:g2ch}.\newline

\noindent We interpret these results, as well as the ones obtained in section \ref{sec:spngv} for $G= Sp(1)$ and $Sp(2)$, as an indication that the 6d gauge symmetry, at least for the theories with $n\neq h^\vee_G$, is realized in the chiral algebra of the \hh{n}{G} CFTs as an affine Kac-Moody algebra.\footnote{ We thank J. de Boer and M. Dedushenko for discussions on this point.}\,\footnote{ In section \ref{sec:excep}, on the other hand, will see that the gauge symmetry for the theories \hh{2}{SU(2)} and \hh{3}{SU(3)}, for which $n=h^\vee_G$, appears to be realized differently.} A similar phenomenon has recently been observed in a related context by Gukov and Dedushenko \cite{Dedushenko:2017osi}, who argued that the chiral algebra of the $\mathcal{N}=(0,2)$ gauge theory with SU(2) gauge group and four chiral multiplets is the $SO(8)$ affine Kac-Moody algebra at level $-2$; note that the \hh{4}{SO(8)} CFT admits a similar UV realization as a $(0,4)$ gauge theory with SU(2) gauge group and four hypermultiplets. Whether an extension of the techniques of \cite{Dedushenko:2017osi} can be used to obtain the chiral algebra of the \hh{4}{SO(8)} CFT from first principles is currently under investigation \cite{wip}.\newline

\noindent The $SU(2)_v$ symmetry should also in principle be realized in terms of an affine Kac-Moody algebra; however, the generators of the reduced moduli space of one $G$ instanton $\widetilde{\mathcal{M}}_{G,1}$ are simultaneously charged under $G$ and $SU(2)_v$, and the $SU(2)_v$ affine symmetry is broken by the expansion \eqref{eq:FGv} which makes the affine $G$ symmetry manifest. In the expansion one nevertheless still clearly recognizes the characters of the Verma modules of the $U(1)^v_{n-h^\vee_G}$ affine Kac-Moody algebra associated to the Cartan subgroup of $SU(2)_R$. Moreover, the value 
\begin{equation}
R^2 = h^\vee_G-n
\end{equation}
is precisely the coefficient multiplying the $\epsilon_+^2$ term of the anomaly polynomial of the string worldsheet theory upon substituting $c_2(R)\to -\epsilon_+^2$ and $c_2(I)\to -\epsilon_+^2$, see the discussion in section \ref{sec:modul} below. \newline

\noindent Unlike the \hh{4}{SO(8)} CFT, many of the \hh{n}{G} theories do not have a known UV  (0,4) gauge theory realization; nevertheless, our results suggest that the chiral algebras of these theories all have a similar structure which is strongly constrained by the global symmetries of the CFTs. A more detailed study of the chiral algebras of these theories is also currently under investigation and will be addressed in \cite{wip}. Finally, we remark that a different realization of chiral algebras of $\mathcal{N}=(0,4)$ CFTs along the lines of \cite{Beem:2013sza} has been proposed in \cite{Beem:2014kka}, which may also shed light on the chiral algebras of this class of CFTs. 

\subsection{NS--R elliptic genus, spectral flow, and low energy spectrum}\label{sec:genspec}
In section \ref{sec:spnns} we found of a spectral flow which relates the states that contribute to the NS--R and R--R elliptic genera for the \hh{1}{Sp(N)} theories. We conjecture that the same transformation 
\begin{equation}
v\to q^{1/2}/v
\label{eq:sftransf}
\end{equation}
leads to a spectral flow for arbitrary \hh{n}{G} theories, and as a consequence, up to a possible overall sign,
\begin{align}\label{eq:genspecflow}
\boxed{\mathcal{E}_n^{G}(\mass_{G},\mass_{F},v,q) = q^{\frac{k_{\epsilon_+}}{4}}v^{-k_{\epsilon_+}} \mathbb{E}_n^{G}(\mass_{G},\mass_{F},q^{1/2}/v,q),}
\end{align}
where $k_{\epsilon_+} = n-h^\vee_G$.\newline

\noindent Of course, this spectral flow symmetry cannot be a property of the full spectrum of the \hh{n}{G} CFT. In particular, the $SU(2)_v$ global symmetry which is used to perform the spectral flow acts nontrivially on the bosonic superfields in the NLSM and therefore is not a chiral symmetry. Nevertheless, in light of the discussion of the previous section we are led naturally to propose that the spectral flow symmetry is a symmetry of the \emph{chiral algebra} of the \hh{n}{G} CFTs, which in particular explains the relation between elliptic genera with NS--R and R--R boundary conditions on the left-moving degrees of freedom.\newline

\noindent We are able to verify the symmetry under spectral flow explicitly whenever a UV Lagrangian description for \hh{n}{G} is known. Rather than providing an extensive list of checks, for brevity we content ourselves here with demonstrating relation \eqref{eq:genspecflow} for the infinite family of theories \hh{4}{SO(8+2N)}. The R--R elliptic genus is given by \cite{Haghighat:2014vxa}:
\begin{align} &\mathbb{E}_4^{SO(8+2N)}(\mass_{SO(8+2N)},\mass_{Sp(2N)},v,q)=\nonumber\\
&\int \frac{dz}{2\pi i z}\frac{\theta_1(v^2,q)}{\eta(q)}\prod_{s=\pm1}\left[\theta_1(z^{2\,s},q)\theta_1(v^2z^{2\,s},q)\frac{\eta(q)^{7}\prod_{j=1}^{N}\theta_1(z^s m_{Sp(N)}^j,q)\theta_1(z^s \tilde{m}_{Sp(N)}^j,q)}{\prod_{j=1}^{4+N}\theta_1(v z^s m_{SO(8+4N)}^j,q)\theta_1(v z^s/m_{SO(8+4N)}^j),q)}\right],
\end{align}
where the integral is along a suitable contour determined by the Jeffrey-Kirwan prescription \cite{Benini:2013nda}, and $m^j_{Sp(N)}$ and $\tilde m^j_{Sp(N)}$ are exponentiated chemical potentials for the regular maximal $Sp(N)\times Sp(N)$ subalgebra of the $Sp(2N)$ flavor symmetry. The shift $v\to q^{1/2}/v$ only affects the terms
\begin{equation}
\prod_{s=\pm 1}\frac{\prod_{j=1}^{N}\theta_1(z^s m_{Sp(N)}^j,q)\theta_1(z^s \tilde{m}_{Sp(N)}^j,q)}{\prod_{j=1}^{4+N}\theta_1(v z^s m_{SO(8+4N)}^j,q)\theta_1(v z^s/m_{SO(8+4N)}^j),q)}
\end{equation}
in the integrand, since under this shift the remainder of the integrand only transforms up to an overall phase. Furthermore, it is possible to absorb the shift by redefining $z\to q^{1/2}z$. Then, the denominator is also only transformed up to a phase, while the numerator transforms to
\begin{equation}
\prod_{s=\pm 1}\prod_{j=1}^{N}\theta_3(z^s m_{Sp(N)}^j,q)\theta_3(z^s \tilde{m}_{Sp(N)}^j,q)
\end{equation}
which corresponds to imposing NS boundary conditions on the chiral fermions appearing in (0,4) the Fermi multiplets, or in other words to computing the NS--R elliptic genus.\newline

\noindent As a further check of the spectral flow symmetry, the strings of the minimal SCFTs with $n>3$ and no matter content, whose spectrum does not include chiral fermions, should have an elliptic genus invariant under the transformation \eqref{eq:sftransf}. Indeed, invariance under this transformation was already observed in \cite{DelZotto:2016pvm}.\newline

\noindent The same considerations that were made in section \ref{sec:spnns} for the \hh{1}{Sp(N)} case on the low energy spectrum in the NS-R sector also apply in general: imposing NS boundary conditions on the chiral fermions implies that the set of vacua of the theory are in one-to-one correspondence with the unique fermionic vacuum, tensored with gauge neutral combinations of the zero-modes of the bosonic fields of the SCFT, modulo relations; these operators are counted by the Hilbert series of the reduced moduli space of one $G$-instanton, $\widetilde{\mathcal{M}}_{G,1}$, leading to the prediction 
\begin{align}\label{eq:ellg0}
\boxed{\mathcal{E}_n^{G}(\mass_{G},\mass_{F},v,q)\bigg\vert_{q^{-\frac{c_L}{24}}} = v^{h^\vee_G-1} \sum_{k=0}^\infty v^{2k}\chi_{k\cdot \theta_G}^{G}(\mass_{G}),}
\end{align}
which indeed always turns out to be case in examples whose elliptic genus is known. Here, $\theta_G$ indicates the highest weight of the adjoint representation of $G$; the overall factor of $v^{h^\vee_G-1}$ multiplying the Hilbert series will be discussed in section \ref{sec:5dlim}.\newline

\noindent At the next excited level $L_0=1/2$, if a UV Lagrangian description is available one can again in analogy with the \hh{1}{Sp(N)} theories construct the states with $U(1)_v$ charge 
\begin{equation}
1+(h^\vee_G-1)
\end{equation}
by taking the 2d gauge group-invariant combinations of zero modes of bosonic fields and  $L_0=1/2$ modes of chiral fermionic fields; such fields transform in the representations of $G\times F$ of the 6d matter hypermultiplets (see the last column in tables \ref{tab:flavor1} and \ref{tab:flavor2}). If the 6d matter transforms in a direct sum of representations 
\begin{equation}
\bigoplus_{i=1}^r\, (R^G_i,R^F_i),
\end{equation}
we observe that the $L_0=1/2$ sector in the NS-R elliptic genus is captured by a straightforward generalization of equation \eqref{eq:1stexcspn}:
\begin{equation}\label{eq:ellg1}
\boxed{\mathcal{E}_n^{G}(\mass_{G},\mass_{F},v,q)\bigg\vert_{q^{-\frac{c_L}{24}+\frac{1}{2}}} =-v^{h^\vee_G-1}\sum_{k=0}^\infty v^{2k+1}\sum_{i=1}^r\chi_{\lambda^G_i+k\cdot\th_G}^{G}(\mass_{G})\chi_{\lambda^F_i}^{F}(\mass_F),}
\end{equation}
which we have verified hold for a large number of theories \hh{n}{G}.\footnote{ To be precise, in the majority of cases we have performed checks after setting $\massG=\massF=\bf{1}$ for semplicity, and for a smaller number of cases we have also performed checks for arbitrary $\massG$ and $\mass_F$.\label{ft:precise}} Here, $\lambda^G_i$ and $\lambda^F_i$ denote respectively the highest weights of the irreducible representation $R^G_i$.\newline

\noindent We find that the central charge of $c_L$ for the left moving degrees of freedom of the \hh{n}{G} CFT can be expressed as follows in terms of the central charges of the $F,$ $G,$ and $U(1)_v$ current algebras:
\begin{equation}\label{eq:ccrel}
\boxed{c_L = c_F + c_G + c_v-24\frac{(h^\vee_G-1)^2}{4(n-h^\vee_G)}.}
\end{equation}
which we have verified to hold for all the theories whose elliptic genus is known. The last term on the right hand side of equation \eqref{eq:ccrel} has the following explanation: the NS-R ground state does not belong to the vacuum Verma module of the $U(1)_v$ WZW model, but rather to the one with
\begin{equation}
h^v_{\ell,n} = \frac{(h^\vee_G-1)^2}{4(n-h^\vee_G)}.
\end{equation}
The values of $\ell$ and $m$ that correspond to this module are $\ell=2n-h^\vee_G-1$ and $m=-1$ for all theories \hh{n}{G}, with the exceptions of \hh{3}{G_2} for which $\ell=-1$ and $m=-2$ and of \hh{4}{SO(8)} for which $\ell=-3$ and $m=-2$. For precisely these two theories the generic assignment $\ell=2n-h^\vee_G-1$ is outside of the bound
\begin{equation}
2(n-h^\vee_G)+1\leq \ell \leq 0,
\end{equation}
and therefore one must shift $\ell\to\ell-2(n-h^\vee_G)$ (and $m\to m-1$) in order to bring $\ell$ within the appropriate range in equation \eqref{eq:FGv}.\newline

\noindent Equation \eqref{eq:ccrel} also provides an independent way to compute the central charge $c_F$ given $G$ and $n$, which in section \ref{sec:wzwunivii} we will use as a consistency check that the flavor symmetry $F$ has been correctly identified. \newline

\noindent The currents of the flavor symmetry of the chiral algebra are chiral operators of conformal weight 1, which we expect to see in the elliptic genera of the \hh{n}{G} CFTs; indeed, expanding the NS-R elliptic genera and extracting the states that contribute at energy $(1,0)$ above the NS-R vacuum,\footnote{ In the NS-R sector we still we have an infinitely-degenerate set of vacua. The vacua contribute to the elliptic genus as the Hilbert series of one $Sp(N)$ instanton, $q^{-\frac{c_L}{24}}v^{h^\vee_G-1}\sum_{k=0}^\infty v^{2k}\chi^{G}_{k\cdot \th_G}(\mass_{G})$. Here we focus on the ground state which is in the trivial representation of $G$.\label{ft:sp}} we find that
\begin{equation}\label{eq:dims}
\boxed{\mathcal{E}_n^{G}(\mass_{G},\mass_{F},v,q)\bigg\vert_{q^{-\frac{c_L}{24}+1}v^{h^\vee_G-1}}=
\chi_{\th_F}^{F}(\mass_{F})+\chi_{\th_G}^{G}(\mass_{G})+1;}
\end{equation}
the last term is the contribution from the current of the $U(1)_v\subset SU(2)_R$ algebra. We find that this expression holds for almost all \hh{n}{G} theories; notably, however, the \hh{1}{Sp(N)} theories with $N\geq 2$ are an exception: for them turns out that
\begin{equation}
\chi_{\th_{Sp(N)}}^{Sp(N)}(\mass_{Sp(N)}) \to \chi_{\th_{Sp(N)}}^{Sp(N)}(\mass_{Sp(N)})+\chi_{(010\dots0)}^{Sp(N)}(\mass_{Sp(N)}),
\end{equation}
in equation \eqref{eq:dims}. The second term on the right hand side is the character of the antisymmetric, $2N^2-N-1$ dimensional representation of $Sp(N)$; notice that the two representations combine into the adjoint representation of $SU(2N)$, of which $Sp(N)$ is a maximal special subalgebra. We will provide more details about this special case in section \ref{sec:5dlimnis1}.\newline

\noindent The 6d SCFT $\six{2}{SU(2)}{}$ enjoys a flavor symmetry $F=SO(8)$ on the tensor branch, but at the superconformal fixed point this symmetry is reduced to a $SO(7)$ subgroup \cite{Ohmori:2015pia}. One can verify (as we do in section \ref{sec:su2case}) that indeed for the \hh{2}{SU(2)} theory equation \ref{eq:dims} holds with $F=SO(7)$ rather than $SO(8)$. This also suggests that the BPS string elliptic genus conveys nontrivial information about the superconformal fixed point.\footnote{ Nevertheless, we still find that the elliptic genus is written more naturally in terms of level 1 $SO(8)$ characters, with a certain specialization of fugacities -- see equations \eqref{eq:spec87a}-\eqref{eq:spec87d}.}

\subsection{Current algebra realization of $F$ (II)}\label{sec:wzwunivii}
\noindent In this section we provide more detailed information about the flavor symmetry $F$ of the \hh{n}{G} CFTs and discuss the subtleties that arise for some of these theories. The matter content of rank 1 6d (1,0) SCFTs on the tensor branch is given in table \ref{tab:sapori}; the flavor symmetry of the SCFTs that is suggested by the matter content is also provided there, and is in general a product of simple and abelian factors
\begin{equation}
F=\prod_i F_i,
\end{equation}
to which we associate a current algebra of total central charge
\begin{equation}
c_F = \sum_{i|\,F_i\text{ simple}}\frac{\text{dim}(F_i)k_i}{h^\vee_{F_i}+k_i}+\sum_{i|\,F_i\text{ abelian}}1.
\end{equation}
From equation \eqref{eq:ccrel} we also obtain an independent prediction for the central charge of the $F$ sector:
\begin{equation}\label{eq:ccrel2}
c_F=c_L + \frac{n\,\text{dim}(G)}{h^\vee_G-n} - 1+6\frac{(h^\vee_G-1)^2}{n-h^\vee_G},
\end{equation}
which provides a consistency check on the flavor symmetry of the SCFT. The two expressions match for most theories, but we encounter a few outliers which we discuss at the end of this section. We collect the information about the flavor symmetry current algebras of the \hh{n}{G} theories in tables \ref{tab:flavor1} and \ref{tab:flavor2}. It is interesting to remark that our computations confirm the flavor symmetries one would compute from field theory for these models. In particular, since the levels correspond to intersection numbers, whenever the former are different from 1 we predict that the corresponding noncompact flavor divisor intersects the gauge theory divisor in the F-theory base in a non-transverse fashion, which has to be contrasted with the assumptions in \cite{Bertolini:2015bwa} where transverse intersections were assumed.\footnote{ The transverse intersection requirement is a necessary condition for the corresponding flavor symmetry to be gauged, but not all flavor symmetries need to give rise to consistent models upon gauging hence there is no contradiction in letting the corresponding noncompact flavor curves to intersect non-transversally.}\newline

\begin{table}[p!]
\begin{center}
\scalebox{0.9}{\begin{tabular}{|c|l|l|l|l|}\hline
$n$&G&F&$c_F$&$(R^G,R^F)$\\\hline\hline
12&$E_8$&$-$&0&$-$\\\hline
8&$E_7$&$-$&0&$-$\\\hline
7&$E_7$&$-$&0&$(\textbf{56},\textbf{1})$\\\hline
6&$E_6$&$-$&0&$-$\\
6&$E_7$&$SO(2)_{12}=U(1)_{12}$&1&$(\textbf{56},\textbf{2})$\\\hline
5&$F_4$&$-$&0&$-$\\
5&$E_6$&$SU(1)_6\times U(1)_6 = U(1)_6$&1&$(\textbf{27},\overline{\textbf{1}}_{-1})\oplus \text{c.c.}$\\
5&$E_7$&$SO(3)_{12}$&$\frac{36}{13}$&$(\textbf{56},\textbf 3)$\\\hline
4&$SO(N),\,N\geq 8$&$Sp(N-8)_1$&$\frac{(N-8)(2N-15)}{N-6}$&$(\textbf{N},\textbf{2(N-8)})$\\
4&$F_4$&$Sp(1)_3$&$\frac{9}{5}$&$(\textbf{26},\textbf{2})$\\
4 &$E_6$&$SU(2)_{6}\times U(1)_{12}$ &$\frac{13}{4}$&$(\textbf{27},\overline{\textbf{2}}_{-1})\oplus \text{c.c.}$\\
4&$E_7$&$SO(4)_{12}$&$\frac{36}{7}$&$(\textbf{56},\textbf 4)$\\\hline
3&$SU(3)$&$-$&0&$-$\\
3&$SO(7)$&$-\times Sp(2)_1$&$\frac{5}{2}$&$(\textbf 8,\textbf 4)$\\
3&$SO(8)$&$Sp(1)_1^a\times Sp(1)_1^b\times Sp(1)_1^c$&3&$(\textbf 8^{\bf{v}},\textbf 2^a)\oplus(\textbf 8^{\bf{s}}, \textbf 2^b)\oplus(\textbf 8^{\bf{c}},\textbf 2^c)$\\
3&$SO(9)$&$Sp(2)_1^a\times Sp(1)_2^b$&4&$(\textbf 9,\textbf 4^a)\oplus(\textbf{16},\textbf 2^b)$\\
3&$SO(10)$&$Sp(3)_1^a\times (SU(1)_4\times U(1)_4)^b$&$\frac{26}{5}$&$(\textbf{10},\textbf 6^a)\oplus[(\textbf{16}^{\bf{s}},\textbf 1_1^b)\oplus \text{c.c.}]$\\
3&$SO(11)$&$Sp(4)_1^a\times \text{Ising}^b$&$\frac{13}{2}$&$(\textbf{11},\textbf 8^a)\oplus(\textbf{32},\textbf{1}_{\textbf{s}}^b)$\\
3&$SO(12)$&$Sp(5)_1$&$\frac{55}{7}$&$(\textbf{12},\textbf{10})\oplus(\textbf{32}^{\bf{s}},\textbf{1})$\\
3&$G_2$&$Sp(1)_1$&1&$(\textbf{7},\textbf{2})$\\
3&$F_4$&$Sp(2)_3$&5&$(\textbf{26},\textbf{4})$\\
3 &$E_6$&$SU(3)_{6}\times U(1)_{18}$ &$\frac{19}{3}$&$(\textbf{27},\overline{\textbf{3}}_{-1})\oplus \text{c.c.}$\\
3&$E_7$&$SO(5)_{12}$&8&$(\textbf{56},\textbf 5)$\\\hline
\end{tabular}}
\end{center}
\caption{Flavor symmetry $F$ of the  theories \hh{n}{G} for $n\geq 3$. For each theory, we denote the simple non-abelian factors in the current algebra by $F_k$, where $k$ is the level. The abelian factors are denoted by $U(1)_{R^2}$, where $R$ is the radius of the compact boson that realizes the current algebra. When $F$ includes a product of several non-abelian factors, we use superscripts $a,b,\dots$ to distinguish more easily between them and their representations. The last column displays the representation $(R^G,R^F)$ of $G\times F$ in which the matter of the parent 6d SCFT transforms. If $F$ includes a $U(1)$ factor, we indicate the charge with respect to it by a subscript. For the Ising model, $\bf{1}_{\bf{s}}$ indicates that the Virasoro primary with $h=\frac{1}{16}$ (whose lowest order component is just a single state).}
\label{tab:flavor1}
\end{table}

\begin{table}[p!]
\begin{center}
\scalebox{0.9}{\begin{tabular}{|c|l|l|l|l|}\hline
$n$&G&F&$c_F$&$(R^G,R^F)$\\\hline\hline
2&$SU(2)$&$SO(8)_1\to SO(7)_1\times\text{Ising}$&$4$&$(\textbf 2,\textbf 8^{\bf{v}})\to (\textbf 2,\textbf 8^{\bf{s}}\times \textbf{1}_{\bf s})$\\
2&$SU(N), N>2$&$SU(2N)_1$&2N-1&$(\textbf N,\overline{\textbf{2N}})\oplus \text{c.c.}$\\
2&$SO(7)$&$Sp(1)_1^a\times Sp(4)_1^b$&7&$(\textbf{7},\textbf{2}^a)\oplus(\textbf{8},\textbf{8}^b)$\\
2&$SO(8)$&$Sp(2)_1^a\times Sp(2)_1^b\times Sp(2)_1^c$&$\frac{15}{2}$&$(\textbf{8}^{\bf{v}},\textbf{4}^a)\oplus(\textbf{8}^{\bf{s}},\textbf{4}^b)\oplus(\textbf{8}^{\bf{c}},\textbf{4}^c)$\\
2&$SO(9)$&$Sp(3)_1^a\times Sp(2)_2^b$&$\frac{41}{5}$&$(\textbf 9,\textbf 6^a)\oplus(\textbf{16},\textbf 4^b)$\\
2&$SO(10)$&$Sp(4)_1^a\times (SU(2)_4\times U(1)_{8})^b$&9&$(\textbf{10},\textbf 8^a)\oplus[(\textbf{16}^{\bf{s}},\textbf 2_1^b)\oplus \text{c.c.}]$\\
2&$SO(11)$&$Sp(5)_1^a\times ???^b$&$\frac{69}{7}$&$(\textbf{11},\textbf{10}^a)\oplus(\textbf{32},\textbf 2^b)$\\
2&$SO(12)_a$&$Sp(6)_1^a\times SO(2)_8$&$\frac{43}{4}$&$(\textbf{12},\textbf{12}^a)\oplus(\textbf{32}^{\bf{s}},\textbf 2^b)$\\
2&$SO(12)_b$&$Sp(6)_1^a\times \text{Ising}^b\times \text{Ising}^c$&$\frac{43}{4}$&$(\textbf{12},\textbf{12}^a)\oplus(\textbf{32}^{\bf{s}},\textbf 1_{\bf{s}}^b)\oplus(\textbf{32}^{\bf{c}},\textbf1_{\bf{s}}^c)$\\
2&$SO(13)$&$Sp(7)_1$&$\frac{35}{3}$&$(\textbf{13},\textbf{14})\oplus(\textbf{64},\textbf{1})$\\
2&$G_2$&$Sp(4)_1$&6&$(\textbf{7},\textbf{8})$\\
2&$F_4$&$Sp(3)_3$&9&$(\textbf{26},\textbf{6})$\\
2 &$E_6$&$SU(4)_{6}\times U(1)_{24}$ &10&$(\textbf{27},\overline{\textbf{4}}_{-1})\oplus \text{c.c.}$\\
2&$E_7$&$SO(6)_{12}$&$\frac{45}{4}$&$(\textbf{56},\textbf{6})$\\\hline
1&$Sp(N)$&$SO(4N+16)_1$&$2N+8$&$(\textbf{2N},\textbf{4N+16})$\\
1&$SU(3)$&$SU(12)_1$&11&$(\textbf 3,\overline{\textbf{12}})\oplus \text{c.c.}$\\
1&$SU(4)$&$SU(12)_1^a\times SU(2)_1^b$&12&$[(\textbf 4,\overline{\textbf{12}}^a)\oplus \text{c.c.}]\oplus (\textbf{6},\textbf{2}^b)$\\
1 &$SU(N),N\geq 4$&$SU(N\!+\!8)_1\!\times\! U(1)_{2N(N-1)(N+8)}$&$N+8$&$[(\textbf N,\overline{\textbf{N+8}}_{-N+4})\oplus(\mathbf{\Lambda^2},\textbf{1}_{-N-8})]\oplus \text{c.c.}$\\
1&$SU(6)_*$&$SU(15)_1\times SO(1)_6 = SU(15)_1$&14&$[(\textbf 6,\overline{\textbf{15}})\oplus \text{c.c.}]\oplus(\textbf{20},\textbf{1})$\\
1&$SO(7)$&$Sp(2)_1^a\times Sp(6)_1^b$&$\frac{49}{4}$&$(\textbf{7},\textbf{4}^a)\oplus(\textbf{8},\textbf{12}^b)$\\
1&$SO(8)$&$Sp(3)_1^a\times Sp(3)_1^b\times Sp(3)_1^c$&$\frac{63}{5}$&$(\textbf{8}^{\bf{v}},\textbf{6}^a)\oplus(\textbf{8}^{\bf{s}},\textbf{6}^b)\oplus(\textbf{8}^{\bf{c}},\textbf{6}^c)$\\
1&$SO(9)$&$Sp(4)_1^a\times Sp(3)_2^b$&13&$(\textbf 9,\textbf 8^a)\oplus(\textbf{16},\textbf 6^b)$\\
1&$SO(10)$&$Sp(5)_1^a\times (SU(3)_4\times U(1)_{12})^b$&$\frac{94}{7}$&$(\textbf{10},\textbf {10}^a)\oplus[(\textbf{16}^{\bf{s}},\textbf 3_{1}^b)\oplus \text{c.c.}]$\\
1 &$SO(11)$&$Sp(6)_1^a\times ???^b$&$\frac{111}{8}$&$(\textbf{11},\textbf{12}^a)\oplus(\textbf{32},\textbf 3^b)$\\
1&$SO(12)_a$&$Sp(7)_1^a\times SO(3)_8^b$&$\frac{43}{3}$&$(\textbf{12},\textbf{14}^a)\oplus(\textbf{32}^{\bf{s}},\textbf 3^b)$\\
1&$SO(12)_b$&$Sp(7)_1^a\times ???^b\times???^c$&$\frac{43}{3}$&$(\textbf{12},\textbf{14}^a)\oplus(\textbf{32}^{\bf{s}},\textbf 2^b)\oplus(\textbf{32}^{\bf{c}},\textbf 1^c)$\\
1&$G_2$&$Sp(7)_1$&$\frac{35}{3}$&$(\textbf{7},\textbf{14})$\\
1&$F_4$&$Sp(4)_3$&$\frac{27}{2}$&$(\textbf{26},\textbf{8})$\\
1 &$E_6$&$SU(5)_{6}\times U(1)_{30}$ &$\frac{155}{11}$&$(\textbf{27},\overline{\textbf{5}}_{-1})\oplus \text{c.c.}$\\
1&$E_7$&$SO(7)_{12}$&$\frac{252}{17}$&$(\textbf{56},\textbf{7})$\\\hline
\end{tabular}}
\end{center}
\caption{Flavor symmetry $F$ of the  theories \hh{n}{G} for $n=1$ and 2. For the \hh{1}{SU(N)} theories, $\bf{\Lambda^2}$ symbolizes the $\frac{N(N-1)}{2}$-dimensional antisymmetric representation of $SU(N)$. We indicate by $???$ the cases where we do not have a good understanding of the worldsheet realization of the flavor symmetry.}
\label{tab:flavor2}
\end{table}

\noindent We can perform a simple consistency check on the central charges listed in tables \ref{tab:flavor1} and \ref{tab:flavor2}: since the current algebras are realized in terms of the chiral fermions of the NLSM, their central charges cannot be greater than the contribution to the left-moving central charge from the chiral fermions, which according to table \ref{tab:ferbos} is given by
\begin{equation}
c_L^{f} = 2h^\vee_G +6(2-n).
\end{equation}
Indeed, for all \hh{n}{G} CFTs listed in tables \ref{tab:flavor1} and \ref{tab:flavor2} we find:
\begin{equation}
c_F\leq c_L^f,
\end{equation}
the bound being saturated only for the \hh{1}{Sp(N)} theories (for which we have indeed shown in section \ref{sec:1spn} that the chiral fermions provide a free field realization of the $SO(16+4N)$ WZW model), for the \hh{2}{SU(2)} theory, and trivially for the theories \hh{3}{SU(3)}, \hh{4}{SO(8)}, \hh{5}{F_4}, \hh{6}{E_6}, \hh{8}{E_7}, and \hh{12}{E_8} corresponding to 6d SCFTs without matter. Moreover, one can check by looking at tables \ref{tab:flavor1} and \ref{tab:flavor2} that the central charge $c_F$ decreases monotonically as one moves along the edges of the Higgsing trees of section \ref{sec:higgsingchains}.\newline

\noindent  A further consistency check is obtained by requiring that the general form of the elliptic genus in terms of affine characters (equations \eqref{eq:elldecomp} and \eqref{eq:FGv}) is consistent with the correct representations appearing at $L_0=1/2$ in the NS-R elliptic genus, as in equation \eqref{eq:ellg1}. This is in particular helpful for checking abelian factors $U(1)_{R^2}$ of the flavor symmetry, for which $c_F\equiv 1$ but the spectrum of primaries depends on the value of $R$.\newline

\noindent In the remainder of this section we comment on subtleties that arise for certain \hh{n}{G} theories. A first subtlety concerns the way $U(N)$ flavor symmetry factors are realized in the CFT. For instance, the $\six{n}{E_6}{}$ SCFTs possess a $U(6-n)$ flavor symmetry which rotates the $6-n$ hypermultiplets in the $\bf 26$ of $E_6$. The flavor symmetry is realized on the string worldsheet in terms of a 
\begin{equation}\label{eq:unwzw}
SU(6-n)_6\times U(1)_{6(6-n)}
\end{equation}
current algebra; interestingly, the levels of the two factors are inconsistent with the ones of the $U(6-n)_k$ current algebra. Indeed, the latter is given by a $\mathbb{Z}_{6-n}$ quotient of the 
\begin{equation}
SU(6-n)_{k}\times U(1)_{(6-n)(k+6-n)}
\end{equation}
 current algebra, which is not compatible with \eqref{eq:unwzw}; moreover, the $U(N)_k$ WZW model is only well defined for odd  $k$ \cite{Gepner:1992kx,Naculich:2007nc}, whereas for the \hh{n}{E_6} CFTs we find an even level. We encounter a similar situation for the $U(4-n)$ flavor symmetry that rotates the spinorial matter in the $\six{n}{SO(10)}{}$ SCFTs. At the level of current algebra, we find that this is captured on the string by a
\begin{equation}
SU(4-n)_{4}\times U(1)_{4(4-n)}
\end{equation}
current algebra, and not by a $U(4-n)_k$ current algebra.\\

\noindent Next, we discuss some peculiarities we encounter for \hh{n}{G} theories whose flavor symmetry includes orthogonal groups of small rank:

\begin{itemize}
\item For the theory \hh{6}{E_7}, we find $F = SO(2)_{12}$, which is equivalent to the $U(1)$ WZW model at radius $R=\sqrt{12}$ (see appendix \ref{sec:WZW}).
\item For certain theories the flavor symmetry $F$ includes a factor of `$SO(1)$', which may or may not correspond to a trivial WZW model according to the following reasoning: Recall that the central charge for a generic $SO(K)_k$ WZW model is given by:
\begin{equation}
c_{SO(K)_k} = \frac{K(K-1)k}{2(K-2+k)}.
\end{equation}
Setting $k$ to be any level except for 1, and analytically continuing the rank to $K=1$ one finds that
\begin{equation}
\lim_{K\to1} c_{SO(K)_k}=0,\qquad\text{for $k\neq 1$},
\end{equation}
so that the CFT for $k\neq 1$ is trivial. On the other hand, setting $k=1$ first and taking the limit gives
\begin{equation}
\lim_{K\to1} c_{SO(K)_1}=\frac{1}{2}.
\end{equation}
This is of course consistent with the well-known fact that the expressions for the characters of the $SO(K)$  WZW model at level 1 reduce to the characters of the Ising model when $K = 1$, which is simply the theory of one free chiral fermion \cite{di1997conformal}.\newline

By computing the central charge \eqref{eq:ccrel2}, we find that the $SO(1)$ factor in $F$ contributes as the trivial $c=0$ theory for the following theories: \hh{1}{SU(6)_*}, \hh{2}{SO(13)}, \hh{3}{SO(12)}, and \hh{7}{E_7}. On the other hand, for \hh{3}{SO(11)} we find that the $SO(1)$ factor contributes $c=1/2$ to the central charge and is therefore captured by the Ising model. By the methods of section \ref{sec:modul} we are able to check that the elliptic genera of these theories are consistent with this identification of the $SO(1)$ factor.

\item For the three theories \hh{3}{SO(12)}, \hh{2}{SO(12)}, and \hh{1}{SO(12)}, there is a priori an ambiguity in the chirality of the $SO(12)$ spinors that contribute to the matter of the 6d SCFT, which may in principle involve $n_s$ hypermultiplets the spinor representation $\textbf{32}^{\bf{s}}$ of $SO(12)$ and $n_c$ hypermultiplets in the conjugate spinor representation $\textbf{32}^{\bf{c}}$. Exchanging $n_s$ and $n_c$ leads to the same elliptic genus up to a redefinition of the $\massG$ fugacities, since the two choices are related by an outer automorphism of $SO(12)$.\newline

\noindent For $n=3$, up to this redefinition there is a unique theory with $n_s=1$ and $n_c=0$; the flavor symmetry rotating the spinors is $SO(1)$, which is realized as the Ising model as discussed above.\newline

\noindent For $n=2$, there are potentially two distinct theories. The first, \hh{2}{SO(12)_a}, has $n_s=2$ and $n_c=0$, and that the flavor symmetry rotating the spinors is expected to be $SO(2)_{8}= U(1)_8$; the second, \hh{2}{SO(12)_b}, has $n_s=1$ and $n_c=1$, and it is expected that the flavor symmetry rotating the spinors is $SO(1)\times SO(1)$. From \eqref{eq:ccrel2}, one expects the central charge of this component of the flavor symmetry to be $c=1$, which is the correct value for $SO(2)_8$ in the former case, and is consistent with $SO(1)$ being realized as the Ising model in the latter. We will look at these theories in  more detail in section \ref{sec:modresults} and will find that both choices lead to seemingly consistent elliptic genera.\newline

\noindent Finally, for $n=1$ there are also two possible choices: a first choice, \hh{1}{SO(12)_a} with $n_s=3$ and $n_c=0$ which contributes a $SO(3)_8 = SU(2)_{16}$ factor to the flavor symmetry, and a second choice, \hh{1}{SO(12)_b} with $n_s=2$ and $n_c=1$, with expected flavor symmetry $SO(2)_8\times SO(1)$. On the other hand, in the second case $SO(2)_8\times SO(1)$ would have central charge 1 or 3/2 depending on whether the $SO(1)$ factor is realized as the trivial theory or as the Ising model. Neither choice is consistent with the prediction from equation \eqref{eq:ccrel2} that the central charge be 8/3, and there does not seem to be an obvious way to recover the correct central charge, even by allowing for enhancements of the flavor symmetry (other than allowing for the possibility that $SO(2)_8\times SO(1)$ enhances to $SO(3)_8$, so that the elliptic genus reduces to  the \hh{1}{SO(12)_a} theory). Therefore we are not able to fit the \hh{1}{SO(12)_b} theory within our framework, possibly indicating an inconsistency in the \hh{1}{SO(12)_b} theory.

\item For the theories \hh{2}{SO(11)} and \hh{1}{SO(11)}, equation \eqref{eq:ccrel2} predicts the central charge of the flavor symmetry chiral algebra, $c_F$, to be respectively $\frac{69}{7}$ and $\frac{111}{8}$. In particular, the factor of the flavor symmetry that rotates the spinorial matter is expected to have central charge respectively $2$ and $\frac{33}{8}$. This is inconsistent with the naive expectation that the flavor symmetry associated to the spinorial matter be respectively $SO(2)_8$ and $SO(3)_8$, for which respectively $c=1$ and $c=8/3$. There also does not appear to be any obvious alternative choice of flavor symmetry that correctly captures these central charges, and therefore somewhat unexpectedly we have not been able to fit these theories within our framework.

\end{itemize}

\noindent Finally, recall from section \ref{sec:6df} that the $\mathcal{O}(-n)\to\mathbb{P}^1$ models with $n=9,$ $10,$ and $11$ have $G=E_8$ gauge algebra coupled respectively to 3, 2, and 1 additional E-string subsectors. Amusingly, one may ask whether these models also admit a description in terms of some effective theory with $n_T=1$ tensor multiplet and an ordinary BPS string sector; however, we immediately run into a problem when calculating $c_F$ for the string of this putative effective theory: from equation \eqref{eq:ccrel2} we find $c_F=-1$ for $n=9$ and $11$, and to $c_F=-\frac{13}{10}$ for $n=10$; suggesting that such an effective description is inconsistent.

\section{Five-dimensional limit}\label{sec:5dlim}
In this section we discuss a relation between the lowest energy states in the elliptic genus and the one-instanton part of the 5d Nekrasov partition function, which arises by taking the zero-size limit of the 6d circle in the $(S^1\times S^1\times \mathbb{R}^4)_{\epsilon_1,\epsilon_2}$ partition function
\begin{equation}\label{eq:z6da}
Z^{6d}_{T^2\times\mathbb{R}^4}(\varphi,\massG,\massF,x,v,q) = Z_{\text{pert}}(\massG,\massF,x,v,q)\times\sum_{k\geq 0} e^{-k\, \Phi}\,\mathbb{E}_{(k)}(\massG,\massF,x,v,q).
\end{equation}
where, as remarked in section \ref{sec:ellgen}, 
\begin{equation}
\Phi = \varphi + \text{ linear combinations of $\tau$, $\massG$, and $\massF$.}
\end{equation}
\noindent As a first step, it is useful to recall the compactification of the 6d theory on a circle of finite size, which gives rise to a five-dimensional theory at finite gauge coupling. Namely, place the 6d SCFT with simple gauge group $G$ and matter content in the representation $\mathcal R$ of $G$ on the tensor branch on the 6d Omega background; this can alternatively be interpreted as a 5d theory with gauge group $U(1)\times G$ on the 5d Omega background $(S^1\times \mathbb{R}^4)_{\epsilon_1,\epsilon_2}$, where the $U(1)$ gauge field $A_{U(1)}$ originates from the reduction of the two-form field $B$ in the tensor multiplet. By a well known argument that can be found in \cite{Witten:2009at} the coupling $\frac{g_{U(1)}^2}{4\pi}$ of this $U(1)$ gauge field is identified with the radius of the 6d circle, $R_{6d}$. The KK momentum along the 6d circle is identified with the instanton charge with respect to the gauge $U(1)$; the $U(1)_C$ isometry of the 6d circle becomes a global symmetry in 5d,  to which a background $U(1)$ gauge field $C$ is associated. Notice that the latter couples to the gauge $U(1)$ field strength via an interaction $\int C\wedge F_{U(1)}\wedge F_{U(1)}$. Hence the KK charge, which is the conserved charge associated to the $U(1)_C$ isometry in 6d, is also identified with the $U(1)$ gauge field instanton charge. Finally, the 6d gauge multiplet for $G$ reduces trivially to a 5d gauge multiplet. We denote the real adjoint-valued scalar in the 5d gauge multiplet by $a_G$. The 5d gauge coupling is given by
\begin{equation}
\frac{8\pi^2}{g_{G,5d}^2} = R_{6d}\frac{8\pi^2}{g_{G,6d}^2}= R_{6d}\,\phi_{6d}=\sqrt{R_{6d}}\phi_{5d},
\end{equation}
where $\phi_{6d}$ is the vev of the scalar in the 6d tensor multiplet, which reduces in 5d to the scalar partner $\phi_{5d}$ of the $U(1)$ gauge field. In equation \eqref{eq:z6da} the combination $\varphi = \phi_{6d} + i\int_{T^2}B$ appears.\newline 

\noindent The complex modulus of the torus gets identified with
\begin{equation}
\tau = \frac{4\pi i}{g^2_{U(1)}}R_{5d}+\int_{S^1_{5d}} C,
\end{equation}
where $R_{5d}$ is the radius of the circle in the 5d Omega background. \newline

The 5d $U(1)$ vector multiplet scalar, $\phi_{5d}$, combines with the $A_{U(1)}$ holonomy on the 5d circle into a complex scalar:
\begin{equation}
\varphi_{5d} = \phi_{5d}+2\pi i\int_{S^1_{5d}} A_{U(1)}.
\end{equation}
 Notice that the imaginary component can also be viewed as a flux $\int_{T^2} B$ for the 6d tensor multiplet two-form field, and appears here due to the 6d $\int B\wedge \text{Tr }F_{G}\wedge F_{G}$ interaction term. Finally, the fugacities $\mass_G,\mass_F$ are interpreted respectively as Coulomb branch parameters for $G$ and as hypermultiplet masses, complexified respectively by Wilson lines for the $G$ gauge potential and of the $F$ background potential.\\

\noindent In the limit of zero size of the 6d circle, $\tau\to i\infty$ and the $U(1)$ gauge field decouples. The $U(1)$ symmetry then gets identified with the 5d topological $U(1)$ global symmetry associated to the $G$ instantons. In this limit we project the spectrum of the 6d theory to the subsector with lowest $U(1)_C$ charge. At a naive level, one expects the 5d limit of the 6d theory to be a 5d theory with gauge group $G$ and matter content in the same representation $\mathcal{R}$ as in 6d, plus the decoupled $U(1)$ vector multiplet. However, we will see that the 5d limit in fact depends on whether $n\geq 3$, $n=2$, or $n=1$. We discuss these three cases in turn, after providing a rationale for these different limits from the perspective of geometric engineering.

\subsection{Remarks on 5d theories and M-theory geometry}\label{sec:5dm}

\noindent The reduction of six-dimensional $\cn=(1,0)$ SCFTs on $S^1$ discussed above is essentially correct, provided the self-intersection of the curve in the F-theory base is sufficiently high. At the level of Nekrasov partition functions, we find the following trichotomy:
\be\label{eq:5dbehaviors}
(\six{n}{G}{\mathcal R})_{6d} \quad\longrightarrow\quad \begin{cases} (G,\mathcal R)_{5d} & n\geq 3\\ (G,\mathcal R)_{5d} \oplus {\mathfrak R}_{2,G} & n = 2 \\ \text{Free hyper} & n =1\end{cases}
\ee
where the notation is as follows: the six-dimensional theory is denoted by $\six{n}{G}{\mathcal R}$ where $\mathcal R$ is a representation of $G$ corresponding to the matter content from table \ref{tab:sapori}, the five-dimensional theory $(G,\mathcal R)_{5d}$ is a 5d gauge theory with gauge group $G$ and matter in the $\mathcal R$ representation, and the five-dimensional theory ${\mathfrak R}_{2,G}$ is a residual decoupled system that arises only when $n=2$.\footnote{ The $n=2$ case overlaps with the class of models studied in \cite{Ohmori:2015pia}, as these theories are all Higgsable to 6d $A_1$ $(2,0)$ theories. It would be interesting to probe the conjecture of \cite{Ohmori:2015pia} which predicts that ${\mathfrak R}_{2,G}$ should consist of an $SU(2)$ SYM sector. This is out of the scope of the present work, but can be addressed with our techinques and we leave it for the future.} For $n=1$ the lowest energy degrees of freedom consist simply of a free hypermultiplet, and any degrees of freedom charged under $G$ only appear at higher levels in the KK tower and therefore decouple in the $R_{6d}\to 0$ limit.\\

\begin{figure}[t]
\begin{center}
\includegraphics[scale=0.8]{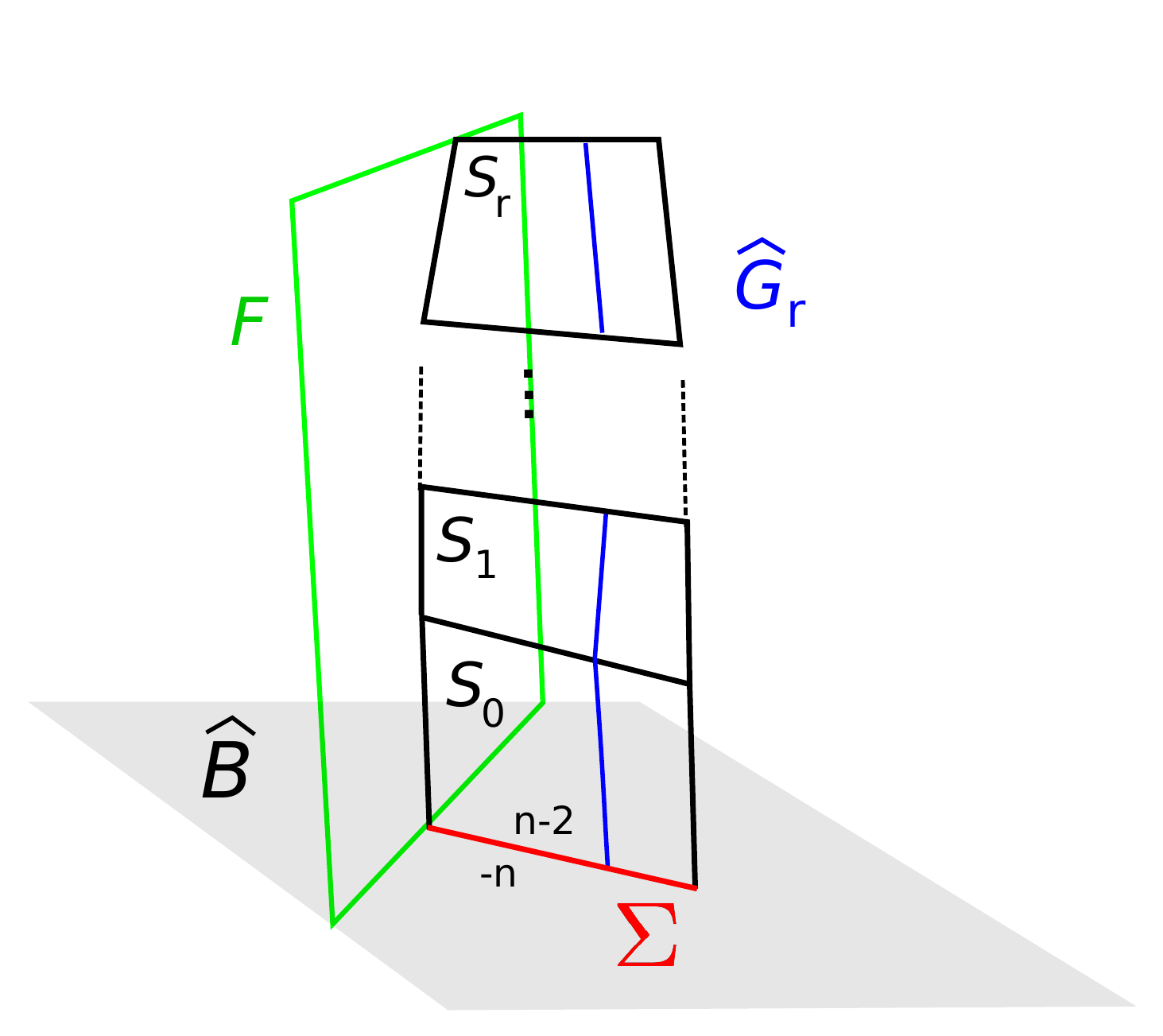}
\end{center}
\caption{Schematic description of the generic geometry for rank-one models.}\label{fig:5dgeom}
\end{figure}

\noindent This has the following geometrical interpretation. The reduction from 6d to 5d is realized within geometric engineering by F-theory/M-theory duality. F-theory on $S^1_{6d} \times \widehat X$, where $\widehat X$ is the elliptic Calabi-Yau associated to the tensor branch of the SCFT, is equivalent to M-theory on $\widehat X$.  In the M-theory picture, M2 branes wrapped on holomorphic two-cycles in $\widehat X$ give rise to BPS particles in 5d. The radius of $S^1_{6d}$ in the F-theory background is inversely proportional to the volume of the elliptic fiber of $\widehat X$ on the M-theory side, so one reaches a genuine 5d theory when the volume of the elliptic fiber is sent to infinite size. For rank-one models with a nontrivial gauge symmetry the elliptic fiber along the compact curve in the base $\widehat B$ has a nontrivial Kodaira type. The relevant geometry is shown schematically in figure \ref{fig:5dgeom} --- see e.g.\cite{Intriligator:1997pq}. Within M-theory, the geometry involves a collection of ruled surfaces $S_i$, $i=0,..,r$. This gives rise to a fibration which is given by a collection of $-2$ curves that combine into an elliptic fiber of Kodaira type, drawn in blue in the figure. We label this collection of curves ${\widehat G}_r$, as it can be viewed as an affinization of the Dynkin graph of the corresponding gauge group $G$. The combination of 2-cycles associated to the imaginary root $\delta$ of ${\widehat G}_r$ gives rise in the M-theory geometric engineering to the extra $U(1)$ gauge symmetry with gauge field $A_{U(1)}$ which becomes global in the limit $R_{6d}\to 0$. As discussed above, this is the global $U(1)$ corresponding to the current
\be
j_{\text{top}} = \ast_{5d} \,\text{tr } F_G\wedge F_G
\ee
whose charge is the $G$ instanton number. The surfaces $S_0,\dots , S_r$ are all compact as long as the radius of the 6d F-theory circle $S^1_{6d}$ is finite. There are several inequivalent ways to  send the volume of the elliptic fiber $E$ of $\widehat X$ to infinity, which are related by flop transitions in the extended K\"ahler cone of $\widehat X$, see e.g. \cite{DelZotto:2017pti}. The limit we discuss in this section involves taking the volume $\tau$ of the elliptic fiber to infinity, while keeping the parameters $\massG$ associated to the 2-cycles corresponding to the real roots of the Dynkin diagram finite. This is achieved by sending to infinite volume the $-2$ curve that provides the ruling of the surface $S_0$ while keeping the volumes of the analogous curves for the $S_1,\dots, S_r$ surfaces finite. \newline

\noindent Let us also comment on how to read off the $U(1)_C$ charge from geometry. It is clear that it is proportional to the class of the elliptic fiber $E$. Notice that $E = F + \cdots$ and therefore we can use $F$ to read off the $U(1)_C$ charge of a wrapped M2 brane.\newline

\noindent The intersection of the base $\widehat B$ with $S_0$ is precisely the rational curve $\Sigma$, which controls the tensor branch vev in 6d and the 6d gauge coupling (in red in Figure \ref{fig:5dgeom}). Within the base $\widehat B$, this curve has self intersection
\be
\Sigma\cdot_{\widehat B} \Sigma = -n\,.
\ee
Locally, the CY condition amounts to the requirement that the normal bundle to $\Sigma$ has the form
\be
\mathcal O(-n) \oplus \mathcal O(n-2) \to \PP^1 \qquad \Sigma \simeq \PP^1\,,
\ee
so that within $S_0$ the curve $\Sigma$ has self-intersection
\be
\Sigma \cdot_{S_0} \Sigma = n-2\,.
\ee
Moreover, $S_0$ viewed as a ruled surface has a generic fiber $F\simeq \PP^1$ which is also a rational curve with
\be
F \cdot_{S_0} F = 0
\ee
with normal bundle
\be
\mathcal O(0) \oplus \mathcal O(-2) \to \PP^1\,;
\ee
finally, we have the following intersection between the two curves
\be\label{eq:ipotesi}
F \cdot_{S_0} \Sigma = 1.
\ee
The curve $F$ is the $-2$ curve of the Kodaira fiber of $\widehat X$ that corresponds to the affine node, so, based on the previous discussion, the 5d limit is achieved by sending the volume of the curve $F$ of the $S_0$ divisor to infinity.\\

\noindent In the case $n>  2$, the curve $\Sigma$ has a positive self-intersection in $S_0$, and therefore it is not contractible. As the volume of $F$, and hence of $S_0$, is scaled to infinity the volume of $\Sigma$ necessarily also scales to infinite size. Moreover, any state with nonzero $U(1)_C$ charge arises from M2 branes wrapped on the elliptic fiber $E$, and by sending $\vol(F)\to\infty$ acquires an infinite mass. In this way, we are left with the collection of contractible surfaces $S_1, \dots, S_r$ that correspond to the 5d gauge theory with gauge group $G$ and matter in a representation $\mathcal R$. This is precisely the naive behavior we have argued for above. An additional subtlety arises in the case $n=2$: because in this case $\Sigma \cdot_{S_0} \Sigma = 0$, the curve $\Sigma$ itself provides a second ruling of the surface $S_0$, which means that the curve can be moved within $S_0$. Therefore there is a moduli space of solutions involving M2 branes wrapping $\Sigma$, that are neutral with respect to $U(1)_C$ and, as we will see in section \ref{sec:5dlimnis2}, lead to extra degrees of freedom in the 5d Nekrasov partition function.\\

\noindent The behavior is markedly different in the case
\be
\Sigma \cdot_{\widehat B} \Sigma = -1.
\ee
By the same argument as above, in this case $\Sigma$ has negative self-intersection within $S_0$ and therefore is not constrained to scale up when $S_0$ does. For definiteness, consider the case in which $S_0$ is a Hirzebruch surface of type $\mathbb F_{1}$. This is not always the case, as in the presence of matter one typically encounters more general surfaces, e.g. blow-ups of Hirzebruch surfaces, but an analogous argument carries over in that case. The divisors $S_0$ and $S_1$ intersect along the curve $\Sigma'=\Sigma\,+\,F$ which has self-intersection $+1$. Therefore, the curve's volume scales to infinity along with $F$, and any M2 brane that is wrapped on a two-cycle which has a component along this curve becomes infinitely heavy in the limit $\text{vol}(F) \to \infty$; 
This in particular means that any degree of freedom corresponding to the gauge group and matter in 6d which also carries BPS string charge becomes infinitely massive (this does not include the W-bosons of $G$ which arise from M2 branes wrapped on the curves in $S_1$,\dots, $S_r$ associated to the finite nodes of $\widehat G_r$, which however contribute to $Z_{pert}$ in equation \eqref{eq:z6da}). An analogous way to rephrase this is that this class of BPS M2 branes carry positive $U(1)_C$ charge since they wrap on curves with a component in $\Sigma'=\Sigma+F$, and therefore differently from the case $n\neq 1$ they do not contribute to lowest energy in the KK tower of states. On the other hand, the local geometry of  $\Sigma$  is the total space of the bundle ${\mathcal O(-1)} \oplus {\mathcal O(-1)} \to \PP^1$ which gives rise to the 5d theory of one free hypermultiplet. This analysis is also compatible with the results of \cite{Jefferson:2017ahm} that enlarges the list of consistent models found in  \cite{Intriligator:1997pq}. Indeed, for the gauge groups and matter content that arise for 6d $n=1$ SCFTs, a putative 5d gauge theory obtained by naive dimensional reduction would possess too much matter to be consistent; this class of models would possess a negative definite 5d metric everywhere along the Coulomb branch to start with, and therefore cannot have a good UV definition in 5d.

\subsection{5d Nekrasov partition functions}
\subsubsection{Circle reduction of 6d models with $n\geq 3$} \label{sec:n35d}

In 5d limit $q\to 0$ the coupling of the $U(1)$ gauge field is turned off, and the 6d Nekrasov partition function reduces to the partition function of $\mathcal{N}=1$ $G$ Yang-Mills theory in the 5d Omega background, with matter content determined by KK reduction of the 6d (1,0) multiplets.  The $g_{G}$-independent term $Z_{pert}$ in equation \eqref{eq:z6da} captures the perturbative piece of this partition function (as well as additional decoupled degrees of freedom), while the sum over elliptic genera captures the instanton terms:
\begin{equation}\label{eq:5dln3}
\sum_{k\geq 0} e^{-k\, \Phi}\,\mathbb{E}_{(k)}(\massG,\massF,x,v,q) \xrightarrow{q\to 0} \sum_{k\geq 0} Q_{5d}^k\,Z_{k}(\massG,\massF,x,v).
\end{equation}
The precise identification of the 5d limit involves some subtleties to which we now turn.\newline

\noindent In the zero-size limit of the 6d circle, the NLSM for $k$ strings is expected to reduce to an ADHM-like quantum mechanics on the moduli space of $k$ instantons in the presence of matter. The states of the string that contribute to the quantum mechanics are the ground states in the Ramond-Ramond sector of the BPS string CFT, which as discussed in section \ref{sec:sigma}, have
\begin{equation}
H_L = -k\frac{n-2}{2}
\end{equation}
(where now, unlike in \eqref{eq:eramondr2}, we include the center of mass hypermultiplet). In other words, we expect that
\begin{equation}
\boxed{\mathbb{E}_{(k)}(\massG,\massF,x,v,q)\bigg\vert_{q^{-k\frac{n-2}{2}}} = Z_{k}(\massG,\massF,x,v),}\label{eq:5delim}
\end{equation}
where $Z_{k}(\massG,\massF,x,v)$ is proportional to the 5d, or K-theoretic, Nekrasov partition function \cite{Nekrasov:2002qd}, up to an overall factor of $v$ which is theory dependent. For instance, for the 6d theories with no matter content \hh{3}{SU(3)}, \hh{4}{SO(8)}, \hh{5}{F_4}, \hh{6}{E_6}, \hh{8}{E_7}, and \hh{12}{E_8}, one finds that \cite{DelZotto:2016pvm}
\begin{equation}
Z_k(\massG,\massF,x,v) = v^{k\, h^\vee_G}Z^{}_{k-inst}(\massG,\massF,x,v)= v^{k\, h^\vee_G}\,\mathcal{H}_{\mathcal{M}_{G,k}}(\massG,\massF,x,v),
\end{equation}
where the $k$-instanton piece of the Nekrasov partition function for 5d $\mathcal{N}=1$ $G$ SYM, $Z_{k-inst}$, is the same as the Hilbert series $\mathcal{H}_{\mathcal{M}_{G,k}}$ of the moduli space of $k$ $G$-instantons \cite{Keller:2011ek,Keller:2012da,Hanany:2012dm}. The overall factor of $v$ is natural from various points of view. First of all, it follows from the fact that
\begin{equation}
\mathbb{E}_{(k)}(\massG,\massF,x,v^{-1},q)=\mathbb{E}_{(k)}(\massG,\massF,x,v,q)
\end{equation}
(up to a possible overall minus sign), which is a manifestation of the $\text{Weyl}[SU(2)_R] = \mathbb{Z}_2$ symmetry of the elliptic genus and is a basic property of Jacobi forms (see appendix \ref{sec:appmod}). Moreover, in \cite{Nakajima:2005fg}, it was observed for the 5d $\mathcal{N}=1$ super-Yang-Mills theory with $G= SU(N)$ that it is natural to include this factor  (with $h^\vee_{SU(N)} = N$) in the definition of the $k$-instanton piece of the 5d Nekrasov partition function; there, it has the interpretation of the contribution from half of the canonical bundle of $\mathcal{M}_{SU(N),k}$ to the $k$-instanton piece of the partition function, which arises naturally if one defines this as the index of the Dirac operator on $\mathcal{M}_{SU(N),k}$ rather than the index of the Dolbeault operator. See also \cite{Rodriguez-Gomez:2013dpa} for related comments on the appearance of a $v$-dependent prefactor. \newline

\noindent The identification \eqref{eq:5delim} also implies the relation
\begin{equation}
e^{-\Phi} = q^{\frac{ n-2}{2}}Q_{5d}\label{eq:normallimit},
\end{equation}
which was observed in the context of minimal SCFTs in \cite{Haghighat:2014vxa}, see also \cite{Hayashi:2017jze}. The 5d limit is taken by keeping $Q_{5d}$ fixed while sending $q\to 0$.

\subsubsection{Circle reduction of 6d models with $n=2$}\label{sec:5dlimnis2}

We next turn to the theories with $n=2$. As anticipated in section \ref{sec:5dm}, we expect extra degrees of freedom to arise from M2 branes wrapping a curve in the homology class of $\Sigma$ in $S_0$: a representative corresponding to a generic nonzero section clearly has vanishing interesection number with all compact divisors in the Calabi-Yau, and  M2 branes wrapped on these holomorphic curves lead to BPS states which are neutral with respect to $G$. These BPS states give extra contributions to the topological string partition function, which appear alongside the Nekrasov partition function for $G$ with matter content determined from KK reduction of the 6d hypermultiplets. For $G=SU(N), N_f=2N$ this effect has been explained in \cite{Hayashi:2013qwa}, where it was argued that the topological string partition function factorizes as the $SU(N),\, N_f=2N$ Nekrasov partition function times a decoupled extra factor:
\begin{equation}
Z_{top} = Z_{extra}\times Z_{SU(N),\, N_f=2N}.
\end{equation}
Therefore, we expect the following slight modification of equation \eqref{eq:5dln3} to hold:
\begin{equation}
\sum_{k\geq 0} e^{-k\, \Phi}\,\mathbb{E}_{(k)}(\massG,\massF,x,v,q) \xrightarrow{q\to 0} \sum_{k\geq 0} Q_{5d}^k\,Z'_{k-inst}(\massG,\massF,x,v),
\end{equation}
where $Q_{5d}=e^{-\Phi}$ in accordance with equation \eqref{eq:normallimit}, and $Z'_{k-inst}(\massF,\massG,x,v)$ differs from the naive $k$-instanton piece of the Nekrasov partition function due to the presence of extra, gauge neutral states.
\newline

\noindent For instance, for the 6d SCFT $\six{2}{SU(2)}{}$ the sum over elliptic genera reduces to the instanton piece of the 5d Nekrasov partition function of $G=SU(2)$ SYM with $N_f=4$ flavors, times an extra factor of \footnote{ This factor differs slightly from the one given in equation (4.69) of \cite{Hayashi:2013qwa}. Namely, the 5d partition function has flavor symmetry $SO(8)$, and therefore depends on an additional fugacity in comparison with the 6d SCFT. Setting this extra fugacity to zero gives \eqref{eq:extrasu2}.}
\begin{equation}\label{eq:extrasu2}
\prod_{j,k\geq 1} (1-Q_{5d}\,e^{2\pi i\, (j-1) \epsilon_1}e^{2\pi i\, k\, \epsilon_2})^{-1}(1-Q_{5d}\,e^{2\pi i\, j\, \epsilon_1}e^{2\pi i\, (k-1) \epsilon_2})^{-1},
\end{equation}
where the two infinite products are associated with the two $\mathcal{O}(0)\to\mathbb{P}^1$ line bundles which one obtains after sending the volume of the adjoint node in the Kodaira fiber to infinity. At the level of the one-instanton partition function, one has:
\begin{equation}
Z'_{1-inst}(\massG,\massF,x,v) = Z_{1-inst}(\massG,\massF,x,v)+(v^{-1}+v)\frac{v}{(1-v\, x)(1-v/x)},
\end{equation}
where
\begin{align} Z_{1-inst}(\massG,\massF,x,v) &= v\frac{-(1+\chi^F_{(100)})\left(\sum_{n=0}^\infty \chi^G_{(2n)}v^{2n+1}\right)+\chi^F_{(001)}\left(\sum_{n=0}^\infty\chi^G_{(2n+1)}v^{2n+2}\right)}{(1-v\, x)(1-v/x)}
 \end{align}
is the 5d Nekrasov partition function, expressed here as an infinite sum in terms of the characters of the 6d flavor and gauge symmetries, $F=SO(7)$ and $G=SU(2)$.

\subsubsection{Circle reduction of 6d models with $n=1$}\label{sec:5dlimnis1}
Finally, let us turn to the case $n=1$; we have seen in section \ref{sec:5dm} that in the limit $q\to 0$ the local geometry reduces simply to the conifold geometry $\mathcal{O}(-1)\oplus \mathcal{O}(-1)\to \mathbb{P}^1$ over the base curve, and all exceptional divisors in the elliptic fiber decouple. In other words, the geometry suggests that the strings of the $n=1$ SCFTs in the 5d limit should simply contribute as a free hypermultiplet to the 5d theory. In the following we discuss how this  phenomenon is realized from the perspective of the strings' CFTs. It is convenient to begin by discussing the more familiar case of the E-string, and then generalize to arbitrary \hh{1}{G} theories.\newline

\noindent The 6d E-string SCFT is realized in M-theory as the theory of one M5 brane probing an M9 plane, where the M9 brane provides the $E_8$ flavor symmetry of the E-string SCFT.\footnote{ To be more precise, the M5-M9 brane system also includes a decoupled theory of one free 6d hypermultiplet that reflects the freedom to move the M5 brane along the M9 plane.} It famously also admits a realization as the UV fixed point of the 5d $Sp(1)$ gauge theory with 8 hypermultiplets. As explained in \cite{Ganor:1996mu}, in order to arrive at this realization one must turn on a Wilson line along the compactification circle, which breaks $E_8$ to $SO(16)$. In the presence of the Wilson line, the 5d $U(1)$ vector multiplet arising from compactification of the tensor multiplet enhances at the origin of the Coulomb branch to $Sp(1)$ and the W-bosons arise from wrapped strings \cite{Seiberg:1996bd,Ganor:1996mu} (see also \cite{Tachikawa:2015mha}). The $Sp(1)$ gauge theory with $N_f\leq 7$ hypermultiplets has flavor symmetry $SO(2N_f)$ on the Coulomb branch, which enhances to the exceptional group $E_{N_f+1}$ at the fixed point; for $N_f=8$, the enhancement is to the affine $E_8$ algebra. The $Sp(1)$ gauge theory on the Coulomb branch is the worldvolume theory supported of the D4 brane in the Type I' configuration of table \ref{tab:type1sp}, where the D8 branes all sit on top of the $O8^-$ plane.\newline

\begin{table}[t!]
\begin{center}
\begin{tabular}{c|cccccccccc}
&0&1&2&3&4&5&6&7&8&9\\\hline
D4&X&X&X&X&X&--&--&--&--&--\\
$O8^{-}$&X&X&X&X&X&X&X&X&X&--\\
8 D8&X&X&X&X&X&X&X&X&X&--
\end{tabular}
\end{center}
\caption{Type I' brane configuration for the 5d Sp(1) gauge theory with 8 flavors.}
\label{tab:type1sp}
\end{table}%

\noindent However, in the limit we are considering in the present work the Wilson line is not turned on. In the Type I' framework, varying the value of the Wilson line back to 0 corresponds to moving one of the D8 branes away from the O8 plane to a critical point; as a consequence, the profile of the dilaton field varies in the interval between the displaced D8 brane and the remaining configuration of O8+D8 branes. The enhancement to the exceptional algebra occurs when the position of the displaced D8 brane is tuned so that the dilaton diverges at the orientifold plane \cite{Polchinski:1996fm}; from the point of view of the D4 brane worldvolume theory this leads to the enhancement of the flavor symmetry group \cite{Seiberg:1996bd}. In this strongly coupled background the description of the D4 brane worldvolume theory as a $Sp(1)$ gauge theory breaks down. It turns out that in order to obtain a finite partition function the sensible limit to take is
\begin{equation}\label{eq:5dlime}
Z_\varphi(\phi,\mass_{SO(16)},x,v,q)\to Z^{(0)}_\varphi(\varphi,\mass_{SO(16)},x,v)=\sum_{k=0}^\infty \mathcal{Q}_{5d}^k \lim_{q\to 0} q^{k/2}\mathbb{E}_{(k)}(\mass_{SO(16)},x,v,q),
\end{equation}
where
\begin{equation}
\mathcal{Q}_{5d} = q^{-1/2}e^{-\Phi}.
\end{equation}
As expected from the geometry, this limit turns out to be very simple: $Z^{(0)}_\varphi(\varphi,\mass_{SO(16)},x,v)$ is just the 5d Nekrasov partition function for a free hypermultiplet of mass $m=-\log(\mathcal{Q}_{5d})$:
\begin{equation}\label{eq:estrhyper}
Z^{(0)}_\varphi(\varphi,\mass_{SO(16)},x,v) = PE\left[\frac{v\,\mathcal{Q}_{5d}}{(1-v\,x)(1-v/x)}\right],
\end{equation}
and any $\mass_{SO(16)}$-dependent states that couple to the $E_8$ degrees of freedom only appear at higher order in the $q$ expansion.\newline

\noindent It is instructive to look at this behavior from the perspective of the BPS string CFT. We adopt the description of the CFT of one E-string (without center of mass piece) as eight chiral fermions with Ramond boundary conditions, coupled to a $O(1)$ gauge field. From this perspective, the elliptic genus receives contributions from the states with $\overline{L}_0 = 0$; the level matching condition for a string wrapped once around the 6d circle is simply
\begin{equation}\label{eq:lmn}
L_0 - \overline{L}_0 = n,
\end{equation}
where $n\in \mathbb{Z}$ reflects the possibility of giving strings momentum around the 6d circle. The $L_0=0$  states in the Ramond sector include the degenerate Ramond vacuum, which provides a set of states in the $\mathbf{128}^{\bf{s}}$ spinor representation of $SO(16)$. However, due to the coupling to $O(1)$ which leads to summing over different periodicities for the fermions, one is also lead to include states built out of the NS vacuum, which from this perspective has $L_0=-1$ and we therefore denote as  $\vert -1\rangle$. The appearance of this state is allowed due to the level matching condition \eqref{eq:lmn}; if 6d circle were decompactified, on the other hand, this state would be interpreted as a tachyon but would be projected out by the more stringent level-matching condition $L_0-\overline{L}_0=0$. The five-dimensional limit \ref{eq:5dlime} singles out precisely the $L_0=-1$ and decouples all states with $L_0\geq 0$.\newline

\noindent Besides the Ramond ground states, then, at $L_0=0$ one also finds the $\textbf{120}$ vector representation of $SO(16)$ which arises from acting on the NS vacuum with a pair of fermionic operators:
\begin{equation}
\vert \textbf{120}\rangle = \left\{\psi^{i}_{1/2}\psi^{j}_{1/2}\vert -1\rangle\right\}^{i,j=1,\dots,16}_{j\neq i}.
\end{equation}

\noindent In other words, the at $L_0=0$ one finds the adjoint representation $\bf{248} = \bf{128}^{\bf{s}} + \bf{120} $ of $E_8$, but the lowest energy state at $L_0=-1$ is just a singlet, which gives rise to the $\mathcal{Q}_{5d}^1$ term in the series expansion of equation \eqref{eq:estrhyper}.\newline

\noindent Similar considerations can be applied to the \hh{1}{Sp(N)} theories. On the one hand, upon turning on a suitable Wilson line it is expected that the 6d SCFT with $n=1$ and $G=Sp(N)$ can be realized in 5d as the $Sp(N+1)$ theory with $8+2N$ flavors \cite{Gaiotto:2015una}. This equivalence has been analyzed in detail for the case $G = Sp(1)$ in \cite{Hayashi:2016abm}, where it was shown that the $T^2\times\mathbb{R}^4 $ partition function of the 6d SCFT agrees with the partition function of the 5d $\mathcal{N}=1$ $Sp(2)$ theory with 10 flavors. On the other hand, if one does not turn on any Wilson lines one finds again a single free hypermultiplet in the spectrum of the theory in the zero size limit of the 6d circle; this can again be studied from the perspective of 
a single string by exploiting the free field realization of the \hh{1}{Sp(N)} CFTs detailed in section \ref{sec:1spn}. Recall that the elliptic genus \eqref{eq:spnel} is given by the following sum over $O(1)$ holonomies:
\begin{align}
&\mathbb{E}_1^{Sp(N)}(\mass_{Sp(N)},\mass_{SO(16+4N)},v,q) =\nonumber\\
&= \frac{1}{2}\sum_{a\in\{0,\frac{1}{2},\frac{\tau}{2},\frac{1+\tau}{2}\}}\left(\prod_{i=1}^{8+2N}\frac{\th_1(e^{2\pi i a}m^i_{SO(16+4N)},q)}{\eta(q)}\right)\left(\prod_{i=1}^{N}\frac{\eta(q)^2}{\th_1(e^{2\pi i a}v\,m^i_{Sp(N)},q)\th_1(e^{2\pi i a}v/m^i_{Sp(N)},q)}\right).\label{eq:spn2}
\end{align}
For the $a=0$ and $a=1/2$ terms the $O(1)$ holonomy does not affect the boundary conditions of the periodic fermions. Their contribution to the $L_0=0$ sector in the elliptic genus is
\begin{equation}
Z^{Sp(N)}_{R}=\frac{1}{2}\sum_{s=\pm1}\frac{\prod_{i=1}^{8+2N}((m_{SO(16+4N)}^i)^{-1/2}+s (m_{SO(16+4N)}^i)^{1/2})}{\prod_{i=1}^N((v\, m_{Sp(N)}^i)^{-1/2}+s(v\,m_{Sp(N)}^i)^{1/2})((v^{-1}m_{Sp(N)}^i)^{-1/2}+s(v^{-1}m_{Sp(N)}^i)^{1/2}))}.
\end{equation}
This coincides with what would be the one-instanton partition function of the 5d $Sp(N)$ theory with $8+2N$ flavors, which can easily be obtained from \cite{Kim:2012gu}: 
\begin{equation}
Z^{Sp(N)}_{R}= \left[\frac{v}{(1-v\,x)(1-v/x)}\right]^{-1}Z_{1-inst}.
\end{equation}
Of course, genuine 5d theories with such a high number of multiplets are inconsistent, as the metric on the 5d Coulomb phase would become negative definite. It is reassuring that the corresponding one-instanton partition function appears here as a component of the 6d Nekrasov partition function and cannot be isolated by taking the 5d limit as is the case for the 6d SCFTs considered in section \ref{sec:n35d}.\newline

\noindent The $a=\frac{\tau}{2}$ and $a=\frac{1+\tau}{2}$ terms in equation \eqref{eq:spn2} can be thought of contributions from $16+4N$ antiperiodic fermions $\psi^i$ in the vector representation of $F=SO(16+4N)$ and $2N$ antiperiodic bosons $\phi^{\pm,i}$ transforming in the fundamental representations of $G=Sp(N)$ and $SU(2)_R$. Their lowest-energy contributions to the elliptic genus consist of a singlet state $\vert -1\rangle$ at $L_0=-1$, which is the only surviving state in the limit \ref{eq:5dlime} and can be viewed as being a component of the Nekrasov partition function of a free hypermultiplet. At $L_0=0$, one finds a number of states obtained by acting on the singlet state with the $L_0=1/2$ components $\phi^{\pm,i}_{1/2},\psi^i_{1/2}$ of the free fermionic and bosonic operators. Specifically, one finds: a set of states in the adjoint representation of $F$ coming from states of the form 
\begin{equation}
\psi_{1/2}^{[i}\psi_{1/2}^{j]}\vert -1\rangle;
\end{equation}
a set of states in the bifundamental representation of $G\times F$ and in the $j=1/2$ representation of $SU(2)_R$, of the form 
\begin{equation}
\psi^i_{1/2}\phi^{\pm,j}_{1/2}\vert -1\rangle;
\end{equation}
a set of states in the adjoint of $G$ and in the $j=1$ representation of $SU(2)_R$ coming from symmetric combinations
\begin{equation}
\phi^{+,(i}_{1/2}\phi^{+,j)}_{1/2}\vert -1\rangle,\qquad \phi^{-,(i}_{1/2}\phi^{-,j)}_{1/2}\vert -1\rangle, \text{and}\qquad \phi^{+,(i}_{1/2}\phi^{-,j)}_{1/2}\vert -1\rangle,
\end{equation}
and finally a set of states in the trivial representation of $SU(2)_R$, in the anti-symmetric representation of $G$ plus a singlet, coming from the following states:
\begin{equation}
\phi^{+,[i}_{1/2}\phi^{-,j]}_{1/2}\vert -1\rangle.
\end{equation}
All in all, these states contribute the following terms to the $L_0=0$ sector of the elliptic genus:
\begin{align}
Z_{extra}&=\chi^{SU(2)_R}_{(2)}(v)\chi^{Sp(N)}_{(20\dots0)}(\mass_{Sp(N)})-\chi^{SU(2)_R}_{(1)}(v)\chi^{Sp(N)}_{(10\dots0)}(\mass_{Sp(N)})\chi^{SO(16+4N)}_{(10\dots0)}(\mass_{SO(16+4N)})\nonumber\\
&+(\chi^{Sp(N)}_{(01\dots0)}(\mass_{Sp(N)})+1).
\end{align}
If on the other hand we expand the $SU(2)_R$ characters in powers of $v$, we find that the $v^0$ coefficient is given by
\begin{equation}
\chi^{Sp(N)}_{(20\dots0)}(\mathbf{m}_{Sp(N)})+\chi^{Sp(N)}_{(01\dots0)}(\mathbf{m}_{Sp(N)})+1
\end{equation}
as already remarked in section \ref{sec:gengauge}.\\

\noindent For other \hh{1}{G} CFTs we do not have a weakly coupled description, but we still expect the elliptic genus to take the same form:
\begin{align}
&\mathbb{E}_{1}^G(\massG,\massF,v,q) =\nonumber\\
& q^{\frac{1}{2}-\frac{1}{6}}\bigg(q^{-1}+q^0\bigg(Z_{extra}(\massG,\massF,v)+\left[\frac{v}{(1-v\,x)(1-v/x)}\right]^{-1}Z_{1-inst}(\massG,\massF,v)\bigg)+\mathcal{O}(q)\bigg),
\end{align}
where $Z_{1-inst}(\massG,\massF,v)$ is the Nekrasov partition function of the would-be 5d $\mathcal{N}=1$ theory with gauge group $G$ and matter content obtained by compactification from the 6d hypermultiplets, whereas $Z_{extra}(\massG,\massF,v)$ comes from acting on the $L_0=-1$ singlet with various combinations of operators as in the $G=Sp(N)$ case.\newline

\noindent The appearance of a singlet at $L_0=-1$ follows also from the spectral flow symmetry \eqref{eq:genspecflow} that maps its contribution to the R--R elliptic genus to contribution of the vacuum in the NS--R elliptic genus \eqref{eq:ellg0} that is neutral under $G$. Summing up the contributions from arbitrary numbers of strings, one again expects the instanton piece of the 5d Nekrasov partition function to coincide with that of a free hypermultiplet of mass $m=-\log(\mathcal{Q}_{5d})$, in accord with the expectations from geometric engineering.

\section{Modular bootstrap of the elliptic genera}\label{sec:modul}

In this section we exploit modularity of the BPS strings' elliptic genera to formulate a general Ansatz in terms of Jacobi forms that captures the elliptic genus for all \hh{n}{G} theories. In section \ref{sec:Ansatz} we formulate the Ansatz, which depends only on $n$ and $G$ and fixes the elliptic genus up to a finite number of undetermined coefficients. In section \ref{sec:constr} we state a number of constraints on the elliptic genus that arise from the universal properties of section \ref{sec:univ}. In section \ref{sec:modresults} we then use these constraints to fix the undetermined coefficients in the elliptic genera for the theories with $\text{rank}(G)\leq 7$; this restriction on the rank covers all exceptional theories whose elliptic genus is not yet known. In most cases, we are able to completely fix the elliptic genus, while in a few cases we are left with a small number of undetermined coefficients. In section \ref{sec:mod5d} we use these results to compute the one-instanton piece of the 5d partition functions that arise from circle compactification of the 6d theory as described in section \ref{sec:5dlim}.

\subsection{The Ansatz}\label{sec:Ansatz}
In this section we aim to express the elliptic genus of an arbitrary \hh{n}{G} theory as an element of a finite-dimensional vector space of meromorphic Jacobi of the correct weight and index. The Ansatz we obtain is a straightforward generalization of the one presented in \cite{DelZotto:2016pvm} for BPS strings of 6d SCFTs without matter.\newline

\noindent In order to make the computations tractable and to keep the discussion succint, we make some simplifying choices. Namely, we only focus on the case of a single BPS string, and we choose to turn off the chemical potentials for $G$ and $F$, so that the elliptic genus of the \hh{n}{G} theory only depends on the parameters $v$ and $q$. The dependence of the elliptic genus on these fugacities is captured in terms of Weyl-invariant Jacobi forms, which have been employed recently in \cite{DelZotto:2017mee} to study the elliptic genera of the strings of the 6d SCFT without matter, and in \cite{Kim:2018gak} for the case at hand of theories with matter.\footnote{ The authors of \cite{Kim:2018gak} were in particular recently able to compute the elliptic genera of the \hh{3}{G_2} and \hh{3}{SO(7)} theories by these techniques, which also fall into the discussion of this section.} Turning off these fugacities has the benefit of greatly reducing the vector space of Jacobi forms one needs to consider.\newline

\noindent The elliptic genus is a torus partition function, which by its behavior under large diffeomorphisms is expected to transform as a Jacobi form of modular weight zero, that is, as a function of the variables\footnote{ In this section, unlike in the rest of the paper, we find it more convenient to write all expressions in terms non-exponentiated variables.} $(\tau,\epsilon_+)\in\mathbb{H}\times\mathbb{C}$ such that (up to possibly an overall sign)
\begin{equation}
\mathbb{E}_{n}^G(\epsilon_+/\tau,-1/\tau) = e^{2\pi i\,\epsilon_+^2\, \rho/\tau}\,\mathbb{E}_{n}^G(\epsilon_+,\tau),
\end{equation}
where the index $\rho$ is the coefficient of $\epsilon_+^2$ in the modular anomaly polynomial. The latter is obtained from the anomaly polynomial of the 2d $(0,4)$ theory in Equation \eqref{eq:anomal2d} by an equivariant integral which amounts to performing the substitution
\begin{equation}
c_2(L) \mapsto -\varepsilon_-^2, \quad c_2(R)\mapsto -\varepsilon_+^2, \quad c_2(I) \mapsto -\varepsilon_+^2, \quad p_1(M_2) \mapsto 0.
\end{equation}
The resulting modular anomaly (for a generic bound state $ \strich=\sum Q^I\Sigma_I$ of BPS strings) is 
\begin{equation}
\begin{aligned}
f &= \, \frac{1}{2} A_{IJ} Q^I Q^J \Big(-\varepsilon_-^2+\varepsilon_+^2\Big)\\& \qquad + Q^I \Big(\frac{1}{4} A_{Ia} \text{Tr} (F^{(a)})^2 -  \frac{1}{4} (2 - A_{II}) \big( 2 \varepsilon_-^2 + 2 \varepsilon_+^2 \big) - h^\vee_{G_I}\varepsilon_+^2 \Big)\\
& = \,  - \frac{\varepsilon_-^2}{2}  \left((2 - A_{II})Q^I + A_{IJ} Q^I Q^J\right) \\
& \quad - \frac{\varepsilon_+^2}{2} \left( \left((2 - A_{II}) + 2 h^\vee_{G_I}\right) Q^I - A_{IJ} Q^I Q^J \right) \\
& \quad + Q^I A_{Ia} \frac{1}{4}  \text{Tr} (F^{(a)})^2
\end{aligned}
\end{equation}
Specializing the above to the case of a rank-one theory with $A_{11} = n$, we obtain
\begin{equation}
\begin{aligned}
f &= \,  - \frac{\varepsilon_-^2}{2}  \left((2 - n)Q + n Q^2\right) \\
& \quad - \frac{\varepsilon_+^2}{2} \left( \left((2 - n) + 2 h^\vee_{G_i}\right) Q^i - n Q^2 \right) \\
& \quad + Q^i A_{ia} \frac{1}{4}  \text{Tr} (F^{(a)})^2\,.
\end{aligned}
\end{equation}
In particular, for $Q=1$, we have
\begin{equation}
\begin{aligned}
f = \,  - \varepsilon_-^2 + \varepsilon_+^2 (n - h^\vee_{G} - 1)  - \frac{1}{4}  A_{1a} \text{Tr} (F^{(a)})^2\,.
\end{aligned}
\end{equation}
\noindent Setting to zero the flavor fugacities and subtracting the contribution of the center of mass hypermultiplet one is left with 
\be
f = \, \varepsilon_+^2 (n - h^\vee_{G}),
\ee
\noindent so that for the \hh{n}{G} theories $\rho$ is given by
\begin{equation}
\rho = n-h^\vee_G.
\end{equation}

\noindent In order to motivate our Ansatz, let us start by briefly recalling the case of the BPS strings for 6d SCFTs without matter that was considered in \cite{DelZotto:2016pvm}, that is: \hh{3}{SU(3)}, \hh{4}{SO(8)}, \hh{5}{F_4}, \hh{6}{E_6}, \hh{8}{E_7}, and \hh{12}{E_8}. For these theories $h^\vee_G=3(n-2)$, so that the index is given by $6-2n$. Moreover the elliptic genus takes the following form:
\begin{equation}
\frac{\mathcal{N}(2\,\epsilon_+,\tau)}{\eta(\tau)^{4(h^\vee_G-1)}\varphi_{-2,1}(2\,\epsilon_+,\tau)^{\frac{\text{dim}(G)-\text{rank}(G)}{2}}},
\end{equation}
where $\eta(\tau)$ is the weight--1/2 Dedekind eta function and $\varphi_{-2,1}(2\epsilon_+,\tau)$ is a holomorphic Jacobi form of weight $-2$ and index 1 with respect to $2\,\epsilon_+$ (i.e. of index 4 with respect to $\epsilon_+$), whose explicit formula is given by:
\begin{equation}
\varphi_{-2,1}(2\epsilon_+,\tau) = -\frac{\theta_1(2\epsilon_+,\tau)^2}{\eta^6}.
\end{equation}

\noindent The factors of $\eta(\tau)$ are responsible for the leading order singularity in the $q$-expansion of the elliptic genus, which is $q^{-\frac{h^\vee_G-1}{6}}$. Furthermore, the factors of $\varphi_{-2,1}(2\epsilon_+,\tau)$ in the denominator can be interpreted as contributions to the elliptic genus from the bosonic generators of the moduli space of one $G$-instanton. The denominator terms are responsible for the meromorphic behavior of the elliptic genus, which diverges when $2\,\epsilon_+$ takes values at the lattice points
\begin{equation}
p_1\cdot 1/2+p_2\cdot\tau/2, \,\,(p_1,p_2)\in\mathbb{Z}^2.
\end{equation}
The numerator is a holomorphic Jacobi form of even weight 
\begin{equation}
m=2(h^\vee_G-1)-(\text{dim}(G)-\text{rank}(G))\quad \in\quad2\,\mathbb{Z}
\end{equation}
and index 
\begin{equation}
k=\frac{1-h^\vee_G/3+\text{dim}(G)-\text{rank}(G)}{2}\quad \in\quad \mathbb{Z}_+/2
\end{equation}
with respect to $2\,\epsilon_+$. In \cite{DelZotto:2016pvm} it was observed that only even powers of $v=e^{2\pi i \epsilon_+}$ appear in the elliptic genera. Therefore, the numerator can also be captured by a holomorphic Jacobi function of integer or half-integer index with respect to $2\, \epsilon_+$, as follows from a lemma of Gritsenko \cite{Gritsenko:1999fk}. The  numerator is then an element of the finite-dimensional vector space of holomorphic Jacobi forms $J^{1/2}_{m,k}(2\,\epsilon_+,\tau)$ of weight $m$ and index $k$, which is in turn a component of the bi-graded ring 
\begin{equation}
J^{1/2}_{*,*}(2\,\epsilon_+,\tau)=\bigoplus_{M\in\mathbb{Z}}\bigoplus_{K\in\mathbb{Z}_+/2}J^{1/2}_{M,K}(2\,\epsilon_+,\tau).
\end{equation}
of Jacobi forms of integer weight and integer or half-integer index. As we review in section \ref{sec:appmod}, $J^{1/2}_{*,*}(2\,\epsilon_+,\tau)$ is a polynomial ring over the ring of $SL(2,\mathbb{Z})$ modular forms (generated by the weight 4 and 6 Eisenstein series $E_4(\tau),E_6(\tau)$), which itself is freely generated by the three Jacobi forms
\begin{equation}
\varphi_{M,\,K}(2\,\epsilon_+,\tau)\in J^{1/2}_{M,\,K}(2\,\epsilon_+,\tau), \qquad (M,K)=(0,1),(-2,1),\text{ or } (0,3/2).
\end{equation}
As a matter of fact, when gauge fugacities $\massG$ are switched off it follows from the structure of the ring of Jacobi forms that cancellations must occur between the numerator and denominator, and the elliptic genus takes the form
\begin{equation}
\frac{\mathcal{N}'(2\,\epsilon_+,\tau)}{\eta(\tau)^{4(h^\vee_G-1)}\varphi_{-2,1}(2\,\epsilon_+,\tau)^{h^\vee_G-1}},
\end{equation}
where $\mathcal{N}'(2\,\epsilon_+,\tau)$ is now a holomorphic Jacobi form of weight 0 and index
\begin{equation}
k'=\frac{5}{6}h^\vee_G-\frac{1}{2}
\end{equation}
with respect to $2\,\epsilon_+$. As is the case for the Hilbert series of the moduli space of one instanton \cite{Benvenuti:2010pq}, the presence of a factor of $\varphi_{-2,1}(2\,\epsilon_+,\tau)^{h^\vee_G-1}$ in the denominator is consistent with the fact that the quaternionic dimension of $\widetilde{\mathcal{M}}_{G,1}$ is $h^\vee_G-1$.\newline

\noindent The generalization to arbitrary \hh{n}{G} theories is simple. The $\eta(\tau)$ factor is fixed by the leading order singularity of the elliptic genus, equation \eqref{eq:ellimo}, to be:
\begin{equation}
\eta(\tau)^{-12(n-2)+4+24\,\delta_{n,1}}.
\end{equation}
The remaining terms in the denominator of the elliptic genus again capture the bosonic degrees of freedom of the CFT, which from the discussion in section \ref{sec:univ} is a sigma model with target space the moduli space of one $G$-instanton. Therefore, in analogy with the theories without matter, one can write the following Ansatz:
\begin{equation}\label{eq:ellansgen}
\boxed{\mathbb{E}_n^G(\epsilon_+,\tau) = \frac{\mathcal{N}_n^G(\epsilon_+,\tau)}{\eta(\tau)^{12(n-2)-4+24\,\delta_{n,1}}\varphi_{-2,1}(2\,\epsilon_+,\tau)^{h^\vee_G-1}}.}
\end{equation}
The numerator factor $\mathcal{N}_n^G(\epsilon_+,\tau)$ is now a holomorphic Jacobi form of even weight
\begin{equation}
m_n^G = 6(n-2)-2+12\,\delta_{n,1}-2(h^\vee_G-1)\quad\in\quad 2\,\mathbb{Z}
\end{equation}
and index
\begin{equation}
k_n^G=n+3\,h^\vee_G-4\quad\in\quad \mathbb{Z}_+
\end{equation}
with respect to $\epsilon_+$. The presence of matter implies that it is no longer possible to express the numerator in terms of even powers of $v$ only, as can be seen already in the case of the M-string, for which
\begin{equation}
\mathbb{E}_1^{SU(2)}(\epsilon_+,\tau) = \eta(\tau)^4\varphi_{-2,1}(\epsilon_+,\tau).
\end{equation}
Therefore the numerator must be expressed in terms of Jacobi forms with elliptic parameter $\epsilon_+$ and no longer $2\,\epsilon_+$.
\newline

\noindent The bi-graded ring of such Jacobi forms of arbitrary integral weight and positive integral index,
\begin{equation}\label{eq:Jring}
J_{*,*}(\epsilon_+,\tau)=\bigoplus_{m\in\mathbb{Z}}\bigoplus_{k\in\mathbb{Z}_+}J_{m,k}(\epsilon_+,\tau).
\end{equation}
 is the subring of the polynomial ring $J^{1/2}_{*,*}(\epsilon_+,\tau)$ over modular forms of $SL(2,\mathbb{Z})$ which is freely generated by
\begin{equation}
\varphi_{-2,0}(\epsilon_+,\tau)\qquad \text{and}\qquad\varphi_{0,0}(\epsilon_+,\tau).
\end{equation}
For fixed weight and index the vector space of Jacobi forms is finite-dimensional.  It follows from the structure of the ring \eqref{eq:Jring} of Jacobi forms that the numerator can always be written in the form
\begin{equation}
\mathcal{N}_n^G(\epsilon_+,\tau)=\varphi_{-2,1}(\epsilon_+,\tau)^{h^\vee_G-3\,n+6(1-\delta_{n,1})}p_{n}^G(E_4(\tau),E_6(\tau),\varphi_{0,1}(\tau),\varphi_{-2,1}(\tau)),
\end{equation}
where
\begin{equation}
p_{n}^G(E_4(\tau),E_6(\tau),\varphi_{0,1}(\tau),\varphi_{-2,1}(\tau))\in J_{0,4n+2h^\vee_G-10+6\,\delta_{n,1}}(\epsilon_+,\tau)
\end{equation}
is a Jacobi form of integer weight and integer index with respect to $\epsilon_+$, which can be expressed as the following finite sum:
\begin{equation}
p_{n}^G(E_4(\tau),E_6(\tau),\varphi_{0,1}(\tau),\varphi_{-2,1}(\tau)) = \sum_{\{a,b,c,d\}\in \mathcal{I}_n^G} r_{a,b,c,d}\, E_4(\tau)^aE_6(\tau)^b\varphi_{-2,1}(\epsilon_+,\tau)^c\varphi_{-2,1}(\epsilon_+,\tau)^d,
\end{equation}
where $r_{a,b,c,d}$ are a set of undetermined numerical coefficients, and $\mathcal{I}_{n}^G$ is the set of tuples of non-negative integers $a,b,c,d$ such that
\begin{align}
4\,a+6\,b-2\,c&=0,\\
c+d&=4n+2h^\vee_G-10+6\,\delta_{n,1},
\end{align}
so that the Jacobi form possesses the correct weight and index.\newline

\noindent In section \ref{sec:constr} we will make use of the universal features of the \hh{n}{G} CFTs discussed in section \ref{sec:univ} to impose constraints on the numerical coefficients $r_{a,b,c,d}$. For the 6d SCFTs with no matter, $\mathcal{N}_n^G(\epsilon_+,\tau)$ can also be obtained easily from the results presented in appendix A.1 of \cite{DelZotto:2016pvm}. There, the numerator factor was determined as a Jacobi form with elliptic parameter $2\, \epsilon_+$; one can express such a Jacobi form in terms of Jacobi forms with elliptic parameter $\epsilon_+$ by making use of identities \eqref{eq:v2vsva}\,--\,\eqref{eq:v2vsvc}.\newline

\noindent We end this section with a brief remark on the flavor symmetry. The $F$ dependence, which is suppressed in equation \eqref{eq:ellansgen}, is realized in terms of chiral fermions and therefore can only enter in the numerator of the elliptic genus. From our discussion in section \ref{sec:wzwuniv}, it follows that the numerator should be given by a linear combination of Kac-Moody characters of $F$, at the level given in tables \ref{tab:flavor1} and \ref{tab:flavor2}, multiplied by the components of a vector-valued Jacobi form of suitable weight and index that captures the dependence of the numerator on $\epsilon_+$ (provided that we do not turn on the chemical potentials $\massG$). The components of the vector-valued Jacobi must transform under a modular transformation in such a way that the combination with the Kac-Moody characters for $F$ is as an ordinary (scalar) Jacobi form. It would be interesting to see whether this additional structure on the numerator can be used to further simplify the computation of elliptic genera.

\subsection{Constraints on the Ansatz}\label{sec:constr}
The general features of the \hh{n}{G} theories that were discussed in section \ref{sec:univ} can be used to impose constraints on the coefficients of the elliptic genus and to fix the undetermined coefficients $r_{a,b,c,d}$ in the Ansatz \eqref{eq:ellansgen}. In particular, one can easily formulate the following constraints:

\begin{itemize}
\item \textbf{Constraint \textbf{C1}:} In section \ref{sec:genspec} we found that the lowest energy and next-to-lowest energy states in the elliptic genus with NS boundary conditions take a universal form, given respectively in equations \eqref{eq:ellg0} and \eqref{eq:ellg1}, which is completely determined in terms of the gauge symmetry, flavor symmetry, and matter content of the theory. Thanks to the spectral flow relation \eqref{eq:genspecflow}, this implies that an infinite number of coefficients in the Ramond elliptic genus are also known. In order to illustrate this it is convenient to express the elliptic genus as the following sum:
\begin{equation}\label{eq:bjkexp}
\mathbb{E}_n^G(v,q) = q^{\frac{1}{6}-\frac{n-2}{2}-\delta_{1,n}}v^{1-n}\sum_{j,k\geq 0} b_{j k} (q/v^2)^jv^k.
\end{equation}
The condition $j\geq 0$ arises from the fact that the energy of the states in the Ramond elliptic genus is bounded from below as in equation \eqref{eq:ellimo} and is automatically satisfied by the Ansatz. On the other hand, the condition $k\geq 0$ is a consequence spectral flow and of the fact that the ground state energy in the NS sector is $-\frac{c_L}{24}$. This condition is not satisfied automatically for a generic choice of coefficients in the Ansatz, and therefore one can obtain a first set of constraints by imposing
\begin{equation}\label{eq:cons1a}
\boxed{b_{jk}=0 \qquad\text{ for } k<0.}
\end{equation}

\noindent The coefficients $b_{jk}$ are bilinear in the dimensions of representations of $F$ and $G$. In particular, from equation \eqref{eq:ellg0} it follows that
\begin{equation}\label{eq:cons1b}
\boxed{b_{n-2+\delta_{1,n}+\ell,0} = \text{dim}(R^G_{\ell\cdot \theta_G}) \qquad\text{ for } \ell\geq0,}
\end{equation}
where $\lambda^G_{\ell\cdot \theta_G}$ is the irreducible representation of $G$ whose highest weight is $\ell$ times the highest weight of the adjoint representation, $\theta_G$. Additionally, from \eqref{eq:ellg1} it follows that
\begin{equation}\label{eq:cons1c}
\boxed{b_{n-1+\delta_{1,n}+\ell,1} = -\sum_{i=1}^r\text{dim}(R^G_{\lambda^G_i+\ell\cdot\bf{\th}})\text{ dim}(R^{F}_i) \qquad\text{ for } \ell\geq0,}
\end{equation}
where we denote by $\lambda^G_i$ the highest weights of the representation $R^G_i$ of $G$. Equations \eqref{eq:cons1a}--\eqref{eq:cons1c} lead to an infinite number of constraints, a finite number of which are independent from each other.
\item \textbf{Constraint \textbf{C2}:} In sections \ref{sec:wzwuniv} and \ref{sec:gengauge} we have seen that the elliptic genus organizes itself in terms of characters of Kac-Moody algebras, according to equations \eqref{eq:elldecomp} and \eqref{eq:FGv}. Turning off the chemical potentials $\massF$ and $\massG$, this implies that the elliptic genus takes the form
\begin{align}\label{eq:FGvnom}
&\mathbb{E}_n^G(v,q)=\nonumber\\ 
&\,\,\sum_{\lambda}\sum_{R_\nu\in \text{Rep}(G)}\,\,\,\sum_{\ell=2(n-h^\vee_G)+1}^{0}\,\sum_{m\in\mathbb{Z}} n^\lambda_{\nu,\ell,m}\, \widehat\chi^F_\lambda(\mathbf{1},q)\,  \frac{q^{-\frac{c_G}{24}+h^G_\nu}\text{dim}(R_\nu)}{\prod_{j=1}^\infty(1-q^j)^{\text{dim}(G)}} \frac{q^{-\frac{c_v}{24}+h^v_{\ell,m}}v^{\ell+2(n-h^\vee_G)m}}{\prod_{j=1}^\infty(1-q^j)^{}}.
 \end{align}
 Multiplying the elliptic genus by a factor of $\prod_{j=1}^\infty(1-q^j)^{\text{dim}(G)+\text{dim}(F)+1}$ removes all denominators in the right hand side that arise from bosonic oscillators (including the ones that appear in the characters of the Kac-Moody algebra for $F$). After multiplying by this factor and expanding equation \eqref{eq:FGvnom} in the same fashion as in equation \eqref{eq:bjkexp}, it turns out that for most \hh{n}{G} elliptic genera several of the $b_{jk}$ vanish, since none of the terms appearing in the sums on the right hand side of equation \eqref{eq:FGvnom} can contribute to them. We use the vanishing of these coefficients to impose an additional set of constraints on the elliptic genera; for an explicit example of this procedure, see appendix \ref{app:xiexample}.\newline
 
\noindent Incidentally, if one pursues this approach for the elliptic genera of the 6d SCFTs without matter, one finds that necessarily
\begin{equation}
b_{jk} = 0 \qquad \text{ if }\quad  j< h^\vee_G/3 \quad \text{ or } \quad k < h^\vee_G/3,
\end{equation}
a fact which was observed in \cite{DelZotto:2016pvm} and employed there to compute the elliptic genera for these theories.\newline

\item \textbf{Constraint \textbf{C3}:} For $n=1$, we can impose additional constraints on all \hh{1}{G} theories by requiring that 
\begin{equation}
\boxed{b_{00} = 1 \qquad \text{and} \qquad b_{0j} =0\quad\text{ for }\quad j>0,}
\end{equation}
which follows from the results of section \ref{sec:5dlimnis1}.
\end{itemize}

\begin{table}[p!]
\begin{center}
\begin{multicols}{2}
\scalebox{0.9}{
\begin{tabular}{|rl|ccc|}
\hline
\multirow{ 2}{*}{$n$}&\multirow{ 2}{*}{G}&\multicolumn{3}{c|}{Number of unfixed coefficients}\\&&Ansatz & \textbf{C1} &\textbf{C1}+\textbf{C2} \\\hline\hline
*7&$E_7$&271&48&--\\\hline
6&$E_6$&140&21&--\\
*6&$E_7$&234&40&--\\\hline
5&$F_4$&80&10&--\\
*5&$E_6$&140&16&--\\
*5&$E_7$&200&33&--\\\hline
4&$SO(8)$&37&3&--\\
*4&$SO(9)$&44&4&--\\
4&$SO(10)$&52&5&--\\
*4&$SO(11)$&81&7&--\\
4&$SO(12)$&70&8&--\\
*4&$SO(13)$&80&10&--\\
4&$SO(14)$&91&12&--\\
*4&$SO(15)$&102&14&--\\
*4&$F_4$&61&7&--\\
*4 &$E_6$&91&12&--\\
*4&$E_7$&169&27&9\\\hline
3&$SU(3)$&10&--&--\\
3&$SO(7)$&19&1&--\\
*3&$SO(8)$&24&1&--\\
*3&$SO(9)$&30&2&--\\
*3&$SO(10)$&37&3&--\\
*3&$SO(11)$&44&4&--\\
*3&$SO(12)$&52&5&--\\\hline
\end{tabular}}
\columnbreak
\scalebox{0.9}{
\begin{tabular}{|rl|ccc|}
\hline
\multirow{ 2}{*}{$n$}&\multirow{ 2}{*}{G}&\multicolumn{3}{c|}{Number of unfixed coefficients}\\&&Ansatz & \textbf{C1} &\textbf{C1}+\textbf{C2}\\\hline\hline
3&$G_2$&14&--&--\\
*3&$F_4$&44&4&--\\
*3 &$E_6$&70&8&--\\
*3&$E_7$&140&21&11\\\hline
2&$SU(2)$&2&--&--\\
2&$SU(3)$&4&--&--\\
2&$SU(4)$&7&--&--\\
2&$SU(5)$&10&--&--\\
2&$SU(6)$&14&--&--\\
2&$SU(7)$&19&1&--\\
2&$SU(8)$&24&1&--\\
*2&$SO(7)$&10&--&--\\
*2&$SO(8)$&14&--&--\\
*2&$SO(9)$&19&1&--\\
*2&$SO(10)$&24&1&--\\
*2&$SO(11)$&30&2&\color{red}2\\
*2&$SO(12)_a$&37&3&2\\
*2&$SO(12)_b$&37&3&1\\
*2&$SO(13)$&44&4&--\\
*2&$G_2$&7&--&--\\
*2&$F_4$&30&2&--\\
*2 &$E_6$&52&5&3\\
*2&$E_7$&114&16& 9\\\hline
\end{tabular}}
\end{multicols}
\end{center}
\caption{Fixing the Ansatz for $n\geq 2$. For each theory, we first list the total number of unfixed coefficients in the Ansatz, and in the remaining columns the number of coefficients left unfixed after imposing successively the constraints \textbf{C1} and \textbf{C2}. For the theory \hh{2}{SO(11)} we cannot impose constraint \textbf{C2} since we do not understand the flavor symmetry $F$. We highlight in red the affected table entry. We prefix by an asterisk the theories for which our results are novel.}
\label{tab:constr1}
\end{table}

\begin{table}[t!]
\begin{center}
\begin{multicols}{2}
\scalebox{0.83}{
\begin{tabular}{|rl|cccc|}
\hline
\multirow{ 2}{*}{$n$}&\multirow{ 2}{*}{G}&\multicolumn{4}{c|}{Number of unfixed coefficients}\\&&Ansatz & \!\!\!\textbf{C1} &\!\!\!\textbf{C1}+\textbf{C2} & \!\!\!\textbf{C1}+\textbf{C2}+\textbf{C3}\\\hline\hline
1&$SU(2)$&2&\!\!\!--&\!\!\!--&\!\!\!--\\
1&$SU(3)$&6&\!\!\!--&\!\!\!--&\!\!\!--\\
1&$SU(4)$&9&\!\!\!--&\!\!\!--&\!\!\!--\\
1&$SU(5)$&13&\!\!\!1&\!\!\!--&\!\!\!--\\
1&$SU(6)$&18&\!\!\!2&\!\!\!--&\!\!\!--\\
1&$SU(7)$&23&\!\!\!3&\!\!\!--&\!\!\!--\\
1&$SU(8)$&29&\!\!\!4&1&\!\!\!--\\
*1&$SU(6)_*$&18&\!\!\!2&\!\!\!--&\!\!\!--\\
1&$Sp(2)$&6&\!\!\!--&\!\!\!--&\!\!\!--\\
1&$Sp(3)$&9&\!\!\!--&\!\!\!--&\!\!\!--\\
1&$Sp(4)$&13&\!\!\!1&\!\!\!--&\!\!\!--\\
1&$Sp(5)$&18&\!\!\!2&\!\!\!--&\!\!\!--\\
1&$Sp(6)$&23&\!\!\!3&\!\!\!--&\!\!\!--\\\hline
\end{tabular}}
\columnbreak
\scalebox{0.83}{
\begin{tabular}{|rl|cccc|}
\hline
\multirow{ 2}{*}{$n$}&\multirow{ 2}{*}{G}&\multicolumn{4}{c|}{Number of unfixed coefficients}\\&&Ansatz & \!\!\!\textbf{C1} &\!\!\!\textbf{C1}+\textbf{C2} & \!\!\!\textbf{C1}+\textbf{C2}+\textbf{C3}\\\hline\hline
1&$Sp(7)$&29&\!\!\!4&\!\!\!--&\!\!\!--\\
1&$G_2$&9&\!\!\!--&\!\!\!--&\!\!\!--\\
*1&$F_4$&36&\!\!\!6&\!\!\!4&\!\!\!1\\
*1 &$E_6$&60&\!\!\!11&\!\!\!6&\!\!\!3\\
*1&$E_7$&126&\!\!\!26&\!\!\!19&\!\!\!15\\
*1&$SO(7)$&13&\!\!\!1&\!\!\!--&\!\!\!--\\
*1&$SO(8)$&18&\!\!\!2&\!\!\!--&\!\!\!--\\
*1&$SO(9)$&23&\!\!\!3&\!\!\!2&\!\!\!--\\
*1&$SO(10)$&29&\!\!\!4&\!\!\!--&\!\!\!--\\
*1&$SO(11)$&36&\!\!\!6&\!\!\!\color{red}6&\!\!\!\color{red}1\\
*1&$SO(12)_a$&43&\!\!\!7&\!\!\!3&\!\!\!1\\
*1&$SO(12)_b$&43&\!\!\!7&\color{red}\!\!\!7&\!\!\!\color{red}1\\\hline
\end{tabular}}
\end{multicols}
\end{center}

\caption{Fixing the Ansatz for $n= 1$. For each theory, we list the total number of unfixed coefficients in the Ansatz, and in the remaining columns the number of coefficients left unfixed after imposing successively the constraints \textbf{C1},\textbf{C2}, and \textbf{C3}. For the \hh{1}{SO(11)} and \hh{2}{SO(12)_b} theories we cannot impose condition \textbf{C2} since we do not have a good enough understanding of the flavor symmetry $F$. We highlight in red the affected table entries.  We prefix by an asterisk the theories for which our results are novel.}
\label{tab:constr2}
\end{table}

\subsection{Computational results} \label{sec:modresults}
One can employ the constraints of the previous section to reduce the number of unfixed coefficients of the Ansatz \eqref{eq:ellansgen} for any \hh{n}{G} theory. As there are infinite numbers of \hh{n}{G} theories, we choose to restrict our analysis to a suitable subset of theories. From the review of section \ref{sec:catalogue}, the elliptic genera of almost all of the \hh{n}{G} theories for which $\text{rank}(G)\geq 8$ have been computed by localization. The only exceptions are the theories \hh{4}{SO(9+2N)} whose elliptic genus we compute by localization in section \ref{sec:soodd}, and the theory \hh{12}{E_8} which, although it does not have a known Lagrangian realization, has also been computed in \cite{DelZotto:2016pvm}. We therefore restrict our attention to the theories with $\text{rank}(G)\leq 7$. We also choose not to compute the elliptic genus for the \hh{8}{E_7} theory by our techniques, since it has already been computed in \cite{DelZotto:2016pvm} and the Ansatz involves a very large number of unfixed coefficients. Finally, we omit the M-string and E-string theories which have trivial gauge group and are already well studied.\newline

\noindent This leaves an additional 72 \hh{n}{G} theories, which we analyze in this section. For 42 of them, to the best of our knowledge the elliptic genus has not yet appeared in the literature. For all theories with $n\geq2$, we successively employ constraints \textbf{C1} and \textbf{C2} to fix the Ansatz, while for the theories with $n=1$ we successively employ constraints \textbf{C1}, \textbf{C2}, and \textbf{C3}. We present the results of our computation in tables \ref{tab:constr1} and \ref{tab:constr2}. The constraints we identified are sufficient to completely fix the elliptic genus for 59 out of the 72 theories, including 31 theories whose elliptic genus was not previously known.\newline

\noindent We have performed several consistency checks on the coefficients of the elliptic genera thus determined. First, in case the elliptic genus has already been calculated (see appendix \ref{sec:catalogue}), we can directly check whether our answer is consistent with the results available in the literature. Second, we can verify whether the condition
\begin{equation}\label{eq:bcoefdim2}
b_{n-1+\delta_{1,n},2} = \text{dim}(G)+\text{dim}(F)+1,
\end{equation}
which follows from equation \eqref{eq:dims}, is satisfied. We have already remarked that this condition does not hold for the theories \hh{1}{Sp(N)} with $N>2$, for which $\text{dim}(Sp(N))$ is replaced by $\text{dim}(SU(2N))$ in equation \eqref{eq:bcoefdim2}; for all other elliptic genera we compute we find that this relation holds. Finally, the coefficients $b_{jk}$ of the elliptic genus must be linear combinations with integer coefficients  of specific products of dimensions of representations of $G$ and $F$ that can be read off from the expansion \eqref{eq:FGvnom}. The coefficients count sets of states transforming in given representations of $G\times F\times U(1)_v$, with a sign; from inspection on the known elliptic genera, moreover, these coefficients always turn out to be small integers: we have not encountered any example where any such coefficient is greater than 3. For instance, for \hh{2}{F_4} the first few $b_{jk}$ we find are given by:
\begin{center}

\begin{tabular}{|l||l|l|l|l|l|l|l|}\hline $k\backslash j$&0&1&2&3&4&5\\\hline\hline 
0& 1 & 0 & 0 & 0 & 0 & -14  \\
 1&52 & -156 & 74 & 56 & 90 & 126 \\\hline
\end{tabular}

\end{center}

\noindent whereas \eqref{eq:bcoefdim2} leads to the prediction that these $b_{jk}$ coefficients are given by:

\begin{center}
\scalebox{0.76}{
\begin{tabular}{|r||r|r|r|r|r|r|r|}\hline $\!\!k\backslash j\!\!$&0&1&2&3&4&5\\\hline\hline 
0& $\!n_{0,0}^{(1)}\!\!$ & 0 & $\!n_{0,2}^{(1)}\!\!$ & 0 & 0 & \!14\,$n_{0,5}^{(1)}\!\!$  \\
 1&$\!52\, n_{1,0}^{(1)}\!+\!364\, n_{1,0}^{(2)}\!+\!84\, n_{1,0}^{(3)}\!\!$ & $\!156\, n_{1,1}^{(1)}\!+\!126\, n_{1,1}^{(2)}\!\!$ & $\!21\, n_{1,2}^{(1)}\!+\!90\, n_{1,2}^{(2)}\!+\!52\, n_{1,2}^{(3)}\!+\!n_{1,2}^{(4)}\!\!$ & $\!56\, n_{1,3}^{(1)}\!\!$ & $\!21\, n_{1,4}^{(1)}\!+\!90\, n_{1,4}^{(2)}\!\!$ & $\!156\, n_{1,5}^{(1)}\!+\!126\, n_{1,5}^{(2)}\!\!$  \\\hline
\end{tabular}}
\end{center}

\noindent There is essentially a unique solution for which all undetermined coefficients $n_{jk}^{(m)}$ are small integers:
\begin{align}
n_{0,0}^{(1)}&=-n_{0,0}^{(1)}=n_{1,0}^{(1)}=-n_{1,1}^{(1)}=n_{1,2}^{(1)}=n_{1,2}^{(3)}=n_{1,2}^{(4)}=n_{1,3}^{(1)}=n_{1,4}^{(2)}=n_{1,5}^{(2)}=1;\nonumber\\
n_{0,2}^{(1)}&=n_{1,0}^{(2)}=n_{1,0}^{(3)}=n_{1,1}^{(2)}=n_{1,2}^{(2)}=n_{1,4}^{(1)}=n_{1,5}^{(1)}=0.
\end{align}

\noindent We have performed similar checks for the entire set of 59 theories; for many theories it turns out that several $b_{jk}$ only involve linear combinations of small numbers of terms, which makes this check easy to perform. On the other hand, for theories where $F$ includes abelian factors (which as Kac-Moody algebras have large numbers of integrable highest weight representations) it turns out that the $b_{jk}$ tend to involve large numbers of terms, and it is more challenging to perform this check except for a small set of $b_{jk}$ coefficients.\newline

\noindent Of the set of 72 theories we considered, 13 remain whose elliptic genus cannot be fixed by applying the constraints \textbf{C1}, \textbf{C2}, and \textbf{C3} alone. One might still wonder if it is possible to fix the remaining undetermined coefficients in the Ansatz by applying additional constraints, such as:
\begin{description}
\item[(a)] Imposing the condition 
\begin{equation}\label{eq:bcoefdim}
b_{n-1+\delta_{1,n},2} = \text{dim}(G)+\text{dim}(F)+1,
\end{equation}
which follows from equation \eqref{eq:dims};
\item[(b)] Looking for relations between coefficients $b_{jk}$ that follow from the fact that the $F$ dependence is captured by affine Kac-Moody algebra characters as in equation \eqref{eq:FGvnom};
\item[(c)] Requiring that the coefficients $b_{jk}$ can be written in terms of small linear combinations of $n_{j,k}^{(m)}$ coefficients as in the \hh{2}{F_4} example given above.
\end{description}
On the other hand, we do not obtain additional constraints from assuming that the dependence of the elliptic genus on the 6d gauge symmetry is captured in terms of the irreducible characters of the $G$ affine Kac-Moody algebra, as opposed to just the Verma modules.\newline

\noindent We now discuss the various remaining theories in turn:
\begin{itemize}
\item For \hh{1}{SO(12)_a},  \hh{2}{SO(12)_a}, and  \hh{1}{F_4}, only one coefficient in the Ansatz was left undetermined after applying constraints \textbf{C1}, \textbf{C2}, and \textbf{C3}. By applying a combination of constraints \textbf{(a)}, \textbf{(b)}, and \textbf{(c)}, we are able to fix the remaining coefficient with a good degree of confidence; namely, after imposing constraint \textbf{(a)}, we can verify that conditions \textbf{(b)} and \textbf{(c)} are also satisfied by the elliptic genera of these theories.
\item For \hh{2}{SO(12)_b}, imposing condition \textbf{(a)} to fix the single undetermined coefficient leads to non-integer values of certain $b_{jk}$ coefficients. Notably, however, we find that if we slightly modify condition \textbf{(a)} and take the $b_{1,2} $ coefficient in the expansion of the elliptic genus to be equal to $\text{dim}(G)+\text{dim}(F)=144$ instead of $145$ the elliptic genus we find is completely consistent with conditions \textbf{(b)} and \textbf{(c)}. Moreover, the elliptic genus we find differs from the one of the \hh{2}{SO(12)_a} theory, supporting the possibility that the parent 6d SCFT $\six{2}{SO(12)_b}{}$ might be a consistent theory distinct from the $\six{2}{SO(12)_a}{}$ SCFT. 
\item For \hh{1}{SO(11)} and \hh{1}{SO(12)_b}, by setting $b_{n-1+\delta_{1,n},2} =\text{dim}(G)+\text{dim}(F)+1$ we are able to completely fix the elliptic genus, but since we do not fully understand the flavor symmetry we are not able to check the validity of our result to a satisfying degree of confidence.
\item For  \hh{2}{SO(11)} the number of undetermined  coefficients is two, and we do not understand the flavor symmetry; we could not find a way to fix the elliptic genus to a sufficient level of confidence.
\item For the theories \hh{1}{E_7}, \hh{2}{E_7}, \hh{3}{E_7}, and \hh{4}{E_7} the number of unfixed coefficients, respectively 15, 9, 11, and 9, is so high that we have not been able to find a criterion to fix them. 
\item Finally, for the theories \hh{2}{E_6} and \hh{1}{E_6} only 3 coefficients are unfixed, and condition \textbf{(a)} can be used to fix one. Unfortunately, however, the number of distinct affine characters of the flavor symmetry $F$ that are allowed to appear in the elliptic genus is extremely high (respectively 1092 and 3360), and so conditions \textbf{(b)} and \textbf{(c)} are not very predictive. As a consequence, we are unable to fix the $E_6$ elliptic genera for these theories with a good degree of confidence.
\end{itemize}
It is of course possible that one may be able to identify additional constraints on the elliptic genus that may suffice to completely fix the yet undetermined elliptic genera.\newline

\noindent In appendix \ref{app:bjk} we display our results for the elliptic genera of the 59 theories which we fixed with conditions \textbf{C1}, \textbf{C2}, \textbf{C3}, as well as the ones for the \hh{1}{SO(12)_a}, \hh{1}{F_4}, \hh{2}{SO(12)_a}, and \hh{2}{SO(12)_b} theories. In order to present our computational results in a concise and useful way, in appendix \ref{app:bjk} we have chosen to tabulate several of the $b_{jk}$ coefficients in their series expansion \eqref{eq:bjkexp}. The tables of coefficients make it easy to compare our results with computations performed by other techniques, such as localization; furthermore, it is easy to verify directly from the tables that various of the universal properties of BPS string CFTs discussed in section \ref{sec:univ} are satisfied for these theories.\newline

\noindent On the other hand, the expressions we obtain for the elliptic genera as meromorphic Jacobi forms are quite unwieldy; rather than including them in the text, we have chosen to attach a Mathematica file to this paper, where we include these expressions for all $59 + 4$ theories whose elliptic genus is completely fixed, as well as for the remaining 9 theories for which some of the coefficients are still undetermined.

\subsection{5d Nekrasov partition functions}\label{sec:mod5d}
In section \ref{sec:5dlim} we argued that the elliptic genus for the \hh{n}{G} theories for $n\geq 2$ reduces in the $q\to 0$ limit to
\begin{equation}\label{eq:5dlimez}
\lim_{q\to 0} q^{-1/6+\frac{n-2}{2}}\mathbb{E}_{n}^G(\massG,\massF,v,q) = \left[\frac{v}{(1-v\, x)(1-v/x)}\right]^{-1}Z_{1-inst}(\massG,\massF,x,v),
\end{equation}
where we have removed the center of mass contribution on the right hand side. For $n\geq 3$, the term appearing on the right hand side is the one-instanton piece of the Nekrasov partition function for the $\mathcal{N}=1$ theory with gauge group $G$ and matter hypermultiplets in the same representation as the 6d matter. For instance, the \hh{7}{E_7} theory gives in the $q\to 0$ limit the one-instanton piece of the Nekrasov partition function of the 5d $G=E_7$ theory with one half-hypermultiplet in the  $\mathbf{56}$ representation of $E_7$. For $n=2$, as discussed in section \ref{sec:5dlimnis2}, the right hand side contains extra states in comparison to the Nekrasov partition function coming from a sector decoupled from the 5d gauge theory.\newline

\noindent Our results for the 6d elliptic genus include many cases that reduce in 5d to theories whose 1-instanton partition function is not yet known. By taking the limit \eqref{eq:5dlimez} we can straightforwardly obtain the 5d one-instanton partition functions from the elliptic genera, in the limit where $\massG\to \bf{1}$ and $\massF\to \bf{1}$. For instance, for $G=F_4$ with two hypermultiplets in the   $\mathbf{26}$ representation, which descends from the 6d SCFT with $n=3$ and $G=F_4$, we obtain:
\begin{equation}\label{eq:3f4v}
Z_{1-inst}(\mathbf{1},\mathbf{1},x,v) = \frac{v^7 \left(5+80 v+268 v^2-1232 v^3+2142 v^4-1232 v^5+268 v^6+80 v^7+5 v^8\right)}{(1-v\, x)(1-v/x)(-1+v)^4 (1+v)^{16}},
\end{equation}
where we only expect our answer to differ from the equivariant calculation of the index by an overall factor of $v$, as remarked in section \ref{sec:5dlim}. We have collected our results for the 5d one-instanton partition functions, with $\massG$ and $\massF$ fugacities switched off, in the tables of appendix \ref{sec:5dappfull}.\newline

\noindent It is also of course desirable to recover the full dependence of the 5d partition function on all chemical potentials. For several theories, we are able to achieve this by comparing the expression for $Z_{1-inst}$ with $F$ and $G$ chemical potentials switched off to the leading order terms in the $q$-expansion of equation \eqref{eq:elldecomp}. We explain this approach with a detailed example in section \ref{sec:5dappgen}. In many instances have been able to infer all-order expansions in $v$ for the elliptic genus, which capture the full dependence on the $\massF$ and $\massG$ fugacities. These results are presented in appendix \ref{sec:5dappfull}.

\section{The exceptional cases \hh{3}{SU(3)} and \hh{2}{SU(2)}} \label{sec:excep}

In section \ref{sec:gengauge} we found evidence that the $G$ dependence of the elliptic genera of the \hh{n}{G} theories is captured in terms of a Kac-Moody algebra at level $-n$, and for $n\neq h^\vee_G$ we found that the elliptic genus admits an expansion given by equation \eqref{eq:FGv}. In this section we discuss the two nontrivial cases \hh{3}{SU(3)} and \hh{2}{SU(2)} for which $n=h^\vee_G$ (the remaining, trivial case being the E-string for which $G=Sp(0)$). We find an expansion of the elliptic genus of these theories similar to equation \eqref{eq:FGv}, and comment on the lack of an obvious relation to the irreducible characters of the affine Kac-Moody algebra for $G$ at critical level.

\subsection{The \hh{3}{SU(3)} case}
\noindent In the case of the \hh{3}{SU(3)} CFT the flavor symmetry $F$ is trivial, and the elliptic genus is given by a single function
\begin{equation}
\mathbb{E}_{3}^{SU(3)}(\mass_{SU(3)},v,q) = \xi^{3,SU(3)}_0(\mass_{SU(3)},v,q).
\end{equation}
Starting from the explicit expression for the elliptic genus obtained by localization in \cite{Kim:2016foj}, we find the following expansion:
\begin{equation}\label{eq:su3expan}
\xi^{3,SU(3)}_0(\mass_{SU(3)},v,q) =\left[q^{8/24}\widetilde\Delta_{SU(3)}(\mass_{SU(3)},q)\right]^{-1}\sum_{n=1}^\infty \chi^{SU(3)}_{(n-1,n-1)}(\mass_{SU(3)})\frac{v^{2n}+q^{n}v^{-2n}}{1-q^n},
\end{equation}
(a similar expression with fugacities $\mass_{SU(3)}\to\mathbf{1}$ was found in \cite{DelZotto:2016pvm}). Note that this expression is explicitly invariant under $v\to q^{1/2}/v$. \newline

\noindent Unlike in the noncritical cases, the $SU(3)$ dependence does not appear to be captured in terms of characters of the $SU(3)$ Kac-Moody algebra at the critical level $k=-3$ in a straightforward way. For an affine Lie algebra at critical level, a formula for the irreducible characters is available \cite{2007arXiv0706.1817A}. In particular, the irreducible character corresponding to the $SU(3)$ representation with Dynkin label $(n_1n_2)$ is given (up to an overall power of $q$) by:
\begin{align}
\label{eq:critcharsu3}\widehat\chi^{SU(3)}_{(n_1n_2)}(\mass_{SU(3)},q)&=\frac{\prod_{j=1}^\infty(1-q^{j})^2}{\widetilde\Delta_{SU(3)}(\mass_{SU(3)},q)}\cdot\frac{\chi_{(n_1n_2)}^{SU(3)}(\mass_{SU(3)},q)}{(1-q^{n_1+1})(1-q^{n_2+1})(1-q^{n_1+n_2+2})},\end{align}
which bears no clear relation to the $SU(3)$ dependence of the elliptic genus \eqref{eq:su3expan}.

\subsection{The \hh{2}{SU(2)} case}\label{sec:su2case}

\noindent The elliptic genus of the \hh{2}{SU(2)} theory is known from \cite{Haghighat:2014vxa}. Although the flavor symmetry of the 6d SCFT is $F=SO(7)$ instead of the naive $SO(8)$ \cite{Ohmori:2015pia}, it turns out that it is still more natural to write the elliptic genus in terms of level 1 $SO(8)$ characters. The embedding of $SO(7)$ into $SO(8)$ is such that the $SO(7)$ spinor representation is identified with the $SO(8)$ vector representation. At the level of fugacities, embedding $SO(7)$ into $SO(8)$ implies the following relations:
\begin{align}\label{eq:spec87a}
m_{SO(8)}^1 &=\frac{m_{SO(7)}^1+m_{SO(7)}^2+m_{SO(7)}^3}{2};\\
m_{SO(8)}^2 &=\frac{-m_{SO(7)}^1+m_{SO(7)}^2+m_{SO(7)}^3}{2};\\
m_{SO(8)}^3 &=\frac{m_{SO(7)}^1-m_{SO(7)}^2+m_{SO(7)}^3}{2};\\
m_{SO(8)}^4 &=\frac{m_{SO(7)}^1+m_{SO(7)}^2-m_{SO(7)}^3}{2}.\label{eq:spec87d}
\end{align}
With this assignment of $SO(8)$ fugacities, the affine level 1 $SO(8)$  characters decompose as follows:
\begin{align}
\widehat\chi^{SO(8)}_{\bf{0}}(\mass_{SO(8)},q)&=\widehat\chi^{SO(7)}_{\bf{0}}(\mass_{SO(7)},q)\widehat\chi^{SO(1)}_{\bf{0}}(q)+\widehat\chi^{SO(7)}_{\bf{v}}(\mass_{SO(7)},q)\widehat\chi^{SO(1)}_{\bf{v}}(q),\\
\widehat\chi^{SO(8)}_{\bf{c}}(\mass_{SO(8)},q)&=\widehat\chi^{SO(7)}_{\bf{0}}(\mass_{SO(7)},q)\widehat\chi^{SO(1)}_{\bf{v}}(q)+\widehat\chi^{SO(7)}_{\bf{v}}(\mass_{SO(7)},q)\widehat\chi^{SO(1)}_{\bf{0}}(q),\\
\widehat\chi^{SO(8)}_{\bf{v}}(\mass_{SO(8)},q)&=\widehat\chi^{SO(8)}_{\bf{s}}(\mass_{SO(7)},q)=\widehat\chi^{SO(7)}_{\bf{s}}(\mass_{SO(7)},q)\widehat\chi^{SO(1)}_{\bf{s}}(q),
\end{align}
where $\widehat\chi^{SO(1)}_{\bf{0}}(q),$ $\widehat\chi^{SO(1)}_{\bf{v}}(q)$, and $\widehat\chi^{SO(1)}_{\bf{s}}(q)$ are the three Ising model characters given in equations \eqref{eq:ising}.\newline

\noindent We find that the \hh{2}{SU(2)} elliptic genus can be written as follows:
\begin{align}\label{eq:esu2ch}
\mathbb{E}_{2}^{SU(2)}(\mass_{SU(2)},v,q) &= \widehat\chi^{SO(8)}_{\bf{0}}(\mass_{SO(8)},q)\xi^{2,SU(2)}_{\bf{0}}(\mass_{SU(2)},v,q)\nonumber\\
&+\widehat\chi^{SO(8)}_{\bf{c}}(\mass_{SO(8)},q)\xi^{2,SU(2)}_{\bf{c}}(\mass_{SU(2)},v,q)\nonumber\\
&+\widehat\chi^{SO(8)}_{\bf{v}}(\mass_{SO(8)},q)\xi^{2,SU(2)}_{\bf{v}}(\mass_{SU(2)},v,q),
\end{align}
where
\begin{align}
\xi^{2,SU(2)}_{\bf{0}}(\mass_{SU(2)},v,q)&=\nonumber\\
\label{eq:xisu2a}&\hspace{-0.4in}\frac{1}{q^{4/24}\prod_{j=0}^\infty(1\!-\!q^j)\widetilde\Delta_{SU(2)}(\mass_{SU(2)},q)} \,\sum_{k\geq 0}\frac{q^{k+1/2}(v^{2k+1}\!+\!v^{-2k-1})}{1-q^{2k+1}}\chi^{SU(2)}_{(2k)}(\mass_{SU(2)}),\\
\xi^{2,SU(2)}_{\bf{c}}(\mass_{SU(2)},v,q)&=\nonumber\\
\label{eq:xisu2b}&\hspace{-0.45in} -\frac{1}{q^{4/24}\prod_{j=0}^\infty(1\!-\!q^j)\widetilde\Delta_{SU(2)}(\mass_{SU(2)},q)} \,\sum_{k\geq 0}\frac{v^{2k+1}+q^{2k+1}v^{-2k-1}}{1-q^{2k+1}}\chi^{SU(2)}_{(2k)}(\mass_{SU(2)}),\\
\xi^{2,SU(2)}_{\bf{v}}(\mass_{SU(2)},v,q)&=\nonumber\\
\label{eq:xisu2c}&\hspace{-0.2in}\frac{1}{q^{4/24}\prod_{j=0}^\infty(1\!-\!q^j)\widetilde\Delta_{SU(2)}(\mass_{SU(2)},q)} \,\sum_{k\geq 0}\frac{v^{2k+2}-q^{k+1}v^{-2k-2}}{1+q^{k+1}}\chi^{SU(2)}_{(2k+1)}(\mass_{SU(2)}).
\end{align}
As a side remark, note that under the spectral flow transformation $v\to q^{1/2}/v$
\begin{equation}
\xi^{2,SU(2)}_{\bf{0}}(\mass_{SU(3)},v,q)\leftrightarrow -\xi^{2,SU(2)}_{\bf{c}}(\mass_{SU(2)},v,q)
\end{equation}
while 
\begin{equation}
\xi^{2,SU(2)}_{\bf{v}}(\mass_{SU(3)},v,q)\to -\xi^{2,SU(2)}_{\bf{v}}(\mass_{SU(3)},v,q).
\end{equation}\\
\noindent Let us now look more in detail at how the flavor symmetry is realized at the level of the elliptic genus. Recall that the $b_{1,2}$ coefficient in the series expansion \eqref{eq:bjkexp} of the elliptic genus is given (with a few exceptions noted in section \ref{sec:modresults}) by
\begin{equation}
\chi^G_{\th_G}(\mass_G)+\chi^F_{\th_F}(\mass_F)+1.
\end{equation}
Computing this coefficient for the \hh{2}{SU(2)} theory, we find
\begin{align}
b_{1,2} = \chi^{SO(8)}_{\th_{SO(8)}}(\mass_{SO(8)})+\chi_{\th_{SU(2)}}^{SU(2)}(\mass_{SU(8)})+2-\chi^{SO(8)}_{\mathbf{c}}(\mass_{SO(8)}).
\end{align}
Using 
\begin{equation}
\chi^{SO(8)}_{\th_{SO(8)}}(\mass_{SO(8)})=\chi^{SO(7)}_{\th_{SO(7)}}(\mass_{SO(7)})+\chi^{SO(7)}_{\mathbf{v}}(\mass_{SO(7)})
\end{equation}
and
\begin{equation}
\chi^{SO(8)}_{\mathbf{c}}(\mass_{SO(8)})=\chi^{SO(7)}_{\mathbf{v}}(\mass_{SO(7)})+1,
\end{equation}
we obtain
\begin{align}
b_{1,2} = \chi^{SO(7)}_{\th_{SO(7)}}(\mass_{SO(7)})+\chi_{\th_{SU(2)}}^{SU(2)}(\mass_{SU(2)})+1,
\end{align}
indicating that the BPS string CFT indeed knows about the reduction of flavor symmetry from $SO(8)$ to $SO(7)$ at the 6d superconformal fixed point!\newline

\noindent Finally, we remark that the irreducible characters of the affine $SU(2)$ Kac-Moody algebra at critical level $k=-2$ are given by:
\begin{align}
\label{eq:critcharsu2}\widehat\chi^{SU(2)}_{(n)}(\mass_{SU(2)},q)&=\frac{\prod_{j=1}^\infty(1-q^{j})}{\widetilde\Delta_{SU(2)}(\mass_{SU(2)},q)}\cdot\frac{1}{1-q^{n+1}}.
\end{align}
One may ask whether the elliptic genus of the \hh{2}{SU(2)} CFT is captured in terms of these characters, as was the case for the theories with $n\neq h^\vee_G$ studied in section \ref{sec:univ}. On the one hand, the $SU(2)$ dependence in equations \eqref{eq:xisu2a} and \eqref{eq:xisu2b} bears some similarity to equation \eqref{eq:critcharsu2} (in fact, they just differ by a factor of $\prod_{j=1}^\infty(1-q^j)^2$); on the other hand, the $SU(2)$ dependence in equation \eqref{eq:xisu2c} cannot easily be captured in terms of the irreducible characters \eqref{eq:critcharsu2}. Also, expressing the dependence of the elliptic genus in terms of $SO(7)$ characters rather than $SO(8)$ characters does not resolve this mismatch. It would be interesting to understand in more detail the chiral algebra of this CFT as well as the \hh{3}{SU(3)} CFT and the way the flavor symmetries are realized in it \cite{wip}.

\section{Conclusions and future directions}\label{sec:concl}

In this paper we have studied the properties of a BPS string of an arbitrary six-dimensional SCFT on the tensor branch from the point of view of the worldsheet $\mathcal{N}=(0,4)$ CFT that lives on it. The upshot is that a uniform description of these CFTs emerges, various features of which are fixed by a minimal amount of data: the Dirac pairing of the string $-n$, the 6d gauge symmetry $G$, and the 6d flavor symmetry $F$, which with a few exceptions is uniquely determined once $n$ and $G$ are specified.\newline

\noindent In the first half of the paper we found a uniform realization of the string worldsheet CFTs in the context of F-theory. We argued that the strings' CFTs can be obtained by compactifying the worldvolume theory of a D3 brane in the presence of defects preserving (0,4) supersymmetry in 2 dimensions; the compactification involves a topological twist which is a hybrid of the $\beta$-twist \cite{Kapustin:2006hi} and the topological duality twist \cite{Martucci:2014ema}. Two kinds of defects appear: generalized chiral defects and folding defects, which are related respectively to the matter and gauge content of the 6d SCFT; from the details of the compactification we are able to infer certain properties of these defects, including their contribution to an effective shift of the R-charge in the curved rigid supersymmetric backgrounds. A precise study of such curved supersymmetric backgrounds requires a more detailed analysis which we leave to future work \cite{DZLfuture}. The compactified theory flows in the IR to the CFT on the string, whose anomalies and central charges we compute by relating them via 2d $\cn=(0,4)$ Ward identities to the coefficients of the 2d anomaly polynomial determined by anomaly inflow from 6d after \cite{Kim:2016foj,Shimizu:2016lbw}.\newline

\noindent In  \cite{Beem:2017ooy}  a deep interconnection was found between the chiral algebras of 4d $\cn=2$ SCFTs  \cite{Beem:2013sza,Beem:2014rza,Lemos:2014lua}  and the geometry of their Higgs branches, which provides a physical realization of Arakawa's results associating a vertex operator algebra to any hyperk\"ahler variety ${\mathcal M}$, see \cite{Arakawa:2017fdq} for a review. In particular, the Schur index of the 4d SCFT \cite{Gadde:2011uv} is identified with the vacuum character of the chiral algebra. However, not all instances studied by Arakawa appear to be realized directly in the context of $\mathcal{N}=2$ SCFT; in particular the cases where ${\mathcal M}$ is the moduli space of one $F_4$ or one $G_2$ instanton appear to be excluded \cite{Argyres:2015ffa,Argyres:2015gha,Argyres:2016xmc,Caorsi:2018ahl}. For the class of $(0,4)$ theories studied in our previous work \cite{DelZotto:2016pvm}, which arise from  compactification of 4d $\cn=2$ theories without the insertion of generalized chiral defects, we were able to recover the Schur index of the 4d $\mathcal{N}=2$ SCFT out of a particular function, $L_G(v,q)$, which arose in the computation of the elliptic genus. The Schur index is obtained by specializing this function to the value $L_G(q^{1/4},q)$; for the cases corresponding to simply-laced gauge group, the Schur index in turn also corresponds to the vacuum character of the chiral algebra of the 4d $\cn=2$ SCFT \cite{Beem:2013sza}. As we have discussed here, one can realize the corresponding BPS string starting from the 4d $\mathcal{N}=2$ theory $H^{Q=1}_{E_6}$ (\emph{i.e.}, the $E_6$ Minahan-Nemeschansky theory) by inserting folding defects. This construction does give rise to a 2d $(0,4)$ theory whose infrared limit is a NLSM with target the moduli space of one $F_4$ instanton. From the elliptic genus of the 2d theory, one can determine the $L_{F_4}(v,q)$ function, which was done in \cite{DelZotto:2016pvm}. When specialized to $L_{F_4}(q^{1/4},q)$, the function coincides with $\widehat\chi^{(F_4)_{-5/2}}_0$, the vacuum character of the chiral algebra of $F_4$ type that in the construction of Arakawa is associated to the hyperk\"ahler manifold ${\mathcal M}_{F_4,1}$ . This suggests a potential generalization of the results of  \cite{Beem:2017ooy} to more general classes of chiral algebras by considering $\beta$-twisted compactifications on $T^2 \times S^2$ backgrounds of 4d $\cn=2$ theories with or without the insertions of 2d $(0,4)$ surface defects; note in particular that the twisted theory flows to a NLSM with target space the Higgs branch of the corresponding 4d $\cn=2$ theory \cite{Kapustin:2006hi,Putrov:2015jpa} for models that admit a $\beta$-twist.\\

\noindent In the second half of the paper we studied the realization of the BPS string CFTs as $\mathcal{N}=(0,4)$ nonlinear sigma models on the moduli space of one instanton of the 6d gauge group. This description correctly accounts for the central charges of the CFT and implies that the 6d flavor symmetry group is realized on the string as a chiral current algebra, whose level we were able to infer from the anomaly polynomial. We also found a simple, uniform realization for the $G$ dependence of the elliptic genus. Namely, we found that (at least for theories for which $n\neq h^\vee_G$) this symmetry is realized as an affine Kac-Moody algebra at level $-n$, and the elliptic genus is expressed in terms of the irreducible characters of this algebra; the theories \hh{2}{SU(2)} and \hh{3}{SU(3)}, for which the level of the Kac-Moody algebra is the critical one, $k= -h^\vee_G$, appear to be an exception to this statement and are worthy of further study. We also found that at the level of chiral algebra the $SU(2)_v$ global symmetry can be exploited to perform a spectral flow that interpolates between different boundary conditions for the chiral fermions. This implies a relation between the elliptic genera computed with R--R and NS--R boundary conditions. From simple considerations about the structure of the NLSM we were able to determine the spectrum of low energy operators contributing to the NS-R elliptic genus, which in turn provides information about the R-R elliptic genus. \newline

\noindent Of course, ultimately the motivation for studying the string CFTs is that they provide an original angle to probe the 6d SCFT to which they belong. In this paper we have encountered several instances which demonstrate the usefulness of this approach: first of all, imposing $c_L\geq 0$, which is required by the unitarity of the string CFT, implies that the gauge algebra of the 6d SCFT must be nontrivial for $n\geq 3$, consistent with the finding in F-theory that tensor multiplets with $n\geq 3$ are always paired to a nontrivial gauge group \cite{Morrison:2012np}. We have also found that the string CFT is a useful tool for understanding the flavor symmetry of the 6d SCFT. For example, the flavor symmetry of the 6d $\six{2}{SU(2)}{}$ SCFT is $SO(8)$ on the tensor branch but is reduced to SO(7) at the superconformal fixed point \cite{Ohmori:2015pia}; in section \ref{sec:excep} we found that this is  reflected explicitly in the elliptic genus of the \hh{2}{SU(2)} CFT: the spectrum organizes itself naturally in terms of $SO(8)$ characters, but among the operators of conformal dimensions $(h_L,h_R)=(1,0)$ above the NS--R vacuum one finds a set of operators transforming in the adjoint representation of $SO(7)$ rather than $SO(8)$. Moreover, in section \ref{sec:wzwunivii} we have encountered some difficulties in realizing the flavor symmetry of the \hh{1}{SO(11)}, \hh{2}{SO(11)}, \hh{1}{SO(12)_b}, and \hh{2}{SO(12)_b} CFTs in terms of WZW models, which might point to potential subtleties with the parent 6d SCFTs.\\

\noindent The most direct connection to 6d physics however comes from the fact that the R--R elliptic genus of the \hh{n}{G} string captures the one-instanton piece of the Nekrasov partition function of the 6d $\six{n}{G}{}$ SCFT. By exploiting modularity and various constraints that arise from the universal features of the string CFTs we were able to determine the elliptic genera of a number of string CFTs of 6d SCFTs with exceptional gauge symmetry or matter content, leading to a wealth of novel results for the associated 6d Nekrasov partition functions (albeit with gauge symmetries fugacities switched off). Moreover, by considering the 5d limit of 6d SCFTs with $n\geq 2$ we also obtained new results for the one-instanton pieces of the Nekrasov partition functions of the resulting 5d $\mathcal{N}=1$ theories, and using the geometric engineering of the 6d SCFTs in F-theory, we were also able to clarify their behavior upon compactification to 5d and explain the different limits obtained for theories with $n=1$, $n=2$, and $n>2$.\\

\noindent The work presented in this paper lends itself naturally to a number of extensions. At the computational level, we found that we could determine the elliptic genera of several 2d CFTs based on simple considerations on their low energy spectrum in the NS--R sector. Understanding the spectrum of low energy operators in more detail may lead to additional constraints and help determine the elliptic genera of the remaining CFTs. It would also be desirable to obtain expressions for the elliptic genera with the fugacities for the 6d gauge symmetry $G$ reinstated, extending the results of \cite{DelZotto:2017mee,Kim:2018gjo}. Moreover, while in this paper we have largely focused on the CFTs describing a single BPS string, several of our results can naturally be extended to study bound states of strings associated to the same gauge factor or to different gauge factors; the CFT of such bound states will be described by nonlinear sigma models on the moduli spaces of higher numbers of instantons. The calculation of central charges and anomaly polynomials easily generalizes to bound states of strings, and likewise one still expects the 6d flavor symmetry to be captured by current algebras at higher level fixed by the anomaly polynomials we computed in this paper. It would be interesting to understand whether other universal features we found for single-string CFTs carry over to bound states. For instance, it would be desirable to know whether the spectral flow symmetry observed in the single string CFT carries over to bound states of strings and leads to relations between R-R and NS-R elliptic genera. Understanding the large $N$ behavior of bound states of strings might also be worthwile. It would also be interesting to obtain a clearer picture of the mechanism by which the strings bind to each other from the point of view of their CFTs, and study whether the classification of 6d SCFTs obtained in \cite{Heckman:2013pva,DelZotto:2014hpa,Heckman:2015bfa} from geometric considerations also follows from \emph{e.g.} unitarity constraints on the CFTs describing bound states of strings.\newline

\noindent One of the salient features of the backgrounds corresponding to frozen singularities in F-theory is that the one-to-one correspondence between simple factors of the gauge group and the tensor multiplets is lost. Therefore the BPS strings corresponding to these theories have features that are slightly different from the ones that we have considered in this paper, in particular the target space of the corrersponding NLSM is no longer the one-instanton moduli space for a simple gauge group. Generalizing our results to this class of geometries is an obvious future direction.\\

\noindent Lastly, it should be possible to interpret various properties of the BPS strings, including the appearance of a $G$ Kac-Moody algebra at negative level and the existence of a spectral flow symmetry, most naturally as features of their chiral algebra. Some of the string CFTs are essentially free theories or admit UV $N=(0,4)$ supersymmetric Lagrangian realizations, and it may therefore be possible to study them by an extension of the techniques of \cite{Dedushenko:2017osi} or \cite{Beem:2014kka}. While most theories are not known to admit such simple realizations, the results of this paper suggest that their chiral algebras bear many similarities to the chiral algebras of the ones that do. One may therefore hope for a uniform description of the chiral algebras of all 6d string CFTs. Among these, it would be important to understand better the exceptional cases for which $n=h^\vee_G$, which as discussed in section \ref{sec:excep} appear to have different properties from the rest. These matters are currently under investigation \cite{wip}.

\section*{Acknowledgements} We thank Chris Beem, Nikolai Bobev, Miranda Cheng, Cyril Closset, Jan de Boer, Mykola Dedushenko, Thomas Dumitrescu, Mathias Gaberdiel, Amihay Hanany, Hirotaka Hayashi, Jonathan J. Heckman, Zohar Komargodski, Peter Koroteev, Kimyeong Lee, Luca Martucci, Sameer Murthy, Nikita Nekrasov, Leonardo Rastelli, Shu-Heng Shao, Jan Troost, Cumrun Vafa and Erik Verlinde for useful conversations. We would like to acknowledge Tsinghua University and the Tsinghua Sanya International Mathematics Forum for hospitality during the workshop ``Superconformal Field Theories in 6 and Lower Dimensions''. This project has received funding from the European Union's Horizon 2020 research and innovation programme under the Marie Sklodowska-Curie grant agreement No 708045.

\newpage

\appendix

\section{Lie algebras and representations}\label{sec:Lie}
\subsection{Simple Lie algebras}\label{sec:finit}
\begin{figure}[t]
\begin{center}
\includegraphics[width=\textwidth]{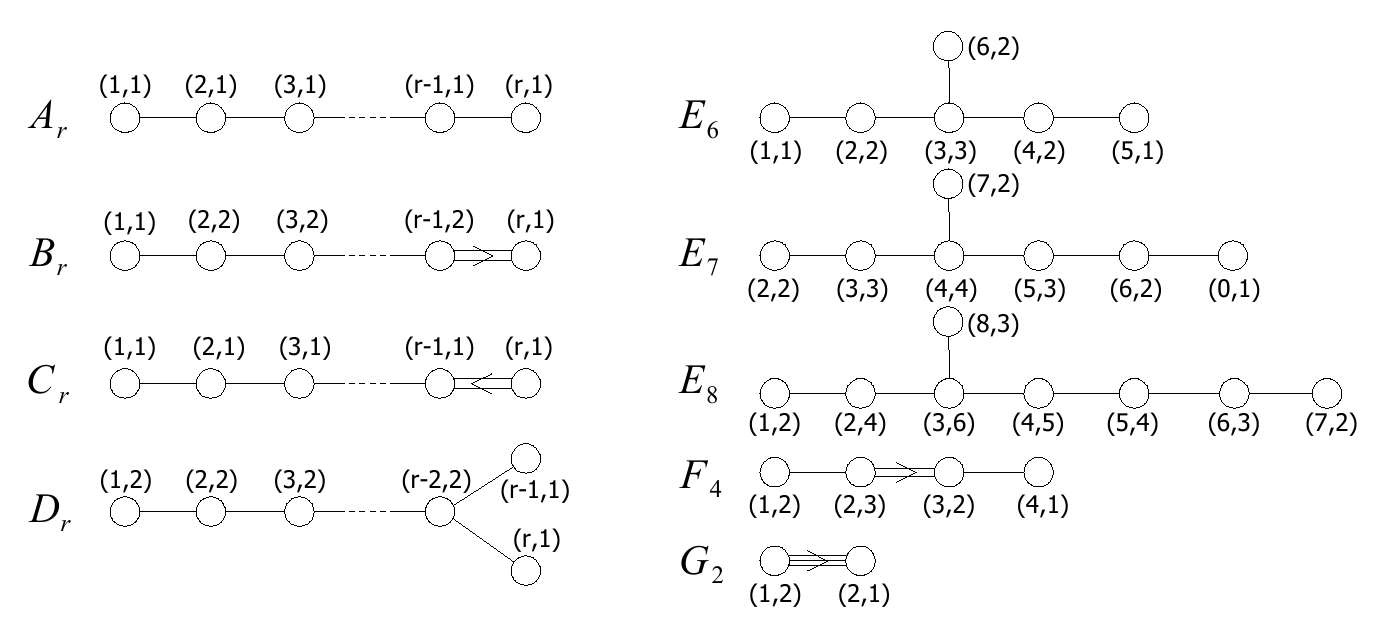}
\caption{Dynkin diagrams associated to simple Lie algebras. We label the $i$-th node by $(i,a_i^\vee)$, where $a_i^\vee$ are the \emph{comarks}.}
\label{fig:simple}
\end{center}
\end{figure}

In this paper we use the standard labeling for the nodes of the Dynkin diagrams associated to simple Lie algebras indicated in figure \ref{fig:simple}. There, we also indicate the comark $a_i^\vee$ associated to each node. The dual Coxeter number of a simple Lie algebra $\mathfrak{g}$ is defined as the sum of all the comarks:
\begin{equation} 
h^\vee_{\mathfrak{g}} = \sum_{i=1}^{\text{rank}(\mathfrak{g})}a_i^\vee.
\end{equation}

\noindent To the $i$-th node of the Dynkin diagram of any Lie algebra $\mathfrak{g}$ one can associate a simple root $\alpha_i$; the total number of nodes equals the rank of the algebra, $\text{rank}(\mathfrak{g})$. We denote the set of positive roots by $\Delta^{\mathfrak{g}}_+$. The Killing form induces a scalar product $\langle\cdot,\cdot\rangle$ on the complexification of the dual Cartan subalgebra of $\mathfrak{g}$, $\mathfrak{h}^*_{\mathbb{C}}\simeq \mathbb{C}^{\text{rank}(\mathfrak{g})}$. By making use of this scalar product one can define the coroots $\alpha_i^\vee$:
\begin{equation}
\alpha_i^\vee=\frac{2 \alpha_i}{\langle\alpha_i,\alpha_i\rangle},
\end{equation}
where the scalar product is normalized so that it equals $2$ for long roots. One furthermore defines the fundamental weights via the relation
\begin{equation}
\langle\omega_i,\alpha_j^\vee\rangle=\delta_{ij};
\end{equation}
the fundamental weights form a basis of the weight space of the Lie algebra.\newline

\noindent We denote the set of irreducible representations of $\mathfrak{g}$ as $\text{Rep}(\mathfrak{g})$. An irreducible representation $R_\lambda$ of a Lie algebra can be labeled in terms of its highest weight $\lambda$, which we expand in terms of the fundamental weights:
\begin{equation}
\lambda = \sum_{i=1}^{\text{rank}(\mathfrak{g})}a_i\, \omega_i.
\end{equation}
In this paper we denote such a representation in terms of its Dynkin labels $a_i$:
\begin{equation}
\lambda \simeq  (a_1a_2\dots a_{\text{rank}(\mathfrak{g})}).
\end{equation}
We separate the entries by commas in cases where ambiguity might arise from this notation.\newline

\noindent In some parts of the paper we also find it convenient to label representations in terms of their dimension. Thus for example the adjoint representation of $\mathfrak{e}_8$ is also denoted by $\mathbf{248}$. For $\mathfrak{g}=\mathfrak{so}_{2N}$ use a superscript $\mathbf{s},\mathbf{c}$ to distinguish between spinor and conjugate spinor representations, and for $\mathfrak{g}=\mathfrak{so}_{8}$ we denote by $\mathbf{8^v},$ $\mathbf{8^s},$ and $\mathbf{8^c}$ respectively the vector, spinor and conjugate spinor representations. Alternatively, we sometimes for convenience use the notation $\textbf{v},\textbf{s},\textbf{c}$ to denote the vector, spinor and conjugate spinor representations of $\mathfrak{so}_{2N}$. Also for $\mathfrak{g}=\mathfrak{su}_N$ we denote respectively by $\mathbf{N}$ and $\mathbf{\overline{N}}$ the $(10\dots0)$ representation and its conjugate representation $(0\dots01)$, and by $\mathbf{\Lambda^2}$ and $\overline{\mathbf{\Lambda^2}}$ respectively the representations $(010\dots0)$ and $(0\dots010)$.\newline

\noindent To any irreducible representation one can associate its quadratic Casimir invariant
\begin{equation}
C_2(R_\lambda) = \langle\lambda,\lambda+2\rho_{\mathfrak{g}}\rangle,
\end{equation}
where the Weyl vector $\rho_{\mathfrak{g}}$ is defined as one half of the sum of positive roots, and has Dynkin labels $(11\dots1)$. Another relevant quantity associated to a representation is its character, which we always evaluate at a specific point $\mathbf{m}_\mathfrak{g}=e^{2\pi i \mathbf{\mu}_\mathfrak{g}}$, where $ \mathbf{\mu}_\mathfrak{g}\in\mathfrak{h}^*_{\mathbb{C}}$: 
\begin{equation}
\chi^{\mathfrak{g}}_\lambda(\mathbf{m}_\mathfrak{g}) = \sum_{\nu \in R_\lambda} \text{mult}(\nu)\,e^{2\pi i (\mathbf{\mu}_{\mathfrak{g}},\nu)},
\end{equation}
where the sum runs over the weights $\nu$ in the representation $R_\lambda$, which we view here as points in $\mathfrak{h}^*_\mathbb{C}$, and $\text{mult}(\nu)$ denotes the multiplicity with which $\nu$ appears in the representation. In the limit $\mathbb{\mu}_{\mathfrak{g}}\to 0$ (which we sometimes denote alternatively by $\mathbf{m}_{\mathfrak{g}}\to\mathbf{1}$), the character reduces to the dimension of the representation:
\begin{equation}
\chi^{\mathfrak{g}}_\lambda(\mathbf{1})=\text{dim}(R_\lambda).
\end{equation}
At various points in the paper we specialize the parameter $\mathbf{m}_{\mathfrak{g}}$ to specific values. For a root $\alpha$ of $\mathfrak{g}$, we find it convenient to define
\begin{equation}\label{eq:malpha}
m_\alpha = e^{2\pi i \langle \mathbf{\mu}_{\mathfrak{g}},\alpha \rangle}.
\end{equation}
Furthermore, in computations we find it convenient to expand $\mathbf{m}_{\mathfrak{g}}$ in the following way:
\begin{itemize}
\item For $\mathfrak{g}=\mathfrak{su}_N$ , define
\begin{equation}
e_1=\omega_1,\quad e_2=\omega_2-\omega_1,\quad e_3=\omega_3-\omega_2,\quad \dots,\qquad e_{N-1}=\omega_{N-1}-\omega_{N-2},\quad e_{N}=-\omega_{N-1},
\end{equation}
and define
\begin{equation}
m_{\mathfrak{su}_N}^i = e^{2\pi i \langle \mathbf{\mu}_{\mathfrak{su}_N},e_i\rangle},
\end{equation}
so that the $\mathbf{N}$ representation has character
\begin{equation}
\chi^{\mathfrak{su}_N}_{(10\dots0)}(\mathbf{m}_{\mathfrak{su}_N})=\sum_{i=1}^Nm_{\mathfrak{su}_N}^i,
\end{equation}
where
\begin{equation}
m_{\mathfrak{su}_N}^N=\frac{1}{\prod_{i=1}^{N-1}m_{\mathfrak{su}_N}^i}.
\end{equation}
\item For $\mathfrak{g}=\mathfrak{so}_{2N}$ , define
\begin{equation}
e_1=\omega_1,\,\,\, e_2=\omega_2-\omega_1,\,\,\, e_3=\omega_3-\omega_2,\,\, \dots,\,\, e_{N-1}=\omega_{N-1}+\omega_{N}-\omega_{N-2},\,\,\, e_N=\omega_{N}-\omega_{N-1},
\end{equation}
and define
\begin{equation}\label{eq:mso2n}
m_{\mathfrak{so}_{2N}}^i = e^{2\pi i \langle \mathbf{\mu}_{\mathfrak{so}_{2N}},e_i\rangle},
\end{equation}
so that the $\mathbf{2\,N}$ representation has character
\begin{equation}
\chi^{\mathfrak{so}_{2N}}_{(10\dots0)}(\mathbf{m}_{\mathfrak{so}_{2N}})=\sum_{i=1}^N\left[m_{\mathfrak{so}_{2N}}^i+\frac{1}{m_{\mathfrak{so}_{2N}}^i}\right].
\end{equation}
\item For $\mathfrak{g}=\mathfrak{so}_{2N+1}$ , define
\begin{equation}
e_1=\omega_1,\,\,\, e_2=\omega_2-\omega_1,\,\,\, e_3=\omega_3-\omega_2,\,\, \dots,\,\, e_{N-1}=\omega_{N-1}-\omega_{N-2},\,\,\, e_N=2\omega_{N}-\omega_{N-1},
\end{equation}
and define
\begin{equation}\label{eq:mso2nplus1}
m_{\mathfrak{so}_{2N+1}}^i = e^{2\pi i \langle \mathbf{\mu}_{\mathfrak{so}_{2N+1}},e_i\rangle},
\end{equation}
so that the $\mathbf{2\,N+1}$ representation has character
\begin{equation}
\chi^{\mathfrak{so}_{2N+1}}_{(10\dots0)}(\mathbf{m}_{\mathfrak{so}_{2N+1}})=1+\sum_{i=1}^N\left[m_{\mathfrak{so}_{2N+1}}^i+\frac{1}{m_{\mathfrak{so}_{2N+1}}^i}\right].
\end{equation}
\item For $\mathfrak{g}=\mathfrak{sp}_N$ , define
\begin{equation}
e_1=\omega_1,\quad e_2=\omega_2-\omega_1,\quad e_3=\omega_3-\omega_2,\quad \dots,\qquad e_{N}=\omega_{N}-\omega_{N-1},
\end{equation}
and define
\begin{equation}\label{eq:mspn}
m_{\mathfrak{sp}_N}^i = e^{2\pi i \langle \mathbf{\mu}_{\mathfrak{sp}_N},e_i\rangle},
\end{equation}
so that the $2\mathbf{N}$ representation has character
\begin{equation}
\chi^{\mathfrak{sp}_N}_{(10\dots0)}(\mathbf{m}_{\mathfrak{sp}_N})=\sum_{i=1}^N\left[m_{\mathfrak{sp}_N}^i+\frac{1}{m_{\mathfrak{sp}_N}^i}\right].
\end{equation}

\end{itemize}

\noindent In this paper representations of the abelian algebra $\mathfrak{u}(1)$ also appear. An irreducible representation of $U(1)$ is determined by its charge $q$, and we choose a normalization such that the vector representation of $\mathfrak{so}(2)$ is equivalent to the sum of two $\mathfrak{u}(1)$ representations with charge respectively $+1$ and $-1$.\newline

\noindent Finally, we remark that in several parts of the text we find it convenient to phrase our discussion in terms of Lie groups $G$ rather than their Lie algebras $\mathfrak{g}$, and we adapt the notation there such that for example by $\Delta_+^G$ we intend the set of positive roots of the Lie algebra, $\Delta_+^\mathfrak{g}$.
\subsection{Affine Lie algebras}\label{sec:appaff}
\begin{figure}[h]
\begin{center}
\includegraphics[width=\textwidth]{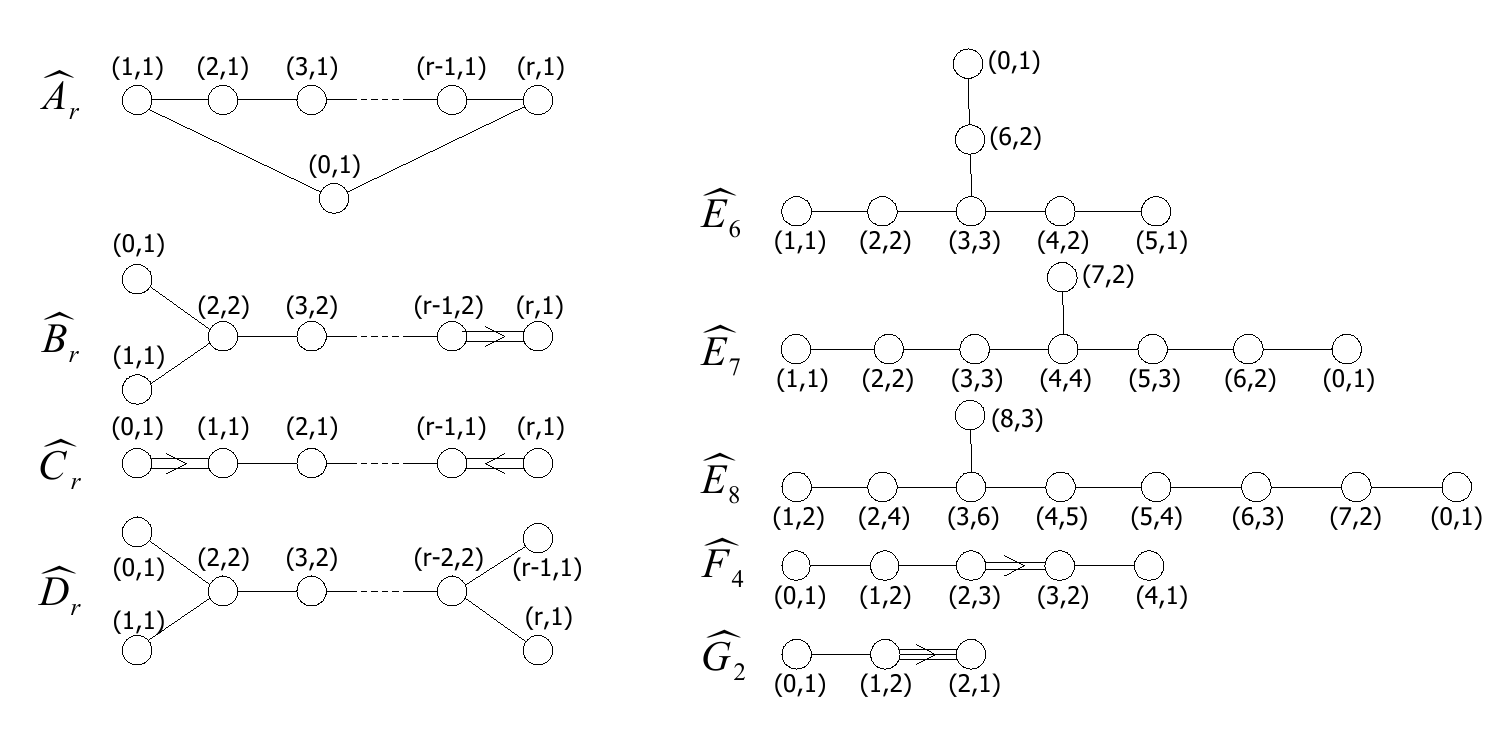}
\caption{Dynkin labels associated to simple Lie algebras. We label the $i$-th node by $(i,a_i^\vee)$, where $a_i^\vee$ are the comarks.}
\label{fig:affine}
\end{center}
\end{figure}
\noindent In the case of affine Lie algebras, we label the additional simple root by $\alpha_0$, corresponding to the $0$ node in figure \ref{fig:affine}. 
Recall that for any simple lie algebra one can construct an affine Lie algebra $\widehat{\mathfrak{g}}$ given by a $\mathfrak{u}(1)\otimes \mathfrak{u}(1)$ extension of the loop algebra $L(\mathfrak{g})=\mathfrak{g}\otimes \mathbb{C}[t,t^{-1}]$:
\begin{equation} \widehat{\mathfrak{g}} = L(\mathfrak{g})\oplus\mathbb{C}\widehat{k}\oplus\mathbb{C}L_0.
\end{equation}
Affine weights are labeled by a weight of $\mathfrak{g}$ as well as its eigenvalues under operators $\widehat k$ and $-L_0$:
\begin{equation} \widehat \lambda = (\lambda;k;n).
\end{equation}
In this notation, the simple roots of the $\mathfrak{g}$ subalgebra are identified with $(\alpha;0;0)$. Furthermore, defining
\begin{equation}
\delta = (0;0;1)
\end{equation}
to be the imaginary root,  one has
\begin{equation}
\alpha_0=(-\theta_\mathfrak{g};0;0)+\delta,
\end{equation}
where $\theta_\mathfrak{g}$ is the highest root of $\mathfrak{g}$. The corresponding affine coroots are given by:
\begin{equation}
\alpha_0^\vee=\alpha_0,\qquad \alpha_i^\vee = (\alpha_i^\vee;0;0).
\end{equation}

\noindent The scalar product between affine weights is defined as:
\begin{equation}
\langle(\lambda;k_\lambda;n_\lambda),(\mu;k_\mu;n_\mu)\rangle=\langle\lambda,\mu\rangle+k_\lambda n_{\mu}+k_\mu n_\lambda.
\end{equation}
Fundamental weights are chosen to have zero $L_0$ eigenvalue and are again fixed by the requirement
\begin{equation} \langle\alpha_i,\widehat\omega_j\rangle= \delta_{ij}.
\end{equation}
Explicitly, one has:
\begin{equation}
\widehat\omega_0 = (0;1;0),\qquad \widehat \omega_i= (\omega_i;a_i^\vee;0), \qquad i =1,\dots,r.
\end{equation}
The positive integers $a_i^\vee$ are the comarks associated to the nodes of the Dynkin diagrams, which are given in figure \ref{fig:affine}.\newline

\noindent An affine representation $R_{\widehat\lambda}$ is characterized in terms by its highest weight $\widehat\lambda = (\lambda;k;n)$, which has an expansion
\begin{equation} \widehat \lambda = \sum_{i=0}^r \lambda_i \widehat\omega_i +n\, \delta.
\end{equation}
Its level $k$ is given by
\begin{equation}\label{eq:aflev}
k = \sum_{i=0}^r a_i^\vee\lambda_i.
\end{equation}
Another common notation (which does not keep track of the value of $n$) is
\begin{equation}
\widehat\lambda = (\lambda_0\lambda_1\dots\lambda_r).
\end{equation}
The level $k$ integrable highest-weight representations are those representations for which the Dynkin labels are all non-negative integers subject to the condition \eqref{eq:aflev}. Moreover, as the value of $n$ is inessential we specialize $n=0$.  \newline

\noindent We define the normalized character of $R_{\widehat\lambda}$ (evaluated at a point $(\mathbf{\mu}_\mathfrak{g};\tau;0)$, where $\mathbf{\mu}_\mathfrak{g}\in\mathfrak{h}^*_{\mathbb{C}}$ and $\tau\in\mathbb{H}$) as
\begin{equation} \widehat\chi^{\mathfrak{g}}_{\lambda}(\mass_{\mathfrak{g}},q)=q^{-\frac{1}{24}\frac{\text{dim}(\mathfrak{g})k}{h^\vee_{\mathfrak{g}}+k}+\frac{C_2(R_\lambda)}{2(h^\vee_{\mathfrak{g}}+k)}}\sum_{\widehat{\nu}}\text{mult}(\widehat{\nu})e^{-2\pi i\tau\langle\widehat\nu,\widehat{\omega}_0\rangle}e^{2\pi i\langle\nu,m\rangle},
\end{equation}
where $q = e^{2\pi i \tau}$ and the sum runs over the weights $\widehat{\nu}$ in the representation, whose finite part we denote by $\nu$. We choose to label the character in terms of the finite component $\lambda$ of the highest weight, keeping in mind that for a given level the remaining component $\lambda_0$ is determined from equation \eqref{eq:aflev}.\newline

\noindent For an integrable highest-weight representation, the character is given by the Weyl-Kac character formula:
\begin{equation}\label{eq:afcha}
\widehat\chi^{\mathfrak{g}}_{\lambda}(\mass_{\mathfrak{g}},q)=q^{-\frac{1}{24}\frac{\text{dim}(\mathfrak{g})k}{h^\vee_{\mathfrak{g}}+k}+\frac{C_2(R_\lambda)}{2(h^\vee_{\mathfrak{g}}+k)}}\frac{\sum_{w\in W_{\widehat{\mathfrak{g}}}}\epsilon(w)e^{2\pi i\langle w(\hat\rho+\widehat\lambda),(\mathbf{\mu}_{\mathfrak{g}};\tau;0)\rangle}}{\sum_{w\in W_{\widehat{\mathfrak{g}}}}\epsilon(w)e^{2\pi i\langle w(\hat\rho),(\mathbf{\mu}_{\mathfrak{g}};\tau;0)\rangle}},
\end{equation}
where $W_{\widehat{\mathfrak{g}}}$ is the affine Weyl group of $\widehat{\mathfrak{g}}$, $\epsilon(w)=(-1)^{\ell(w)}$ is the length of the Weyl group element $w$, and $\hat\rho$ is the affine Weyl vector, defined as the element of the affine weight lattice with Dynkin labels $(111\dots1)$. Thanks to the Macdonald-Weyl identity, the character can be rewritten in the form
\begin{equation}
\widehat\chi^{\mathfrak{g}}_{\lambda}(\mass_{\mathfrak{g}},q)=q^{-\frac{1}{24}\frac{\text{dim}(\mathfrak{g})k}{h^\vee_{\mathfrak{g}}+k}+\frac{C_2(R_\lambda)}{2(h^\vee_{\mathfrak{g}}+k)}}\frac{\sum_{w\in W_{\widehat{\mathfrak{g}}}}\epsilon(w)e^{2\pi i\langle w(\hat\rho+\widehat\lambda)-\hat\rho,(\mathbf{\mu}_{\mathfrak{g}};\tau;0)\rangle}}{\Delta_{\mathfrak{g}}(\mass_{\mathfrak{g}},q)},
\end{equation}
where
\begin{equation}\label{eq:kdet}
\Delta_{\mathfrak{g}}(\mass_{\mathfrak{g}},q) = \prod_{j=1}^
\infty (1-q^j)^{\text{rank}(\mathfrak{g})}\prod_{\alpha\in \Delta^{\mathfrak{g}}_+}(1-q^{j-1} m_\alpha)(1-q^j m_\alpha^{-1}).
\end{equation}
Moreover, the sum in the numerator admits an expansion in positive powers of $q$, with coefficients given in terms of linear combinations of characters of the finite part $\mathfrak{g}$ of the affine Lie algebra. In particular, to leading order in the $q$ expansion,\newline
\begin{equation}
\widehat\chi^{\mathfrak{g}}_{\lambda}(\mass_{\mathfrak{g}},q)=q^{-\frac{1}{24}\frac{\text{dim}(\mathfrak{g})k}{h^\vee_{\mathfrak{g}}+k}+\frac{C_2(R_\lambda)}{2(h^\vee_{\mathfrak{g}}+k)}}(\chi^{\mathfrak{g}}_\lambda(\mass_{\mathfrak{g}})+\mathcal{O}(q)).
\end{equation}

\noindent In the limit $\mass_{\mathfrak{g}}\to \mathbf{1}$, the characters associated to the integrable representations at a given level become elements of a vector-valued, weight zero modular form \cite{Kac:1984mq}.

\section{Modular and Jacobi forms}\label{sec:appmod}
In this appendix we collect some useful information about various modular and Jacobi forms we make use of in the text. First of all, the Dedekind eta function is defined as
\begin{equation}
\eta(\tau) = q^{1/24}\prod_{j=1}^\infty (1-q^j),
\end{equation}
where $\tau\in\mathbb{H}$ and $q=e^{2\pi i\tau}$, is a weight-1/2 modular form with nontrivial multiplier system, which under an $S$ transformation behaves as follows:
\begin{equation}
\eta(-1/\tau) = \sqrt{-i \tau}\,\eta(\tau).
\end{equation}
Another important class of functions are the Jacobi theta functions:
\begin{align} \theta_1(\zeta,\tau) &= i q^{1/8}z^{-1/2}(1-z)\prod_{j=1}^\infty(1-q^j)(1-q^j z)(1-q^j z^{-1}),\\
\theta_2(\zeta,\tau) &= q^{1/8}z^{1/2}(1+z)\prod_{j=1}^\infty(1-q^j)(1+q^j z)(1+q^j z^{-1}),\\
\theta_3(\zeta,\tau) &= \prod_{j=1}^\infty(1-q^j)(1+q^{j-1/2} z)(1+q^{j-1/2} z^{-1}),\\
\theta_4(\zeta,\tau) &= \prod_{j=1}^\infty(1-q^j)(1-q^{j-1/2} z)(1-q^{j-1/2} z^{-1}),
\end{align}
where $z=e^{2\pi i \zeta}$.\footnote{ In most of the text, for notational ease we use the exponentiated variables $z,q$ as the arguments of modular and Jacobi forms, rather than $\zeta$ and $\tau$.} The function $\theta_1(\zeta,\tau)$ is a Jacobi form of weight 1/2 and index 1/2, also with a nontrivial multiplier system:
\begin{equation}
\theta_1(\zeta/\tau,-1/\tau) = -i\sqrt{-i \tau}e^{2\pi i \frac{1}{2}\frac{\zeta^2}{\tau}}\theta_1(\zeta,\tau);
\end{equation}
on the other hand, the functions $\theta_2(\zeta,\tau),$ $\theta_3(\zeta,\tau),$ and $\theta_4(\zeta,\tau)$ transform as components of a vector-valued Jacobi form, also of weight 1/2 and index 1/2:
\begin{equation}
\begin{pmatrix}\theta_2(\zeta/\tau,-1/\tau)\\\theta_3(\zeta/\tau,-1/\tau)\\\theta_4(\zeta/\tau,-1/\tau)\end{pmatrix} = \sqrt{-i \tau}e^{2\pi i \frac{1}{2}\frac{\zeta^2}{\tau}}\begin{pmatrix}0&0&1\\0&1&0\\1&0&0\end{pmatrix}\begin{pmatrix}\theta_2(\zeta,\tau)\\\theta_3(\zeta,\tau)\\\theta_4(\zeta,\tau)\end{pmatrix}.
\end{equation}
Under a shift $\zeta\to\zeta+\tau$ of the elliptic parameter, the functions $\theta_3(\zeta,\tau)$ and $\theta_4(\zeta,\tau)$ are invariant, while $\theta_1(\zeta,\tau)$ and $\theta_2(\zeta,\tau)$ transform back to themselves up to a minus sign. Also, $\theta_1(-\zeta,\tau)=-\theta_1(\zeta,\tau)$, while for $\ell=2,$ $3,$ and $4$ one has $\theta_\ell(-\zeta,\tau)=\theta_\ell(\zeta,\tau)$. Moreover, under a shift $\zeta \to \frac{\tau}{2}-\zeta$, one has:
\begin{equation}
\begin{pmatrix}\theta_1(\tau/2-\zeta,\tau)\\\theta_2(\tau/2-\zeta,\tau)\\\theta_3(\tau/2-\zeta,\tau)\\\theta_4(\tau/2-\zeta,\tau)\end{pmatrix} = e^{-\pi i \tau/4}e^{\pi i z}\begin{pmatrix}0&0&0&i\\0&0&1&0\\0&1&0&0\\-i&0&0&0\end{pmatrix}\begin{pmatrix}\theta_1(\zeta,\tau)\\\theta_2(\zeta,\tau)\\\theta_3(\zeta,\tau)\\\theta_4(\zeta,\tau)\end{pmatrix}.
\end{equation}

We also resort to structure theorems on the spaces of holomorphic modular forms. First of all, a holomorphic  $SL(2,\mathbb{Z})$ modular form $f(\tau)$ of weight $m$ is a function from $\mathbb{H}$ to $\mathbb{C}$ that satisfies
\begin{equation}
f\left(\frac{a \tau+b}{c\tau+d}\right) = (c\tau+d)^m\,f(\tau)\qquad \forall \begin{pmatrix}a&b\\c&d\end{pmatrix}\in SL(2,\mathbb{Z})
\end{equation}
and admits a Fourier expansion
\begin{equation}
f(\tau)=\sum_{n\geq 0} a_n\, q^n.
\end{equation}
The ring of holomorphic modular forms $M_*=\bigoplus_{m\geq 0} M_m$, which admits an obvious grading in terms of the weight $m$, is generated freely by the two Eisenstein series
\begin{equation} E_4(\tau) =1+240\sum_{k=1}^\infty\frac{n^3q^k}{1-q^k},\end{equation}
\begin{equation} E_6(\tau) =1-504\sum_{k=1}^\infty\frac{n^5q^k}{1-q^k},\end{equation}
which have respectively weight $4$ and $6$. \newline

\noindent On the other hand, a holomorphic Jacobi form of index $k\geq 0$ and weight $m\in2\mathbb{Z}$  is defined as a holomorphic function $\varphi(\zeta,\tau)$ from $\mathbb{C}\times \mathbb{H}$ to $\mathbb{C}$ such that
\begin{equation}
\varphi\left(\frac{\zeta}{c\tau+d},\frac{a \tau+b}{c\tau+d}\right) = (c\tau+d)^m\,e^{2\pi i k\frac{ c\,\zeta^2}{c\tau+d}}\varphi(\zeta,\tau)\qquad \forall \begin{pmatrix}a&b\\c&d\end{pmatrix}\in SL(2,\mathbb{Z})
\end{equation}
and 
\begin{equation}
\varphi\left(\zeta+a\tau+b,\tau\right)= e^{-2\pi i k(a^2\tau+2 a z)}\varphi(\zeta,\tau)\qquad \forall a,b\in\mathbb{Z},
\end{equation}
which admits a Fourier expansion
\begin{equation}
\varphi\left(\zeta,\tau\right) =\sum_{n\geq 0}\sum_{r\in\mathbb{Z}} c_{n,r} q^n z^r,
\end{equation}
such that 
\begin{equation}
c_{n,r}= 0 \qquad \text{if }\qquad r^2\geq 4\,k\, n.
\end{equation}
The ring of all such functions, $J^{even}_{*,*}=\bigoplus_{m,k}J^{even}_{m,k}$, is a polynomial ring over the ring of holomorphic modular forms which is freely generated by the two functions $\varphi_{-2,1}(\zeta,\tau),$ and $\varphi_{0,1}(\zeta,\tau)$. These functions are given in terms of the Dedekind eta function and Jacobi theta functions defined above as:
\begin{align}
\varphi_{-2,1}(\zeta, \tau) &= -\frac{\theta_1(\zeta,\tau)^2}{\eta(\tau)^6},\\
\varphi_{0,1}(\zeta, \tau) &= 4\sum_{j=2}^4\frac{\theta_j(\zeta,\tau)^2}{\theta_j(0,\tau)^2}.
\end{align}
In the paper we also make use of the larger ring $J^{even,1/2}_{*,*}=\bigoplus_{m,k}J^{even,1/2}_{m,k}$, of weak Jacobi forms of even weight and half/integer index. It has been shown in \cite{Gritsenko:1999fk} that this ring is again a polynomial ring over the ring of holomorphic modular forms, which is freely generated by the functions
\begin{align}
\varphi_{-2,1}(\zeta, \tau),\qquad \varphi_{0,1}(\zeta, \tau), \quad \text{and } \qquad\varphi_{0,3/2}(\zeta, \tau)=\frac{\theta_1(2\,\zeta,\tau)}{\theta_1(\zeta,\tau)}.
\end{align}
Note also that the functions $\varphi_{-2,1}(\zeta,\tau)$ and $\varphi_{0,1}(\zeta,\tau)$ are even functions of $\zeta$, while $\varphi_{0,3/2}(\zeta,\tau)$  is odd.\newline

\noindent Finally, one can show that the following relations hold between the generators of weak Jacobi forms with elliptic variable $2\,\zeta$ and those with elliptic variable $\zeta$:
\begin{align}
\varphi_{-2,1}(2\,\zeta,\tau) &= \frac{1}{216}\varphi_{-2,1}(\zeta,\tau)^4E_6(\tau)-\frac{1}{144}\varphi_{-2,1}(\zeta,\tau)^3\varphi_{0,1}(\zeta,\tau)E_4(\tau)\nonumber\\
&+\frac{1}{432}\varphi_{-2,1}(\zeta,\tau)\varphi_{0,1}(\zeta,\tau)^3;\label{eq:v2vsva}\\
\varphi_{0,1}(2\,\zeta,\tau) &=\frac{1}{192}\varphi_{-2,1}(\zeta,\tau)^4E_4(\tau)^2-\frac{1}{108}\varphi_{-2,1}(\zeta,\tau)^3\varphi_{0,1}(\zeta,\tau)E_6(\tau)\nonumber\\
&+\frac{1}{288}\varphi_{-2,1}(\zeta,\tau)^2\varphi_{0,1}(\zeta,\tau)^2E_4(\tau)+\frac{1}{1728}\varphi_{0,1}(\zeta,\tau)^4;\label{eq:v2vsvb}\\
\varphi_{0,\frac{3}{2}}(2\,\zeta,\tau) &= \varphi_{-2,1}(\zeta,\tau)^6\left(\frac{1}{55296}E_4(\tau)^3-\frac{1}{46656}E_6(\tau)^2\right)\nonumber\\
&+\frac{1}{62208}\varphi_{-2,1}(\zeta,\tau)^5\varphi_{0,1}(\zeta,\tau)E_4(\tau)E_6(\tau)-\frac{5}{165888}\varphi_{-2,1}(\zeta,\tau)^4\varphi_{0,1}(\zeta,\tau)^2E_4(\tau)^2\nonumber\\
&+\frac{5}{186624}\varphi_{-2,1}(\zeta,\tau)^3\varphi_{0,1}(\zeta,\tau)^3-\frac{5}{497664}\varphi_{-2,1}(\zeta,\tau)^2\varphi_{0,1}(\zeta,\tau)^4\nonumber\\
&+\frac{1}{1492992}\varphi_{0,1}(\zeta,\tau)^6.\label{eq:v2vsvc}
\end{align}

\section{A catalogue of BPS string elliptic genera}\label{sec:catalogue}

In this appendix we collect a list of references to other works where the elliptic genera of various BPS string CFTs are presented in a form which may be readily compared to our results. We warn the reader that in what follows we make no attempt to provide a comprehensive review of the literature.

\begin{figure}[t!]
\begin{center}
\subfloat[\phantom{x}]{\includegraphics[width=0.45\textwidth]{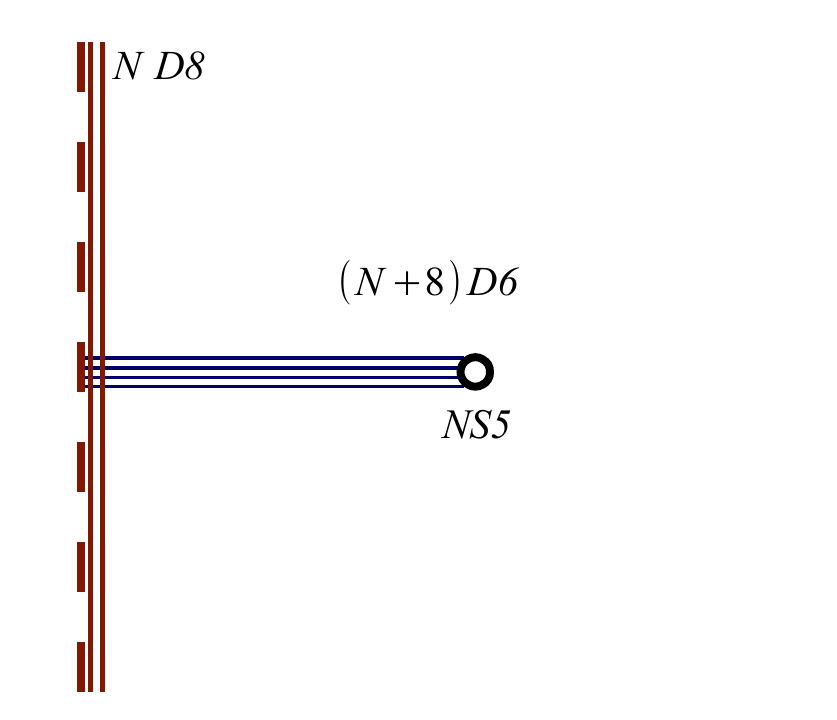}}
\subfloat[\phantom{x}]{\includegraphics[width=0.4\textwidth]{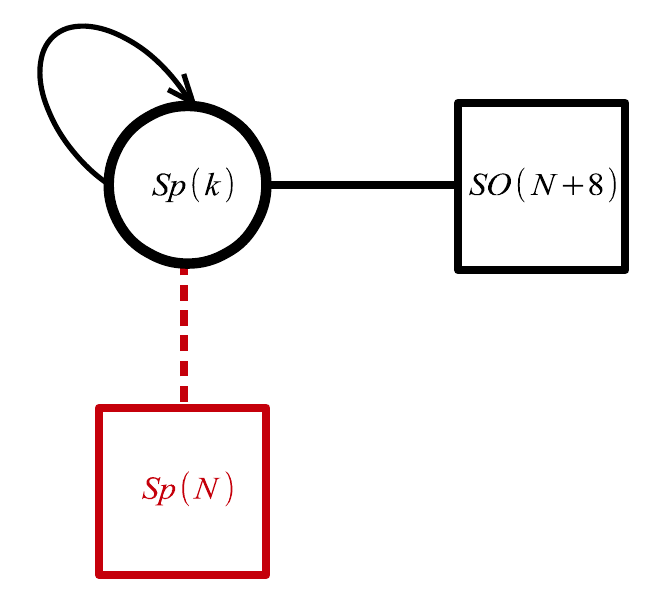}}
\caption{(a): Brane construction and (b): 2d (0,4) quiver for $k$ BPS strings of the 6d (1,0) SCFT with $(n,G) = (4,SO(8+N))$. The dashed and continuous edge symbolizes respectively bifundamental Fermi and hypermultiplets, while the gauge node supports an adjoint vector multiplet and a hypermultiplet in the antisymmetric representation of $Sp(k)$, symbolized by the arrow.}
\label{fig:oddSON}
\end{center}
\end{figure}

\begin{itemize}
\item The CFT for one E-string has been understood since the 1990s to consist of the $E_8$ level 1 current algebra and a decoupled center of mass hypermultiplet \cite{Ganor:1996mu,Seiberg:1996vs,Klemm:1996hh}. In \cite{Minahan:1998vr} it was argued that the CFT for $k$ strings included a subsector consisting of a $E_8$ current algebra at level $k$. A UV gauge theory realization of the E-string CFT for arbitrary numbers of strings, with full dependence on the Omega background parameters, was ultimately given in \cite{Kim:2014dza} and was employed by the authors to compute the elliptic genera of the CFTs using localization. A UV quiver gauge theory description for the infinite sequence of theories \hh{1}{Sp(N)} which Higgs to the E-string was given implicitly in \cite{Gadde:2015tra}; the elliptic genus (for one and two strings) was computed in \cite{Yun:2016yzw}. On the other hand, the UV gauge theories for the \hh{1}{SU(N)} theories and their multiple-string generalizations, as well as their elliptic genera, were obtained in \cite{Kim:2015fxa}.

\item The CFT of M-strings was first studied in \cite{Haghighat:2013gba}, where the elliptic genus was computed using the refined topological vertex \cite{Iqbal:2007ii} as well as by an index computation. The CFTs of the infinite sequence of \hh{2}{SU(N)} theories (the $A_n$ orbifolds of the M-string) were studied in \cite{Haghighat:2013tka} and \cite{Hohenegger:2013ala} and their elliptic genera computed via the topological vertex \cite{Haghighat:2013tka,Hohenegger:2013ala}, by supersymmetric localization in \cite{Haghighat:2013tka} and with an index computation in \cite{Hohenegger:2013ala}.
\item The elliptic genera for the \hh{3}{SU(3)} CFT and its multiple-string generalization have been computed via supersymmetric localization in \cite{Kim:2016foj}.
\item The elliptic genera for the \hh{4}{SO(8)} CFT and its multiple-string generalization  have been computed via supersymmetric localization in \cite{Haghighat:2014vxa}, and so have the elliptic genera for the \hh{4}{SO(8+2N)} CFTs. The ones for the \hh{4}{SO(9+2N)} CFTs have not appeared explicitly in the literature but may be obtained for instance from the brane configuration in figure \ref{fig:oddSON}(a), which leads to the 2d $(0,4)$ quiver gauge theory of figure \ref{fig:oddSON}(b), from which one may compute the elliptic genera via localization;\footnote{ The possibility of constructing a Lagrangian model for the $n=4$, $G=SO(2N+9)$ strings by a simple modification of the $G=SO(2N+8)$ model was already remarked to us by C. Vafa in 2014.} this computation is performed in appendix \ref{sec:soodd}.

\item The elliptic genera for the \hh{5}{F_4},\hh{6}{E_6},\hh{8}{E_7}, and \hh{12}{E_8} CFTs (for one string) have been computed in \cite{DelZotto:2016pvm} exploiting modularity of the elliptic genus, in the limit where the fugacities $\massF,\massG$ are switched off and for a single string. The elliptic genera for \hh{6}{E_6} and \hh{8}{E_7} have also been computed by localization, with all fugacities turned on, respectively in \cite{Putrov:2015jpa} and \cite{Agarwal:2018ejn}.

\item UV quiver gauge theories and elliptic genera for the \hh{3}{G_2} and \hh{3}{SO(7)} CFTs and their multiple-string generalizations have recently been computed by supersymmetric localization \cite{Kim:2018gjo}; the elliptic genus for the \hh{3}{G_2} string has also been computed in \cite{Kim:2018gak} by exploiting modularity.
\item For the remaining theories ($G=SU(6)_*$ for $n=1$, $G=SO(8),\dots, SO(12)$ for $n=1,2,3$, $G=SO(13)$ for $n=2$, $G=F_4$ for $n=1,\dots,4$, $G=E_6$ for $n=1,\dots,5$, and $G=E_7$ for $n=1,\dots,7$) no approach to computing the elliptic genera at all orders in $q$ is has been found at the time of writing of this paper.
\end{itemize}

\section{Elliptic genera of the $n=4$, $G=SO(2M+1)$ BPS strings}\label{sec:soodd}
In this appendix we compute the elliptic genera of the strings of the $\bf{4}_{SO(2M+1)}$ SCFTs (with $M\geq 4$). The brane configuration corresponding to this class of 6d SCFTs is given in figure \ref{fig:oddSON}(a) (with $N=2M-7$), while the 2d (0,4) quiver of their strings is shown in figure \ref{fig:oddSON}(b). The computation of the elliptic genus is a slight modification of the one for $G=SO(2M)$ in \cite{Haghighat:2014vxa}, which we can easily adapt to the present case.\newline 

\noindent In \cite{Haghighat:2014vxa} the elliptic genus for $G=SO(2M)$ was constructed starting from a brane configuration in which only a $Sp(M-4)\times Sp(M-4)$ is visible, but the quiver theory displays the full $F=Sp(2M-8)$ flavor symmetry. For the case $G=SO(2M+1)$ under consideration here, the field content is essentially the same as for $G=SO(2M)$, the only difference between the two cases being that for $k$ strings the hypermultiplets that couple to $G$ transform in the bifundamental of $Sp(k)\times SO(2M+1)$ rather than $Sp(k)\times SO(2M)$, and the Fermi multiplets transform in the bifundamental of $Sp(k)\times Sp(2M-7)$ rather than $Sp(k)\times Sp(2M-8)$. This allows us to straightforwardly write down the elliptic genus for $k$ strings of the $G=SO(2M+1)$ SCFT:
\begin{align}\label{eq:soodd}
&\mathbb{E}_{(k)}(\mass_{SO(2M+1)},\mass_{Sp(2M-7)},x,v,q) =\nonumber\\
& \frac{1}{2^n\,n!}\left[\frac{\eta(q)^2}{\theta_1(v x,q)\theta_1(v/x,q)}\right]^k\int \prod_{i=1}^k \left[d\zeta_i\,\eta(q)^2\frac{\theta_1(v^2,q)}{\eta(q)}\prod_{s=\pm}\frac{\theta_1(z_i^{2s},q)\theta_1(v^2z_i^{2s},q)}{\eta(q)^2}\right]\nonumber\\
&\times \prod_{i<j}^k\prod_{s_1=\pm,\,s_2=\pm}\left[\frac{\th_1(z_i^{s_1}z_j^{s_2},q)\th_1(v^2z_i^{s_1}z_j^{s_2},q)}{\th_1(v\, x\, z_i^{s_1}z_j^{s_2},q)\th_1(v\,x^{-1}z_i^{s_1}z_j^{s_2},q)}\right]\nonumber\\
&\times\prod_{i=1}^k\frac{\prod_{j=1}^{2M-7}\prod_{s=\pm}\th_1(z_i^s\,m_{Sp(2M-7)}^j,q)/\eta(q)}{\th_1(v\, z_i,q)\th_1(v\, z_i^{-1},q)\prod_{j=1}^{M}\prod_{s_1=\pm}\prod_{s_2=\pm}\th_1(v\,z_i^{s_1}\,(m_{SO(2M+1)}^j)^{s_2},q)/\eta(q)},
\end{align}
where $\zeta_i=\frac{\log(z_i)}{2\pi i}$ are holonomies of the 2d $Sp(k)$ gauge field, the fugacities $m_{Sp(2M-7)}^i$ and $m_{SO(2M+1)}^i$ are defined as in appendix \ref{sec:finit}, and the integral is computed by the Jeffrey-Kirwan prescription \cite{Benini:2013nda,Benini:2013xpa}. The only difference from the $G=SO(2M)$ case, given in equations (3.13)-(3.19) of \cite{Haghighat:2014vxa}, is in the last line of equation \eqref{eq:soodd}.\newline 

\noindent For the case of $k=1$ string, the integration reduces to summing over residues at
\begin{align}
z_1 &= v\, m_{SO(2M+1)}^i\qquad\text{and}\qquad z_1 = v/m_{SO(2M+1)}^i,
\end{align}
while the residue at $z_1=v$ vanishes. This leads to the following result:
\begin{align}
&\mathbb{E}_{4}^{SO(2M+1)}(\mass_{SO(2M+1)},\mass_{Sp(2M-7)},v,q)\! =\! \bigg[\!\sum_{i=1}^M\frac{\th_1(v^2(m_{SO(2M+1)}^i)^2,q)\th_1(v^4(m_{SO(2M+1)}^i)^2,q)}{\th_1(m_{SO(2M+1)}^i,q)\th_1(v^2\,m_{SO(2M+1)}^i,q)}\times\nonumber\\
&\frac{\eta(q)^{10}\prod_{j=1}^{2M-7}\prod_{s=\pm}\th_1(v\, m_{SO(2M+1)}^i\,(m_{Sp(2M-7)}^j)^s,q)}{\prod_{\substack{j=1\\j\neq i}}^{M}\prod_{s_1=\pm}\prod_{s_2=\pm}\th_1(v\, (v\,m_{SO(2M+1)}^i)^{s_1}(m_{SO(2M+1)}^j)^{s_2},q)}\!+\!\bigg(m_{SO(2M+1)}^i\to1/m_{SO(2M+1)}^i\bigg)\bigg],
\end{align}
which indeed agrees with our results obtained exploiting modularity in section \ref{sec:modul} for $G=SO(9)$, $\dots$, $SO(15)$.

\section{WZW models}\label{sec:WZW}
In this section we collect a few basic properties of the WZW models, mostly following the discussion in \cite{di1997conformal}. The spectrum of the (chiral half of the) unitary WZW CFT associated to a Lie group $F$ (which we assume to be simple) includes a set of chiral currents $j^a_F(z)$ of conformal dimension $1$, with $a=1,\dots,\text{dim}(F)$, which satisfy the OPE
\begin{equation}
j^a_F(z)j^b_F(w) \sim \frac{k \delta^{ab}}{(z-w)^2} +\sum_c i f_{abc} \frac{j_F^c(w)}{z-w}.
\end{equation}
The modes that appear in the Laurent expansion of the currents,
\begin{equation}
j^a_F(z) \sum_{n\in\mathbb{Z}}z^{-n-1}j^a_F[n],
\end{equation}
satisfy the commutation relations of the affine Lie algebra $\widehat{\mathfrak{f}}$ of $F$ at level $k$:
\begin{equation}
[j^a_{F}[n],j^b_{F}[m]] =\sum_c i f_{abc} J^c_{F}[n+m]+ k\, n\, \delta_{ab}\delta_{n+m,0},
\end{equation}
where $f_{abc}$ are the structure constants of $\mathfrak{f}$.\\

\noindent The energy-momentum tensor of the theory takes the Sugawara form
\begin{equation}
T(z) = \frac{1}{2(h^\vee_{F}+k)}\sum_a :j^aj^a:(z),
\end{equation}
and its OPE with itself is given by:
\begin{equation}
T(z)T(w) \sim \frac{c_F/2}{(z-w)^4}+\frac{2T(w)}{(z-w)^2}+\frac{\partial T(w)}{z-w},
\end{equation}
where the central charge is given by
\begin{equation} c_{F} = \frac{\text{dim}(F)k}{h^\vee_{F}+k},\end{equation}
where $\mathfrak{f}$ is the Lie algebra of $F$.\newline

\noindent The states of the CFT can be organized in terms of the extended chiral algebra $\widehat{\mathfrak{f}}$. of the theory. In particular, one can define a set of WZW primary fields $\vert\lambda\rangle$ such that
\begin{align}
j^a_F[0]\, \vert\lambda\rangle &= -t^a_\lambda\, \vert\lambda\rangle,\nonumber\\
j^a_F[n]\, \vert\lambda\rangle &= 0 \text{ for } n>0,
\end{align}
where $t^a_\lambda$ are the matrices in the representation $R_\lambda$ of $\widehat{\mathfrak{f}}$ corresponding to the roots of the Lie algebra $\mathfrak{f}$, so that WZW primaries organize themselves in terms of representations $R_\lambda$. It can furthermore be shown that for a given $k$ only the choices of $\lambda$ which correspond to integrable highest-weight representations of $\widehat{\mathfrak{f}}$ give rise to states in the CFT, in the sense that any other choice of $\lambda$ leads to states whose correlators with other states in the CFT all vanish. The conformal dimension of the primary field $\vert\lambda\rangle$ is
 \begin{equation}
 h_{\widehat{\lambda}}=\frac{C_2(R_\lambda)}{2(h^\vee_F+k)},
 \end{equation}
and the extended chiral algebra descendants of the WZW primary are states of the form
 \begin{equation}
j^a_F[-n_1]j^b_F[-n_2]\dots\vert\lambda\rangle,\qquad n_1,n_2,\dots \geq 0.
 \end{equation}
The chiral component of the Hilbert space of the WZW model organizes itself into a number of sectors $H^{WZW_F}_\lambda$, corresponding to the families of states associated to each WZW primary. For each sector one can define a flavored character which counts the states in that sector:
\begin{equation}
\widehat{\chi}^F_\lambda(\mass_F,q) = \text{Tr}_{H^{WZW_F}_\lambda} q^{L_0-\frac{c_F}{24}}\prod_{j=1}^{\text{rank}(F)}e^{2\pi i\langle \mathbf{\mu}_F,K^j_F\rangle},
\end{equation}
where $K^j_F$ are generators of the Cartan subalgebra of $\mathfrak{f}$. These characters coincide with the characters \eqref{eq:afcha} of the corresponding integrable highest-weight representations $R_\lambda$ of $\widehat{\mathfrak{f}}$.\newline

\noindent For certain WZW models it is possible to give closed formulas for the characters in terms of Jacobi theta functions. For instance, the $SO(2N)$ WZW model at level 1 has four conformal primaries: the trivial, vector, spinor and conjugate spinor representations, whose highest weights have Dynkin labels $ (10\dots0),(01\dots0),(0\dots10),$ and $ (0\dots01)$ respectively; we denote these representations concisely as \textbf{1}, \textbf{v}, \textbf{s}, and \textbf{c}. The corresponding characters are given by:
\begin{align}
\widehat\chi^{SO(2N)}_{\bf{1}}(\mass_{SO(2N)},q) &= \frac{1}{2}\left(\prod_{i=1}^{N}\frac{\theta_3(m_{SO(2N)}^i,q)}{\eta(q)}+\prod_{i=1}^{N}\frac{\theta_4(m_{SO(2N)}^i,q)}{\eta(q)}\right),\nonumber\\
\widehat\chi^{SO(2N)}_{\bf{v}}(\mass_{SO(2N)},q) &= \frac{1}{2}\left(\prod_{i=1}^{N}\frac{\theta_3(m_{SO(2N)}^i,q)}{\eta(q)}-\prod_{i=1}^{N}\frac{\theta_4(m_{SO(2N)}^i,q)}{\eta(q)}\right),\nonumber\\
\widehat\chi^{SO(2N)}_{\bf{s}}(\mass_{SO(2N)},q) &= \frac{1}{2}\left(\prod_{i=1}^{N}\frac{\theta_2(m_{SO(2N)}^i,q)}{\eta(q)}+e^{\pi i n/2}\prod_{i=1}^{N}\frac{\theta_1(m_{SO(2N)}^i,q)}{\eta(q)}\right),\nonumber\\
\widehat\chi^{SO(2N)}_{\bf{c}}(\mass_{SO(2N)},q) &= \frac{1}{2}\left(\prod_{i=1}^{N}\frac{\theta_2(m_{SO(2N)}^i,q)}{\eta(q)}-e^{-\pi i n/2}\prod_{i=1}^{N}\frac{\theta_1(m_{SO(2N)}^i,q)}{\eta(q)}\right),
\end{align}
where the fugacities $m_{SO{2N}}^i$ are given in equation \eqref{eq:mso2n}.\newline

\noindent Analogous formulas can also be obtained for $F=SO(2N+1)$. Here we just remark that the characters of the `SO(1)' WZW model at level 1, which has $c=1/2$, are given by:
\begin{align}\label{eq:ising}
\widehat\chi^{SO(1)}_{\bf{1}}(\mass_{SO(1)},q) &= \frac{1}{2}\left(\sqrt{\frac{\theta_3(1,q)}{\eta(q)}}+\sqrt{\frac{\theta_4(1,q)}{\eta(q)}}\right),\nonumber\\
\widehat\chi^{SO(1)}_{\bf{v}}(\mass_{SO(1)},q) &= \frac{1}{2}\left(\sqrt{\frac{\theta_3(1,q)}{\eta(q)}}-\sqrt{\frac{\theta_4(1,q)}{\eta(q)}}\right),\nonumber\\
\widehat\chi^{SO(1)}_{\bf{s}}(\mass_{SO(1)},q) &= \sqrt{\frac{\theta_2(1,q)}{2\eta(q)}}.
\end{align}
The unique level 1 character for $F=E_8$ also has a simple closed form expression: it is given by the following sum of $SO(16)$ level 1 characters:
\begin{equation}
\widehat\chi^{E_8}_{\bf{1}}(\mass_{E_8},q) = \widehat\chi^{SO(16)}_{\bf{1}}(\mass_{SO(16)},q)+\widehat\chi^{SO(16)}_{\bf{s}}(\mass_{SO(16)},q) = \frac{\frac{1}{2}\sum_{\ell=1}^4\prod_{i=1}^8\theta_\ell(m^i_{SO(16)},q)}{\eta(q)^8},
\end{equation}
where the regular embedding of $SO(16)$ into $E_8$ fixes the precise relation between $\mass_{E_8}$ and $\mass_{SO(16)}$ fugacities. \newline

\noindent The two level 1 characters of $SU(2)$ are given in terms of the following theta series:
\begin{align}
\widehat\chi_{\bf 1}^{SU(2)}(\mathbf{m}_{SU(2)},q) &=\frac{\sum_{l\in\mathbb{Z}}q^{l^2}(m^1_{SU(2)})^{2l}}{\eta(q)};\label{eq:su2ch}\\
\widehat\chi_{\bf 2}^{SU(2)}(\mathbf{m}_{SU(2)},q) &=\frac{\sum_{l+1/2\in\mathbb{Z}}q^{l^2}(m^1_{SU(2)})^{2l}}{\eta(q)}.\label{eq:su2ch2}
\end{align}

\noindent The abelian case  $F=U(1)$, which coincides with the $c=1$ theory of a boson compactified on a circle or radius $R$, falls slightly outside of the discussion above. We denote the CFT at radius $R$ as $U(1)_{R^2}$. When the square of the radius of the circle is a rational number
\begin{equation}
R=\sqrt{\frac{2p'}{p}},
\end{equation}
the CFT is rational and the spectrum can be organized into a finite number of characters of an extended algebra. The $2\,p\, p'+1$ primaries under the extended algebra can be labeled by an integer $\ell$ in the range $-pp'+1\leq \ell\leq pp'$, and have conformal dimension
\begin{equation}
h_\ell = \frac{\ell^2}{4p p'}
\end{equation}
and character 
\begin{equation}
\widehat{\chi}^{U(1)}_{\ell}(m_{U(1)},q) = \frac{1}{\eta(\tau)}\sum_{k\in\mathbb{Z}}q^{\frac{(2\,p\,p'k+\ell)^2}{4\,p\,p'}}m_{U(1)}^{2\,p\, p' k+\ell}.
\end{equation}
At $R^2 = 2$, the $U(1)_{R^2} $ CFT coincides with the $SU(2)$ WZW model at level 1.\newline

\noindent For the several cases for which no simple closed formula for the characters is available, we have computed the characters to sufficiently high powers in $q$ using the \texttt{Sage} platform.\newline

\noindent We conclude this appendix by remarking that for low rank exceptional isomorphisms of Lie algebras lead to equivalences between WZW models:
\begin{align} 
Sp(1)_{k} &\simeq SU(2)_k,\nonumber\\
SO(3)_{k} &\simeq SU(2)_{2k},\nonumber\\
SO(4)_k &\simeq SU(2)_k\times SU(2)_k,\nonumber\\
SO(5)_k &\simeq Sp(2)_k,\nonumber\\
SO(6)_k &\simeq SU(4)_k. \end{align}
Moreover, the $SO(2)_k$ WZW model coincides with the $U(1)_k$ model.

\section{An explicit example of $\xi^{n,G}_\lambda$ functions}\label{app:xiexample}
In this appendix we provide some details about the functions $\xi^{n,G}_\lambda(\mass_G,v,q)$ that we compute for a nontrivial example: the \hh{3}{F_4} CFT. We also show how the form of these functions leads to constraints on the elliptic genus.\newline

\noindent The elliptic genus is expected, according to equation \eqref{eq:bjkexp}, to take the form
\begin{equation}\label{eq:FGVF43}
\mathbb{E}_3^{F_4}(v,q) = q^{-\frac{1}{3}}v^{-2}\sum_{j,k\geq 0} b_{j k} (q/v^2)^jv^k=\sum_\lambda\widehat{\chi}^{Sp(2)}_\lambda(\mass_{Sp(2)},q)\xi^{3,F_4}_\lambda(\mass_{F_4},v,q),
\end{equation}
where $\widehat{\chi}^{Sp(2)}_\lambda(\mass_{Sp(2)},q)$ are the characters corresponding to the ten integrable representations of $F=Sp(2)$ at level 3:
\begin{equation}
R_{(00)}^{Sp(2)},\quad R_{(01)}^{Sp(2)},\quad R_{(10)}^{Sp(2)},\quad R_{(03)}^{Sp(2)},\quad R_{(30)}^{Sp(2)},\quad R_{(12)}^{Sp(2)},\quad R_{(02)}^{Sp(2)},\quad R_{(20)}^{Sp(2)},\quad R_{(21)}^{Sp(2)}, \quad R_{(11)}^{Sp(2)}.
\end{equation}
We keep the terms on the right hand side of equation \eqref{eq:FGVF43} which may contribute to the $b_{jk}$ coefficients for a suitable range of $j$ and $k$, which we choose to be $0\leq j\leq 3$ and $0\leq k\leq 6$.\newline

\noindent First, we compute the $Sp(2)=SO(5)$ characters to the desired power in the $q$ expansion. We find:
\begin{equation}
\widehat{\chi}^{Sp(2)}_\lambda(\mass_{Sp(2)},q) = \frac{1}{\widetilde\Delta_{Sp(2)}(\mass_{Sp(2)},q)}\widetilde{\chi}^{Sp(2)}_\lambda(\mass_{Sp(2)},q),
\end{equation}
where
\begin{align}
\frac{1}{\widetilde\Delta_{Sp(2)}(\mass_{Sp(2)},q)}&=1+q \chi^{Sp(2)}_{(20)}+q^2(\chi^{Sp(2)}_{(00)}+\chi^{Sp(2)}_{(01)}+\chi^{Sp(2)}_{(20)}+\chi^{Sp(2)}_{(02)}+\chi^{Sp(2)}_{(40)})\nonumber\\
&+q^3(\chi^{Sp(2)}_{(00)}+\chi^{Sp(2)}_{(01)}+4\chi^{Sp(2)}_{(20)}+\chi^{Sp(2)}_{(02)}+2\chi^{Sp(2)}_{(21)}+\chi^{Sp(2)}_{(40)}+\chi^{Sp(2)}_{(22)}+\chi^{Sp(2)}_{(60)})
\end{align}
plus higher order terms, and the rescaled characters $\widetilde{\chi}^{Sp(2)}_\lambda(\mass_{Sp(2)},q)$ are given to the desired order in the $q$ expansion by:
\begin{align}
\widetilde{\chi}^{Sp(2)}_{(00)}(\mass_{Sp(2)},q)&=q^{-5/24};\\
\widetilde{\chi}^{Sp(2)}_{(01)}(\mass_{Sp(2)},q)&=q^{1/8}\chi^{Sp(2)}_{(01)};\\
\widetilde{\chi}^{Sp(2)}_{(10)}(\mass_{Sp(2)},q)&=\chi^{Sp(2)}_{(10)};\\
\widetilde{\chi}^{Sp(2)}_{(03)}(\mass_{Sp(2)},q)&=q^{31/24}(\chi^{Sp(2)}_{(03)}-q\,\chi^{Sp(2)}_{(23)});\\
\widetilde{\chi}^{Sp(2)}_{(30)}(\mass_{Sp(2)},q)&=q^{2/3}(\chi^{Sp(2)}_{(30)}-q\,\chi^{Sp(2)}_{(50)});\\
\widetilde{\chi}^{Sp(2)}_{(12)}(\mass_{Sp(2)},q)&=q^{1}(\chi^{Sp(2)}_{(12)}-q\,\chi^{Sp(2)}_{(32)});\\
\widetilde{\chi}^{Sp(2)}_{(02)}(\mass_{Sp(2)},q)&=q^{5/8}\chi^{Sp(2)}_{(02)};\\
\widetilde{\chi}^{Sp(2)}_{(20)}(\mass_{Sp(2)},q)&=q^{7/24}(\chi^{Sp(2)}_{(20)}-q^2\,\chi^{Sp(2)}_{(60)});\\
\widetilde{\chi}^{Sp(2)}_{(21)}(\mass_{Sp(2)},q)&=q^{19/24}(\chi^{Sp(2)}_{(21)}-q\,\chi^{Sp(2)}_{(41)});\\
\widetilde{\chi}^{Sp(2)}_{(11)}(\mass_{Sp(2)},q)&=0.
\end{align}

We also need to expand the $ \xi^{3,F_4}_\lambda(\mass_{F_4},v,q)$ functions according to \eqref{eq:FGv} and keep all terms that may contribute to equation \eqref{eq:FGVF43} for $0\leq j\leq 3$ and $0\leq k\leq 6$.\newline 

\noindent The range of $\ell$ in the sum of equation \eqref{eq:FGv} is $-11\leq \ell\leq 0$. We define the rescaled functions
\begin{equation}
\widetilde\xi^{\,3,F_4}_\lambda(\mass_{F_4},v,q)=\xi^{\,3,F_4}_\lambda(\mass_{F_4},v,q)\cdot \widetilde\Delta_{F_4}(\mass_{F_4},q)\prod_{j=1}^\infty(1-q^j)
\end{equation}
in order to remove the terms in the denominator of equation \eqref{eq:FGv}.  Explicitly, we find:
\begin{align}&\widetilde\xi^{3,F_4}_{(00)}(\mass_{F_4},v,q) =  q^{7/8}\bigg((v^{-2}n^{(00)}_{(0000),-2,0}+v^2n^{(00)}_{(0000),-10,-1}\chi^{F_4}_{(0000)}+\dots)\nonumber\\
&\quad\quad\quad\quad\quad\quad\quad\quad+q(v^{-2}n^{(00)}_{(1000),-4,0}\chi^{F_4}_{(1000)}+n^{(00)}_{(0001),-2,0}\chi^{F_4}_{(0001)}+v^4n^{(00)}_{(0001),-10,-1}\chi^{F_4}_{(0001)}+\dots)\nonumber\\
&\quad\quad\quad\quad\quad\quad\quad\quad+q^2(v^{-2}n^{(00)}_{(2000),-6,0}\chi^{F_4}_{(2000)}+n^{(00)}_{(0010),-2,0}\chi^{F_4}_{(0010)}+\dots)+\dots\bigg);\label{eq:bm1}
\end{align}
\begin{align}
&\widetilde\xi^{3,F_4}_{(01)}(\mass_{F_4},v,q) =  q^{-11/24}\bigg(v^{4}n^{(01)}_{(0000),-6,-1}\chi^{F_4}_{(0000)}\nonumber\\
&\quad\quad\quad\quad\quad\quad+q(v^{-2}n^{(01)}_{(0100),-6,0}\chi^{F_4}_{(0100)}+n^{(00)}_{(0002),-4,0}\chi^{F_4}_{(0002)}+v^4n^{(01)}_{(1000),0,0}\chi^{F_4}_{(0001)}+\dots)+\dots\bigg);
\end{align}
\begin{align}
&\widetilde\xi^{3,F_4}_{(10)}(\mass_{F_4},v,q) =  q^{2/3}\bigg((v^{3}n^{(10)}_{(0000),-9,-1}+\dots)+q(v^{-1}n^{(10)}_{(0001),-3,0}\chi^{F_4}_{(0001)}+\dots)\quad\quad\nonumber\\
&\quad\quad\quad\quad\quad\quad\quad\quad\quad\quad\quad\quad\quad+q^2(v^{-1}n^{(10)}_{(1001),-5,0}\chi^{F_4}_{(1001)}+vn^{(10)}_{(0010),-3,0}\chi^{F_4}_{(0010)}+\dots)+\dots\bigg);
\end{align}
\begin{align}
&\widetilde\xi^{3,F_4}_{(03)}(\mass_{F_4},v,q) =  q^{3/8}\bigg((v^{-2}n^{(03)}_{(0000),-4,0}+\dots)+q(n^{(03)}_{(0001),-4,0}\chi^{F_4}_{(0001)}+\dots)+\dots\bigg);\quad\quad\quad\quad\quad
\end{align}
\begin{align}
&\widetilde\xi^{3,F_4}_{(30)}(\mass_{F_4},v,q) =  q^{1}\bigg((v^{1}n^{(30)}_{(0000),-1,0}+v^{3}n^{(30)}_{(0000),-11,-1}+\dots)\nonumber\\
&\quad\quad\quad\quad\quad\quad\quad\quad\quad\quad\quad\quad\quad+q(v^{-1}n^{(30)}_{(0010),-5,0}\chi^{F_4}_{(0010)}+v^{3}n^{(30)}_{(0001),-1,0}\chi^{F_4}_{(0001)}+\dots)+\dots\bigg);
\end{align}
\begin{align}
&\widetilde\xi^{3,F_4}_{(12)}(\mass_{F_4},v,q) =  q^{2/3}\bigg((v^{-1}n^{(12)}_{(0000),-3,0}+\dots)+q(vn^{(12)}_{(0001),-3,0}\chi^{F_4}_{(0001)}+\dots)+\dots\bigg);\quad\quad\quad\quad\quad
\end{align}
\begin{align}
&\widetilde\xi^{3,F_4}_{(02)}(\mass_{F_4},v,q) =  q^{25/24}\bigg((v^{2}n^{(02)}_{(0000),0,0}+\dots)+q(v^4n^{(02)}_{(0001),0,0}\chi^{F_4}_{(0001)}+\dots)+\dots\bigg);\quad\quad\quad\quad\quad
\end{align}
\begin{align}
&\widetilde\xi^{3,F_4}_{(20)}(\mass_{F_4},v,q) =  q^{3/8}\bigg((v^{4}n^{(02)}_{(0000),-8,-1}+\dots)+q(v^{-2}n^{(20)}_{(0001),-4,0}\chi^{F_4}_{(0001)}+\dots)\nonumber\\
&\quad\quad\quad\quad\quad\quad\quad\quad\quad\quad\quad\quad\quad\quad\quad\quad\quad+q^2(n^{(20)}_{(0010),-4,0}\chi^{F_4}_{(0010)}+v^{2}n^{(20)}_{(1000),-2,0}\chi^{F_4}_{(1000)})+\dots\bigg);
\end{align}
\begin{align}
&\widetilde\xi^{3,F_4}_{(21)}(\mass_{F_4},v,q) =  q^{7/8}\bigg((n^{(21)}_{(0000),-2,0}+v^4n^{(21)}_{(0000),-10,-1}+\dots)\nonumber\\
&\quad\quad\quad\quad\quad\quad\quad\quad\quad\quad\quad\quad\quad\quad\quad+q(n^{(21)}_{(1000),-4,0}\chi^{F_4}_{(1000)}+v^2n^{(21)}_{(0001),-4,0}\chi^{F_4}_{(0001)}+\dots)+\dots\bigg);
\end{align}
\begin{align}\label{eq:bm10}
&\widetilde\xi^{3,F_4}_{(11)}(\mass_{F_4},v,q) =q^{13/4}\bigg(\dots\bigg),\quad\quad\quad\quad\quad\quad\quad\quad\quad\quad\quad\quad\quad\quad\quad\quad\quad\quad\quad\quad\quad\quad\quad\quad\quad\quad\quad\quad\quad
\end{align}
where the $\dots$ indicate higher order terms which do not contribute to the elliptic genus for the specified range of $j,k$. There are overall 33 undetermined integral coefficients $n^{\lambda}_{\nu,\ell,m}$. \newline

\noindent As reported in table \ref{tab:constr1}, after imposing constraint $\bf{C1}$ the modular Ansatz still has 4 unfixed coefficients which we would like to fix by exploiting equation \eqref{eq:FGv}. In order to impose constraint $\bf{C2}$, we first rescale the elliptic genus as:
\begin{equation}
\widetilde{\mathbb{E}}_3^{F_4}(v,q)=\widetilde\Delta_{Sp(2)}(\mass_{Sp(2)},q)\widetilde\Delta_{F_4}(\mass_{F_4},q)\prod_{j=1}^\infty(1-q^j)\times \mathbb{E}_3^{F_4}(v,q)
\end{equation}
and expand it as
\begin{equation}\label{eq:FGVF43a}
\widetilde{\mathbb{E}}_3^{F_4}(v,q)=q^{-\frac{1}{3}}v^{-2}\sum_{j,k\geq 0} \widetilde b_{j k} (q/v^2)^jv^k.\end{equation}
Equation \eqref{eq:FGv} then implies that 
\begin{equation}\label{eq:FGVF43b}
\sum_{j,k\geq 0} \widetilde b_{j k} (q/v^2)^jv^k = \sum_\lambda\widetilde{\chi}^{F_4}_\lambda(\mass_{Sp(2)},q) \widetilde{\xi}^{3,F_4}_\lambda(\mass_{F_4},v,q).
\end{equation}
Expanding the right hand side in terms of the functions \eqref{eq:bm1}-\eqref{eq:bm10}, we find that certain coefficients $\widetilde{b}_{jk}$ are necessarily zero; we indicate these in the table \ref{tab:0star} by a 0, while we indicate by an asterisk the ones which are not necessarily vanishing.\newline

\begin{table}[t!]
\begin{center}\begin{tabular}{|l||l|l|l|l|l|l|l|l|l|}\hline $k\backslash j$&0&1&2&3&4&5&6\\\hline\hline 0&
 0 & 0 & 0 & 0 & 0 & 0 & *  \\
1& * & 0 & 0 & 0 & * & * & *  \\
 2&* & * & * & * & * & * & *  \\
3& * & * & * & * & * & * & *  \\\hline
\end{tabular}\end{center}
\caption{Coefficients $\widetilde{b}_{jk}$ in equation \eqref{eq:FGVF43}. The coefficients which are identically zero are indicated by a `0', while the rest are indicated by an asterisk.}
\label{tab:0star}
\end{table}
\noindent In the modular Ansatz, the coefficients $b_{00},$ $b_{01}, $ and $b_{10}$ are already guaranteed to vanish as a consequence of imposing constraint $\bf{C1}$. On the other hand, the six coefficients $b_{02},\dots, b_{05}$, $b_{12}$, and $b_{13}$ do not automatically vanish. This gives six constraints on the four undetermined coefficients of the Ansatz, which is sufficient to uniquely fix them. After fixing these undetermined coefficients, one finds the following values of $\widetilde{b}_{jk}$ coefficients in the expansion \eqref{eq:FGVF43a} of the elliptic genus:

\begin{center}\begin{tabular}{|l||l|l|l|l|l|l|l|l|l|}\hline $k\backslash j$&0&1&2&3&4&5&6\\\hline\hline 0&
 0 & 0 & 0 & 0 & 0 & 0 & 5  \\
1& 1 & 0 & 0 & 0 & 0 & -4 & -10  \\
 2&52 & -104 & 0 & 20 & 14 & 0 & -9  \\
3& 1053 & -4212 & 4350 & 0 & -910 & -56 & 364  \\\hline
\end{tabular}\end{center}

\noindent Having determined the elliptic genus with fugacities turned off, we can compare the numerical values for $\widetilde{b}_{jk}$ with equation \eqref{eq:FGVF43b} to determine the values of the 33 $n^\lambda_{\nu,\ell,m}$ coefficients in equations \eqref{eq:bm1}--\eqref{eq:bm10}. \newline

\noindent For instance, for $j=3,k=6$ one finds:
\begin{equation}
364 =260 n^{(01)}_{(1000),0,0} +364n^{(02)}_{(0001),0,0}, 
\end{equation}
from which the only solution for which the $n^\lambda_{\nu,\ell,m}$ take reasonably small values is:
\begin{equation}
n^{(01)}_{(1000),0,0} = 0 \text{ and } n^{(02)}_{(0001),0,0} =1.
\end{equation}
Proceeding similarly for the remaining coefficients, we are easily able to fix all remaining coefficients in equations \eqref{eq:bm1}--\eqref{eq:bm10}. This allows us to fix the functions $\xi^{3,F_4}_{\lambda}(\mass_{F_4},v,q)$ completely:
\begin{align}&\xi^{3,F_4}_{(00)}(\mass_{F_4},v,q) =  q^{7/8}\bigg(v^{-2}+q(v^{-2}\chi^{F_4}_{(1000)}+v^4\chi^{F_4}_{(0001)})+q^2v^{-2}\chi^{F_4}_{(2000)}\bigg);\\
&\xi^{3,F_4}_{(01)}(\mass_{F_4},v,q) =  q^{-11/24}\bigg(v^{4}\chi^{F_4}_{(0000)}+q(v^{-2}\chi^{F_4}_{(0100)}+\chi^{F_4}_{(0002)})\bigg);\\
&\xi^{3,F_4}_{(10)}(\mass_{F_4},v,q) =  -q^{2/3}\bigg(v^{3}+q\,v^{-1}\chi^{F_4}_{(0001)}+q^2v^{-1}\chi^{F_4}_{(1001)}\bigg);\\
&\xi^{3,F_4}_{(30)}(\mass_{F_4},v,q) =  q\,v;\\
&\xi^{3,F_4}_{(02)}(\mass_{F_4},v,q) =  q^{49/24}v^4\chi^{F_4}_{(0001)}\\
&\xi^{3,F_4}_{(20)}(\mass_{F_4},v,q) =  q^{3/8}\bigg(-v^{4}+q^2\chi^{F_4}_{(0010)}\bigg);\\
&\xi^{3,F_4}_{(21)}(\mass_{F_4},v,q) =  q^{7/8}\bigg(-v^4-q\, v^2\chi^{F_4}_{(0001)}\bigg);\\
&\xi^{3,F_4}_{(03)}(\mass_{F_4},v,q) = \xi^{3,F_4}_{(12)}(\mass_{F_4},v,q) = \xi^{3,F_4}_{(11)}(\mass_{F_4},v,q) =0.\end{align}
Note that in this example all coefficients turn out to be either $0$, $1$, or $-1$.\newline

\noindent Via equation \eqref{eq:FGVF43}, this calculation allows us to determine the $\mass_{F_4}$ and $\mass_{Sp(2)}$ dependence of the elliptic genus up to a reasonable order in its series expansion.

\section{Tables of elliptic genus coefficients}\label{app:bjk}

By exploiting modularity, in section \ref{sec:modul} we have obtained explicit expressions for the one-string elliptic genera of 63 out of the 72 known six-dimensional SCFTs with one tensor multiplet and $\text{rank}(G)$ up to rank 7, with fugacities $\mass_F,\mass_G$ turned off. Except for the cases of smallest rank, these expressions tend to be quite lengthy. Rather than including them here, we provide them them in an attached Mathematica file. In this appendix, on the other hand, we provide tables of coefficients of the elliptic genera, which display several of the universal features that are outlined in section \ref{sec:univ} and can be easily employed to compare our results with other calculations of the elliptic genera. \newline

\noindent The tables are to be read as follows: for any theory \hh{n}{G} we expand the elliptic genus as a power series:
\begin{equation}\mathbb{E}_{n}^{G}(v,q) = q^{\frac{1}{6}-\frac{n-2}{2}-\delta_{1,n}}v^{1-n}\sum_{j,k\geq 0} b_{j k} (q/v^2)^jv^k.
\end{equation}
The tables list the coefficients $b_{j k}$ for a suitable range of $j$ and $k$.\newline

\noindent Of particular relevance are the following sequences of coefficients:
\begin{itemize}
\item The $j=0$ row gives the $v$-expansion of the one-instanton piece of the 5d Nekrasov partition function for the theory obtained from the 6d SCFT in the zero radius limit of the 6d circle (see section \ref{sec:5dlim}); note that for certain theories the $j=0$ coefficients are only nonzero for large values of $k$, so all the coefficients displayed in the tables below vanish. Nevertheless, in appendix \ref{sec:5dapp} we provide expressions for the $j=0$ coefficients which are valid to all orders in $v$.
\item The $k=0$ column gives the one-instanton piece of the Nekrasov partition function of the 5d pure $G$ $\mathcal{N}=1$ SYM theory, as in equation \eqref{eq:cons1b}, while the $k=1$ column captures information about the bifundamental matter of the 6d SCFT as in equation \eqref{eq:cons1c}.
\item For the majority of theories, the $j=n-1+\delta_{1,n}$, $k=2$ coefficient is equal to $\text{dim}(G)+\text{dim}(F)+1$ as in equation \eqref{eq:bcoefdim}. The exceptions are the theories \hh{1}{Sp(N)} for $N\geq 2$ and \hh{1}{SO(12)_b} (see the discussion in section \ref{sec:modresults}).
\end{itemize}
\hspace{0.3in}
\begin{table}[H]
\begin{center}\scalebox{0.702}{
\begin{tabular}{|l||l|l|l|l|l|l|l|l|l|}\hline $k\backslash j$&0&1&2&3&4&5&6&7&8\\\hline\hline 0&
1 & 0 & 0 & 0 & 0 & 0 & 0 & 0 & 0 \\
1& 3 & -40 & 194 & 472 & -1021 & 1536 & -2048 & 2560 & -3072 \\
2& 5 & -80 & 582 & -2480 & 4998 & 9808 & -23994 & 36784 & -49147 \\
3& 7 & -120 & 970 & -4960 & 18057 & -46952 & 61836 & 108696 & -293239 \\
4& 9 & -160 & 1358 & -7440 & 30095 & -96976 & 255442 & -515968 & 532246 \\
5& 11 & -200 & 1746 & -9920 & 42133 & -145464 & 429150 & -1102944 & 2427308 \\
6& 13 & -240 & 2134 & -12400 & 54171 & -193952 & 600810 & -1658256 & 4124353 \\
7& 15 & -280 & 2522 & -14880 & 66209 & -242440 & 772470 & -2211008 & 5778395 \\
8& 17 & -320 & 2910 & -17360 & 78247 & -290928 & 944130 & -2763760 & 7429365 \\\hline
\end{tabular}}
\end{center}
\caption{table of $b_{jk}$ coefficients for \hh{1}{SU(2)}.}
\end{table}
\begin{table}[H]
\begin{center}\scalebox{0.702}{
\begin{tabular}{|l||l|l|l|l|l|l|l|l|l|}\hline $k\backslash j$&0&1&2&3&4&5&6&7&8\\\hline\hline 0&
1 & 0 & 0 & 0 & 0 & 0 & 0 & 0 & 0 \\
1& 8 & -72 & 152 & 440 & 520 & -4608 & 13312 & -28160 & 50688 \\
2& 27 & -360 & 2008 & -5400 & 2870 & 17128 & 6104 & -129384 & 401435 \\
3& 64 & -1008 & 7272 & -31400 & 86464 & -134568 & 13424 & 330328 & -20032 \\
4& 125 & -2160 & 17648 & -91000 & 331506 & -890200 & 1716080 & -1956032 & -224024 \\
5& 216 & -3960 & 34840 & -197640 & 818720 & -2644712 & 6863760 & -14236480 & 22201384 \\
6& 343 & -6552 & 60552 & -364760 & 1628830 & -5797656 & 17159008 & -43080128 & 91730079 \\
7& 512 & -10080 & 96488 & -605800 & 2842560 & -10748584 & 34317680 & -95072832 & 231514304 \\
8& 729 & -14688 & 144352 & -934200 & 4540634 & -17897048 & 60055632 & -176828608 & 464904985 \\\hline
\end{tabular}
}
\end{center}
\caption{table of $b_{jk}$ coefficients for \hh{1}{SU(3)}.}
\end{table}
\begin{table}[H]
\begin{center}\scalebox{0.702}{
\begin{tabular}{|l||l|l|l|l|l|l|l|l|l|}\hline $k\backslash j$&0&1&2&3&4&5&6&7&8\\\hline\hline 0&
 1 & 0 & 0 & 0 & 0 & 0 & 0 & 0 & 0 \\
 1&15 & -108 & 162 & 276 & 783 & 128 & -16896 & 78720 & -235264 \\
 2&84 & -992 & 4710 & -10080 & 3880 & 11424 & 29286 & -2016 & -638892 \\
3& 300 & -4440 & 29568 & -114500 & 270420 & -340548 & 48128 & 246204 & 565332 \\
4& 825 & -13920 & 110160 & -542016 & 1838865 & -4440512 & 7371600 & -7028640 & 668338 \\
5& 1911 & -35140 & 309210 & -1740060 & 7041927 & -21715464 & 52206882 & -96895180 & 130916670 \\
6& 3920 & -76608 & 724542 & -4447264 & 20024268 & -70656000 & 202708158 & -480625344 & 939588888 \\
 7&7344 & -150192 & 1496880 & -9766260 & 47260200 & -181837516 & 580179072 & -1573241940 & 3667470216 \\
8& 12825 & -271680 & 2817648 & -19236480 & 98071385 & -400941408 & 1374258288 & -4061540576 & 10524196914 \\\hline
\end{tabular}
}
\end{center}
\caption{table of $b_{jk}$ coefficients for \hh{1}{SU(4)}.}
\end{table}
\begin{table}[H]
\begin{center}\scalebox{0.702}{
\begin{tabular}{|l||l|l|l|l|l|l|l|l|l|}\hline $k\backslash j$&0&1&2&3&4&5&6&7&8\\\hline\hline 0&
 1 & 0 & 0 & 0 & 0 & 0 & 0 & 0 & 0 \\
1& 24 & -150 & 194 & 202 & 408 & 1824 & -3328 & -55840 & 394368 \\
 2&200 & -2170 & 9296 & -17670 & 7308 & 12538 & -1968 & 117126 & -120632 \\
 3&1000 & -14130 & 88900 & -320572 & 693592 & -798690 & 158856 & 599838 & -1024680 \\
4& 3675 & -60830 & 469520 & -2232060 & 7219992 & -16360848 & 25145904 & -22328790 & 2401696 \\
5& 10976 & -202020 & 1774150 & -9903410 & 39366600 & -117671424 & 270049808 & -471489438 & 594085944 \\
6& 28224 & -560700 & 5385184 & -33433218 & 151129508 & -529617726 & 1489166800 & -3410975826 & 6353004528 \\
7& 64800 & -1363740 & 13990536 & -93720760 & 462937608 & -1801792122 & 5749054872 & -15391101594 & 34958597160 \\
8& 136125 & -2997060 & 32344800 & -229438392 & 1209644592 & -5074442320 & 17667293616 & -52433918862 & 134819632032 \\\hline
\end{tabular}
}
\end{center}
\caption{table of $b_{jk}$ coefficients for \hh{1}{SU(5)}.}
\end{table}
\begin{table}[H]
\begin{center}\scalebox{0.652}{
\begin{tabular}{|l||l|l|l|l|l|l|l|l|l|}\hline $k\backslash j$&0&1&2&3&4&5&6&7&8\\\hline\hline 0&
 1 & 0 & 0 & 0 & 0 & 0 & 0 & 0 & 0 \\
1& 35 & -198 & 232 & 214 & -21 & 1876 & 1680 & -14196 & -183464 \\
2& 405 & -4128 & 16464 & -29024 & 11706 & 23104 & -36848 & 86720 & 260373 \\
3& 2695 & -36750 & 221724 & -760940 & 1557101 & -1701634 & 329808 & 1476798 & -2348563 \\
4& 12740 & -207648 & 1571724 & -7283392 & 22792635 & -49588416 & 72864988 & -62207872 & 6191064 \\
5& 47628 & -876708 & 7680764 & -42580510 & 167016850 & -488713026 & 1088867280 & -1833431804 & 2223625278 \\
6& 149940 & -3015936 & 29288196 & -183247456 & 830327589 & -2895984224 & 8037320060 & -18018269536 & 32605072278 \\
7& 413820 & -8906436 & 93392628 & -637920024 & 3198608868 & -12557442856 & 40106780532 & -106586786534 & 238358008812 \\
 8&1029105 & -23364000 & 260031948 & -1898849760 & 10266844476 & -43920568384 & 154842653924 & -461713031136 & 1183204470747 \\\hline
\end{tabular}
}
\end{center}
\caption{table of $b_{jk}$ coefficients for \hh{1}{SU(6)}.}
\end{table}
\begin{table}[H]
\begin{center}\scalebox{0.702}{
\begin{tabular}{|l||l|l|l|l|l|l|l|l|l|}\hline $k\backslash j$&0&1&2&3&4&5&6&7\\\hline\hline 0&
 1 & 0 & 0 & 0 & 0 & 0 & 0 & 0  \\
1& 48 & -252 & 274 & 252 & -228 & 952 & 4536 & -5400 \\
2& 735 & -7140 & 27012 & -45108 & 16998 & 39708 & -61212 & 29708 \\
3& 6272 & -83160 & 485898 & -1608180 & 3164712 & -3327324 & 554116 & 3147540 \\
4& 37044 & -596232 & 4443488 & -20194020 & 61696002 & -130515228 & 186131376 & -154246344 \\
5& 169344 & -3116400 & 27239436 & -150165288 & 583139508 & -1680840084 & 3671120772 & -6038635992 \\
6& 640332 & -13006224 & 127370376 & -801499944 & 3638271504 & -12649513968 & 34805722028 & -76951883316 \\
 7&2090880 & -45821160 & 488830188 & -3389732136 & 17195257296 & -67982653816 & 217486465392 & -575678320620 \\
8& 6073353 & -141381240 & 1612755936 & -12049911720 & 66458923968 & -288778622328 & 1028844925920 & -3082991552840 \\\hline
\end{tabular}
}
\end{center}
\caption{table of $b_{jk}$ coefficients for \hh{1}{SU(7)}.}
\end{table}
\begin{table}[H]
\begin{center}\scalebox{0.702}{
\begin{tabular}{|ll|l|l|l|l|l|l|l|l|}\hline $k\backslash j$&0&1&2&3&4&5&6&7\\\hline\hline 0&
 1 & 0 & 0 & 0 & 0 & 0 & 0 & 0  \\
1& 63 & -312 & 320 & 296 & -317 & 312 & 3384 & 7128 \\
2& 1232 & -11520 & 41844 & -67072 & 23760 & 60160 & -78332 & -13312 \\
3& 13104 & -169848 & 967648 & -3115184 & 5955360 & -6081608 & 894144 & 5790064 \\
4& 94500 & -1505280 & 11078496 & -49586432 & 148778157 & -308381440 & 430533600 & -349049088 \\
5& 518364 & -9535680 & 83176620 & -456460824 & 1759089020 & -5014874080 & 10800110640 & -17482263048 \\
6& 2317392 & -47443968 & 467719164 & -2956466688 & 13440933264 & -46636945152 & 127582632164 & -279460228864 \\
7& 8833968 & -196499160 & 2125800864 & -14921076984 & 76406098800 & -303879888192 & 974147978784 & -2573467183328 \\
8& 29630601 & -704301312 & 8198155680 & -62406567168 & 349797709668 & -1539653616384 & 5535251144160 & -16668756397056 \\\hline
\end{tabular}
}
\end{center}
\caption{table of $b_{jk}$ coefficients for \hh{1}{SU(8)}.}
\end{table}
\begin{table}[H]
\begin{center}\scalebox{0.702}{
\begin{tabular}{|l||l|l|l|l|l|l|l|l|}\hline $k\backslash j$&0&1&2&3&4&5&6&7\\\hline\hline 0&
 1 & 0 & 0 & 0 & 0 & 0 & 0 &  0 \\
1& 35 & -200 & 260 & 20 & 875 & -1260 & 10640 & -36148 \\
 2&405 & -4140 & 16660 & -30580 & 19826 & -8660 & 63700 & -184140 \\
 3&2695 & -36792 & 222480 & -767560 & 1595125 & -1864780 & 894008 & -176780 \\
4& 12740 & -207760 & 1573880 & -7303652 & 22917795 & -50166500 & 75016480 & -68986440 \\
5& 47628 & -876960 & 7685860 & -42630980 & 167346410 & -490325716 & 1095236580 & -1854745160 \\
6& 149940 & -3016440 & 29298780 & -183356600 & 831071605 & -2899794380 & 8053101980 & -18073750672 \\
7& 413820 & -8907360 & 93412620 & -638132880 & 3200110228 & -12565415380 & 40141093720 & -106712365860 \\
8& 1029105 & -23365584 & 260067060 & -1899233400 & 10269626220 & -43935780840 & 154910202524 & -461968529260 \\\hline
\end{tabular}
}
\end{center}
\caption{table of $b_{jk}$ coefficients for \hh{1}{SU(6)_*}.}
\end{table}
\begin{table}[H]
\begin{center}\scalebox{0.702}{
\begin{tabular}{|l||l|l|l|l|l|l|l|l|l|}\hline $k\backslash j$&0&1&2&3&4&5&6&7&8\\\hline\hline 0&
 1 & 0 & 0 & 0 & 0 & 0 & 0 & 0 & 0 \\
1& 10 & -96 & 292 & -96 & 2058 & -8192 & 20480 & -40960 & 71680 \\
 2&35 & -480 & 2856 & -9248 & 15750 & -17440 & 84776 & -287200 & 688163 \\
3& 84 & -1344 & 9964 & -45280 & 139132 & -294112 & 420936 & -580832 & 1810300 \\
4& 165 & -2880 & 23888 & -126112 & 477778 & -1378080 & 3088720 & -5331200 & 7208280 \\
5& 286 & -5280 & 46900 & -269664 & 1139320 & -3793888 & 10352600 & -23514880 & 44314066 \\
6& 455 & -8736 & 81272 & -493856 & 2231390 & -8074272 & 24479904 & -63786240 & 144368959 \\
7& 680 & -13440 & 129276 & -816608 & 3861620 & -14751968 & 47758440 & -134923008 & 338101988 \\
8& 969 & -19584 & 193184 & -1255840 & 6137642 & -24359712 & 82476016 & -245743872 & 656643625 \\\hline
\end{tabular}
}
\end{center}
\caption{table of $b_{jk}$ coefficients for \hh{1}{Sp(2)}.}
\end{table}
\begin{table}[H]
\begin{center}\scalebox{0.702}{
\begin{tabular}{|l||l|l|l|l|l|l|l|l|l|}\hline $k\backslash j$&0&1&2&3&4&5&6&7&8\\\hline\hline 0&
 1 & 0 & 0 & 0 & 0 & 0 & 0 & 0 & 0 \\
1& 21 & -168 & 414 & -168 & 21 & 8192 & -49152 & 172032 & -458752 \\
 2&126 & -1568 & 8310 & -23520 & 35380 & -23520 & -40842 & 522720 & -2457474 \\
3& 462 & -7056 & 49476 & -209720 & 590850 & -1132152 & 1433624 & -960120 & -1866750 \\
4& 1287 & -22176 & 180900 & -931392 & 3393705 & -9244928 & 19233468 & -30427680 & 35281510 \\
5& 3003 & -56056 & 503250 & -2910600 & 12242019 & -39948720 & 104819022 & -224759080 & 394299510 \\
6& 6188 & -122304 & 1173354 & -7335328 & 33832986 & -123332160 & 370343706 & -938417088 & 2030226408 \\
7& 11628 & -239904 & 2416680 & -15975960 & 78591240 & -308819560 & 1013000076 & -2851819800 & 7005006096 \\
8& 20349 & -434112 & 4539816 & -31301760 & 161497175 & -670004832 & 2339316720 & -7080221792 & 18930661794 \\\hline
\end{tabular}
}
\end{center}
\caption{table of $b_{jk}$ coefficients for \hh{1}{Sp(3)}.}
\end{table}
\begin{table}[H]
\begin{center}\scalebox{0.702}{
\begin{tabular}{|l||l|l|l|l|l|l|l|l|l|}\hline $k\backslash j$&0&1&2&3&4&5&6&7&8\\\hline\hline 0&
 1 & 0 & 0 & 0 & 0 & 0 & 0 & 0 & 0 \\
1& 36 & -256 & 560 & -256 & 36 & 0 & 32768 & -262144 & 1179648 \\
2& 330 & -3840 & 18880 & -49408 & 70152 & -49408 & 18880 & -265984 & 3146058 \\
3& 1716 & -25344 & 170976 & -692736 & 1854572 & -3379968 & 4158912 & -3379968 & 3034220 \\
4& 6435 & -109824 & 884928 & -4477440 & 15915024 & -41951232 & 83982912 & -128196864 & 147998752 \\
5& 19448 & -366080 & 3310736 & -19219200 & 80612532 & -260032512 & 667972992 & -1390912768 & 2361125988 \\
6& 50388 & -1018368 & 9992320 & -63734528 & 298296856 & -1094429952 & 3275080128 & -8186561280 & 17317499520 \\
7& 116280 & -2480640 & 25867200 & -176743424 & 894572796 & -3590016768 & 11914381632 & -33590839040 & 81815642556 \\
8& 245157 & -5457408 & 59659392 & -429646848 & 2306412576 & -9889224704 & 35372882496 & -108612850944 & 291735314784 \\\hline
\end{tabular}
}
\end{center}
\caption{table of $b_{jk}$ coefficients for \hh{1}{Sp(4)}.}
\end{table}
\begin{table}[H]
\begin{center}\scalebox{0.702}{
\begin{tabular}{|l||l|l|l|l|l|l|l|l|l|}\hline $k\backslash j$&0&1&2&3&4&5&6&7\\\hline\hline 0&
 1 & 0 & 0 & 0 & 0 & 0 & 0 & 0 \\
1& 55 & -360 & 730 & -360 & 55 & 0 & 0 & 131072  \\
2& 715 & -7920 & 36950 & -91920 & 126426 & -91920 & 36950 & -7920 \\
3& 5005 & -72072 & 472670 & -1855920 & 4807235 & -8507400 & 10318308 & -8507400 \\
4& 24310 & -411840 & 3287570 & -16425552 & 57414335 & -148243920 & 290209230 & -434652960 \\
5& 92378 & -1750320 & 15915900 & -92646840 & 388001900 & -1243035816 & 3154283110 & -6461977200 \\
 6&293930 & -6046560 & 60363940 & -390904800 & 1850251975 & -6828272880 & 20425933880 & -50725384512 \\
7& 817190 & -17907120 & 191829700 & -1344351840 & 6954608778 & -28379585520 & 95172862070 & -269335998360 \\
8& 2042975 & -47070144 & 532878940 & -3969489120 & 21972846970 & -96681303840 & 352734033924 & -1097369959920 \\\hline
\end{tabular}
}
\end{center}
\caption{table of $b_{jk}$ coefficients for \hh{1}{Sp(5)}.}
\end{table}
\begin{table}[H]
\begin{center}\scalebox{0.702}{
\begin{tabular}{|l||l|l|l|l|l|l|l|l|l|}\hline $k\backslash j$&0&1&2&3&4&5&6&7\\\hline\hline 0&
 1 & 0 & 0 & 0 & 0 & 0 & 0 & 0 \\
1& 78 & -480 & 924 & -480 & 78 & 0 & 0 & 0  \\
2& 1365 & -14560 & 65352 & -156960 & 211682 & -156960 & 65352 & -14560 \\
3& 12376 & -174720 & 1121484 & -4303520 & 10894872 & -18917856 & 22752072 & -18917856 \\
4& 75582 & -1272960 & 10087584 & -49926240 & 172447626 & -439115872 & 847529088 & -1255045824 \\
5& 352716 & -6718400 & 61351368 & -357913920 & 1497819674 & -4778545824 & 12035999976 & -24419800768 \\
 6&1352078 & -28217280 & 285599184 & -1871746240 & 8940106812 & -33168338112 & 99330679448 & -245966456736 \\
 7&4457400 & -99845760 & 1093039752 & -7816186560 & 41146639248 & -170223959360 & 576190947696 & -1638255055008 \\
8& 13037895 & -309046400 & 3599439648 & -27550297920 & 156304791772 & -702349866816 & 2605397698176 & -8201957116288 \\\hline
\end{tabular}
}
\end{center}
\caption{table of $b_{jk}$ coefficients for \hh{1}{Sp(6)}.}
\end{table}
\begin{table}[H]
\begin{center}\scalebox{0.702}{
\begin{tabular}{|l||l|l|l|l|l|l|l|l|}\hline $k\backslash j$&0&1&2&3&4&5&6&7\\\hline\hline 0&
 1 & 0 & 0 & 0 & 0 & 0 & 0 & 0 \\
1& 105 & -616 & 1142 & -616 & 105 & 0 & 0 & 0  \\
 2&2380 & -24640 & 107366 & -251328 & 334520 & -251328 & 107366 & -24640 \\
3& 27132 & -376992 & 2379272 & -8972040 & 22337812 & -38280088 & 45787152 & -38280088 \\
4& 203490 & -3410880 & 26872104 & -132030976 & 452040449 & -1139797120 & 2178876280 & -3202899392 \\
5& 1144066 & -21886480 & 200550700 & -1172148912 & 4904012610 & -15604903776 & 39120480630 & -78892132088 \\
6& 5200300 & -109830336 & 1124196332 & -7439811456 & 35803341980 & -133468775616 & 400437497964 & -990651116224 \\
7& 20058300 & -457626400 & 5100038216 & -37076890928 & 198003321740 & -828627305968 & 2828123668328 & -8080816527696 \\
8& 67863915 & -1647455040 & 19644950440 & -153762470400 & 890244245282 & -4070880504768 & 15316806055432 & -48733092166528 \\\hline
\end{tabular}
}
\end{center}
\caption{table of $b_{jk}$ coefficients for \hh{1}{Sp(7)}.}
\end{table}
\begin{table}[H]
\begin{center}\scalebox{0.702}{
\begin{tabular}{|l||l|l|l|l|l|l|l|l|l|}\hline $k\backslash j$&0&1&2&3&4&5&6&7&8\\\hline\hline 0&
 1 & 0 & 0 & 0 & 0 & 0 & 0 & 0 & 0 \\
1& 14 & -98 & 120 & 350 & 910 & -1216 & -11520 & 63168 & -198016 \\
2& 77 & -896 & 4110 & -7840 & -1370 & 17248 & 40974 & -89472 & -335795 \\
3& 273 & -4004 & 26250 & -98630 & 217015 & -208614 & -182788 & 447258 & 970679 \\
4& 748 & -12544 & 98370 & -477120 & 1579725 & -3639776 & 5394622 & -3128800 & -5100524 \\
5& 1729 & -31654 & 276870 & -1544970 & 6175245 & -18676588 & 43438192 & -75584530 & 87019530 \\
6& 3542 & -68992 & 649740 & -3965920 & 17722815 & -61876640 & 174768900 & -404326720 & 757815968 \\
7& 6630 & -135240 & 1343580 & -8731310 & 42038360 & -160673842 & 508042238 & -1360145630 & 3111214236 \\
8& 11571 & -244608 & 2530620 & -17225600 & 87500420 & -356097504 & 1213415216 & -3558427040 & 9123119434 \\\hline
\end{tabular}
}
\end{center}
\caption{table of $b_{jk}$ coefficients for \hh{1}{G_2}.}
\end{table}
\begin{table}[H]
\begin{center}\scalebox{0.702}{
\begin{tabular}{|l||l|l|l|l|l|l|l|l|}\hline $k\backslash j$&0&1&2&3&4&5&6&7\\\hline\hline 0&
 1 & 0 & 0 & 0 & 0 & 0 & 0 & 0  \\
1& 52 & -208 & 89 & 120 & 308 & 792 & 3072 & 1704 \\
 2&1053 & -8424 & 23204 & -18512 & -11350 & -10704 & 6820 & 54936 \\
3& 12376 & -143208 & 689364 & -1685272 & 1866152 & -39608 & -765216 & -1659000 \\
4& 100776 & -1478048 & 9711260 & -36937160 & 86395300 & -116741552 & 59692724 & 41427000 \\
5& 627912 & -10883808 & 87746010 & -432815080 & 1437418628 & -3300129352 & 5065163264 & -4423934120 \\
6& 3187041 & -62581896 & 584202348 & -3435586440 & 14178101676 & -43106676088 & 97983239544 & -163533566872\\
7& 13748020 & -297457160 & 3102350628 & -20743270824 & 99578268180 & -363361928088 & 1037554927552 & -2340760733472 \\
8& 51949755 & -1214825040 & 13822964948 & -101985799864 & 547895844084 & -2278100342760 & 7591910100408 & -20670151476576  \\\hline
\end{tabular}
}
\end{center}
\caption{table of $b_{jk}$ coefficients for \hh{1}{F_4}.}
\end{table}
\begin{table}[H]
\begin{center}\scalebox{0.702}{
\begin{tabular}{|l||l|l|l|l|l|l|l|l|l|}\hline $k\backslash j$&0&1&2&3&4&5&6&7&8\\\hline\hline 0&
 1 & 0 & 0 & 0 & 0 & 0 & 0 & 0 & 0 \\
1& 21 & -124 & 110 & 260 & 789 & 1152 & -8704 & -14464 & 223488 \\
2& 168 & -1764 & 7024 & -10588 & -4116 & 10916 & 50032 & 46364 & -503640 \\
3& 825 & -11412 & 69276 & -233688 & 434927 & -283380 & -389832 & 226572 & 1676015 \\
4& 3003 & -48972 & 369840 & -1700184 & 5202264 & -10629024 & 12838416 & -3438780 & -12574496 \\
5& 8918 & -162344 & 1404746 & -7682004 & 29644377 & -84687744 & 180329328 & -273616588 & 246274245 \\
6& 22848 & -450072 & 4276384 & -26181428 & 116169412 & -396796620 & 1075184112 & -2327385492 & 3947607216 \\
7& 52326 & -1093848 & 11130168 & -73807664 & 359942835 & -1377924132 & 4300502952 & -11166392996 & 24266871171 \\
8& 109725 & -2402664 & 25763232 & -181348656 & 947203248 & -3927734240 & 13475440272 & -39235439724 & 98330686560 \\\hline
\end{tabular}
}
\end{center}
\caption{table of $b_{jk}$ coefficients for \hh{1}{SO(7)}.}
\end{table}
\begin{table}[H]
\begin{center}\scalebox{0.702}{
\begin{tabular}{|l||l|l|l|l|l|l|l|l|l|}\hline $k\backslash j$&0&1&2&3&4&5&6&7&8\\\hline\hline 0&
 1 & 0 & 0 & 0 & 0 & 0 & 0 & 0 & 0 \\
1& 28 & -144 & 92 & 216 & 588 & 1848 & -2464 & -29176 & 56976 \\
2& 300 & -2880 & 10094 & -11904 & -8104 & 4416 & 41454 & 144384 & -260564 \\
3& 1925 & -25200 & 142134 & -431280 & 670691 & -218856 & -657052 & -299112 & 1736931 \\
4& 8918 & -141120 & 1023484 & -4444992 & 12511959 & -22394112 & 20688612 & 3535296 & -21767636 \\
5& 32928 & -592704 & 5039944 & -26818680 & 99212344 & -265762776 & 512485572 & -658521264 & 396776184 \\
6& 102816 & -2032128 & 19302276 & -117375552 & 512196412 & -1696845696 & 4376235864 & -8782492416 & 13252883352 \\
 7&282150 & -5987520 & 61719588 & -412790112 & 2015818938 & -7649920824 & 23361506350 & -58392693816 & 119647476102 \\
8& 698775 & -15681600 & 172164168 & -1236951360 & 6558107850 & -27385792896 & 93662349144 & -268572416448 & 653514465071 \\\hline
\end{tabular}
}
\end{center}
\caption{table of $b_{jk}$ coefficients for \hh{1}{SO(8)}.}
\end{table}
\begin{table}[H]
\begin{center}\scalebox{0.702}{
\begin{tabular}{|l||l|l|l|l|l|l|l|l|}\hline $k\backslash j$&0&1&2&3&4&5&6&7\\\hline\hline 0&
 1 & 0 & 0 & 0 & 0 & 0 & 0 & 0 \\
1& 36 & -168 & 94 & 168 & 468 & 1456 & 2016 & -16560  \\
2& 495 & -4440 & 14244 & -14712 & -9066 & -1848 & 28452 & 134504 \\
3& 4004 & -50160 & 267438 & -752280 & 1044204 & -232104 & -770540 & -778344 \\
4& 22932 & -353808 & 2482496 & -10310520 & 27272082 & -44523144 & 34688784 & 11092752 \\
5& 102816 & -1831200 & 15328908 & -79686672 & 284855316 & -726010296 & 1301494452 & -1488757776  \\
6& 383724 & -7593696 & 71975304 & -434396496 & 1866929904 & -6028515168 & 14949532076 & -28319747544 \\
7& 1241460 & -26634960 & 276957516 & -1861197264 & 9079256232 & -34150691824 & 102358483944 & -248103044904 \\
8& 3581721 & -81921840 & 915368256 & -6673637520 & 35737932576 & -149790689136 & 510200930592 & -1443615699152 \\\hline
\end{tabular}
}
\end{center}
\caption{table of $b_{jk}$ coefficients for \hh{1}{SO(9)}.}
\end{table}
\begin{table}[H]
\begin{center}\scalebox{0.702}{
\begin{tabular}{|l||l|l|l|l|l|l|l|l|l|}\hline $k\backslash j$&0&1&2&3&4&5&6&7\\\hline\hline 0&
 1 & 0 & 0 & 0 & 0 & 0 & 0 & 0 \\
1& 45 & -196 & 110 & 120 & 391 & 1152 & 2124 & -792 \\
2& 770 & -6560 & 19842 & -19680 & -7110 & -7872 & 17290 & 115648 \\
3& 7644 & -92484 & 472780 & -1264044 & 1660818 & -441172 & -694716 & -1197432 \\
4& 52920 & -799680 & 5465616 & -21949344 & 55603779 & -85982720 & 63374700 & 13067584 \\
5& 282744 & -4989600 & 41220300 & -210290508 & 732226820 & -1800471036 & 3076437870 & -3303090628 \\
6& 1241460 & -24577344 & 232368444 & -1392936192 & 5911557018 & -18710825952 & 45065511608 & -81985256736 \\
 7&4671810 & -101080980 & 1057754016 & -7129660956 & 34722608970 & -129616856208 & 382781764860 & -906421301260 \\
8& 15520791 & -360456096 & 4083433068 & -30103638048 & 162385463460 & -682190154720 & 2315054813436 & -6481392386112  \\\hline
\end{tabular}
}
\end{center}
\caption{table of $b_{jk}$ coefficients for \hh{1}{SO(10)}.}
\end{table}
\begin{table}[H]
\begin{center}\scalebox{0.652}{
\begin{tabular}{|l||l|l|l|l|l|l|l|l|}\hline $k\backslash j$&0&1&2&3&4&5&6&7\\\hline\hline 0&
 1 & 0 & 0 & 0 & 0 & 0 & 0 & 0 \\
1& 66 & -264 & 175 & 0 & 459 & 160 & 3030 & 1704 \\
2& 1638 & -13024 & 37125 & -39480 & 10059 & -35688 & 49689 & -37040 \\
3& 23100 & -266112 & 1290081 & -3286304 & 4346958 & -2380224 & 1097942 & -3954696 \\
4& 222156 & -3256704 & 21474365 & -82748424 & 200841537 & -303068832 & 251722053 & -82582152  \\
5& 1613898 & -28087488 & 227685777 & -1132610688 & 3818763138 & -9040729320 & 14897234001 & -15981701160 \\
6& 9447438 & -187098912 & 1762140237 & -10462560680 & 43669190334 & -134850391440 & 314361285071 & -550479506568 \\
7& 46562373 & -1020323304 & 10776662897 & -72955273248 & 354577272357 & -1310825156872 & 3800990387739 & -8759070969360 \\
 8&199377750 & -4740241792 & 54822956331 & -410877059736 & 2240474903238 & -9448968696288 & 31934354072709 & -88265081159848 \\\hline
\end{tabular}
}
\end{center}
\caption{table of $b_{jk}$ coefficients for \hh{1}{SO(12)_a}.}
\end{table}
\begin{table}[H]
\begin{center}\scalebox{0.702}{
\begin{tabular}{|l||l|l|l|l|l|l|l|l|l|}\hline $k\backslash j$&0&1&2&3&4&5&6&7&8\\\hline\hline 0&
 1 & 0 & -7 & 16 & -24 & 32 & -40 & 48 & -56 \\
1& 3 & -16 & 25 & 0 & -71 & 176 & -285 & 384 & -480 \\
2& 5 & -32 & 96 & -176 & 189 & 0 & -467 & 1136 & -1872 \\
3& 7 & -48 & 160 & -384 & 756 & -1136 & 1069 & 0 & -2323 \\
4& 9 & -64 & 224 & -576 & 1300 & -2592 & 4323 & -5648 & 4859 \\
5& 11 & -80 & 288 & -768 & 1820 & -3936 & 7640 & -13184 & 19674 \\
6& 13 & -96 & 352 & -960 & 2340 & -5248 & 10752 & -20304 & 35430 \\
7& 15 & -112 & 416 & -1152 & 2860 & -6560 & 13824 & -27136 & 50218 \\
8& 17 & -128 & 480 & -1344 & 3380 & -7872 & 16896 & -33920 & 64638 \\\hline
\end{tabular}
}
\end{center}
\caption{table of $b_{jk}$ coefficients for \hh{2}{SU(2)}.}
\end{table}
\begin{table}[H]
\begin{center}\scalebox{0.702}{
\begin{tabular}{|l||l|l|l|l|l|l|l|l|l|}\hline $k\backslash j$&0&1&2&3&4&5&6&7&8\\\hline\hline 0&
 1 & 0 & -1 & -20 & 90 & -232 & 470 & -828 & 1330 \\
1& 8 & -36 & 44 & 20 & -44 & -340 & 1664 & -4448 & 9160 \\
2& 27 & -180 & 532 & -864 & 629 & 376 & -719 & -3400 & 17416 \\
3& 64 & -504 & 1908 & -4720 & 8344 & -10024 & 5732 & 4264 & -7404 \\
4& 125 & -1080 & 4616 & -13496 & 30681 & -56120 & 80498 & -82280 & 39853 \\
5& 216 & -1980 & 9100 & -29160 & 74816 & -162148 & 299772 & -466628 & 587968 \\
6& 343 & -3276 & 15804 & -53680 & 148015 & -351756 & 735192 & -1357888 & 2204377 \\
7& 512 & -5040 & 25172 & -89024 & 257544 & -648824 & 1458180 & -2955744 & 5429088 \\
8& 729 & -7344 & 37648 & -137160 & 410669 & -1077232 & 2540628 & -5461016 & 10790865 \\\hline
\end{tabular}
}
\end{center}
\caption{table of $b_{jk}$ coefficients for \hh{2}{SU(3)}.}
\end{table}
\begin{table}[H]
\begin{center}\scalebox{0.702}{
\begin{tabular}{|l||l|l|l|l|l|l|l|l|l|}\hline $k\backslash j$&0&1&2&3&4&5&6&7&8\\\hline\hline 0&
 1 & 0 & -1 & 0 & -70 & 448 & -1610 & 4352 & -9870 \\
1& 15 & -64 & 79 & 0 & -9 & 64 & -2297 & 13312 & -46520 \\
2& 84 & -576 & 1725 & -2816 & 2343 & -448 & -61 & 2816 & -42223 \\
3& 300 & -2560 & 10381 & -26624 & 47165 & -57216 & 42406 & -13312 & -298 \\
4& 825 & -8000 & 38091 & -120320 & 283087 & -517504 & 733480 & -773888 & 551586 \\
5& 1911 & -20160 & 106139 & -378880 & 1038969 & -2316544 & 4288284 & -6584320 & 8226740 \\
6& 3920 & -43904 & 247689 & -958720 & 2896091 & -7268480 & 15604884 & -28983296 & 46563419 \\
7& 7344 & -86016 & 510425 & -2093056 & 6760273 & -18359936 & 43330056 & -90327040 & 167511165 \\
8& 12825 & -155520 & 959191 & -4107264 & 13933995 & -40043136 & 100932416 & -227385344 & 462669790 \\\hline
\end{tabular}
}
\end{center}
\caption{table of $b_{jk}$ coefficients for \hh{2}{SU(4)}.}
\end{table}
\begin{table}[H]
\begin{center}\scalebox{0.702}{
\begin{tabular}{|l||l|l|l|l|l|l|l|l|l|}\hline $k\backslash j$&0&1&2&3&4&5&6&7&8\\\hline\hline 0&
 1 & 0 & -1 & 0 & 0 & -252 & 2100 & -9648 & 32550 \\
1& 24 & -100 & 124 & 0 & -124 & 352 & -24 & -12648 & 91200 \\
2& 200 & -1400 & 4251 & -7000 & 5849 & 0 & -7949 & 19648 & -4251 \\
3& 1000 & -9000 & 38176 & -100800 & 180876 & -220000 & 156700 & 9648 & -247900 \\
4& 3675 & -38500 & 196625 & -655200 & 1592224 & -2951200 & 4191075 & -4382800 & 2833250 \\
5& 10976 & -127400 & 733524 & -2818200 & 8140100 & -18716904 & 35134624 & -54080096 & 67220100 \\
6& 28224 & -352800 & 2209949 & -9354800 & 30258051 & -79590896 & 175861625 & -331388496 & 534403300 \\
 7&64800 & -856800 & 5714176 & -25956000 & 90975724 & -262542800 & 646529076 & -1385477104 & 2608959500 \\
8& 136125 & -1881000 & 13168251 & -63117600 & 234934576 & -725674500 & 1931317325 & -4526691104 & 9463184100 \\\hline
\end{tabular}
}
\end{center}
\caption{table of $b_{jk}$ coefficients for \hh{2}{SU(5)}.}
\end{table}
\begin{table}[H]
\begin{center}\scalebox{0.702}{
\begin{tabular}{|l||l|l|l|l|l|l|l|l|l|}\hline $k\backslash j$&0&1&2&3&4&5&6&7&8\\\hline\hline 0&
 1 & 0 & -1 & 0 & 0 & 0 & -924 & 9504 & -53130 \\
1& 35 & -144 & 179 & 0 & -179 & 144 & 889 & 0 & -66198 \\
2& 405 & -2880 & 8848 & -14688 & 12329 & 0 & -12329 & 5184 & 57350 \\
3& 2695 & -25200 & 110400 & -298240 & 542332 & -663824 & 473955 & 0 & -420825 \\
4& 12740 & -141120 & 757625 & -2626400 & 6558060 & -12349568 & 17672137 & -18531808 & 12124497 \\
5& 47628 & -592704 & 3639181 & -14757120 & 44403755 & -104978608 & 200384283 & -311062016 & 387847460 \\
6& 149940 & -2032128 & 13734112 & -62102208 & 211781479 & -579406720 & 1315368329 & -2521066912 & 4103016876 \\
7& 413820 & -5987520 & 43500960 & -213190656 & 795881912 & -2413262544 & 6166189029 & -13563009536 & 25980535867 \\
8& 1029105 & -15681600 & 120559275 & -628689600 & 2513967480 & -8230614528 & 22928202815 & -55635665568 & 119281776315 \\\hline
\end{tabular}
}
\end{center}
\caption{table of $b_{jk}$ coefficients for \hh{2}{SU(6)}.}
\end{table}
\begin{table}[H]
\begin{center}\scalebox{0.702}{
\begin{tabular}{|l|l|l|l|l|l|l|l|l|}\hline $k\backslash j$&0&1&2&3&4&5&6&7\\\hline\hline 0&
 1 & 0 & -1 & 0 & 0 & 0 & 0 & -3432 \\
1& 48 & -196 & 244 & 0 & -244 & 196 & -48 & 3432 \\
2& 735 & -5292 & 16416 & -27440 & 23127 & 0 & -23127 & 27440 \\
3& 6272 & -60368 & 271020 & -745584 & 1371952 & -1690696 & 1211084 & 0 \\
4& 37044 & -428064 & 2386349 & -8529136 & 21779225 & -41624520 & 60092816 & -63314272 \\
5& 169344 & -2222640 & 14329168 & -60571056 & 188335812 & -456080240 & 884841460 & -1387541232 \\
6& 640332 & -9236304 & 66157056 & -314791680 & 1119685084 & -3165617364 & 7362858187 & -14351595104 \\
7& 2090880 & -32443488 & 251776944 & -1308916224 & 5137991040 & -16228172800 & 42806145984 & -96426865392 \\
8& 6073353 & -99891792 & 825805035 & -4599601776 & 19474888644 & -66881396736 & 193668842592 & -484485941600 \\\hline
\end{tabular}

}
\end{center}
\caption{table of $b_{jk}$ coefficients for \hh{2}{SU(7)}.}
\end{table}
\begin{table}[H]
\begin{center}\scalebox{0.702}{
\begin{tabular}{|l||l|l|l|l|l|l|l|l|}\hline $k\backslash j$&0&1&2&3&4&5&6&7\\\hline\hline 0&
 1 & 0 & -1 & 0 & 0 & 0 & 0 & 0 \\
1& 63 & -256 & 319 & 0 & -319 & 256 & -63 & 0 \\
2& 1232 & -8960 & 28017 & -47104 & 39839 & 0 & -39839 & 47104 \\
3& 13104 & -129024 & 590513 & -1648640 & 3064689 & -3800576 & 2731504 & 0 \\
4& 94500 & -1128960 & 6482476 & -23740416 & 61757311 & -119587840 & 174133728 & -184414208 \\
5& 518364 & -7096320 & 47542572 & -207728640 & 663462436 & -1639975680 & 3229234547 & -5115265024 \\
6& 2317392 & -35126784 & 263461716 & -1305722880 & 4806514188 & -13969951744 & 33191990195 & -65722116096 \\
7& 8833968 & -144967680 & 1185759828 & -6463328256 & 26431333236 & -86379245568 & 234193247568 & -538964647936 \\
8& 29630601 & -518259456 & 4541329287 & -26674053120 & 118344386556 & -422961896448 & 1266010130448 & -3253134753792 \\\hline
\end{tabular}
}
\end{center}
\caption{table of $b_{jk}$ coefficients for \hh{2}{SU(8)}.}
\end{table}
\begin{table}[H]
\begin{center}\scalebox{0.702}{
\begin{tabular}{|l||l|l|l|l|l|l|l|l|l|}\hline $k\backslash j$&0&1&2&3&4&5&6&7&8\\\hline\hline 0&
 1 & 0 & 0 & -8 & -35 & 336 & -1317 & 3688 & -8519 \\
1& 14 & -56 & 51 & 48 & 7 & -344 & -630 & 8512 & -35036 \\
 2&77 & -512 & 1443 & -2016 & 753 & 1552 & 63 & -7024 & -5171 \\
3& 273 & -2288 & 9030 & -22128 & 35799 & -34272 & 5900 & 26176 & 420 \\
4& 748 & -7168 & 33606 & -103872 & 236250 & -406976 & 511339 & -393624 & 17393 \\
5& 1729 & -18088 & 94281 & -332304 & 895446 & -1945024 & 3452071 & -4921600 & 5296513 \\
 6&3542 & -39424 & 220857 & -847776 & 2533905 & -6267968 & 13179345 & -23719936 & 36229900 \\
7& 6630 & -77280 & 456204 & -1859760 & 5964231 & -16050592 & 37418381 & -76689152 & 138787518 \\
8& 11571 & -139776 & 858636 & -3660672 & 12356022 & -35286720 & 88234055 & -196702208 & 394629950 \\\hline
\end{tabular}
}
\end{center}
\caption{table of $b_{jk}$ coefficients for \hh{2}{G_2}.}
\end{table}
\begin{table}[H]
\begin{center}\scalebox{0.702}{
\begin{tabular}{|l||l|l|l|l|l|l|l|l|}\hline $k\backslash j$&0&1&2&3&4&5&6&7\\\hline\hline 0&
 1 & 0 & 0 & 0 & 0 & -14 & -70 & -216  \\
 1&52 & -156 & 74 & 56 & 90 & 126 & 448 & -1348 \\
2& 1053 & -6318 & 14117 & -11544 & -2065 & 616 & 6660 & 12996 \\
 3&12376 & -107406 & 406432 & -838148 & 908054 & -278910 & -260850 & -162008 \\
4& 100776 & -1108536 & 5658107 & -17462484 & 35136029 & -45104518 & 30486813 & 513540 \\
5& 627912 & -8162856 & 50812932 & -200047368 & 550927748 & -1099670598 & 1573235888 & -1487730508 \\
6& 3187041 & -46936422 & 337074079 & -1568026824 & 5274548964 & -13507533022 & 26869094301 & -41293163340 \\
7& 13748020 & -223092870 & 1785727008 & -9392648100 & 36394268948 & -110078146326 & 267922966188 & -531517036124 \\
8& 51949755 & -911118780 & 7943335361 & -45932548116 & 197938648545 & -675704802882 & 1892326957877 & -4433982679452 \\\hline
\end{tabular}
}
\end{center}
\caption{table of $b_{jk}$ coefficients for \hh{2}{F_4}.}
\end{table}
\begin{table}[H]
\begin{center}\scalebox{0.702}{
\begin{tabular}{|l||l|l|l|l|l|l|l|l|l|}\hline $k\backslash j$&0&1&2&3&4&5&6&7&8\\\hline\hline 0&
 1 & 0 & 0 & -2 & -27 & -20 & 1027 & -5958 & 22057 \\
1& 21 & -78 & 61 & 54 & 49 & -118 & -1263 & 368 & 34148 \\
 2&168 & -1106 & 2994 & -3770 & 849 & 2376 & 1799 & -3518 & -29724 \\
3& 825 & -7146 & 28646 & -69034 & 104604 & -85566 & -451 & 54720 & 33333 \\
4& 3003 & -30646 & 151258 & -480798 & 1088896 & -1798160 & 2061990 & -1304906 & -193235 \\
5& 8918 & -101556 & 571431 & -2132702 & 5925554 & -12891132 & 22247190 & -29845102 & 28859157 \\
 6&22848 & -281484 & 1734236 & -7198126 & 22716939 & -57851398 & 122249718 & -216042714 & 316189540 \\
7& 52326 & -684012 & 4505004 & -20174310 & 69522117 & -196427868 & 470616039 & -972117806 & 1740019295 \\
8& 109725 & -1502292 & 10414332 & -49382314 & 181545392 & -552475098 & 1443393448 & -3304130206 & 6697776060 \\\hline
\end{tabular}
}
\end{center}
\caption{table of $b_{jk}$ coefficients for \hh{2}{SO(7)}.}
\end{table}
\begin{table}[H]
\begin{center}\scalebox{0.702}{
\begin{tabular}{|l||l|l|l|l|l|l|l|l|l|}\hline $k\backslash j$&0&1&2&3&4&5&6&7&8\\\hline\hline 0&
 1 & 0 & 0 & 0 & -15 & -64 & 287 & 1728 & -19215 \\
1& 28 & -96 & 59 & 64 & 75 & 96 & -1220 & -2816 & 16768 \\
 2&300 & -1920 & 4865 & -5184 & -196 & 2816 & 5190 & 3456 & -45241 \\
3& 1925 & -16800 & 66525 & -152960 & 207284 & -123744 & -55740 & 58496 & 173940 \\
4& 8918 & -94080 & 473935 & -1505344 & 3304770 & -5059840 & 4944037 & -1892160 & -1639248 \\
5& 32928 & -395136 & 2321410 & -8904000 & 24887830 & -53072672 & 86903151 & -105405184 & 83317010 \\
6& 102816 & -1354752 & 8863302 & -38555776 & 125315359 & -322157568 & 672406592 & -1144825280 & 1563120876 \\
7& 282150 & -3991680 & 28285110 & -134717184 & 486175437 & -1413958112 & 3425954142 & -7027986048 & 12249028858 \\
8& 698775 & -10454400 & 78795090 & -401969280 & 1567387743 & -4980406272 & 13375882455 & -31002012608 & 62669671935 \\\hline
\end{tabular}
}
\end{center}
\caption{table of $b_{jk}$ coefficients for \hh{2}{SO(8)}.}
\end{table}
\begin{table}[H]
\begin{center}\scalebox{0.652}{
\begin{tabular}{|l||l|l|l|l|l|l|l|l|l|}\hline $k\backslash j$&0&1&2&3&4&5&6&7&8\\\hline\hline 0&
 1 & 0 & 0 & 0 & -5 & -44 & -95 & 1456 & 429 \\
1& 36 & -118 & 68 & 60 & 84 & 134 & -340 & -3608 & -4280 \\
2& 495 & -3114 & 7606 & -7520 & -508 & 2680 & 6096 & 8776 & -16074 \\
3& 4004 & -35156 & 138200 & -308904 & 393184 & -199732 & -114164 & 56560 & 223684 \\
4& 22932 & -247884 & 1267072 & -4024108 & 8660300 & -12647208 & 11215914 & -2977932 & -4261158 \\
5& 102816 & -1282680 & 7774740 & -30412512 & 85418344 & -179833016 & 284507040 & -323208660 & 223496044 \\
6& 383724 & -5318280 & 36369804 & -163762680 & 544151025 & -1410138574 & 2921674928 & -4853150808 & 6325588310 \\
7& 1241460 & -18651996 & 139612836 & -696342168 & 2602763528 & -7744363160 & 18952660776 & -38752829440 & 66373180536 \\
8& 3581721 & -57364164 & 460669440 & -2484244224 & 10136320143 & -33326857120 & 91557798580 & -214613330920 & 433765167251 \\\hline
\end{tabular}

}
\end{center}
\caption{table of $b_{jk}$ coefficients for \hh{2}{SO(9)}.}
\end{table}
\begin{table}[H]
\begin{center}\scalebox{0.702}{
\begin{tabular}{|l||l|l|l|l|l|l|l|l|}\hline $k\backslash j$&0&1&2&3&4&5&6&7\\\hline\hline 0&
 1 & 0 & 0 & 0 & -3 & -16 & -103 & -48 \\
1& 45 & -144 & 86 & 48 & 88 & 160 & -243 & -736 \\
2& 770 & -4800 & 11538 & -11312 & 351 & 1968 & 6427 & 12304 \\
3& 7644 & -67536 & 264946 & -584752 & 727803 & -369776 & -146259 & 39264 \\
4& 52920 & -583296 & 3018522 & -9610752 & 20491020 & -29251696 & 24994161 & -6373856 \\
5& 282744 & -3636864 & 22590900 & -89797440 & 253668779 & -530960352 & 824437119 & -904981216 \\
6& 1241460 & -17905536 & 126761316 & -586487136 & 1984425114 & -5184363184 & 10713955577 & -17548934352 \\
7& 4671810 & -73616400 & 575268738 & -2975442624 & 11439020010 & -34693369952 & 85735276641 & -175328820640 \\
8& 15520791 & -262451904 & 2216098170 & -12488323584 & 52844569869 & -178688324304 & 500546068987 & -1186103984576 \\\hline
\end{tabular}
}
\end{center}
\caption{table of $b_{jk}$ coefficients for \hh{2}{SO(10)}.}
\end{table}
\begin{table}[H]
\begin{center}\scalebox{0.652}{
\begin{tabular}{|l|l|l|l|l|l|l|l|l|}\hline $k\backslash j$&0&1&2&3&4&5&6&7\\\hline\hline 0&
 1 & 0 & 0 & 0 & -2 & 0 & -65 & 144 \\
1& 66 & -208 & 146 & 0 & 132 & 64 & 71 & 0 \\
2& 1638 & -10176 & 24541 & -26000 & 8744 & -4992 & 13695 & 4832 \\
3& 23100 & -206976 & 818012 & -1814784 & 2323365 & -1486288 & 262759 & -371200 \\
4& 222156 & -2525952 & 13369697 & -43113840 & 92344902 & -132562176 & 118728676 & -49678624 \\
5& 1613898 & -21745152 & 140310401 & -573281280 & 1645528555 & -3461643088 & 5361177380 & -5889533952 \\
6& 9447438 & -144665664 & 1078909832 & -5205925296 & 18159108739 & -48318072064 & 100506127656 & -163975921808 \\
7& 46562373 & -788186256 & 6569279717 & -35898720000 & 144212523466 & -451656029968 & 1138839165448 & -2348714494976 \\
8& 199377750 & -3659225856 & 33314037614 & -200620394832 & 897738323185 & -3173937344512 & 9188877126670 & -22247675867120 \\\hline
\end{tabular}
}
\end{center}
\caption{table of $b_{jk}$ coefficients for \hh{2}{SO(12)_a}.}
\end{table}
\begin{table}[H]
\begin{center}\scalebox{0.702}{
\begin{tabular}{|l|l|l|l|l|l|l|l|l|}\hline $k\backslash j$&0&1&2&3&4&5&6&7\\\hline\hline 0&
 1 & 0 & 0 & 0 & 0 & -24 & 89 & -560 \\
1& 66 & -208 & 144&24&0&480&-519&-1632 \\
2& 1638 & -10176 &24519&-25712&6892&2800&-10023&57408 \\
3& 23100 & -206976 & 817882&-1812960&2310583&-1426224&49569&233744\\
4& 222156 & -2525952 & 13369147&-43105680&92284010&-132254664&117539906&-45924864\\
5& 1613898 & -21745152 & 140308531&-573252240&1645300811&-3460428560&5356188890&-5872654176\\
6& 9447438 & -144665664 & 1078904398&-5205837648&18158392835&-48314083616&100488952236&-163914717136\\
7& 46562373 & -788186256 & 6569265703&-35898486624&144210551342&-451644637904&1138788176396&-2348525112416\\
8& 199377750 & -3659225856 & 33314004724&-200619831984&897733427591&3173908187384&9188742332120&-22247157728176\\\hline
\end{tabular}
}
\end{center}
\caption{table of $b_{jk}$ coefficients for \hh{2}{SO(12)_b}.}
\end{table}
\begin{table}[H]
\begin{center}\scalebox{0.652}{
\begin{tabular}{|l||l|l|l|l|l|l|l|l|}\hline $k\backslash j$&0&1&2&3&4&5&6&7\\\hline\hline 0&
 1 & 0 & 0 & 0 & 0 & -14 & 13 & 0  \\
 1&78 & -246 & 184 & 0 & 0 & 414 & -78 & -2016 \\
2& 2275 & -14170 & 34491 & -37984 & 14824 & 0 & -12545 & 55428 \\
3& 37400 & -337650 & 1343860 & -3011060 & 3949818 & -2740344 & 625970 & 147420 \\
4& 415701 & -4789070 & 25623440 & -83363256 & 180151060 & -262494856 & 244659096 & -119438520  \\
5& 3461458 & -47509462 & 311410944 & -1288483040 & 3734402980 & -7918507302 & 12377186244 & -13852158216  \\
6& 23046023 & -361266906 & 2750438847 & -13502044556 & 47742957054 & -128332589450 & 268924830830 & -441474787956 \\
7& 128271000 & -2232770540 & 19083881388 & -106586370072 & 435987698822 & -1385009164420 & 3529169247240 & -7331469476188  \\
8& 616258500 & -11677308500 & 109474843920 & -676675858040 & 3096495676326 & -11151643938096 & 32758441385150 & -80174635432992 \\\hline
\end{tabular}
}
\end{center}
\caption{table of $b_{jk}$ coefficients for \hh{2}{SO(13)}.}
\end{table}
\begin{table}[H]
\begin{center}\scalebox{0.702}{
\begin{tabular}{|l||l|l|l|l|l|l|l|l|l|}\hline $k\backslash j$&0&1&2&3&4&5&6&7&8\\\hline\hline 0&
 0 & 0 & 1 & 0 & 8 & 0 & 27 & 0 & 64 \\
1& 1 & 0 & 0 & 0 & 9 & 0 & 64 & 0 & 216 \\
2& 8 & 0 & 9 & 0 & 0 & 0 & 53 & 0 & 360 \\
 3&27 & 0 & 64 & 0 & 53 & 0 & 0 & 0 & 245 \\
4& 64 & 0 & 216 & 0 & 360 & 0 & 245 & 0 & 0 \\
5& 125 & 0 & 512 & 0 & 1188 & 0 & 1600 & 0 & 971 \\
6& 216 & 0 & 1000 & 0 & 2816 & 0 & 5211 & 0 & 6168 \\
7& 343 & 0 & 1728 & 0 & 5500 & 0 & 12288 & 0 & 19818 \\
8& 512 & 0 & 2744 & 0 & 9504 & 0 & 24000 & 0 & 46528 \\\hline
\end{tabular}
}
\end{center}
\caption{table of $b_{jk}$ coefficients for \hh{3}{SU(3)}.}
\end{table}
\begin{table}[H]
\begin{center}\scalebox{0.702}{
\begin{tabular}{|l||l|l|l|l|l|l|l|l|l|}\hline $k\backslash j$&0&1&2&3&4&5&6&7&8\\\hline\hline 0&
 0 & 0 & 0 & 0 & 5 & -32 & 140 & -448 & 1218 \\
1& 1 & 0 & 0 & 0 & -1 & 0 & 132 & -864 & 3610 \\
2& 21 & -32 & 32 & 0 & -5 & 0 & -32 & 0 & 1985 \\
3& 168 & -448 & 820 & -864 & 560 & 0 & -132 & 64 & -560 \\
4& 825 & -2880 & 6936 & -11424 & 14536 & -12800 & 6900 & 0 & -2006 \\
5& 3003 & -12320 & 34884 & -71712 & 120340 & -161056 & 173040 & -136800 & 66927 \\
6& 8918 & -40768 & 128576 & -303072 & 597548 & -986016 & 1391040 & -1644160 & 1595550 \\
7& 22848 & -112896 & 384664 & -995904 & 2185760 & -4110656 & 6791808 & -9824256 & 12442264 \\
8& 52326 & -274176 & 990192 & -2745792 & 6507886 & -13398112 & 24596472 & -40391648 & 59697986 \\\hline
\end{tabular}
}
\end{center}
\caption{table of $b_{jk}$ coefficients for \hh{3}{SO(7)}.}
\end{table}
\begin{table}[H]
\begin{center}\scalebox{0.702}{
\begin{tabular}{|l||l|l|l|l|l|l|l|l|l|}\hline $k\backslash j$&0&1&2&3&4&5&6&7&8\\\hline\hline 0&
 0 & 0 & 0 & 0 & 1 & 8 & -96 & 560 & -2270 \\
1& 1 & 0 & 0 & 0 & -2 & 0 & 38 & 256 & -3369 \\
2& 28 & -48 & 38 & 8 & 0 & -16 & -76 & 48 & 708 \\
3& 300 & -960 & 1736 & -1728 & 764 & 256 & 96 & -512 & -1500 \\
4& 1925 & -8400 & 21480 & -36688 & 44240 & -34104 & 11464 & 3720 & 3369 \\
5& 8918 & -47040 & 147070 & -325760 & 554910 & -730688 & 724192 & -482880 & 145530 \\
6& 32928 & -197568 & 705796 & -1834728 & 3795484 & -6422728 & 8965812 & -10156784 & 8939168 \\
7& 102816 & -677376 & 2662464 & -7727552 & 18169840 & -35873856 & 60568904 & -87710080 & 108129264 \\
8& 282150 & -1995840 & 8430912 & -26536608 & 68393217 & -150204712 & 287427296 & -483941880 & 718911028 \\\hline
\end{tabular}
}
\end{center}
\caption{table of $b_{jk}$ coefficients for \hh{3}{SO(8)}.}
\end{table}
\begin{table}[H]
\begin{center}\scalebox{0.702}{
\begin{tabular}{|l||l|l|l|l|l|l|l|l|l|}\hline $k\backslash j$&0&1&2&3&4&5&6&7&8\\\hline\hline 0&
 0 & 0 & 0 & 0 & 0 & 4 & 6 & -268 & 2116 \\
1& 1 & 0 & 0 & 0 & -1 & -4 & 0 & 212 & 170 \\
2& 36 & -68 & 50 & 12 & 0 & -12 & -56 & -144 & -36 \\
3& 495 & -1788 & 3296 & -3220 & 1155 & 680 & 0 & -412 & -1325 \\
4& 4004 & -20152 & 54730 & -96324 & 114160 & -81396 & 17166 & 19384 & -2116 \\
5& 22932 & -141960 & 483808 & -1133924 & 1977545 & -2594628 & 2464000 & -1453892 & 190621 \\
6& 102816 & -734160 & 2912364 & -8157240 & 17667168 & -30634032 & 42847730 & -47462412 & 39305640 \\
7& 383724 & -3042864 & 13468416 & -42692664 & 106715805 & -219628604 & 379624320 & -553494568 & 674483511 \\
8& 1241460 & -10669032 & 51315660 & -178281720 & 494121760 & -1145255280 & 2275563774 & -3922608352 & 5884380680 \\\hline
\end{tabular}
}
\end{center}
\caption{table of $b_{jk}$ coefficients for \hh{3}{SO(9)}.}
\end{table}
\begin{table}[H]
\begin{center}\scalebox{0.702}{
\begin{tabular}{|l||l|l|l|l|l|l|l|l|l|}\hline $k\backslash j$&0&1&2&3&4&5&6&7&8\\\hline\hline 0&
 0 & 0 & 0 & 0 & 0 & 0 & 14 & -32 & -674 \\
1& 1 & 0 & 0 & 0 & -1 & 0 & -14 & 0 & 1000 \\
2& 45 & -92 & 68 & 12 & 0 & -12 & -68 & 124 & -1045 \\
3& 770 & -3040 & 5784 & -5728 & 2010 & 1152 & 0 & -1152 & -1336 \\
4& 7644 & -42588 & 122056 & -221480 & 263676 & -183280 & 30196 & 50668 & 0 \\
5& 52920 & -366912 & 1342068 & -3298080 & 5902644 & -7798880 & 7284468 & -4017248 & 116592 \\
6& 282744 & -2284128 & 9854040 & -29328852 & 66178359 & -117613440 & 165998096 & -182425360 & 146015652 \\
7& 1241460 & -11233728 & 54650376 & -186062592 & 490131700 & -1046847936 & 1852644768 & -2731167520 & 3321256884 \\
8& 4671810 & -46151820 & 246102516 & -926394084 & 2730649548 & -6631573704 & 13633247000 & -24046382136 & 36526686612 \\\hline
\end{tabular}
}
\end{center}
\caption{table of $b_{jk}$ coefficients for \hh{3}{SO(10)}.}
\end{table}
\begin{table}[H]
\begin{center}\scalebox{0.652}{
\begin{tabular}{|l||l|l|l|l|l|l|l|l|l|}\hline $k\backslash j$&0&1&2&3&4&5&6&7&8\\\hline\hline 0&
 0 & 0 & 0 & 0 & 0 & 0 & 0 & 48 & -255 \\
1& 1 & 0 & 0 & 0 & -1 & 0 & 0 & -48 & 0 \\
2& 55 & -120 & 92 & 8 & 0 & -8 & -92 & 120 & 200 \\
3& 1144 & -4840 & 9540 & -9752 & 3744 & 1488 & 0 & -1488 & -3744 \\
4& 13650 & -82160 & 246848 & -462088 & 560130 & -394984 & 71728 & 100144 & 0 \\
5& 112200 & -847440 & 3286136 & -8415368 & 15467387 & -20735176 & 19424424 & -10568584 & 56403 \\
6& 703494 & -6238320 & 28806624 & -90213120 & 211137472 & -384543328 & 550176564 & -606292456 & 479544108 \\
7& 3586440 & -35864400 & 188273448 & -680093568 & 1874297920 & -4139175384 & 7496610080 & -11204712064 & 13686616296 \\
8& 15520791 & -170450280 & 987592320 & -3972386088 & 12339695160 & -31218639928 & 66197493824 & -119379939624 & 183922406640 \\\hline
\end{tabular}
}
\end{center}
\caption{table of $b_{jk}$ coefficients for \hh{3}{SO(11)}.}
\end{table}
\begin{table}[H]
\begin{center}\scalebox{0.652}{
\begin{tabular}{|l||l|l|l|l|l|l|l|l|l|}\hline $k\backslash j$&0&1&2&3&4&5&6&7&8\\\hline\hline 0&
 0 & 0 & 0 & 0 & 0 & 0 & 0 & 0 & 165 \\
1& 1 & 0 & 0 & 0 & -1 & 0 & 0 & 0 & -165 \\
2& 66 & -152 & 122 & 0 & 0 & 0 & -122 & 152 & -66 \\
3& 1638 & -7328 & 14960 & -15904 & 6910 & 1408 & 0 & -1408 & -6910 \\
4& 23100 & -147840 & 462956 & -893664 & 1111286 & -814040 & 191700 & 158136 & 0 \\
5& 222156 & -1795200 & 7311216 & -19418784 & 36649600 & -50094752 & 47714700 & -26720160 & 1272584 \\
6& 1613898 & -15402816 & 75236832 & -246081792 & 595164394 & -1110311664 & 1615203928 & -1799004312 & 1431701106 \\
7& 9447438 & -102232416 & 571319892 & -2169449568 & 6218563110 & -14154709888 & 26217103880 & -39798057376 & 49064889198 \\
 8&46562373 & -556049208 & 3448909464 & -14667256032 & 47666004672 & -125023810032 & 272698290096 & -502378917392 & 785763764522 \\\hline
\end{tabular}
}
\end{center}
\caption{table of $b_{jk}$ coefficients for \hh{3}{SO(12)}.}
\end{table}
\begin{table}[H]
\begin{center}\scalebox{0.702}{
\begin{tabular}{|l||l|l|l|l|l|l|l|l|l|}\hline $k\backslash j$&0&1&2&3&4&5&6&7&8\\\hline\hline 0&
 0 & 0 & 0 & 2 & -7 & 28 & -64 & 154 & -286 \\
1& 1 & 0 & 0 & 0 & 0 & 32 & -126 & 448 & -1075 \\
2& 14 & -14 & 18 & 0 & -1 & -2 & 0 & 306 & -1274 \\
3&77 & -128 & 252 & -224 & 184 & 0 & -18 & -32 & 0 \\
4& 273 & -572 & 1386 & -1894 & 2576 & -2114 & 1392 & 0 & -177 \\
5& 748 & -1792 & 4914 & -8256 & 13986 & -17120 & 19424 & -15008 & 8638 \\
6& 1729 & -4522 & 13464 & -25578 & 49308 & -73180 & 103908 & -117596 & 119780 \\
7& 3542 & -9856 & 31122 & -64160 & 134706 & -224672 & 363384 & -494400 & 631352 \\
8& 6630 & -19320 & 63756 & -139342 & 310843 & -560794 & 988548 & -1504986 & 2188872 \\\hline
\end{tabular}
}
\end{center}
\caption{table of $b_{jk}$ coefficients for \hh{3}{G_2}.}
\end{table}
\begin{table}[H]
\begin{center}\scalebox{0.652}{
\begin{tabular}{|l||l|l|l|l|l|l|l|l|l|}\hline $k\backslash j$&0&1&2&3&4&5&6&7&8\\\hline\hline 0&
 0 & 0 & 0 & 0 & 0 & 0 & 5 & 20 & -282 \\
1& 1 & 0 & 0 & 0 & 0 & -4 & -10 & 0 & 445 \\
 2&52 & -104 & 63 & 20 & 14 & 0 & -9 & -292 & -630 \\
3& 1053 & -4212 & 7626 & -6552 & 1169 & 1204 & 1246 & 520 & -735 \\
 4&12376 & -71604 & 204724 & -356372 & 382158 & -205296 & -12984 & 38052 & 47996 \\
5& 100776 & -739024 & 2771158 & -6798428 & 11772934 & -14419412 & 11556342 & -4309188 & -1268849 \\
6& 627912 & -5441904 & 24520358 & -74594556 & 168305172 & -291899816 & 389737341 & -386624816 & 254747324 \\
7& 3187041 & -31290948 & 161203486 & -570136684 & 1532228038 & -3279299588 & 5702866566 & -8070216388 & 9126753214 \\
8& 13748020 & -148728580 & 848959748 & -3360036076 & 10246505556 & -25358941140 & 52337158336 & -91211114924 & 134417094168 \\\hline
\end{tabular}
}
\end{center}
\caption{table of $b_{jk}$ coefficients for \hh{3}{F_4}.}
\end{table}
\begin{table}[H]
\begin{center}\scalebox{0.702}{
\begin{tabular}{|l||l|l|l|l|l|l|l|l|}\hline $k\backslash j$&0&1&2&3&4&5&6&7\\\hline\hline 0&
 0 & 0 & 0 & 0 & 0 & 0 & 0 & -2  \\
1& 1 &0 &0 &0 &0 &0 &-8 &-16 \\
2&78 &-162 &88 &20 &27 &20 &2 &108 \\
3&2430 &-10368 &18382 &-14256 &1898 &464 &2430 &3184 \\
4&43758 &-277992 &818262 &-1383868 &1336426 &-559926 &-55982 &-57612 \\
5&537966 &-4447872 &17830722 &-44652160 &75077496 &-83838128 &55183076 &-11828160 \\
6&4969107 &-49691070 &246665562 &-794475864 &1824658484 &-3087757604 &3820192146 &-3254775622 \\
7&36685506 &-424030464 &2466530352 &-9503406640 &26897792168 &-58618543536 &100133981910 &-133439345056 \\
8&225961450 &-2928460392 &19265962144 &-85010356288 &280290468042 &-728384278466 &1533138581700 &-2642650203582 \\\hline
\end{tabular}
}
\end{center}
\caption{table of $b_{jk}$ coefficients for \hh{3}{E_6}.}
\end{table}
\begin{table}[H]
\begin{center}\scalebox{0.702}{
\begin{tabular}{|l||l|l|l|l|l|l|l|l|l|}\hline $k\backslash j$&0&1&2&3&4&5&6&7&8\\\hline\hline 0&
 0 & 0 & 0 & 0 & -1 & 0 & -28 & 0 & -300 \\
 1&0 & 0 & 0 & 0 & 0 & 0 & -29 & 0 & -707 \\
2& 1 & 0 & 0 & 0 & 0 & 0 & -2 & 0 & -464 \\
3& 28 & 0 & 29 & 0 & 2 & 0 & 0 & 0 & -58 \\
4& 300 & 0 & 707 & 0 & 464 & 0 & 58 & 0 & 0 \\
5& 1925 & 0 & 6999 & 0 & 9947 & 0 & 5365 & 0 & 928 \\
 6&8918 & 0 & 42889 & 0 & 92391 & 0 & 101850 & 0 & 49775 \\
7& 32928 & 0 & 193102 & 0 & 544786 & 0 & 894198 & 0 & 843165 \\
8& 102816 & 0 & 699762 & 0 & 2392663 & 0 & 5096487 & 0 & 7032993 \\\hline
\end{tabular}
}
\end{center}
\caption{table of $b_{jk}$ coefficients for \hh{4}{SO(8)}.}
\end{table}
\begin{table}[H]
\begin{center}\scalebox{0.702}{
\begin{tabular}{|l||l|l|l|l|l|l|l|l|l|}\hline $k\backslash j$&0&1&2&3&4&5&6&7&8\\\hline\hline 0&
 0 & 0 & 0 & 0 & 0 & -2 & 9 & -72 & 231 \\
1& 0 & 0 & 0 & 0 & 0 & 0 & 0 & -76 & 360 \\
 2&1 & 0 & 0 & 0 & 0 & 0 & -1 & 0 & 9 \\
3& 36 & -18 & 40 & 0 & 0 & 2 & 0 & 0 & -40 \\
4& 495 & -462 & 1314 & -684 & 855 & -18 & -9 & 76 & -9 \\
5& 4004 & -5148 & 17028 & -15396 & 25740 & -13788 & 12880 & -612 & -324 \\
6& 22932 & -36036 & 132704 & -159192 & 316053 & -278280 & 357390 & -195624 & 152777 \\
 7&102816 & -185640 & 741780 & -1064232 & 2378100 & -2707254 & 4173660 & -3593760 & 3929652 \\
8& 383724 & -767448 & 3271140 & -5319840 & 12980880 & -17413812 & 30382730 & -33166980 & 43785090 \\\hline
\end{tabular}
}
\end{center}
\caption{table of $b_{jk}$ coefficients for \hh{4}{SO(9)}.}
\end{table}
\begin{table}[H]
\begin{center}\scalebox{0.702}{
\begin{tabular}{|l||l|l|l|l|l|l|l|l|l|}\hline $k\backslash j$&0&1&2&3&4&5&6&7&8\\\hline\hline 0&
 0 & 0 & 0 & 0 & 0 & 0 & -5 & 40 & -279 \\
1& 0 & 0 & 0 & 0 & 0 & 0 & 1 & 0 & -245 \\
2& 1 & 0 & 0 & 0 & 0 & 0 & -2 & 0 & 56 \\
3& 45 & -40 & 56 & 0 & -5 & 0 & 10 & 0 & -112 \\
4& 770 & -1280 & 2580 & -2040 & 1617 & 0 & -245 & -40 & 490 \\
5& 7644 & -17640 & 44110 & -58368 & 74235 & -53872 & 32253 & 0 & -5877 \\
 6&52920 & -150528 & 436254 & -752640 & 1222155 & -1403648 & 1443189 & -982352 & 499765 \\
7& 282744 & -931392 & 3009720 & -6156288 & 11723738 & -17082864 & 22655787 & -23612416 & 21452859 \\
8& 1241460 & -4561920 & 16035624 & -37022832 & 79084500 & -134489856 & 209693627 & -272934384 & 320594252 \\\hline
\end{tabular}
}
\end{center}
\caption{table of $b_{jk}$ coefficients for \hh{4}{SO(10)}.}
\end{table}
\begin{table}[H]
\begin{center}\scalebox{0.702}{
\begin{tabular}{|l||l|l|l|l|l|l|l|l|l|}\hline $k\backslash j$&0&1&2&3&4&5&6&7&8\\\hline\hline 0&
 0 & 0 & 0 & 0 & 0 & 0 & 0 & -14 & 154 \\
1& 0 & 0 & 0 & 0 & 0 & 0 & 0 & 6 & -11 \\
2& 1 & 0 & 0 & 0 & 0 & 0 & -1 & -6 & 0 \\
3& 55 & -66 & 77 & 0 & -14 & 0 & 14 & 14 & -66 \\
4& 1144 & -2574 & 4815 & -4466 & 2954 & 154 & -889 & 0 & 735 \\
5& 13650 & -42900 & 104855 & -160006 & 194845 & -153076 & 75845 & 9482 & -29491 \\
 6&112200 & -437580 & 1279047 & -2533102 & 4109862 & -5095904 & 5078920 & -3556784 & 1470238 \\
7& 703494 & -3197700 & 10645448 & -24989690 & 48638496 & -76538462 & 101986291 & -110913242 & 98067660 \\
8& 3586440 & -18290844 & 67269411 & -178497858 & 395474962 & -727883200 & 1160357220 & -1583817950 & 1869442995 \\\hline
\end{tabular}
}
\end{center}
\caption{table of $b_{jk}$ coefficients for \hh{4}{SO(11)}.}
\end{table}
\begin{table}[H]
\begin{center}\scalebox{0.702}{
\begin{tabular}{|l||l|l|l|l|l|l|l|l|l|}\hline $k\backslash j$&0&1&2&3&4&5&6&7&8\\\hline\hline 0&
 0 & 0 & 0 & 0 & 0 & 0 & 0 & 0 & -42 \\
1& 0 & 0 & 0 & 0 & 0 & 0 & 0 & 0 & 27 \\
2& 1 & 0 & 0 & 0 & 0 & 0 & -1 & 0 & -27 \\
3& 66 & -96 & 103 & 0 & -27 & 0 & 27 & 0 & -61 \\
4& 1638 & -4480 & 8382 & -8448 & 5129 & 576 & -2187 & 0 & 2187 \\
 5&23100 & -88704 & 224913 & -371712 & 455928 & -369408 & 164314 & 44544 & -93897 \\
6& 222156 & -1064448 & 3298922 & -7096320 & 11893791 & -15372032 & 15378627 & -10761216 & 3823551 \\
7& 1613898 & -9060480 & 32465070 & -83128320 & 169434325 & -279711744 & 381972903 & -424986624 & 374859639 \\
8& 9447438 & -59799168 & 239370417 & -695844864 & 1629505845 & -3163425408 & 5222290225 & -7355904000 & 8835469032 \\\hline
\end{tabular}
}
\end{center}
\caption{table of $b_{jk}$ coefficients for \hh{4}{SO(12)}.}
\end{table}
\begin{table}[H]
\begin{center}\scalebox{0.702}{
\begin{tabular}{|l||l|l|l|l|l|l|l|l|l|}\hline $k\backslash j$&0&1&2&3&4&5&6&7&8\\\hline\hline 0&
 0 & 0 & 0 & 0 & 0 & 0 & 0 & 0 & 0 \\
1& 0 & 0 & 0 & 0 & 0 & 0 & 0 & 0 & 0 \\
2& 1 & 0 & 0 & 0 & 0 & 0 & -1 & 0 & 0 \\
3& 78 & -130 & 134 & 0 & -44 & 0 & 44 & 0 & -134 \\
4& 2275 & -7150 & 13697 & -14560 & 8464 & 1430 & -4466 & 0 & 4466 \\
5& 37400 & -165750 & 441602 & -772200 & 966322 & -797160 & 330054 & 136864 & -242946 \\
6& 415701 & -2309450 & 7638640 & -17484220 & 30403198 & -40556386 & 41062240 & -28619032 & 8937641 \\
7& 3461458 & -22632610 & 87499854 & -239745220 & 511877560 & -878703202 & 1232023366 & -1396956444 & 1236834410 \\
8& 23046023 & -170600430 & 742996397 & -2323025120 & 5738161356 & -11650207270 & 19899034288 & -28804097088 & 35248051027 \\\hline
\end{tabular}%
}
\end{center}
\caption{table of $b_{jk}$ coefficients for \hh{4}{E_6}.}
\end{table}
\begin{table}[H]
\begin{center}\scalebox{0.702}{
\begin{tabular}{|l||l|l|l|l|l|l|l|l|l|}\hline $k\backslash j$&0&1&2&3&4&5&6&7&8\\\hline\hline 0&
 0 & 0 & 0 & 0 & 0 & 0 & 1 & 0 & 52 \\
1& 0 & 0 & 0 & 0 & 0 & 0 & 0 & 0 & 53 \\
2& 0 & 0 & 0 & 0 & 0 & 0 & 0 & 0 & 0 \\
3& 1 & 0 & 0 & 0 & 0 & 0 & 0 & 0 & -1 \\
4& 52 & 0 & 53 & 0 & 0 & 0 & -1 & 0 & 0 \\
5& 1053 & 0 & 2432 & 0 & 1484 & 0 & 0 & 0 & -53 \\
6& 12376 & 0 & 44980 & 0 & 59996 & 0 & 28824 & 0 & 0 \\
7& 100776 & 0 & 495872 & 0 & 1023464 & 0 & 1034880 & 0 & 434010 \\
8& 627912 & 0 & 3856722 & 0 & 10670660 & 0 & 16410602 & 0 & 13979280 \\\hline
\end{tabular}%
}
\end{center}
\caption{table of $b_{jk}$ coefficients for \hh{5}{E_6}.}
\end{table}
\begin{table}[H]
\begin{center}\scalebox{0.702}{
%
}
\end{center}
\caption{table of $b_{jk}$ coefficients for \hh{7}{E_7}.}
\end{table}

\section{One-instanton component of $Z^{5d}_{S^1\times\mathbb{R}^4}$}
In this section we describe the computation of one-instanton component of 5d Nekrasov partition functions starting from the elliptic genera of the \hh{n}{G} theories. In section \ref{sec:5dappgen} we discuss our approach, and in section \ref{sec:5dappfull} we provide the results of our computations.
\label{sec:5dapp}
\subsection{Computing $Z_{1-inst}$}\label{sec:5dappgen}
From the discussion of section \ref{sec:5dlim} it follows that for $n\geq 3$ the leading order terms in the $q$-expansion of the 6d $T^2\times\mathbb{R}^4$ partition function gives the Nekrasov partition function for the 5d $\mathcal{N}=1$ theory with gauge group $G$ and matter hypermultiplets in the same representations as the 6d matter, up to an overall power of $v$. For $n=2$, the leading order terms give the Nekrasov partition function with the addition of some decoupled states which are neutral under the gauge group.\newline

\noindent By exploiting modularity as in section \ref{sec:modul}, we have computed the one-string elliptic genus for 42 theories with $n\geq 2$; from these computations we can readily obtain the one-instanton piece of the corresponding 5d  partition function, $Z_{1-inst}$, with fugacities $\mass_G$ and $\mass_F$ switched off. To the best of our knowledge, this leads to new results for the 5d theories in table \ref{tab:new5d}.\footnote{ While this work was under completion, the papers \cite{Kim:2018gak,Kim:2018gjo,Hayashi:2018bkd} appeared with similar results for the theories with $G=SO(7) $ and $ G_2$, with respectively two hypermultiplets in the spinor representation and one hypermultiplet in the $\mathbf{7}$ representation.}\newline
\begin{table}[t]
\begin{center}
\begin{tabular}{|l|lr|}\hline $G$&Matter&\\\hline\hline 
$E_7$ & $m\cdot \mathbf{56}$&$\quad m=1^*,2^*,3^*$\\
$E_6$ & $m\cdot (\mathbf{27}+\overline{\mathbf{27}})$&$\quad m=1^*,2,3$\\
$F_4$ & $m\cdot \mathbf{26}$&$\quad m=1^*,2^*,\underline{3}^*$\\
$G_2$ & $m\cdot \mathbf{7}$&$\quad m=\underline{4}^*$\\
$SO(7)$ & $(3-m)\cdot\mathbf{7}\oplus 2(4-m)\cdot \mathbf{8}$&$\quad m=\underline{2}^*$\\
$SO(8)$ & $(4-m)\cdot\mathbf{8^v}\oplus(4-m)\cdot\mathbf{8^s}\oplus(4-m)\cdot\mathbf{8^c}$&$\quad m=\underline{2},3$\\
$SO(9)$ & $(5-m)\cdot\mathbf{9}\oplus(4-m)\cdot\mathbf{16}$&$\quad m=\underline{2},3^*$\\
$SO(10)$ & $(6-m)\cdot\mathbf{10}\oplus(4-m)\cdot\mathbf{16^s}\oplus(4-m)\cdot\mathbf{16^c}$&$\quad m=\underline{2},3^*$\\
$SO(11)$ & $(7-m)\cdot\mathbf{11}\oplus (4-m)\cdot\mathbf{32}$&$\quad m=3^*$\\
$SO(12)_a$ & $(8-m)\cdot\mathbf{12}\oplus (4-m)\cdot\mathbf{32}^s$&$\quad m=\underline{2},3$\\
$SO(12)_b$ & $(8-m)\cdot\mathbf{12}\oplus(\frac{4-m}{2})\cdot\mathbf{32}^s\oplus(\frac{4-m}{2})\cdot\mathbf{32}^c$&$\quad m=\underline{2}$\\
$SO(13)$ & $(9-m)\cdot\mathbf{13}\oplus(\frac{4-m}{2})\cdot\mathbf{64}$&$\quad m=\underline{2}$\\
\hline
\end{tabular}
\end{center}
\caption{Choices of gauge and matter content of 5d theories for which we obtain novel results for the one-instanton partition function. For the values of $m$ which are underlined, the one-instanton partition function contains extra gauge-neutral states in comparison to the Nekrasov partition function (as discussed in section \ref{sec:5dlim}). We denote by an asterisk the theories for which we are able to determine the one-instanton partition function to all orders in $v$ with chemical potentials $\mass_F$ and $\mass_G$ turned on; for the remaining theories we are only able to obtain partial information about the dependence on $\massF$ and $\massG$.}
\label{tab:new5d}
\end{table}

\noindent By using equations \eqref{eq:elldecomp} and \eqref{eq:FGv} as in appendix \ref{app:xiexample}, one can in fact reconstruct the dependence of the one-instanton functions $Z_{1-inst}(\mass_G,\mass_F,v,q)$, written as a power series in $v$. For several theories, we are able to perform such a computation to very high orders in $v$, and this allows us to conjecture an exact expression for the one-instanton piece of the Nekrasov partition function of the corresponding 5d theory, with all fugacities turned on. This also leads to several novel results for the theories listed in table \ref{tab:new5d}.\newline

\noindent We illustrate this idea by means of a specific example: the theory with $G=G_2$ and 4 hypermultiplets in the $\mathbf{7}$ representation, which transform under $F=Sp(4)$. Expanding our result for $Z_{1-inst}$ to a suitable power in $v$, we find
\begin{align}\label{eq:5dex1}
&\left[\frac{v}{(1-v\, x)(1-v/x)}\right]^{-1}\,Z_{1-inst}(\mathbf{1},\mathbf{1},x,v)=\mathbb{E}_{2}^{G_2}(\mathbf{1},\mathbf{1},v,q)\bigg\vert_{q^{1/6}} = \nonumber\\
&\frac{(-1+v)^2 \left(1+8 v+30 v^2+64 v^3+30 v^4+8 v^5+v^6\right)}{v (1+v)^6}= v^{-1}-8 v^2\!-\!35 v^3\!+\!336 v^4\!-\!1317 v^5\!+\!\dots
\end{align}
On the other hand, by expanding equation \eqref{eq:elldecomp} we obtain:
\begin{align}\label{eq:5dex2}
\mathbb{E}_{2}^{G_2}(\mass_{G_2},\mass_{Sp(4)},v,q)\bigg\vert_{q^{1/6}} &= n^{(00)}_{(0000),-1,0}v^{-1}+v\,n^{(00)}_{(0000),-3,-1}+v^2\,n^{(00)}_{(1000),-2,-1}\chi^F_{\mathbf 8}\nonumber\\
&+v^3(n^{(10)}_{(0000),-1,-1}\chi^G_{\mathbf 7}+n^{(00)}_{(0001),-1,-1}\chi^F_{\mathbf 42})+n^{(10)}_{(0010),0,-1}v^3\chi^F_{\mathbf 48}\chi^G_{\mathbf 7}\nonumber\\
&+v^4(n^{(20)}_{(0100),-3,-2}\chi^F_{\mathbf 27}\chi^G_{\mathbf 27}+n^{(01)}_{(0001),-3,-2}\chi^F_{\mathbf 42}\chi^G_{\mathbf 14})+\dots,
\end{align}
and in order to restore the $\mass_{G_2}$, $\mass_{Sp(4)}$ dependence we need to find small integers $n^\lambda_{\nu,\ell,m}$ for which equations \eqref{eq:5dex1} and \eqref{eq:5dex2} agree. Several of these coefficients are uniquely determined:
\begin{equation}
n^{(00)}_{(0000),-1,0}=1;\qquad n^{(00)}_{(0000),-3,-1}=0;\qquad n^{(00)}_{(1000),-2,-1}=-1;\qquad n^{(10)}_{(0010),0,-1}=1.
\end{equation}
We can also easily determine the remaining coefficients with high confidence. For example, for the $v^4$ term we need to impose
\begin{equation}
729\,n^{(20)}_{(0100),-3,-2}+588\,n^{(01)}_{(0001),-3,-2}=-1317.
\end{equation}
The generic solution for integer coefficients is:
\begin{equation}
n^{(20)}_{(0100),-3,-2}=195-196 c;\qquad n^{(01)}_{(0001),-3,-2}=-244+243 c; c\in\mathbb{Z}.
\end{equation}
The only choice that leads to small coefficients is $c=1$, which gives 
\begin{equation}
n^{(20)}_{(0100),-3,-2}=n^{(01)}_{(0001),-3,-2}=-1.
\end{equation}
Likewise, for the $v^3$ coefficient the only reasonable choice is 
\begin{equation}
n^{(10)}_{(0000),-1,-1}=1\quad \text{ and } \quad n^{(00)}_{(0001),-1,-1}=-1.
\end{equation}
One can easily automate such computations, and pushing the calculation to higher order one begins to recognize a pattern in the representations that appear in the one-instanton partition function:
 \begin{align}
 &\left[\frac{v}{(1-v\, x)(1-v/x)}\right]^{-1}\,Z_{1-inst}(\mathbf{1},\mathbf{1},x,v)=\nonumber\\ 
 & v^{-1}-v^2\chi^F_{(1000)}+v^3(\chi^G_{(10)}-\chi^F_{(0001)})+v^4\chi^G_{(10)}\chi^F_{(0010)}-v^5(\chi^G_{(20)}\chi^F_{(0100)}+\chi^G_{(01)}\chi^F_{(0001)})\nonumber\\
 &+v^6(\chi^G_{(30)}\chi^F_{(1000)}+\chi^G_{(11)}\chi^F_{(0010)})
 -v^7(\chi^G_{(40)}+\chi^G_{(21)}\chi^F_{(0100)}+\chi^G_{(02)}\chi^F_{(0001)})\nonumber\\
 &+v^8(\chi^G_{(31)}\chi^F_{(1000)}+\chi^G_{(12)}\chi^F_{(0010)})-v^9(\chi^G_{(41)}+\chi^G_{(22)}\chi^F_{(0100)}+\chi^G_{(03)}\chi^F_{(0001)})+\mathcal{O}(v^{10})
  \end{align}
(in fact one can easily push the computation to higher orders with little effort). This leads us to conjecture that
 \begin{align}\label{eq:g2expa}
  &\left[\frac{v}{(1-v\, x)(1-v/x)}\right]^{-1}\,Z_{1-inst}(\mathbf{1},\mathbf{1},x,v)=\nonumber\\ 
  &= v^{-1}-v^2\chi^F_{(1000)}+v^3\chi^G_{(10)}-\sum_{n=0}^\infty v^{3+2n}\chi^G_{0n}\chi^F_{(0001)}+\sum_{n=0}^\infty v^{4+2n}\chi^G_{(1n)}\chi^F_{(0010)}\nonumber\\
 &-\sum_{n=0}^\infty v^{5+2n}\chi^G_{2n}\chi^F_{(0100)}+\sum_{n=0}^\infty v^{6+2n}\chi^G_{(3n)}\chi^F_{(1000)}-\sum_{n=0}^\infty v^{7+2n}\chi^G_{(4n)}.
  \end{align}
As a consistency check, setting the fugacities $\mass_{G_2},$ and $\mass_{Sp(N)}$ to zero and plugging in the formulas for the dimensions of the representations of $G_2$ and $Sp(4)$ that appear in equation \eqref{eq:g2expa}, that is:
\begin{equation}
\text{dim}(R^{G_2}_{(n_1,n_2)}) = \frac{(n_1+1)(n_2+1)(n_1+n_2+2)(n_1+2n_2+3)(n_1+3n_2+4)(2n_1+3n_2+5)}{120},
\end{equation}
and
\begin{equation}
\text{dim}(R^{Sp(4)}_{(1000)}) = 8;\quad\text{dim}(R^{Sp(4)}_{(0100)}) =27;\quad\text{dim}(R^{Sp(4)}_{(0010)}) = 48;\quad\text{dim}(R^{Sp(4)}_{(0001)}) = 42,
\end{equation}
one finds from equation \eqref{eq:g2expa} that
\begin{equation}
Z_{1-inst}(\mathbf{1},\mathbf{1},v)=v^{-1}+\sum_{n=0}^\infty \frac{1}{15}(-1)^nn(76-70n^2+9n^4)v^n,
\end{equation}
which indeed is the series expansion of 
\begin{equation}
\frac{(-1+v)^2 \left(1+8 v+30 v^2+64 v^3+30 v^4+8 v^5+v^6\right)}{v (1+v)^6}.
\end{equation}
\subsection{One-instanton partition functions}\label{sec:5dappfull}
\begin{table}[p!]
\begin{center}\scalebox{0.85}{
\begin{tabular}{|ll|r|}
\hline
$n$&$G$&$Z_{1-inst}$\\\hline
2 & $SU(2)$ &$v^{-1}(1+v)^{-2}(1-v)^2 [1+4 v+v^2]$\\
2 & $SU(3)$ &$v^{-1} (1+v)^{-4}(1-v)^2 [1+6 v+16 v^2+6 v^3+v^4]$\\
2 & $SU(4)$ &$v^{-1} (1+v)^{-6}(1-v)^2 [1+8 v+29 v^2+64 v^3+29 v^4+8 v^5+v^6]$\\
2 & $SU(5)$ &$v^{-1} (1+v)^{-8}(1-v)^2 [1+10 v+46 v^2+130 v^3+256 v^4+130 v^5+46 v^6+10 v^7+v^8]$\\
2 & $SU(6)$ &$v^{-1} (1+v)^{-10}(1-v)^2 [1+12 v+67 v^2+232 v^3+562 v^4+1024 v^5+562 v^6+232 v^7$\\
&&$+67 v^8+12 v^9+v^{10}]$\\
2 & $SU(7)$ &$v^{-1} (1+v)^{-12}(1-v)^2 [1+14 v+92 v^2+378 v^3+1093 v^4+2380 v^5+4096 v^6+2380 v^7$\\
&&$+1093 v^8+378 v^9+92 v^{10}+14 v^{11}+v^{12}]$\\
2 & $SU(8)$ &$v^{-1} (1+v)^{-14}(1-v)^2 [1+16 v+121 v^2+576 v^3+1941 v^4+4944 v^5+9949 v^6$\\
&&$+16384 v^7+9949 v^8+4944 v^9+1941 v^{10}+576 v^{11}+121 v^{12}+16 v^{13}+v^{14}]$\\
2 & $SO(7)$ & $v^{-1} (1+v)^{-8}(1-v)^2 [1+10 v+47 v^2+138 v^3+256 v^4+138 v^5+47 v^6+10 v^7+v^8]$\\
2 & $SO(8)$ & $v^{-1} (1+v)^{-10}(1-v)^2 [1+12 v+68 v^2+244 v^3+615 v^4+1024 v^5+615 v^6+244 v^7$\\
&&$+68 v^8+12 v^9+v^{10}]$\\
2 & $SO(9)$ & $v^{-1} (1+v)^{-12}(1-v)^2 [1+14 v+93 v^2+392 v^3+1181 v^4+2658 v^5+4106 v^6+2658 v^7$\\
&&$+1181 v^8+392 v^9+93 v^{10}+14 v^{11}+v^{12}]$\\
2 & $SO(10)$ & $v^{-1} (1+v)^{-14}(1-v)^2 [1+16 v+122 v^2+592 v^3+2060 v^4+5472 v^5+11287 v^6+16496 v^7$\\
&&$+11287 v^8+5472 v^9+2060 v^{10}+592 v^{11}+122 v^{12}+16 v^{13}+v^{14}]$\\
2 & $SO(12)_a$ & $v^{-1} (1+v)^{-18}(1-v)^2 [1+20 v+192 v^2+1180 v^3+5226 v^4+17804 v^5+48575 v^6$\\
&&$+108512 v^7+197370 v^8+267144 v^9+197370 v^{10}+108512 v^{11}+48575 v^{12}$\\
&&$+17804 v^{13}+5226 v^{14}+1180 v^{15}+192 v^{16}+20 v^{17}+v^{18}]$\\
2 & $SO(12)_b$ & $v^{-1} (1+v)^{-18}(1-v)^2 [1+20 v+192 v^2+1180 v^3+5228 v^4+17820 v^5+48633 v^6$\\
&&$+108640 v^7+197566 v^8+267368 v^9+197566 v^{10}+108640 v^{11}+48633 v^{12}$\\
&&$+17820 v^{13}+5228 v^{14}+1180 v^{15}+192 v^{16}+20 v^{17}+v^{18}]$\\
2 & $SO(13)$ & $v^{-1} (1+v)^{-20}(1-v)^2 [1+22 v+233 v^2+1584 v^3+7780 v^4+29466 v^5+89645 v^6$\\
&&$+224944 v^7+471813 v^8+818552 v^9+1077376 v^{10}+818552 v^{11}+471813 v^{12}$\\
&&$+224944 v^{13}+89645 v^{14}+29466 v^{15}+7780 v^{16}+1584 v^{17}+233 v^{18}+22 v^{19}+v^{20}]$\\
2 & $G_2$ & $v^{-1} (1+v)^{-6}(1-v)^2 [1+8 v+30 v^2+64 v^3+30 v^4+8 v^5+v^6]$\\
2 & $F_4$ & $v^{-1} (1+v)^{-16}(1-v)^2 [1+18 v+155 v^2+852 v^3+3369 v^4+10240 v^5+24825 v^6+47834 v^7$\\
&&$+66180 v^8+47834 v^9+24825 v^{10}+10240 v^{11}+3369 v^{12}+852 v^{13}+155 v^{14}+18 v^{15}+v^{16}]$\\\hline

\end{tabular}}
\end{center}
\caption{5d limit of the one-string elliptic genus for the theories \hh{n}{G} with $n=2$, with chemical potentials $\mass_F,$ $\mass_G$ turned off. The partition function is displayed with the center of mass factor $\frac{v}{(1-v\,x)(1-v/x)}$ removed.}
\label{tab:5dlimtab1}
\end{table}

\begin{table}[p!]
\begin{center}\scalebox{0.85}{
\begin{tabular}{|ll|r|}
\hline
$n$&$G$&$Z_{1-inst}$\\\hline
3 & $SU(3)$ &$(1-v)^{-4}(1+v)^{-4}[v^2+4 v^4+v^6]$\\
3 & $SO(7)$ &$(1-v)^{-4}(1+v)^{-8}v^4 [5-12 v+22 v^2-12 v^3+5 v^4]$\\
3 & $SO(8)$ &$(1-v)^{-4}(1+v)^{-10}v^4 [1+14 v-37 v^2+68 v^3-37 v^4+14 v^5+v^6]$\\
3 & $SO(9)$ &$(1-v)^{-4}(1+v)^{-12}2 v^5 [2+19 v-62 v^2+106 v^3-62 v^4+19 v^5+2 v^6]$\\
3 & $SO(10)$ &$(1-v)^{-4}(1+v)^{-14}2 v^6 [7+54 v-210 v^2+344 v^3-210 v^4+54 v^5+7 v^6]$\\
3 & $SO(11)$ &$(1-v)^{-4}(1+v)^{-16}v^7 [48+321 v-1436 v^2+2302 v^3-1436 v^4+321 v^5+48 v^6]$\\
3 & $SO(12)$ &$(1-v)^{-4}(1+v)^{-18}11 v^8 [15+90 v-451 v^2+716 v^3-451 v^4+90 v^5+15 v^6]$\\
3 & $G_2$ &$(1-v)^{-4}(1+v)^{-6}v^3 [2-3 v+8 v^2-3 v^3+2 v^4]$\\
3 & $F_4$ &$(1-v)^{-4}(1+v)^{-16}v^6 [5+80 v+268 v^2-1232 v^3+2142 v^4-1232 v^5+268 v^6+80 v^7+5 v^8]$\\
3 & $E_6$ &$(1-v)^{-4}(1+v)^{-22}2 v^7 [1+28 v+356 v^2+2045 v^3+1583 v^4-19638 v^5+36572 v^6$\\
&&$-19638 v^7+1583 v^8+2045 v^9+356 v^{10}+28 v^{11}+v^{12}]$\\
\hline
4 & $SO(8)$ &$(1-v)^{-10}(1+v)^{-10}v^5 [1+18 v^2+65 v^4+65 v^6+18 v^8+v^{10}]$\\
4 & $SO(9)$ &$(1-v)^{-10}(1+v)^{-12}v^6 [2-5 v+36 v^2-46 v^3+130 v^4-90 v^5+130 v^6-46 v^7$\\
&&$+36 v^8-5 v^9+2 v^{10}]$\\
4 & $SO(10)$ &$(1-v)^{-10}(1+v)^{-14}v^7 [5-20 v+99 v^2-184 v^3+370 v^4-360 v^5+370 v^6$\\
&&$-184 v^7+99 v^8-20 v^9$\\
&&$+5 v^{10}]$\\
4 & $SO(11)$ &$(1-v)^{-10}(1+v)^{-16}v^8 [14-70 v+306 v^2-651 v^3+1180 v^4-1278 v^5+1180 v^6-651 v^7$\\
&&$+306 v^8-70 v^9+14 v^{10}]$\\
4 & $SO(12)$ &$(1-v)^{-10}(1+v)^{-18}v^9 [42-240 v+999 v^2-2264 v^3+3947 v^4-4464 v^5+3947 v^6-2264 v^7$\\
&&$+999 v^8-240 v^9+42 v^{10}]$\\
4 & $SO(13)$ &$(1-v)^{-10}(1+v)^{-20}v^{10} [132-825 v+3366 v^2-7898 v^3+13550 v^4-15642 v^5+13550 v^6$\\
&&$-7898 v^7+3366 v^8-825 v^9+132 v^{10}]$\\
4 & $SO(14)$ &$(1-v)^{-10}(1+v)^{-22}13 v^{11} [33-220 v+891 v^2-2136 v^3+3640 v^4-4248 v^5+3640 v^6$\\
&&$-2136 v^7+891 v^8-220 v^9+33 v^{10}]$\\
4 & $SO(15)$ &$(1-v)^{-10}(1+v)^{-24}2 v^{12} [715-5005 v+20241 v^2-49231 v^3+83680 v^4-98280 v^5$\\
&&$+83680 v^6-49231 v^7+20241 v^8-5005 v^9+715 v^{10}]$\\
4 & $F_4$ &$(1-v)^{-10}(1+v)^{-16}v^7 [1+10 v-49 v^2+266 v^3-549 v^4+1068 v^5-1110 v^6+1068 v^7$\\
&&$-549 v^8+266 v^9-49 v^{10}+10 v^{11}+v^{12}]$\\
4 & $E_6$ &$(1-v)^{-10}(1+v)^{-22}v^9 [3+44 v+33 v^2-1052 v^3+6513 v^4-17404 v^5+31905 v^6-37432 v^7$\\
&&$+31905 v^8-17404 v^9+6513 v^{10}-1052 v^{11}+33 v^{12}+44 v^{13}+3 v^{14}]$\\\hline

\end{tabular}}
\end{center}
\caption{5d limit of the one-string elliptic genus for the theories \hh{n}{G} with $n=3$ and $4$, with chemical potentials $\mass_F,$ $\mass_G$ turned off.}
\label{tab:5dlimtab2}
\end{table}

In tables \ref{tab:5dlimtab1}--\ref{tab:5dlimtab3} we provide the one-instanton partition functions of the 5d $\mathcal{N}=1$ theories we obtain from compactification of the 6d SCFTs, with fugacities $\mass_F$ and $\mass_G$ turned off. In the text that follows we present our results with chemical potentials restored. We omit to discuss the cases without matter and the two families of theories with $n=2$, $G=SU(N)$ and $n=4$, $G=SO(8+N)$ since their 5d partition function is already well studied. For the remaining theories, the matter content of the 5d theory can be read off from tables \ref{tab:flavor1} and \ref{tab:flavor2}. In this entire section we have factored out the center of mass term $\frac{v}{(1-v\,x)(1-v/x)}$ from $Z_{1-inst}$.\newline

\noindent In the paper we have not kept track of the overall sign of the elliptic genus and this also affects the overall sign of the one-instanton partition function we compute. Here we make a choice of sign so that the leading order coefficient in the $v$-expansion of $Z_{1-inst}$ has positive sign when the fugacities $\massF$ and $\massG$ are set to $\mathbf{1}$.\newline

\begin{table}[t]
\begin{center}\scalebox{0.85}{
\begin{tabular}{|ll|r|}
\hline
$n$&$G$&$Z_{1-inst}$\\\hline
5 & $F_4$ &$(1-v)^{-16}(1+v)^{-16}v^8 [1+36 v^2+341 v^4+1208 v^6+1820 v^8+1208 v^{10}+341 v^{12}+36 v^{14}+v^{16}]$\\
5 & $E_6$ &$(1\!-\!v)^{-16}(1\!+\!v)^{-22}v^{10} [1\!+\!8 v\!-\!43 v^2\!+\!456 v^3\!-\!1436 v^4\!+\!5116 v^5\!-\!9848 v^6\!+\!19504 v^7\!-\!24164 v^8$\\
&&$+30016 v^9\!-\!24164 v^{10}\!+\!19504 v^{11}\!-\!9848 v^{12}\!+\!5116 v^{13}\!-\!1436 v^{14}\!+\!456 v^{15}\!-\!43 v^{16}\!+\!8 v^{17}\!+\!v^{18}]$\\
5 & $E_7$ &$(1\!-\!v)^{-16}(1\!+\!v)^{-34}v^{14} [7+126 v+307 v^2-4096 v^3+8014 v^4+121428 v^5-898477 v^6+3512818 v^7$\\
&&$-9100043 v^8+17425312 v^9-25317468 v^{10}+28704184 v^{11}-25317468 v^{12}+17425312 v^{13}$\\
&&$-\!9100043 v^{14}\!+\!3512818 v^{15}\!-\!898477 v^{16}\!+\!121428 v^{17}\!+\!8014 v^{18}\!-\!4096 v^{19}\!+\!307 v^{20}\!+\!126 v^{21}\!+\!7v^{22}]$\\
\hline
6 & $E_6$ &$(1-v)^{-22}(1+v)^{-22}v^{11} [1+56 v^2+945 v^4+6776 v^6+23815 v^8+43989 v^{10}+43989 v^{12}$\\
&&$+23815 v^{14}+6776 v^{16}+945 v^{18}+56 v^{20}+v^{22}]$\\
6 & $E_7$ &$(1-v)^{-22}(1+v)^{-34}v^{15} [2+24 v-43 v^2+52 v^3+8027 v^4-53360 v^5+279039 v^6-950972 v^7$\\
&&$+2698740 v^8-5898532 v^9+10988680 v^{10}-16600348 v^{11}+21616127 v^{12}-23243264 v^{13}$\\
&&$+21616127 v^{14}-16600348 v^{15}+10988680 v^{16}-5898532 v^{17}+2698740 v^{18}-950972 v^{19}$\\
&&$+279039 v^{20}-53360 v^{21}+8027 v^{22}+52 v^{23}-43 v^{24}+24 v^{25}+2v^{26}]$\\\hline
7 & $E_7$ &$(1-v)^{-28}\,\,(1+v)^{-34}\,\,v^{16} \,\,[1+6\, v-13\, v^2+764\, v^3-3200\, v^4+24172\, v^5-76952\, v^6+317380\, v^7$\\
&&$-758576 v^8\!+\!2097116 v^9\!-\!3826888 v^{10}\!+\!7681284 v^{11}\!-\!10844990 v^{12}\!+\!16441672 v^{13}\!-\!18077202 v^{14}$\\
&&$+21130252 v^{15}-18077202 v^{16}+16441672 v^{17}-10844990 v^{12}+7681284 v^{11}-3826888 v^{10}$\\
&&$+2097116 v^9-758576 v^8+317380 v^7-76952 v^6+24172 v^5-3200 v^4+764 v^3-13 v^2+6 v+1]$\\\hline
\end{tabular}}
\end{center}
\caption{5d limit of the one-string elliptic genus for the theories \hh{n}{G} with $n=5,6,7$, with chemical potentials $\mass_F,$ $\mass_G$ turned off.}
\label{tab:5dlimtab3}
\end{table}

\noindent $\bullet\quad \mathbf{n=2,\, G=SO(7),\,F=Sp(1)_a\times Sp(4)_b}$:\newline

\noindent We find that the first 11 terms in the $v$ expansion in $Z_{1-inst}$ are given by:
 \begin{align}&Z_{1-inst} = v^{-1}-v^2\chi^{F}_{(1)_a}-v^3\chi^F_{(0100)_b}+v^4(\chi^G_{(001)}\chi^F_{(1000)_b}-\chi^F_{(1)_a\otimes(0001)_b})\nonumber\\
 &+v^5(\chi^G_{(100)}\chi^F_{(0001)_b}-\chi^G_{(002)}+\chi^G_{(001)}\chi^F_{(1)_a\otimes(0010)_b})\nonumber\\
  &-v^6(\chi^G_{(101)}\chi^F_{(0010)_b}+\chi^G_{(010)}\chi^F_{(1)_a\otimes(0001)_b}+\chi^G_{(002)}\chi^F_{(1)_a\otimes(0100)_b})\nonumber\\
    &+v^7(\chi^G_{(110)}\chi^F_{(0001)_b}+\chi^G_{(102)}\chi^F_{(0100)_b}+\chi^G_{(011)}\chi^F_{(1)_a\otimes(0010)_b}+\chi^G_{(003)}\chi^F_{(1)_a\otimes(1000)_b})\nonumber\\
      &-v^8(\chi^G_{(111)}\chi^F_{(0010)_b}+\chi^G_{(103)}\chi^F_{(1000)_b}+\chi^G_{(020)}\chi^F_{(1)_a\otimes(0001)_b}+\chi^G_{(012)}\chi^F_{(1)_a\otimes(0100)_b}+\chi^G_{(004)}\chi^F_{(1)_a})\nonumber\\
    &+v^9(\chi^G_{(120)}\chi^F_{(0001)_b}+\chi^G_{(112)}\chi^F_{(0100)_b}+\chi^G_{(104)}+\chi^G_{(021)}\chi^F_{(1)_a\otimes(0010)_b}+\chi^G_{(013)}\chi^F_{(1)_a\otimes(1000)_b})\nonumber\\
&+\mathcal{O}(v^{10}).  
 \end{align}
We have pushed this computation to $\mathcal{O}(v^{16})$ and conjecture that the Nekrasov partition function to all orders in $v$ is given by:
 \begin{align}Z_1 &= v^{-1}-v^2\chi^{F}_{(1)_a}-v^3\chi^F_{(0100)_b}+v^4\chi^G_{(001)}\chi^F_{(1000)_b}-v^5\chi^G_{(002)}-\sum_{n=0}^\infty v^{4+2n}\chi^G_{(0n0)}\chi^F_{(1)_a\otimes(0001)_b}\nonumber\\
 &+\sum_{n=0}^\infty v^{5+2n}(\chi^G_{(1n0)}\chi^F_{(0001)_b}+\chi^G_{(0n1)}\chi^F_{(1)_a\otimes(0010)_b})\nonumber\\
 &-\sum_{n=0}^\infty v^{6+2n}(\chi^G_{(1n1)}\chi^F_{(0010)_b}+\chi^G_{(0n2)}\chi^F_{(1)_a\otimes(0100)_b})\nonumber\\
 &+\sum_{n=0}^\infty v^{7+2n}(\chi^G_{(1n2)}\chi^F_{(0100)_b}+\chi^G_{(0n3)}\chi^F_{(1)_a\otimes(1000)_b})\nonumber\\
 &-\sum_{n=0}^\infty v^{8+2n}(\chi^G_{(1n3)}\chi^F_{(1000)_b}+\chi^G_{(0n4)}\chi^F_{(1)_a})+\sum_{n=0}^\infty v^{9+2n}\chi^G_{(1n4)}.
 \end{align}
As a further check, upon specializing $\massG,\massF\to 1$ this expression matches with the result in table \ref{tab:5dlimtab1}.\newline\\

\noindent $\bullet\quad \mathbf{n=2,\, G=SO(8),\,F=Sp(2)_a\times Sp(2)_b\times Sp(2)_c}$:\newline

\noindent In this case we are not able to completely fix the Nekrasov partition function due to the appearance of many representations of $F=Sp(2)^3$ of the same dimension. We make the assumption that the combinations of representations that appear as coefficients in the Nekrasov partition function must be triality invariant. Still, it is sometimes possible to find more than one triality-invariant combination of representations. For instance, the combinations of representations
\begin{equation}
\chi^G_{(1000)}\chi^{F}_{(10)_a\otimes (10)_a}+\chi^G_{(0010)}\chi^{F}_{(10)_b\otimes (10)_b}+\chi^G_{(0001)}\chi^{F}_{(10)_c\otimes (10)_c} 
\end{equation}
and
\begin{equation}
\chi^G_{(1000)}\chi^{F}_{(10)_b\otimes (10)_c}+\chi^G_{(0010)}\chi^{F}_{(10)_a\otimes (10)_c}+\chi^G_{(0001)}\chi^{F}_{(10)_a\otimes (10)_b}
\end{equation}
are both triality invariant and have the same dimension. For cases such as these where we are not able to determine the precise representations appearing we use the following notation:
\begin{equation}
(\chi^G_{(1000)}+\chi^G_{(0010)}+\chi^G_{(0001)})\chi^{F}_{(10)_{(abc)}\otimes (10)_{(abc)}}.
\end{equation}
With this proviso, by studying the partition function up to $\mathcal{O}(v^{15})$ we have found the following expression, which we conjecture to hold at all orders in $v$:
\begin{align} Z_{1-inst} &= v^{-1}-v^3(\chi^{F}_{(01)_a}+\chi^{F}_{(01)_b}+\chi^{F}_{(01)_c})-v^5((\chi^G_{(1000)}+\chi^G_{(0010)}+\chi^G_{(0001)})\chi^{F}_{(10)_{(abc)}\otimes (10)_{(abc)}}\nonumber\\
&+\chi^G_{(0100)}) -v^6(\chi^G_{(1010)}\chi^{F}_{(10)_c}+\chi^G_{(1001)}\chi^{F}_{(10)_b}+\chi^G_{(0011)}\chi^{F}_{(10)_a}) +v^7\chi^G_{(1011)}\nonumber\\
    &-\sum_{n=0}^\infty v^{5+2n}\chi^G_{(0n00)}\chi^F_{(01)_a\otimes (01)_b\otimes (01)_c}\nonumber\\
    &+\sum_{n=0}^\infty v^{6+2n}(\chi^G_{(1n00)}+\chi^G_{(0n10)}+\chi^G_{(0n01)})\chi^F_{(01)_{(abc)}\otimes (01)_{(abc)}\otimes (10)_{(abc)}}\nonumber\\
    &-\sum_{n=0}^\infty v^{7+2n}\bigg((\chi^G_{(1n10)}+\chi^G_{(1n01)}+\chi^G_{(0n11)})\chi^F_{(01)_{(abc)}\otimes (10)_{(abc)}\otimes (10)_{(abc)}}\nonumber\\
    &\qquad\qquad\qquad\qquad +(\chi^G_{(2n00)}+\chi^G_{(0n20)}+\chi^G_{(0n02)})\chi^F_{(01)_{(abc)}\otimes (01)_{(abc)}}\bigg)\nonumber\\
        &+\sum_{n=0}^\infty v^{8+2n}\bigg(\chi^G_{(1n11)}\chi^F_{(10)_a\otimes(10)_b\otimes(10)_c}\nonumber\\
        &\qquad\qquad\qquad+(\chi^G_{(2n10)}+\chi^G_{(2n01)}+\chi^G_{(1n20)}+\chi^G_{(1n02)}+\chi^G_{(0n21)}+\chi^G_{(0n12)})\chi^F_{(10)_{(abc)}\otimes(01)_{(abc)}}\nonumber\\
            &-\sum_{n=0}^\infty v^{9+2n}\bigg((\chi^G_{(2n20)}\chi^F_{(01)_{c}}+\chi^G_{(2n02)}\chi^F_{(01)_{b}}+\chi^G_{(0n22)}\chi^F_{(01)_{a}})\nonumber\\
                &\qquad\qquad\qquad\qquad +(\chi^G_{(2n11)}+\chi^G_{(1n21)}+\chi^G_{(1n12)})\chi^F_{(10)_{(abc)}\otimes (10)_{(abc)}}\bigg)\nonumber\\
 &+\sum_{n=0}^\infty v^{10+2n}(\chi^G_{(2n21)}\chi^F_{(10)_{c}}+\chi^G_{(2n12)}\chi^F_{(10)_{b}}+\chi^G_{(1n22)}\chi^F_{(10)_{a}})-\sum_{n=0}^\infty v^{11+2n}\chi^G_{(2n22)}.
 \end{align}
As a further check, upon specializing $\massG,$ $\massF\to 1$ this expression matches with the result in table \ref{tab:5dlimtab3}.\newline\\

\noindent $\bullet\quad \mathbf{n=2,\, G=SO(9),\,F=Sp(3)_a\times Sp(2)_b}$:\newline

\noindent We are able to determine the first 8 coefficients of $Z_{1-inst}$ unequivocally:
 \begin{align}Z_{1-inst} &= v^{-1}-v^3\chi^{F}_{(01)_b}-v^4(\chi^{F}_{(100)_a\otimes(01)_b}+\chi^{F}_{(001)_a})+v^5(\chi^G_{(1000)}\chi^F_{(01)_b}-\chi^F_{(010)_a\otimes(20)_b})\nonumber\\
&+v^6(\chi^G_{(0100)}\chi^F_{(100)_a\otimes(20)_b}+\chi^G_{(0100)}\chi^F_{(100)_a}+\chi^G_{(0001)}\chi^F_{(010)_a\otimes(10)_b}-\chi^F_{(001)_a\otimes(02)_b})+\mathcal{O}(v^7).
 \end{align}
 At higher powers of $v$, large numbers of representations are allowed to contribute, and it becomes difficult to identify the correct representations.\newline\\

\noindent $\bullet\quad \mathbf{n=2,\, G=SO(10),\,F=Sp(4)_a\times SU(2)_b\times U(1)_c}$:\newline

\noindent In this case we are only able to unambiguously identify the first few coefficients of the Nekrasov partition function:
\begin{align}&Z_{1-inst}=v^{-1}-v^3\chi^F_{(2)_b}-v^{4}\chi^F_{(1000)_a\otimes((2)_c\oplus(-2)_c)}+\mathcal{O}(v^5).\end{align}
 At higher powers of $v$, large numbers of representations are allowed to contribute, and it becomes difficult to identify the correct representations.\newline
 
\noindent $\bullet\quad \mathbf{n=2,\, G=SO(12)_a,\,F=Sp(6)_a\times U(1)_b}$:\newline

\noindent We have determined the Nekrasov partition function up to $\mathcal{O}(v^9)$. We find:
 \begin{align}
 Z_{1-inst} &= v^{-1}-v^3\chi^{F}_{(2)_b\oplus(-2)_b}-v^5\chi^{F}_{(010000)_a}+v^6\chi^G_{(100000)}\chi^{F}_{(100000)_a}\nonumber\\
& -v^7(\chi^G_{(200000)}\chi^F_{(01)_b}+\chi^F_{(000100)_a\otimes((2)_b\oplus(-2)_b)}+\chi^F_{(000001)_a})\nonumber\\
&+v^8(\chi^G_{(100000)}\chi^F_{(001000)_a\otimes((2)_b\oplus(-2)_b))}+\mathcal{O}(v^9).
 \end{align}
\newpage
\noindent $\bullet\quad \mathbf{n=2,\, G=SO(12)_b,\,F=Sp(6)}$:\newline

\noindent We have determined the Nekrasov partition function up to $\mathcal{O}(v^{10})$. We find:
 \begin{align}
 Z_{1-inst} &= v^{-1}-2v^5\chi^{F}_{(100000)}+v^6(2\chi^{G}_{(100000)}+\chi^{F}_{(010000)})-v^7(\chi^G_{(100000)}\chi^{F}_{(100000)}+2\chi^{F}_{(001000)})\nonumber\\
& \!\!\!\!+v^8(\chi^G_{(200000)}\!+\!2\chi^G_{(200000)}\chi^F_{(010000)}\!-\!\chi^F_{(000001)})\!-\!v^9(2\chi^G_{(200000)}\chi^F_{(100000)}\!+\!2\chi^F_{(000010)})\!+\!\mathcal{O}(v^{10}).
 \end{align}

\noindent $\bullet\quad \mathbf{n=2,\, G=SO(13),\,F=Sp(7)}$:\newline

\noindent We have determined the first several coefficients of the Nekrasov partition function:
 \begin{align}&Z_{1-inst} = v^{-1}-v^4\chi^{F}_{(1000000)}+v^5\chi^{G}_{(10000)_a}-v^7\chi^G_{(0001000)}+v^8(\chi^{G}_{(10000)}\chi^F_{(0010000)}-\chi^F_{(0000001)})\nonumber\\
& -v^9\chi^G_{(200000)}\chi^F_{(0100000)}+v^{10}(\chi^G_{(300000)}\chi^F_{(1000000)}+\chi^G_{(000001)}\chi^F_{(0000010)}+\chi^G_{(010000)}\chi^F_{(0000100)}))\nonumber\\
&+v^{11}(\chi^G_{(400000)}+\chi^G_{(100001)}\chi^F_{(0000100)}+\chi^G_{(001000)}\chi^F_{(0000010)}+\chi^G_{(000100)}\chi^F_{(0001000)})\nonumber\\
&+v^{12}(\chi^G_{(101000)}\chi^F_{(0000100)}+\chi^G_{(200001)}\chi^F_{(0001000)}+\chi^G_{(210000)}\chi^F_{(0010000)}-\chi^G_{(000010)}\chi^F_{(0000001)})\nonumber\\
&+\mathcal{O}(v^{13}).
 \end{align}
 
\noindent $\bullet\quad \mathbf{n=2,\, G=G_2,\,F=Sp(4)}$:\newline

\vspace{-0.05in}
\noindent We find the following all-order expression for $Z_{1-inst}$:
\vspace{-0.05in}
 \begin{align}Z_{1-inst} &= v^{-1}-v^2\chi^F_{(1000)}+v^3\chi^G_{(10)}-\sum_{n=0}^\infty v^{3+2n}\chi^G_{0n}\chi^F_{(0001)}+\sum_{n=0}^\infty v^{4+2n}\chi^G_{(1n)}\chi^F_{(0010)}\nonumber\\
 &-\sum_{n=0}^\infty v^{5+2n}\chi^G_{2n}\chi^F_{(0100)}+\sum_{n=0}^\infty v^{6+2n}\chi^G_{(3n)}\chi^F_{(1000)}-\sum_{n=0}^\infty v^{7+2n}\chi^G_{(4n)}.
  \end{align}
\noindent Upon specializing $\massG,\massF\to 1$ this expression matches with the result in table \ref{tab:5dlimtab1}.\newline

\newpage\noindent $\bullet\quad \mathbf{n=2,\, G=F_4,\,F=Sp(3)}$:\newline

\vspace{-0.05in}
\noindent We have determined the coefficients of $Z_{1-inst}$ up to $\mathcal{O}(v^{20})$ and conjecture that the Nekrasov partition function to all orders in $v$ is given by:
\vspace{-0.05in}
 \begin{align}Z_{1-inst} &= v^{-1}-v^4\chi^F_{(001)}-v^5\chi^F_{(101)}-v^6\chi^F_{(201)}+v^7(\chi^G_{(1000)}\chi^F_{(010)}+\chi^G_{(0001)}\chi^F_{(101)}-\chi^F_{(030)})\nonumber\\
 &+v^8(\chi^G_{(1000)}\chi^F_{(300)}-\chi^G_{(0010)}\chi^F_{(001)}+\chi^G_{(0001)}\chi^F_{(120)})\nonumber\\
 &-v^9(\chi^G_{(2000)}+\chi^G_{(1001)}\chi^F_{(200)}+\chi^G_{(0002)}\chi^F_{(020)}+\chi^G_{(0010)}\chi^F_{(210)})\nonumber\\
  &+v^{10}(\chi^G_{(1010)}\chi^F_{(100)}+\chi^G_{(0100)}\chi^F_{(300)}+\chi^G_{(0011)}\chi^F_{(110)})\nonumber\\
   &-v^{11}(\chi^G_{(1100)}+\chi^G_{(0101)}\chi^F_{(200)}+\chi^G_{(0020)}\chi^F_{(010)})+v^{12}\chi^G_{(0110)}\chi^F_{(100)}-v^{13}\chi^G_{(0200)}\nonumber\\
   &-\sum_{n=0}^\infty v^{8+2n}\chi^G_{(n,0,0,0)}\chi^F_{(0,0,3)}+\sum_{n=0}^\infty v^{9+2n}\chi^G_{(n,0,0,1)}\chi^F_{(0,1,2)}\nonumber\\
   &-\sum_{n=0}^\infty v^{10+2n}\left(\chi^G_{(n,0,1,0)}\chi^F_{(0,2,1)}+\chi^G_{(n,0,0,2)}\chi^F_{(1,0,2)}\right)\nonumber\\
   &+\sum_{n=0}^\infty v^{11+2n}\left(\chi^G_{(n,1,0,0)}\chi^F_{(0,3,0)}+\chi^G_{(n,0,0,3)}\chi^F_{(0,0,2)}+\chi^G_{(n,0,1,1)}\chi^F_{(1,1,1)}\right)\nonumber\\
&-\sum_{n=0}^\infty v^{12+2n}\left(\chi^G_{(n,0,1,2)}\chi^F_{(0,1,1)}+\chi^G_{(n,0,2,0)}\chi^F_{(2,0,1)}+\chi^G_{(n,1,0,1)}\chi^F_{(1,2,0)}\right)\nonumber\\
&+\sum_{n=0}^\infty v^{13+2n}\left(\chi^G_{(n,1,0,2)}\chi^F_{(0,2,0)}+\chi^G_{(n,0,2,1)}\chi^F_{(1,0,1)}+\chi^G_{(n,1,1,0)}\chi^F_{(2,1,0)}\right)\nonumber\\
&-\sum_{n=0}^\infty v^{14+2n}\left(\chi^G_{(n,2,0,0)}\chi^F_{(3,0,0)}+\chi^G_{(n,0,3,0)}\chi^F_{(0,0,1)}+\chi^G_{(n,1,1,1)}\chi^F_{(1,1,0)}\right)\nonumber\\
&+\sum_{n=0}^\infty v^{15+2n}\left(\chi^G_{(n,1,2,0)}\chi^F_{(0,1,0)}+\chi^G_{(n,2,0,1)}\chi^F_{(2,0,0)}\right)\nonumber\\
   &-\sum_{n=0}^\infty v^{16+2n}\chi^G_{(n,2,1,0)}\chi^F_{(1,0,0)}+\sum_{n=0}^\infty v^{17+2n}\chi^G_{(n,3,0,0)}.
 \end{align}
 As a check, upon specializing $\massG,\massF\to 1$ this expression matches with the result in table \ref{tab:5dlimtab1}.\newline\\

\noindent $\bullet\quad \mathbf{n=3,\, G=SO(7),\,F=Sp(1)_a\times Sp(4)_b}$:\newline

\noindent We have determined the coefficients in $Z_{1-inst}$ up to $\mathcal{O}(v^{20})$ and conjecture the following all-order expression:
 \begin{align}Z_{1-inst} &= \sum_{n=0}^\infty v^{4+2n}\chi^G_{(0n0)}\chi^F_{(01)}- \sum_{n=0}^\infty v^{5+2n}\chi^G_{(0n1)}\chi^F_{(10)}+ \sum_{n=0}^\infty v^{6+2n}\chi^G_{(0n2)}.
 \end{align}
As a check, upon specializing $\massG,\massF\to 1$ this expression matches with the result in table \ref{tab:5dlimtab2}.\newline\\

\noindent $\bullet\quad \mathbf{n=3,\, G=SO(8),\,F=Sp(1)_a\times Sp(1)_b\times Sp(1)_c}$:\newline

\noindent As in the $n=2,\,\, G=SO(8)$ case, we assume invariance of the elliptic genus spectrum under triality to help us determine the representations that appear in the Nekrasov partition function. However, we are not able to completely fix the answer. We find:
 \begin{align}Z_{1-inst} &= v^4+\sum_{n=0}^\infty v^{5+2n}\chi^G_{(0n00)}\chi^F_{(1)_a\otimes(1)_b\otimes(1)_c}- \sum_{n=0}^\infty v^{6+2n}(\chi^G_{(1n00)}+\chi^G_{(0n10)}+\chi^G_{(0n01)})\chi^F_{(1)_{abc}\otimes (1)_{abc}}\nonumber\\
&+\sum_{n=0}^\infty v^{7+2n}(\chi^G_{(1n10)}\chi^F_{(1)_c}+\chi^G_{(1n01)}\chi^F_{(1)_b}+\chi^G_{(0n11)}\chi^F_{(1)_c})-\sum_{n=0}^\infty v^{8+2n}\chi^G_{(1n11)}.
 \end{align}

\noindent $\bullet\quad \mathbf{n=3,\, G=SO(9),\,F=Sp(2)_a\times Sp(1)_b}$:\newline

\noindent We have determined the coefficients of $Z_{1-inst}$ up to $\mathcal{O}(v^{14})$ and conjecture the following all-order expression:
 \begin{align}Z_{1-inst} &= v^5\chi^F_{(10)_a}-v^6\chi^G_{(1000)}\nonumber\\
&+\sum_{n=0}^\infty v^{6+2n}\chi^G_{(0n00)}\chi^F_{(01)_a\otimes(2)_b}-\sum_{n=0}^\infty v^{7+2n}(\chi^G_{(1n00)}\chi^F_{(10)_a\otimes(2)_b}+\chi^G_{(0n01)}\chi^F_{(01)_a\otimes(1)_b})\nonumber\\
&+\sum_{n=0}^\infty v^{8+2n}(\chi^G_{(0n10)}\chi^F_{(01)_a}+\chi^G_{(2n00)}\chi^F_{(2)_b}+\chi^G_{(1n01)}\chi^F_{(10)_a\otimes(1)_b})\nonumber\\
&-\sum_{n=0}^\infty -v^{9+2n}(\chi^G_{(2n01)}\chi^F_{(1)_b}+\chi^G_{(1n10)}\chi^F_{(10)_a})+\sum_{n=0}^\infty v^{10+2n}\chi^G_{(2n10)}.
 \end{align}
As a check, upon specializing $\massG,\massF\to 1$ this expression matches with the result in table \ref{tab:5dlimtab2}.\newline

\noindent $\bullet\quad \mathbf{n=3,\, G=SO(10),\,F=Sp(3)_a\times U(1)_b}$:\newline

\noindent We have determined the coefficients of $Z_{1-inst}$ up to $\mathcal{O}(v^{14})$ and conjecture the following all-order expression:
\vspace{-0.05in}
 \begin{align}Z_{1-inst} &= v^6\chi^F_{(010)_a}-v^7\chi^G_{(100000)}\chi^F_{(100)_a}+v^8\chi^G_{(20000)}
 +\sum_{n=0}^\infty v^{7+2n}\chi^G_{(0n000)}\chi^F_{(001)_a\otimes((2)_b\oplus(-2)_b)}\nonumber\\
&-\sum_{n=0}^\infty v^{8+2n}(\chi^G_{(1n000)}\chi^F_{(010)_a\otimes((2)_b\oplus(-2)_b)}+\chi^G_{(0n001)}\chi^F_{(001)_a\otimes((1)_b\oplus(-1)_b)})\nonumber\\
&+\sum_{n=0}^\infty v^{9+2n}(\chi^G_{(2n000)}\chi^F_{(100)_a\otimes((2)_b\oplus(-2)_b)}+\chi^G_{(1n001)}\chi^F_{(010)_a\otimes((1)_b\oplus(-1)_b)}+\chi^G_{(0n100)}\chi^F_{(001)_a)})\nonumber\\
&-\sum_{n=0}^\infty v^{10+2n}(\chi^G_{(3n000)}\chi^F_{(2)_b\oplus(-2)_b}+\chi^G_{(2n001)}\chi^F_{(100)_a\otimes((1)_b\oplus(-1)_b)}+\chi^G_{(1n100)}\chi^F_{(010)_a)})\nonumber\\
&+\sum_{n=0}^\infty v^{11+2n}(\chi^G_{(3n001)}\chi^F_{\otimes((1)_b\oplus(-1)_b)}+\chi^G_{(2n100)}\chi^F_{(100)_a)})-\sum_{n=0}^\infty v^{12+2n}\chi^G_{(3n100)}.
 \end{align}
As a check, upon specializing $\massG,\massF\to 1$ this expression matches with the result in table \ref{tab:5dlimtab2}.\newline

\noindent $\bullet\quad \mathbf{n=3,\, G=SO(11),\,F=Sp(4)}$:\newline

\vspace{-0.05in}
\noindent We have determined the coefficients of $Z_{1-inst}$ up to $\mathcal{O}(v^{16})$ and conjecture the following all-order expression:
\vspace{-0.05in}
 \begin{align}Z_{1-inst} &= v^7\chi^F_{(0010)}-v^8\chi^G_{(10000)}\chi^F_{(0100)}+v^9\chi^G_{(20000)}\chi^F_{(1000)}-v^{10}\chi^G_{(30000)}\nonumber\\
&+\sum_{n=0}^\infty v^{8+2n}\chi^G_{(0n000)}\chi^F_{(0001)}-\sum_{n=0}^\infty v^{9+2n}(\chi^G_{(1n000)}\chi^F_{(0010)}+\chi^G_{(0n001)}\chi^F_{(0001)})\nonumber\\
&+\sum_{n=0}^\infty v^{10+2n}(\chi^G_{(2n000)}\chi^F_{(0100)}+\chi^G_{(1n001)}\chi^F_{(0010)}+\chi^G_{(0n100)}\chi^F_{(0001)})\nonumber\\
&-\sum_{n=0}^\infty v^{11+2n}(\chi^G_{(3n000)}\chi^F_{(1000)}+\chi^G_{(2n001)}\chi^F_{(0100)}+\chi^G_{(1n100)}\chi^F_{(0010)})\nonumber\\
&+\sum_{n=0}^\infty v^{12+2n}(\chi^G_{(4n000)}+\chi^G_{(3n001)}\chi^F_{(1000)}+\chi^G_{(2n100)}\chi^F_{(0100)})\nonumber\\
&-\sum_{n=0}^\infty v^{13+2n}(\chi^G_{(4n001)}+\chi^G_{(3n100)}\chi^F_{(1000)})+\sum_{n=0}^\infty v^{14+2n}\chi^G_{(4n100)}.
 \end{align}
As a check, upon specializing $\massG,\massF\to 1$ this expression matches with the result in table \ref{tab:5dlimtab2}.\newline\\

\noindent $\bullet\quad \mathbf{n=3,\, G=SO(12),\,F=Sp(5)}$:\newline

\noindent The matter content consists of one spinor half-hypermultiplet of definite chirality. We have determined the coefficients of $Z_{1-inst}$ up to $\mathcal{O}(v^{18})$, except for the fact that we are not able to determine the spinor labels appearing in the various coefficients. We find the following all-order expression:
 \begin{align}Z_{1-inst} &= v^8\chi^F_{(00010)}-v^9\chi^G_{(100000)}\chi^F_{(00100)}+v^{10}\chi^G_{(200000)}\chi^F_{(01000)}-v^{11}\chi^G_{(300000)}\chi^F_{(10000)}+v^{12}\chi^G_{(400000)}\nonumber\\
&-\sum_{n=0}^\infty v^{10+2n}\chi^G_{(0n00,a,1-a)}\chi^F_{(00001)}+\sum_{n=0}^\infty v^{11+2n}(\chi^G_{(1n00,a,1-a)}\chi^F_{(00010)}+\chi^G_{(0n1000)}\chi^F_{(00001)})\nonumber\\
&-\sum_{n=0}^\infty v^{12+2n}(\chi^G_{(2n00,a,1-a)}\chi^F_{(00100)}+\chi^G_{(1n1000)}\chi^F_{(00010)})\nonumber\\
&+\sum_{n=0}^\infty v^{13+2n}(\chi^G_{(3n00,a,1-a)}\chi^F_{(01000)}+\chi^G_{(2n1000)}\chi^F_{(00100)})\nonumber\\
&-\sum_{n=0}^\infty v^{14+2n}(\chi^G_{(4n00,a,1-a)}\chi^F_{(10000)}+\chi^G_{(3n1000)}\chi^F_{(01000)})\nonumber\\
&+\sum_{n=0}^\infty v^{15+2n}(\chi^G_{(5n00,a,1-a)}+\chi^G_{(4n1000)}\chi^F_{(10000)})-\sum_{n=0}^\infty v^{16+2n}\chi^G_{(5n1000)},
 \end{align}
 where $a=0$ or $1$ for each term.\newline

 \noindent As a check, upon specializing $\massG,\massF\to 1$ this expression matches with the result in table \ref{tab:5dlimtab2}.\newline\\

\noindent $\bullet\quad \mathbf{n=3,\, G=G_2,\,F=Sp(1)}$:\newline

\noindent We have determined the coefficients of $Z_{1-inst}$ up to $\mathcal{O}(v^{20})$ and conjecture the following all-order expression:
 \begin{align}
 Z_{1-inst} &= \sum_{n=0}^\infty v^{3+2n}\chi^G_{(0n)}\chi^F_{(1)}-\sum_{n=0}^\infty v^{4+2n}\chi^G_{(1n)}.
 \end{align}
As a check, upon specializing $\massG,\massF\to 1$ this expression matches with the result in table \ref{tab:5dlimtab2}.\newline\\

\noindent $\bullet\quad \mathbf{n=3,\, G=F_4,\,F=Sp(2)}$:\newline

\vspace{-0.05in}
\noindent We have determined the coefficients of $Z_{1-inst}$ up to $\mathcal{O}(v^{16})$ and conjecture the following all-order expression:
 \begin{align}Z_{1-inst} &= v^{6}\chi^F_{(01)}+v^7\chi^{F}_{(30)}-v^8(\chi^G_{(1000)}+\chi^G_{(0001)}\chi^F_{(20)})+v^9\chi^G_{(0010)}\chi^F_{(10)}-v^{10}\chi^G_{(0100)}\nonumber\\
 &+\sum_{n=0}^\infty v^{8+2n}\chi^G_{(n000)}\chi^F_{(03)}-\sum_{n=0}^\infty v^{9+2n}\chi^G_{(n001)}\chi^F_{(12)}\nonumber\\
 &+\sum_{n=0}^\infty v^{10+2n}(\chi^G_{(n010)}\chi^F_{(21)}+\chi^G_{(n002)}\chi^F_{(02)})\nonumber\\
  &-\sum_{n=0}^\infty v^{11+2n}(\chi^G_{(n100)}\chi^F_{(30)}+\chi^G_{(n011)}\chi^F_{(11)})\nonumber\\
   &+\sum_{n=0}^\infty v^{12+2n}(\chi^G_{(n020)}\chi^F_{(01)}+\chi^G_{(n101)}\chi^F_{(20)})\nonumber\\
    &-\sum_{n=0}^\infty v^{13+2n}\chi^G_{(n110)}\chi^F_{(10)}+\sum_{n=0}^\infty v^{14+2n}\chi^G_{(n200)}.
 \end{align}
As a check, upon specializing $\massG,\massF\to 1$ this expression matches with the result in table \ref{tab:5dlimtab2}.\newline

\vspace{-0.05in}
\noindent $\bullet\quad \mathbf{n=3,\, G=E_6,\,F=SU(3)_a\times U(1)_b}$:\newline

\noindent In this case we are only able to unambiguously identify the first few coefficients of the Nekrasov partition function:
\begin{align}
Z_{1-inst}&=v^7\chi^F_{(3)_b\oplus(-3)_b}-v^{8}(\chi^F_{(30)_a}+\chi^F_{(03)_a})+v^9\chi^F_{(22)_a\otimes((3)_b\oplus(-3)_b)}\nonumber\\
&-v^{10}(\chi^G_{(100000)}\chi^F_{(12)_a\otimes(-2)_b}+\chi^G_{(000010)}\chi^F_{(21)_a\otimes(2)_b}+\chi^G_{(000001)}\chi^F_{(11)_a}\nonumber\\
&-\chi^F_{(30)_a\otimes(\pm 6)_b}-\chi^F_{(03)_a\otimes(\mp 6)_b}-\chi^F_{(33)_a})+\mathcal{O}(v^{12}),
\end{align}
where in the last line we do not have sufficient information to fix all $U(1)$ charges.\newline

\noindent $\bullet\quad \mathbf{n=4,\, G=F_4,\,F=Sp(1)}$:\newline

\noindent We have determined the coefficients of $Z_{1-inst}$ up to $\mathcal{O}(v^{16})$ and conjecture the following all-order expression:
 \begin{align}Z_{1-inst} &= v^{7}-\sum_{n=0}^\infty v^{8+2n}\chi^G_{(n000)}\chi^F_{(3)}+\sum_{n=0}^\infty v^{9+2n}\chi^G_{(n001)}\chi^F_{(2)}-\sum_{n=0}^\infty v^{10+2n}\chi^G_{(n010)}\chi^F_{(1)}\nonumber\\
 &+\sum_{n=0}^\infty v^{11+2n}\chi^G_{(n100)}.
 \end{align}
As a check, upon specializing $\massG,\massF\to 1$ this matches with the result in table \ref{tab:5dlimtab2}.\newline

\noindent $\bullet\quad \mathbf{n=4,\, G=E_6,\,F=SU(2)_a\times U(1)_b}$:\newline

\noindent In this case we are only able to unambiguously identify the first few coefficients of the Nekrasov partition function:
\begin{align}Z_{1-inst}&=v^9\chi^F_{(2)_a}+v^{10}\chi^F_{(3)_a\otimes((3)_b\oplus(-3)_b)}-v^{11}(\chi^G_{(100000)}\chi^F_{(2)_a\otimes(-2)_b}+\chi^G_{(000010)}\chi^F_{(2)_a\otimes(2)_b}\nonumber\\
&+\chi^G_{(000001)}-\chi^F_{(6)_b\oplus(-6)_b}-\chi^F_{(6)_a})+\mathcal{O}(v^{12}).\end{align}

\noindent $\bullet\quad \mathbf{n=5,\, G=E_6,\,F=U(1)}$:\newline

\noindent We have determined the coefficients of $Z_{1-inst}$ up to $\mathcal{O}(v^{20})$ and conjecture the following all-order expression:
  \begin{align}
  Z_{1-inst} &= v^{10}+\sum_{n=0}^\infty v^{11+2n}\chi^G_{(00000n)}\chi^F_{(3)\oplus (-3)}+\sum_{n=0}^\infty v^{12+2n}(\chi^G_{(10000n)}\chi^F_{(-2)}+\chi^G_{(00001n)}\chi^F_{(2)})\nonumber\\
 &+\sum_{n=0}^\infty v^{13+2n}(\chi^G_{(01000n)}\chi^F_{(-1)}+\chi^G_{(00010n)}\chi^F_{(1)})+\sum_{n=0}^\infty v^{14+2n}\chi^G_{(00100n)}
 \end{align}
As a check, upon specializing $\massG,\massF\to 1$ this matches with the result in table \ref{tab:5dlimtab3}.\\

\noindent $\bullet\quad \mathbf{n=5,\, G=E_7,\,F=SO(3)}$:\newline

\noindent We write the representations of $G=SO(3)$ as representations of $SU(2)$. \noindent We have determined the coefficients of $Z_{1-inst}$ up to $\mathcal{O}(v^{30})$ and conjecture the following all-order expression:
 \begin{align}Z_{1-inst} &= v^{14}\chi^F_{(6)}-v^{16}(\chi^G_{(1000000)}\chi^F_{(4)}-\chi^F_{(12)})+v^{17}(\chi^G_{(0000001)}\chi^F_{(4)}-\chi^G_{(0000010)}\chi^F_{(10)})\nonumber\\
 &+v^{18}(\chi^G_{(2000000)}+\chi^G_{(0000100)}\chi^F_{(8)})-v^{19}(\chi^G_{(0001000)}\chi^F_{(6)}+\chi^G_{(1000001)}\chi^F_{(2)})\nonumber\\
 &+v^{20}(\chi^G_{(0010000)}\chi^F_{(4)}+\chi^G_{(1100000)}+\chi^G_{(0000002)}\chi^F_{(2)})-v^{21}\chi^G_{(0100001)}\chi^F_{(2)}+v^{22}\chi^G_{(0200000)}\nonumber\\
 &+\sum_{n=0}^\infty v^{19+2n}\chi^G_{(n000001)}\chi^F_{(12)}-\sum_{n=0}^\infty v^{20+2n}(\chi^G_{(n000011)}\chi^F_{(10)}+\chi^G_{(n100000)}\chi^F_{(12)})\nonumber\\
 &+\sum_{n=0}^\infty v^{21+2n}(\chi^G_{(n000101)}\chi^F_{(8)}+\chi^G_{(n100010)}\chi^F_{(10)})-\sum_{n=0}^\infty v^{22+2n}(\chi^G_{(n001001)}\chi^F_{(6)}+\chi^G_{(n100100)}\chi^F_{(8)})\nonumber\\
  &+\sum_{n=0}^\infty v^{23+2n}(\chi^G_{(n010001)}\chi^F_{(4)}+\chi^G_{(n101000)}\chi^F_{(6)}+\chi^G_{(n000003)})\nonumber\\
  &-\sum_{n=0}^\infty v^{24+2n}(\chi^G_{(n100002)}\chi^F_{(2)}+\chi^G_{(n110000)}\chi^F_{(4)})+\sum_{n=0}^\infty v^{25+2n}\chi^G_{(n200001)}+\sum_{n=0}^\infty v^{26+2n}\chi^G_{(n300000)}.
 \end{align}
Note that as expected only representations of $SO(3)$, corresponding to representations of $SU(2)$ with labels of the type $(2n)$, appear.\newline

\noindent As a check of our result, upon specializing $\massG,\massF\to 1$ this expression matches with the result in table \ref{tab:5dlimtab3}.\newline\\

\noindent $\bullet\quad \mathbf{n=6,\, G=E_7,\,F=SO(2)}$:\newline

\noindent We have determined the coefficients of $Z_{1-inst}$ up to $\mathcal{O}(v^{20})$ and conjecture the following all-order expression:

 \begin{align}Z_{1-inst} &= v^{15}\chi^F_{(2)\oplus (-2)}-v^{17}\chi^G_{(1000000)}+v^{18}\chi^G_{(0000001)}\chi^F_{(1)\oplus(-1)}-v^{19}\chi^G_{(0100000)}\nonumber\\
 &+\sum_{n=0}^\infty v^{17+2n}\chi^G_{(n000000)}\chi^F_{(6)\oplus (-6)}-\sum_{n=0}^\infty v^{18+2n}\chi^G_{(n000010)}\chi^F_{(5)\oplus (-5)}\nonumber\\
 &+\sum_{n=0}^\infty v^{19+2n}\chi^G_{(n000100)}\chi^F_{(4)\oplus (-4)})-\sum_{n=0}^\infty v^{20+2n}\chi^G_{(n001000)}\chi^F_{(3)\oplus (-3)}\nonumber\\
 &+\sum_{n=0}^\infty v^{21+2n}(\chi^G_{(n010000)}\chi^F_{(2)\oplus (-2)}+\chi^G_{(n000002)})	-\sum_{n=0}^\infty v^{22+2n}\chi^G_{(n100001)}\chi^F_{(1)\oplus (-1)}\nonumber\\
 &+\sum_{n=0}^\infty v^{23+2n}\chi^G_{(n200000)}.
 \end{align}
We label the representations of $SO(2)=U(1)$ in terms of their $U(1)$ charges.\newline

\noindent As a check of our result, upon specializing $\massG,\massF\to 1$ this expression matches with the result in table \ref{tab:5dlimtab3}.\newline\\

\noindent $\bullet\quad \mathbf{n=7,\, G=E_7}$:\newline

\noindent For this theory the flavor symmetry is trivial. We find the following all-order expression for the one-instanton Nekrasov partition function:
 \begin{align}
 Z_1 &= v^{16}+\sum_{n=0}^\infty v^{19+2n}\chi^G_{(n00001)}+\sum_{n=0}^\infty v^{20+2n}\chi^G_{(n100000)},
 \end{align}
which agrees in the  $\massG,\massF\to 1$ limit to the result in table \ref{tab:5dlimtab3}.

\bibliography{OEIS}

\end{document}